\documentclass[11pt]{amsart}

\usepackage{tikz}
\usepackage{amsmath}
\usepackage{amssymb}
\usepackage{graphicx}
\usepackage[ps,dvips,xdvi,arrow,frame,matrix,graph,curve,cmtip]{xy}
\usepackage{enumerate}
\usepackage{float}
\usepackage[algo2e,figure,vlined]{algorithm2e}

\newcommand{\N}{\mathbb{N}}
\newcommand{\intO}{{\int_{\Omega}}}

\newcommand{\dV}{{\, dV}}

\newcommand{\dt}{{\, dt}}
	   \newcommand{\bu}{{\mathbf {u}}}
\newcommand{\bpsi}{{\boldsymbol{\psi}}}

\newcommand{\eps}{\varepsilon}
\newcommand{\Gam}{\Gamma}

\newcommand{\lam}{\lambda}

\newcommand{\R}{\mathbb{R}}
\newcommand{\C}{\mathbb{C}}
\newcommand{\Z}{\mathbb{Z}}
\newcommand{\D}{\mathbb{D}}
\newcommand{\s}{\mathbb{S}}
\newcommand{\A}{\mathbb{A}}
\newcommand{\tet}{\mathbb{T}}
\newcommand{\oct}{\mathbb{O}}
\newcommand{\Oh}{\mathbb{O}_h}

\DeclareMathOperator{\nul}{Null}
\DeclareMathOperator{\fix}{Fix}
\DeclareMathOperator{\stab}{Stab}

\DeclareMathOperator{\aut}{Aut}

\DeclareMathOperator{\spn}{span}
\DeclareMathOperator{\sym}{Sym}
\DeclareMathOperator{\MI}{MI}
\DeclareMathOperator{\GL}{GL}
\newcommand{\bigcupdot}{%
  \mathop{%
    \vphantom{\bigcup}%
    \mathpalette\setbigcupdot\cdot}\displaylimits}
\newcommand{\setbigcupdot}[2]{\ooalign{\hfil$#1\bigcup$\hfil\cr\hfil$#2$\hfil\crcr}}

\theoremstyle{plain} 

\theoremstyle{definition}

\theoremstyle{remark}

\floatstyle{boxed}
\newfloat{Algorithm}{tbp}{lop} 
\floatname{Algorithm}{\sc{Algorithm}}

\begin{document}

\thanks{\today}

\title[PDE on the Cube]
{Newton's Method and Symmetry for Semilinear Elliptic PDE on the Cube}

\author{John M. Neuberger}
\author{N\'andor Sieben}
\author{James W. Swift}

\email{
John.Neuberger@nau.edu,
Nandor.Sieben@nau.edu,
Jim.Swift@nau.edu}

\address{
Department of Mathematics and Statistics,
Northern Arizona University PO Box 5717,
Flagstaff, AZ 86011-5717, USA
}


\begin{abstract}
We seek discrete approximations  to solutions $u:\Omega\to \R$ of semilinear elliptic partial
differential equations of the form $\Delta u + f_s(u) = 0$,
where $f_s$ is a one-parameter family of nonlinear functions
and $\Omega$ is a domain in $\R^d$.
The main achievement of this paper is
the approximation of solutions to the PDE
on the cube $\Omega=(0,\pi)^3\subseteq\R^3$.
There are 323 possible isotropy subgroups of functions on the cube, which fall into 99 conjugacy 
classes.
The bifurcations with symmetry in this problem are quite interesting, including many with 
3-dimensional critical eigenspaces.
Our automated symmetry analysis is necessary with so many isotropy subgroups and bifurcations among 
them,
and it allows our code to follow one branch in each equivalence class that
is created at a bifurcation point.
Our most complicated result is the complete analysis of
a degenerate bifurcation with a 6-dimensional critical eigenspace.

This article extends the authors' work in
{\it Automated Bifurcation Analysis for Nonlinear Elliptic Partial Difference Equations on Graphs}
(Int. J. of Bifurcation and Chaos, 2009),
wherein they combined symmetry analysis with modified implementations of the
gradient Newton-Galerkin algorithm (GNGA, Neuberger and Swift) to
automatically generate bifurcation diagrams and solution graphics for small,
discrete problems with large symmetry groups.
The code described in the current paper is efficiently implemented in parallel,
allowing us to investigate a relatively fine-mesh discretization of the cube.
We use the methodology and corresponding library presented in our paper
{\it An MPI Implementation of a Self-Submitting Parallel Job Queue}
(Int. J. of Parallel Prog., 2012).
\end{abstract}

\maketitle

\tolerance=10000

\begin{section}{Introduction}

We are interested in finding and approximating solutions $u:\Omega\to\R$ of
semilinear elliptic equations with zero Dirichlet boundary conditions,
\begin{eqnarray}
\label{pde}
	\left\{ \begin{array}{rl}
			\Delta u + f_s(u) = 0 & \hbox{in } \Omega \\
			u=0 & \hbox{on } {\partial \Omega},
			\end{array}
			\right .
\end{eqnarray}
where $f_s:\R\to\R$ satisfies $f_s(0)=0$,  $f_s'(0) =s$, and
$\Omega$ is a region in $\R^2$ or $\R^3$.
Our code also works for zero Neumann boundary conditions, and
a wide range of nonlinearities.
In this paper we present results for PDE~(\ref{pde}) with
$f_s(t)=st+t^3$
on the square $\Omega=(0,\pi)^2$,
and more challengingly, on the cube $\Omega=(0,\pi)^3$.
By finding and following new, bifurcating branches of (generally) lesser symmetry we are able to approximate, 
within reason,
any solution that is connected by branches to the trivial branch.
The more complicated solutions bifurcating farther from the origin $(u,s)=(0,0)$ are of course progressively 
more challenging to locate and accurately approximate.

Generally, we apply Newton's method to the gradient of an action functional
whose critical points are solutions to our PDE.
For an exposition of our initial development of the gradient Newton-Galerkin algorithm (GNGA) 
and our first application of it to the square, see \cite{NS}.

This article extends the methods for so-called partial difference equations (PdE) from \cite{NSS3} to large graphs, 
that is, fine mesh discretizations for PDE.
For small graphs with possibly large symmetry groups,
the code in \cite{NSS3} automated the analysis of symmetry, isotypic decomposition,
and bifurcation.
We use here the GAP
(Groups, Algorithms, and Programming, see~\cite{GAP})
and Mathematica codes presented in those articles to
automatically generate a wealth of symmetry information for use by our branch-following C++ code.
Some of this information  is summarized in
the bifurcation digraph, which shows the generic symmetry-breaking bifurcations.
In \cite{NSS3} we developed two modified implementations of the
gradient Newton-Galerkin algorithm, namely the
tangent algorithm (tGNGA) for following bifurcation curves and the
cylinder algorithm (cGNGA) for switching branches at bifurcation points.
Together with a secant method for locating these bifurcation points,
in the current PDE setting we are able to handle most difficulties that arise when encountering accidental 
degeneracies and high-dimensional critical eigenspaces.

Since we use here a fine mesh to investigate PDE~(\ref{pde}) on the cube $\Omega=(0,\pi)^3$,
the practical implementation of the algorithms from~\cite{NSS3} requires increased efficiencies.
In particular, we use isotypic decompositions of invariant fixed-point spaces to take advantage of the block diagonal structure of the Hessian matrix.
In doing so,  we substantially reduce the dimension of the Newton search direction linear system, and hence also reduce the number of costly numerical integrations.
The same theory allows for reduced dimensions in many of the search spaces when seeking new,
bifurcating solutions near high-dimensional bifurcation points in the presence of symmetry.

Even with these efficiency improvements, we found it necessary to convert our high-level branch following strategy to use parallel computing.
Most of the details of the parallel implementation can be found in~\cite{NSS4}, where we present
a general methodology using self-submitting job queues to implement many types of mathematical algorithms in parallel.
In particular, therein we develop a light-weight, easy-to-use C++ parallel job queue library, which we have used in obtaining the cube results found in this article.

Our numerical results are summarized in bifurcation diagrams,
which plot a scalar function such as the value of $u$ at a generic point $u(x^*, y^*, z^*)$ versus $s$,
for approximate solutions to Equation~(\ref{pde})
with parameter $s$.
These diagrams indicate by line type the Morse Index (MI) of solutions, which
typically changes at bifurcation and turning points.
We present graphics for individual approximate solutions in several formats.
For the most part, we find that representative graphics using a small, fixed collection of patches (``flags'')
most clearly show the symmetries of real-valued functions of three variables.
We call these {\em flag diagrams}.
We also include some contour plots of actual solution approximations.
A fairly comprehensive collection of graphics and supporting information describing the symmetries of functions on the cube,
possible bifurcations of nonlinear PDE on the cube,
and more example approximate PDE solutions can be found on the companion website \cite{cubeWEB}.

In~\cite{NSS3}, we considered many small graphs where scaling was not used, and hence PDE were not involved.
In the present setting, we approximate a solution $u$ to PDE~(\ref{pde}) with a vector $\bu=(u_n)\in\R^N$ whose components represent $u$ values at $N$ regularly spaced grid points in $\Omega$ located a distance $\Delta x$ apart.
Thus, our approach is equivalent to applying our algorithms to the finite dimensional semilinear elliptic partial
{\it difference} equation (PdE)
\begin{eqnarray}
\label{dpde}
			-L \bu + f_s(\bu) = 0 & \hbox{in } \Omega_N,
\end{eqnarray}
where $\Omega_N$ is a graph with $N$ vertices coming from a grid.
The matrix $L$ is in fact the graph Laplacian on $\Omega_N$,
scaled by $\frac1{(\Delta x)^2}$, and
modified at boundary vertices to enforce a zero-Dirichlet problem boundary condition.
See \cite{NSS} for a discussion of ghost points for enforcing boundary conditions.
For general regions in $\R^2$ and $\R^3$, we approximate eigenvectors of $L$ using standard linear 
techniques,
e.g., Matlab's {\it eigs} or some other easy to use implementation of ARPACK.
For the square and cube the eigenfunctions are of course well known explicitly in terms of sine 
functions, so the consideration of $L$ is not necessary.
Since accurate PDE results require the dimension $N$ to be very large,
in the expansions of our approximate solutions $\bu = \sum_{m=1}^M a_m \bpsi_m$
we use $M\ll N$ discretized eigenfunctions of $-\Delta$ with this boundary condition.

In Section~\ref{theory} we present some theory for the action functional, its gradient and Hessian,
symmetry of functions, the corresponding fixed-point subspaces, and bifurcations with symmetry.
We apply the general symmetry theory to the basis generation process
for the square and cube.
In Section~\ref{algorithms} we outline the algorithms used in our project.
We include a high-level description of our numerical methods and corresponding implementations,
including the new use of self-submitting parallel job queues applied to obtain accurate 
high-resolution solutions for the cube.
We also describe our method for taking advantage of the block structure of the Hessian and 
our procedure for generating contour plots of approximate solutions.
Section~\ref{square} contains the results
from our experiment on the square, essentially an efficient and automatic refinement of the 
computations found in~\cite{NS}.

Our main numerical results are found in Section~\ref{cube}.  Namely, we investigate PDE~(\ref{pde}) 
on the cube,
where it is required to use a large number of eigenmodes and spacial grid points in order to find 
nodally complicated
solutions of high MI, lying in many different fixed-point spaces of the fairly large symmetry 
group.
We present several interesting examples from the companion website.
The website shows examples of a solution with each of the symmetry types that we found in our investigation.
Due to space limitations, we do not show all of these solutions in this paper.
Rather, we concentrate on the first six primary bifurcations, and one of the secondary bifurcations
with $\tet_d$ symmetry (the symmetry of a tetrahedron).
Two of the primary bifurcations that we consider have degenerate bifurcation points,
including one with a six-dimensional eigenspace that is the direct sum of two 3-dimensional 
irreducible
representations of $\Oh$, the symmetry of the cube.

There does not seem to be much in the literature that specifically investigates the bifurcation 
and symmetry of solutions to semilinear elliptic PDE on the cube.
The article~\cite{BuddHumphriesWathen} is interesting for pushing the nonlinearity power to the 
critical exponent in the cube case. 
The interested reader can consult works by Zhou and co-authors for 
alternate but related methods and algorithms for computing solutions to semilinear elliptic PDE, 
e.g., \cite{WangZhou1, WangZhou2, Zhou3} and the recent book~\cite{Zhou4}.

\end{section}

\begin{section}{Symmetry and Invariance}
\label{theory}

For the convenience of the reader, in this section we summarize enough notation and theory
from~\cite{NSS3} to follow our new results. 
We also include square and cube-specific information
required to apply our algorithms in our particular cases.

\begin{subsection}{The Functional Setting.}

Our techniques rely on two levels of approximation,
namely the restriction of functions to a suitably large $M$-dimensional Galerkin subspace 
$B_M\subseteq H = H_0^1(\Omega)$, and the discretization of $\Omega$ to $\Omega_N$.
We call the natural numbers $M$ and $N$ the Galerkin and spacial dimensions of our approximations, respectively.
For the regions $\Omega$ considered in this paper, it suffices to divide the region up into $N$ squares or cubes with edge length $\Delta x$,
and then place a gridpoint $x_n$ in the center of each such cell.
The graph $\Omega_N$ has a vertex $v_i$ corresponding to each gridpoint,
with edges $e_{nj}$ determined by the several neighbors $x_j$ which are at distance $\Delta x$ away from $x_n$.
With this arrangement, the simple numerical integration scheme used below in Equations~\ref{grad} and \ref{hess}
to evaluate the nonlinear terms in our gradient and Hessian computations becomes the midpoint method.

The eigenvalues of the negative Laplacian with the zero Dirichlet boundary condition satisfy
\begin{eqnarray}
\label{evals}
0<\lambda_1<\lambda_2\leq\lambda_3\leq\cdots\to\infty,
\end{eqnarray}
and the corresponding eigenfunctions
$\{\psi_m\}_{m\in\N}$
can be chosen to be an orthogonal basis
for the Sobolev space $H$,
and an orthonormal basis for
the larger Hilbert space $L^2=L^2(\Omega)$, with the usual inner products.
In the cases $\Omega=(0,\pi)^d$ for $d=2$ and $d=3$,
we take the first $M$ eigenvalues (counting multiplicity) from
$$
\{\lam_{i, j} := i^2+j^2\mid i,j\in\N\} \ \hbox{or} \ \{\lam_{i, j, k} := i^2+j^2+k^2\mid i,j,k\in\N\},
$$
and singly index them in a vector $\lambda=(\lambda_1,\ldots,\lambda_M)$.
In these cases, the corresponding eigenfunctions $\psi_m$ that we use are appropriate linear combinations of the
well-known eigenfunctions
$\psi_{i,j}(x,y)= \frac2\pi \sin (i x) \sin (j y)$ and
$\psi_{i,j,k}(x,y,z)=\left(\frac{2}{\pi}\right)^{3/2}\sin (i x) \sin (j y) \sin (k z)$.
We process the eigenfunctions using the projections given in Section~\ref{isoSub}
in order to understand and exploit the symmetry of functions in terms of the nonzero coefficients of their
eigenfunction expansions.
The $\psi_m$ are discretized as
$\bpsi_m\in\R^N$, $m\in \{1,\ldots, M\}$, by evaluating the
functions at the gridpoints, i.e., 
$\bpsi_m =( \psi_m(x_1),\ldots, \psi_m(x_N))$.
The eigenvectors $\bpsi_m$ form an orthonormal basis for an $M$-dimensional subspace of
$\R^N$.

Using variational theory, we define a nonlinear functional $J$ whose critical points are the solutions of PDE~(\ref{pde}).
We use the Gradient-Newton-Galerkin-Algorithm (GNGA, see \cite{NS})
to approximate these critical points,
that is, we seek approximate solutions $u$ lying in the subspace
$$
B_M := {\rm span}\{\psi_1,\ldots,\psi_M\}\subseteq H,
$$
which in turn are discretely approximated in $\R^N$ by
$$
\bu=\sum_{m=1}^{M} a_m \bpsi_m.
$$
The coefficient vectors $a$ in $\R^M$
(simultaneously the approximation vectors $\bu$ in $\R^N$)
are computed by applying Newton's method to the eigenvector expansion coefficients of the approximation
$-L + f_s(\bu)$ of
the gradient $\nabla J_s(u)$.

Let $F_s(p)=\int_0^p f_s(t)\dt$ for all
$p\in{\R}$ define the primitive of $f_s$.
We then define the action functional $J_s:H\to\R$ by
\begin{equation}
\label{Jdef}
J_s( u)=\intO {\textstyle \frac12}|\nabla u|^2 - F_s(u) \dV 
=\sum_{m=1}^M {\textstyle \frac12} a_m^2\lambda_m 
- \intO  F_s(u) \dV.
\end{equation}
The class of nonlinearities $f_s$ found in \cite{AR, CCN} imply that
$J_s$ is well defined and of class $C^2$ on $H$.
Computing directional derivatives and integrating by parts gives
\begin{equation}
\label{JprimeDef}
J_s'(u)(\psi_m) = \intO \nabla u \cdot \nabla \psi_m -  f_s(u) \, \psi_m  \dV
                   = a_m\lambda_m - \intO f_s(u) \, \psi_m \dV,
\end{equation}
for $m\in\{1,\ldots, M\}$.
Replacing the nonlinear integral term with a sum that is in fact the midpoint
method given our specific (square or cube) grid gives the gradient coefficient
vector $g\in\R^M$ defined by
\begin{equation}
\label{grad}
g_m=a_m\lambda_m - \sum_{n=1}^N f_s(u_n) (\bpsi_m)_n \,\Delta V.
\end{equation}
Here, the constant mesh area or volume factor is given by
$\Delta V = \text{Vol}(\Omega)/N=\pi^d/N$.
The functions $P_{B_M} \nabla J_s(u)$ and $\sum_{m=1}^M g_m \psi_m$ are approximately equal and are pointwise
approximated by the vector $-L \bu + f_s(\bu)$.

To apply Newton's method to find a zero of $g$ as a function of $a$,
we compute the coefficient matrix $h$ for the Hessian as well.
A calculation shows that
\begin{equation}
\label{JprimePrimeDef}
J_s''(u)(\psi_l,\psi_m) = \intO \nabla \psi_l \cdot \nabla \psi_m -  f_s'(u) \, \psi_l\,\psi_m \dV
= \lambda_l\delta_{lm} - \intO f_s'(u) \, \psi_l \,\psi_m \dV,
\end{equation}
where $\delta_{lm}$ is the Kronecker delta function.
Again using numerical integration, for $l, m\in\{1,\ldots, M\}$ we compute elements of $h$ by
\begin{equation}
\label{hess}
h_{lm}=\lambda_l\delta_{lm}  -
             \sum_{n=1}^{N} f_s(u_n) (\bpsi_l)_n (\bpsi_m)_n  \, \Delta V.
\end{equation}

The coefficient vector $g\in\R^M$ and the $M\times M$ coefficient matrix $h$ represent suitable projections of the $L^2$ 
gradient and Hessian of $J$, restricted to the subspace $B_M$, where all such quantities are defined.
The least squares solution $\chi$ to the $M$-dimensional linear system $h\chi=g$ always exists and is identified with 
the projection of the search direction
$(D^2_2J_s(u))^{-1}\nabla_2 J_s(u)$ onto $B_M$.
The $L_2$ search direction is not only defined for all points $u\in B_M$ such that the Hessian is invertible, but is 
in that case equal to
$(D^2_H J_s(u))^{-1}\nabla_H J_s(u)$.

The Hessian function $h_s: \R^M \rightarrow \R^M$ or $h: \R^{M+1} \rightarrow \R^M$
is very important for identifying bifurcation points.
If $h(a^*, s^*)$ is invertible at a solution $(a^*, s^*)$, then the Implicit Function Theorem
guarantees that there is locally a unique solution branch through $(a^*, s^*)$.
When $h(a^*, s^*)$ is not invertible,
then $(a^*, s^*)$ is a candidate for a bifurcation point, defined in the next subsection.
The kernel
$$
\tilde E = \nul h(a^*, s^*)
$$
of the Hessian at a bifurcation point
is called the {\em critical eigenspace}.
A Lyapunov-Schmidt reduction of the gradient $g(a^*, s^*)$ can be done to obtain
the {\em Lyapunov-Schmidt reduced gradient} $\tilde g: \tilde E \rightarrow \tilde E$ \cite{GSS}.
Local to the point where $h$ is singular,
there is a one-to-one correspondence between zeros of $g$ and
zeros of $\tilde g$.
The {\em Lyapunov-Schmidt reduced bifurcation equations} are $\tilde g = 0$.
We refer to the reduced gradient or reduced bifurcation equations when the Lyapunov-Schmidt 
reduction is understood.

Newton's method in coefficient space is implemented by fixing $s$, initializing
the coefficient vector $a$ with a guess, and iterating
\begin{equation}
a \leftarrow a - \chi, \quad\text{where} \ h(a, s) \chi = g(a, s).
\label{newtonsMethod}
\end{equation}
When it converges, the algorithm converges to vectors $a$ and $\bu = \sum_{m=1}^M a_m\bpsi_m$
giving $g=0$, and hence an approximate solution to PDE~(\ref{pde}) has been found.
The search direction $\chi$ in Newton's method is found by solving the system in (\ref{newtonsMethod})
without inverting $h(a, s)$.  The solver we use returns the least squares solution for an overdetermined
system and the solution with smallest norm for an underdetermined system.
We observe experimentally that Newton algorithms work well even
near bifurcation points where the Hessian is not invertible.
In Section~\ref{algorithms} we include brief descriptions of the
tGNGA and cGNGA,
the modifications of the GNGA actually implemented in our current code.

\begin{figure}
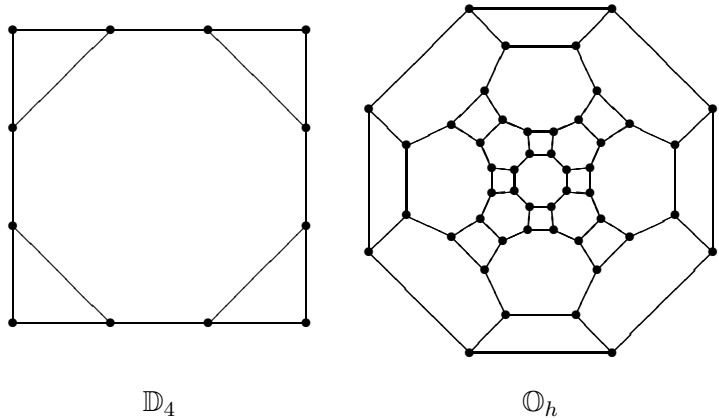

\begin{center}
\begin{tabular}{ccc}
\input{figures/D4graph.tex} & &
\input{figures/octgraph.tex} \\
& & \\
$\D_4$ & &  $\Oh$ \\
\end{tabular}
\end{center}
\caption
{
Small graphs used to generate symmetry information for analyzing functions on the square and cube, respectively.
The graph on the right, with full octahedral symmetry,
is the skeleton graph of the great rhombicuboctahedron, which has
48 vertices, 72 edges, and 26 faces.
This solid can be inscribed in the cube, with 8 vertices in each of the 6 faces.  Thus,
there is a one-to-one correspondence between elements of $\Oh$ and vertices of the skeleton graph,
after one vertex (chosen arbitrarily) has been assigned to the identity element.
}
\label{D4_and_Oct}
\end{figure}

In Sections~\ref{isoSub}~and~\ref{basis} we explain the details of our method for constructing a specific basis of
eigenfunctions that allows our code to exploit and report symmetry information.
Regardless of whether we know a basis for $B_M$ in terms of sines or must approximate one using
numerically computed eigenvectors of the sparse matrix $L$,
we are required to first compute various symmetry quantities relevant to the region $\Omega$ and the possible
symmetry types of expected solutions.
For this we use GAP~\cite{GAP}.
We start with a graph $G$ which has the same symmetry as $\Omega_N$, but with a significantly smaller number of vertices.
In Figure~\ref{D4_and_Oct} we show a 12-vertex graph with the $\D_4$ symmetry of the square,
and a 48-vertex graph with
the $\Oh$ symmetry of the cube.
These graphs are the smallest we found that allow functions on the vertices to have all possible
symmetry types.
For example, the square with four vertices does not allow a function with only four-fold rotational symmetry.
The group information used to analyze the symmetry of all functions
on the square and cube regions was done automatically by our suite of programs.
The automated GAP processes adapted from~\cite{NSS3} would fail if fed instead the exceedingly
large graphs $\Omega_N$.
Our GAP code is applied to the smaller graph $G$ to generate
all the files encoding the bifurcation digraph and the underlying fixed-point space decompositions.
The results from the small-graph analysis inform us and our code about the symmetries of functions on $\Omega$ and the
vectors in $\R^N$ which approximate such functions at the gridpoints.

\end{subsection}

\subsection{Symmetry of functions}
\label{functSym}

We assume that $f_s$ is odd; the case where $f_s$ is not odd is easily inferred.
To discuss the symmetry of solutions to Equation~(\ref{pde}),
we note that $\aut(\Omega)\times \Z_2 \cong \aut(\Omega_N)\times \Z_2 \cong \aut(G)\times \Z_2$, where $G$ is one of the
small graphs depicted in Figure~\ref{D4_and_Oct}.
We define
$$
\Gamma_0=\aut(\Omega_N)\times \Z_2,
$$
where $\Z_2=\{1,-1\}$ is written multiplicatively.

The natural action of $\Gamma_0$ on $\R^N$ is defined by
\begin{equation}
\label{Gamma_action}
(\gamma\cdot \bu)_i=\beta u_{\pi^{-1}(i)},
\end{equation}
where $\gamma=(\alpha_\pi,\beta)\in \Gamma_0$ and $\bu\in \R^N$.
We usually write $\alpha$ for $(\alpha,1)$ and $-\alpha$ for $(\alpha,-1)$.
The \emph{symmetry} of $\bu$ is the isotropy subgroup 
$\sym(\bu):=\stab(\bu,\Gamma_0)=\{\gamma\in\Gamma_0\mid \gamma\cdot \bu=\bu\}$.
The symmetries $\sym(u)$ of functions $u:\Omega\to\R$ are isomorphic to $\sym(\bu)$.
Two subgroups $\Gamma_i$ and $\Gamma_j$ of $\Gamma_0$ are called \emph{conjugate} if 
$\Gamma_i=\gamma\Gamma_j \gamma^{-1}$ for some
$\gamma\in\Gamma_0$. The \emph{symmetry type} of $\bu$ is the conjugacy class $[\sym(\bu)]$ 
of the symmetry of $\bu$;
a similar definition holds for the symmetry type of a function $u:\Omega\to\R$.
We say that two symmetry types are \emph{isomorphic} if they have isomorphic representatives.
We use the notation ${\mathcal G}:=\{\Gamma_0,\ldots,\Gamma_q \}$ for the set of symmetries and
${\mathcal S}:=\{S_0=[\Gamma_0],\ldots ,S_r\}$ for the set of symmetry types.

Let $X$ be the set of all solutions $(\bu,s)$ to (\ref{dpde}) in $\R^N\times \R$.
We define a \emph{branch of solutions} to be a maximal subset of $X$ that is a $C^1$ manifold with constant symmetry.
The \emph{trivial branch} $\{ (0, s) \mid s \in \R\}$ contains the \emph{trivial solution} $\bu = 0$,
which has symmetry $\Gamma_0$
if $f_s$ is odd, and symmetry $\aut(G)$ otherwise.
A \emph{bifurcation point} is a solution in the closure of at least two different solution branches.
We call the branch containing the bifurcation point the {\em mother}, and the other branches,
for which the bifurcation point is a limit point, are called {\em daughters}.
Note that there is not a bifurcation at a fold point, where a branch of constant symmetry is
not monotonic in $s$.

The action of $\Gam_0$ on $\R^N$ induces an action of $\Gam_0$ on $\R^M$, given the correspondence of
functions and coefficient vectors $\bu = \sum_{m=1}^M \bf {\psi}_m$.
The gradient function ${g}_s: \R^M \to \R^M$
is $\Gam_0$-equivariant, i.e.,
$g_s(\gamma \cdot a) = \gamma \cdot g_s(a)$ for all $\gamma\in\Gam_0$, $a \in \R^M$, and $s \in \R$.
As a consequence, if
$(a, s)$ is a solution to~(\ref{dpde})
then
$(\gamma \cdot a, s)$ is also a solution to (\ref{dpde}),
for all $\gamma \in \Gamma_0$.
Following the standard treatment~\cite{GSS, NSS3},
for each $\Gam_i \leq \Gam_0$ we define
the {\em fixed-point subspace} of the $\Gamma_0$ action on $\R^M$ to be
$$
\fix(\Gamma_i, \R^M) = \{ a \in \R^M \mid \gamma \cdot a = a \text{ for all } \gamma \in \Gamma_i\}.
$$
The fixed-point subspaces for any of the function spaces $V = G_M$, $\R^n$, $\tilde E$,
or $H$ is defined as
$$
\fix(\Gamma_i, V) = \{ u \in V \mid \gamma \cdot a = a \text{ for all } \gamma \in \Gamma_i\}.
$$
There is a one-to-one correspondence between $a \in \R^M$ and $u \in G_M$, and we will
often write $\fix(\Gam_i)$ when the equation is valid for any ambient space.
These fixed-point subspaces are important because they are $g_s$-invariant, meaning that
$g_s(\fix(\Gamma_i)) \subseteq \fix(\Gamma_i)$.
For efficiency in our code, we restrict $g_s$ to one of these fixed-point subspaces,
as described in Subsection \ref{blockhessian}.

As mentioned in the previous subsection, the reduced bifurcation equations are $\tilde g = 0$,
where $\tilde g: \tilde E \rightarrow \tilde E$ is the reduced gradient on the critical eigenspace
$\tilde E$.   A fold point $(a^*, s^*)$ is not a bifurcation point even though 
$h(a^*, s^*)$ is singular. 

Consider a bifurcation point $(a^*, s^*)$ in coefficient space, or the
corresponding $(\bu^*, s^*)$, and let $\Gam_i = \sym(u^*)$ be the symmetry of the mother solution.
Then there is a natural action of $\Gam_i$ on $\tilde E$,
and the reduced gradient $\tilde g$ is equivariant.
That is, $\tilde g(\gamma \cdot e, s) = \gamma \cdot \tilde g(e, s)$ for all $\gamma \in \Gam_i$,
and $e \in \tilde E$.
Let $\Gam_i'$ be the kernel of the action of $\Gam_i$ on $\tilde E$.
Then $\Gamma_i/\Gamma_i'$ acts freely on $\tilde E$, and we
say that the mother branch  {\em undergoes a bifurcation with $\Gamma_i/\Gamma_i'$ symmetry}.
This bifurcation is {\em generic}, or {\em non-degenerate}, if the action of $\Gamma_i/\Gamma_i'$
on $\tilde E$ is irreducible, and other non-degeneracy conditions (see \cite{GSS}) are met.
For each $\Gam_j \le \Gam_i$, if $ e \in \fix(\Gam_j, \tilde E)$,
then $\sym(\bu^* + e) = \Gam_j$.  Most of the fixed-point subspaces in $\tilde E$ are empty.
The subgroups $\{\Gam_j \mid \fix(\Gam_j, \tilde E) \neq \emptyset \}$ can be arranged in a
{\em lattice of isotropy subgroups}.
The Equivariant Branching Lemma (EBL), described in \cite{GSS},
states that there is generically an {\em EBL branch} of bifurcating solutions
with symmetry $\Gam_j$ if the fixed-point subspace $\fix(\Gam_j, \tilde E)$ is one-dimensional.
In a gradient system such as PDE~(\ref{pde}), there is generically a branch of
bifurcating solutions with symmetry $\Gam_i$ if $\Gam_i$ is a maximal isotropy subgroup.
See \cite{GSS, NSS3} for details.

The  {\em bifurcation digraph}, defined in \cite{NSS3}, summarizes some information about all of
the generic bifurcations that are possible for a system with a given symmetry.  In particular,
if there is a daughter with symmetry $\Gamma_j$ created at a generic bifurcation
of a mother solution with symmetry $\Gam_i$ in a gradient system,
then there is an arrow in the bifurcation digraph
$$
\xymatrix{
[ \Gamma_i ]
\ar [r]^{\Gamma_i/\Gamma_i'}&
[\Gamma_j].
}
$$
The arrow in the bifurcation digraph is either solid, dashed, or dotted,
as described in \cite{NSS3}.
Roughly speaking,
a solid arrow indicates a pitchfork bifurcation within some one-dimensional subspace of $\tilde E$,
a dashed arrow indicates a transcritical bifurcation within some one-dimensional subspace of $\tilde E$,
and a dotted arrow indicates a more exotic bifurcation.

In~\cite{NSS3} we defined an
{\it anomalous invariant subspace} (AIS) ${\mathcal A} \subseteq V$, with $V = G_M$ or $H$,
to be a $g_s$-invariant subspace that is not a fixed-point subspace.
Consider the PDE (\ref{pde}) on the cube, with $f_s$ odd.
For positive integers $p$, $q$, and $r$,
not all 1, the set
\begin{equation}
\label{AISpqr}
{\mathcal A}_{p,q,r} = \spn \left
\{ \psi_{i,j,k} \mid \textstyle{\frac{i}{p}, \frac{j}{q}, \frac{k}{r} } \in \Z \right \}
\end{equation}
is an AIS.
The function space ${\mathcal A}_{1,1,1}$ is  all of $V$, so it is not an AIS.

The space ${\mathcal A}_{p,q,r}$ is the appropriate function space if one solved the
PDE~(\ref{pde}) in the box $(0, \pi/p) \times (0, \pi/q) \times (0, \pi/r)$.  Any solution
in this box extends to a solution on the cube, with nodal planes dividing the cube into
$p q r$ boxes.
These anomalous invariant subspaces are caused by the so-called hidden symmetry \cite{Gomes, NS}
of the problem
that is related to the symmetry of the PDE on all of $\R^3$.

There are in fact a multitude of AIS for the PDE on the cube.
There are proper subspaces of every AIS ${\mathcal A}_{p,q,r}$ consisting of functions with symmetry
on the domain $(0, \pi/p) \times (0, \pi/q) \times (0, \pi/r)$.  For example, consider the case
where $p = q = r > 1$.
Since there are 323 fixed-point subspaces
for functions on the cube, including the zero-dimensional $\{ 0 \}$,
there are 322 AIS that are subspaces of ${\mathcal A}_{p,p,p}$, for each $p > 1$.
Our code is capable of finding solutions in any AIS, within reason.
The theory of anomalous solutions within AIS is unknown.
The book \cite{GS} is a good reference on invariant spaces of nonlinear operators in general.

\subsection{Isotypic Decomposition}
\label{isoSub}

To analyze the bifurcations of a branch of solutions with symmetry $\Gamma_i$,
we need to understand the isotypic decomposition of the action of $\Gamma_i$ on
$\R^n$.

Suppose a finite group $\Gamma$ acts on $V=\R^n$ according to the representation
$g\mapsto \alpha_g:\Gamma\to \aut(V)\cong {\GL}_{n}(\R)$.
In our applications we choose $\Gamma\in{\mathcal G}$ and the
group action is the one in Equation~(\ref{Gamma_action}).
Let $\{\alpha^{(k)}_{\Gamma}:\Gamma\to {\GL}_{d_\Gamma^{(k)}}(\R)\mid k \in K_\Gamma\}$ be the set of
irreducible representations of $\Gamma$ over $\R$,
where $d_{\Gamma^{(k)}}$ is the dimension of the representation.
We write $\alpha^{(k)}$ and $K$ when the subscript $\Gam$ is understood.
It is a standard result of representation theory that
there is an orthonormal basis $B_{\Gamma}=\bigcup_{k\in K} B_{\Gamma}^{(k)}$
for $V$ such that
$B_\Gamma^{(k)}=\bigcupdot_{l=1}^{\,L_k} B_\Gamma^{(k,l)}$ and
$[\alpha_g|_{V_{\Gamma}^{(k,l)}}]_{B_\Gamma^{(k,l)}}=\alpha^{(k)}(g)$ for all $g\in\Gamma$,
where $V_{\Gamma}^{(k,l)}:=\spn(B_\Gamma^{(k,l)})$.
Each $V_{\Gamma}^{(k,l)}$ is an irreducible subspace of $V$.
Note that $B_\Gamma^{(k)}$ might be empty for some $k$, corresponding to $V_\Gamma^{(k)}=\{0\}$.
The {\em isotypic decomposition of } $V$ under the action of $\Gamma$ is
\begin{equation}
\label{iso_decomp}
V = \bigoplus_{k\in K} V_{\Gamma}^{(k)},
\end{equation}
where $V_{\Gamma}^{(k)}=\bigoplus_{l=1}^{L_k} V_{\Gamma}^{(k,l)}$ are the {\em isotypic components}.

The isotypic decomposition of $V$ under the action of each $\Gamma_i$ is required by
our algorithm.
The decomposition under the action of $\aut (G)$ is the same
as the decomposition under the action of  $\Gamma_0$.
While there are twice as many irreducible representations of  $\Gamma_0 = \aut (G) \times \Z_2$
as there are of $\aut (G)$, if $\alpha^{(k)}_{\Gamma_0}(-1) = I$ then $V_{\Gamma_0}^{(k)} = \{ 0 \}$.
The other half of the irreducible representations have $\alpha^{(k)}_{\Gamma_0}(-1) = -I$.
The irreducible representations of $\Gamma_0$ and of $\aut (G)$ can be labeled so that
$V_{\Gamma_0}^{(k)} = V_{\aut (G)}^{(k)}$ for $k \in K_{\aut (G)}$.

The isotypic components are uniquely determined, but the decomposition into
irreducible spaces is not.
Our goal is to find $B_\Gamma^{(k)}$ for all $k$ by finding the projection
$P_\Gamma^{(k)}:V\to V_\Gamma^{(k)}$.
To do this, we first need to
introduce representations over the complex numbers $\C$ for two reasons.
First,
irreducible representations over $\C$ are better understood than those over $\R$.
Second, our GAP program uses the field $\C$ since
irreducible representations over $\R$ are not
readily obtainable by GAP.

There is a natural action of $\Gamma$ on $W:=\C^n$ given by the representation
$g\mapsto \beta_g:\Gamma\to \aut(W)$ such that $\beta_g$ and $\alpha_g$ have the
same matrix representation. The isotypic decomposition
$W=\bigoplus_{k\in \tilde K} W_{\Gamma}^{(k)}$ is defined as above using the set
$\{\beta^{(k)}:\Gamma\to {\GL}_{\tilde d_\Gamma^{(k)}}(\C)\mid k \in \tilde K_\Gam\}$ of
irreducible representations of $\Gamma$ over $\C$.

The {\em characters} of the irreducible representation $\beta^{(k)}$ are
$\chi^{(k)}(g) := {\rm Tr} \, \beta^{(k)}(g)$. The projection $Q_\Gamma^{(k)}:W\to W_\Gamma^{(k)}$
is known to be
\begin{equation}
\label{projQ}
Q_{\Gamma}^{(k)} ={\frac{\tilde d_\Gamma^{(k)}}{|\Gamma|}}\sum_{g\in \Gamma} \chi^{(k)}(g) \beta_g .
\end{equation}
We are going to get the $P_\Gamma^{(k)}$'s in terms of $Q_\Gamma^{(k)}$'s.
A general theory for constructing these projections can be found in~\cite{NSS3}, but here it is enough that
if $\chi^{(k)} = \overline{ \chi^{(k)} }$, then $P_\Gam^{(k)}=Q_\Gam^{(k)}\mid_V$ and
	$d_\Gamma^{(k)}=\tilde d_\Gamma^{(k)}$, whereas if
$ \chi^{(k)} \neq \overline{ \chi^{(k)} }$, then $P_\Gam^{(k)}=\left(Q_\Gam^{(k)}+\overline{Q_\Gam^{(k)}}\right)\mid_V$ and
	$d_\Gamma^{(k)}=2\tilde d_\Gamma^{(k)}$, for all $k\in K_\Gam$.

\subsection{Basis processing}
\label{basis}

In this subsection, we describe how the package of programs from \cite{NSS3} is modified to generate the
basis needed to approximate solutions to
PDE~(\ref{pde}) on the square or cube.  In principle a brute force method is possible,
wherein the very large graph $\Omega_N$ is used.
However, the GAP and {\it Mathematica} portions of the package in
\cite{NSS3} cannot process such large graphs with current computer systems.
To remedy this,  we wrote specialized {\it Mathematica} basis-generation programs
for the square and cube.
These programs use the known eigenfunctions, $\psi_{i,j}$ or $\psi_{i,j,k}$, together with the
GAP output from the graphs in Figure~\ref{D4_and_Oct} to generate the data needed by
the GNGA program.

It can be shown that for polynomial $f_s$ and the known eigenfunctions
$\psi_m$ defined in terms of sine functions on the square and cube,
the midpoint numerical integration can be made exact, up to the arithmetic precision used in the computation.
In particular, consider the case when $f_s$ is cubic and the eigenfunctions are
$\psi_{i,j}$ (for $d = 2$) or $\psi_{i,j,k}$ (for $d = 3$).
Let
$\tilde M$
be the desired number of sine frequencies in each dimension, that is, $i, j, k \leq \tilde M$.
Then our midpoint numerical integration is exact
if $2{\tilde M} + 1$ gridpoints in each direction are used, giving a total of $N = (2{\tilde M}+1)^d$ total gridpoints.
For example, for analyzing PDE~(\ref{pde}) on $\Omega=(0,\pi)^2$ we used ${\tilde M}=30$ sine frequencies and,
for exact integration,  $N=(2\cdot 30+1)^2 = 3721$ gridpoints.
Similarly, for the cube $\Omega=(0,\pi)^3$, we used ${\tilde M}=15$ sine frequencies and $N=(2\cdot 15+1)^3 = 29,791$
gridpoints.

The only input for the basis generation program is ${\tilde M}$,
the desired number of sine frequencies in each dimension.
We define the bases
$$B_M=\{\bpsi_{i,j} \mid 1\leq i,j\leq {\tilde M} \hbox{ and } \lambda_{i,j}=i^2+j^2<({\tilde M}+1)^2+1\}$$
for the square, and
$$B_M=\{\bpsi_{i,j,k} \mid 1\leq i,j,k \leq {\tilde M} \hbox{ and } \lambda_{i,j,k}=i^2+j^2+k^2<({\tilde M}+1)^2+2\}$$
for the cube. These bases include all eigenfunctions with
eigenvalues less than
$\lambda_{{\tilde M}+1,1}$ and $\lambda_{{\tilde M}+1,1, 1}$ respectively.
This process gives $M=719$ for the square, not ${\tilde M}^2=900$, and $M=1848$ for the cube, as opposed to
${\tilde M}^3=3375$.

The basis generation code  also produces an automorphism file, which is an $8 \times N$
(for the square)
or $48 \times N$ (for the cube) matrix describing how the elements of
$\D_4$ or $\Oh$ permute the vertices in $\Omega_N$.
This file is used by the GNGA program
but not by the square and cube basis generation code. Instead,
the action of $\D_4$ or $\Oh$ on the eigenfunctions
is achieved very efficiently using the standard 2D representation of $\D_4$ acting on the plane
and the standard 3D representation of $\Oh$ acting on $\R^3$.
This requires us to define $\psi_{i,j}$ and $\psi_{i,j,k}$ for negative integers $i, j, k$
in this way:
$\psi_{-i, j}(x,y) = \psi_{i,j}(\pi-x,y)$ and $\psi_{-i, j,k}(x,y,z) = \psi_{i,j,k}(\pi-x,y,z)$.

The GNGA program does not use the basis $B_M$.
Rather, it uses a basis spanning the same $M$-dimensional space, obtained from
projections of the eigenfunctions in $B_M$ onto the isotypic
components identified by the GAP program.  The output of the GAP program for the small graphs shown in
Figure~\ref{D4_and_Oct} is used to achieve these projections.
For the cube, the eigenspaces of the Laplacian for the
eigenvalue $\lam_{i,j,k}$ are 1-dimensional if $i=j = k$, 3-dimensional if $i=j< k $ or $i<j =k$,
and 6-dimensional if $i < j < k$.  Each of these eigenspaces are separated into their isotypic
components automatically by the basis generation programs, using the characters found by GAP
for the graphs in Figure~\ref{D4_and_Oct}.
Then the projections in Equation~(\ref{projQ}) are written as $2 \times 2$
or $3 \times 3$ matrices, where $\beta_g$ is the standard 2D or 3D representation
of $\D_4$ or $\Oh$, respectively.
With these changes, the construction of the bases proceeds as described in \cite{NSS3}.

\subsection{The symmetry of a cube}
\label{symmetryOfCube}

\begin{figure}
\begin{tabular}{ccc}
\scalebox{.8}
{\includegraphics{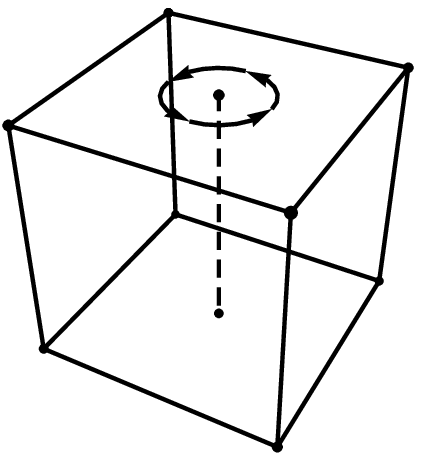}} &  
\scalebox{.8}
{\includegraphics{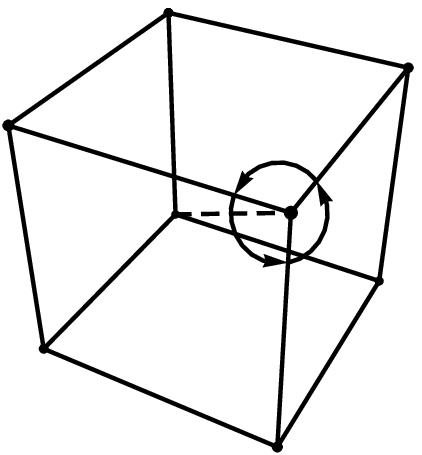}} & 
\scalebox{.8}
{\includegraphics{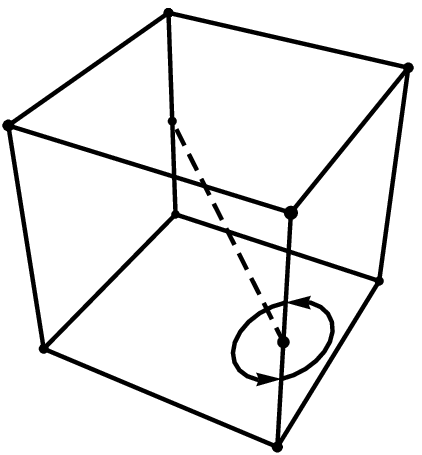}} \\
$R_{90}$& $R_{120}$& $R_{180}$
\end{tabular}
\caption{$R_{90}$, $R_{120}$, and $R_{180}$ are matrices that rotate the cube centered at the origin by
$90^\circ$, $120^\circ$ and $180^\circ$, respectively.
The dashed lines are the axes of rotation.
The 48 element automorphism group of the cube
is identified with $\Oh$.
The matrix group $\Oh$ is generated by these three rotation matrices and $-I_3$,
the inversion through the center of the cube.
}
\label{R90}
\end{figure}

\newcommand{\irredScale}{.21}
\begin{figure}
\begin{tabular}{ccccc}
\scalebox{\irredScale}{\includegraphics{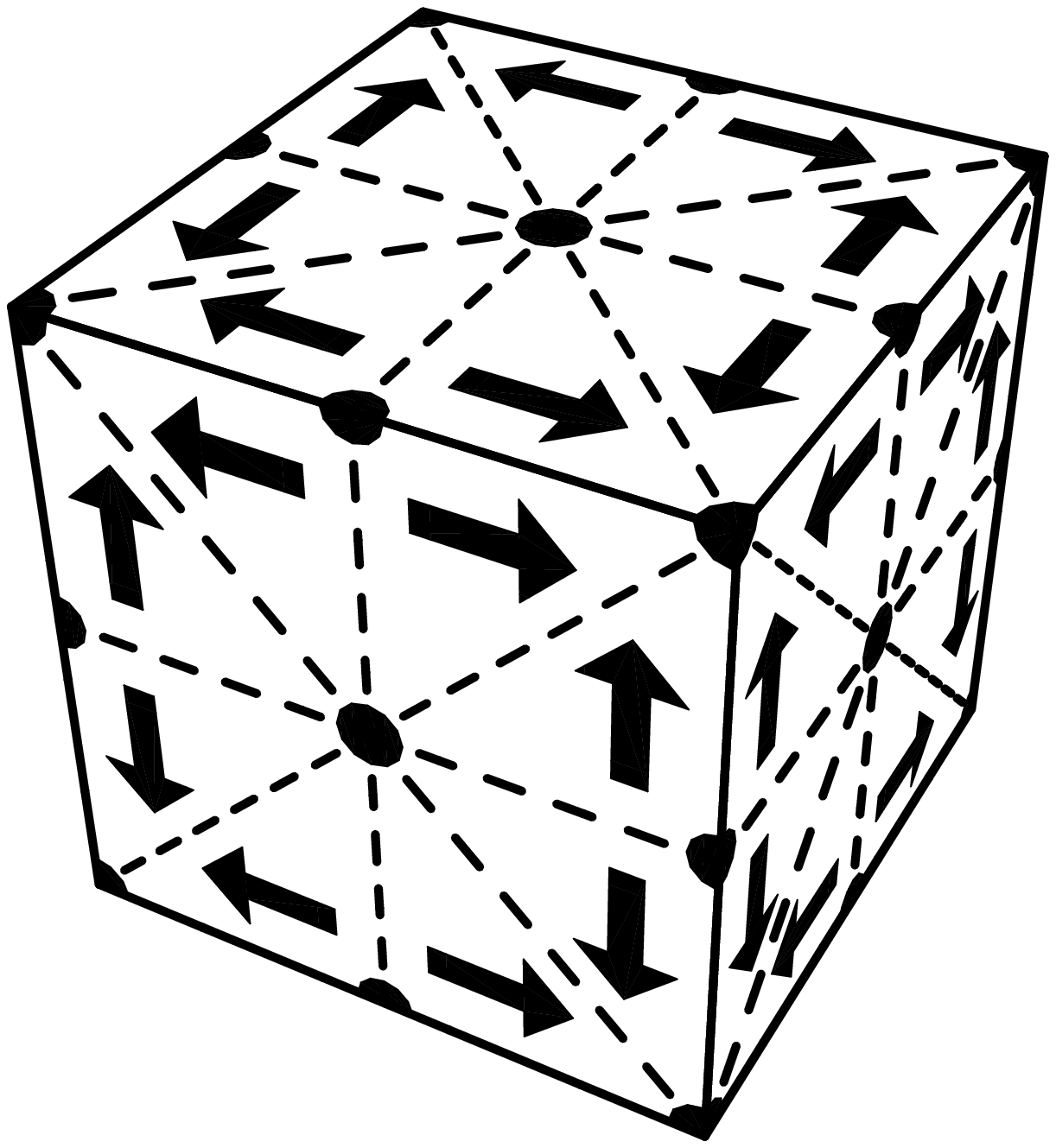}}&
\scalebox{\irredScale}{\includegraphics{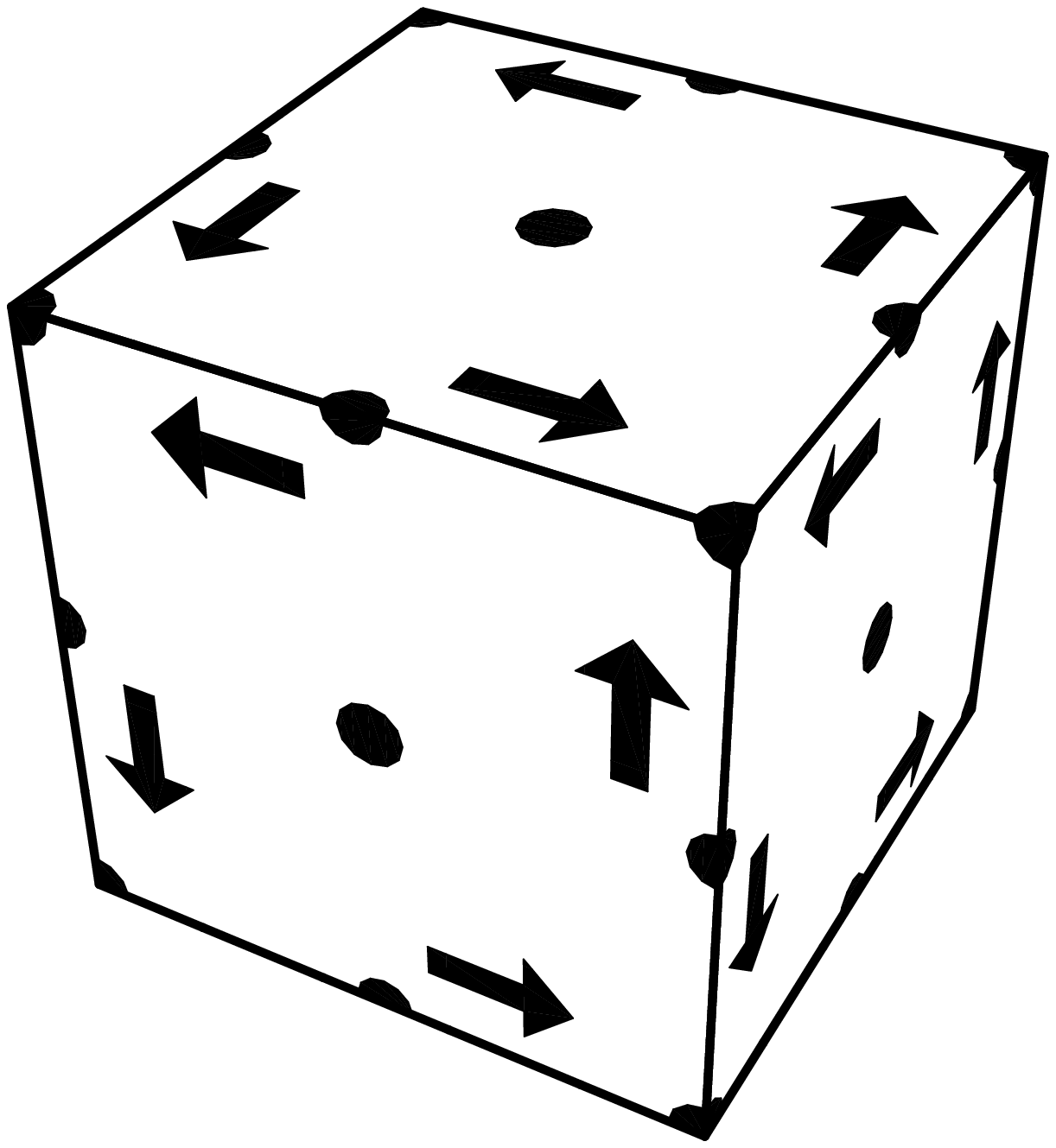}}&
\scalebox{\irredScale}{\includegraphics{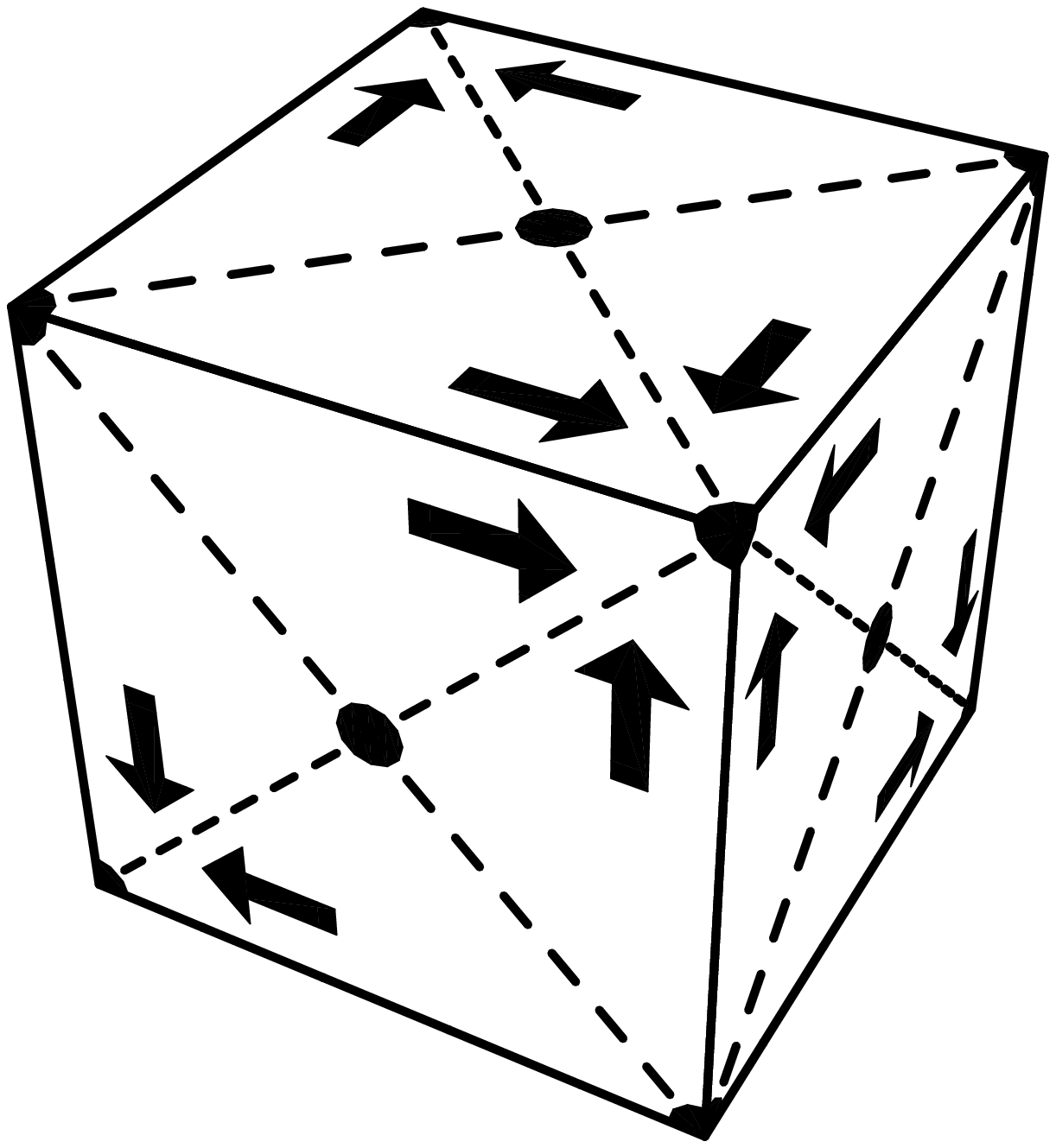}}&
\scalebox{\irredScale}{\includegraphics{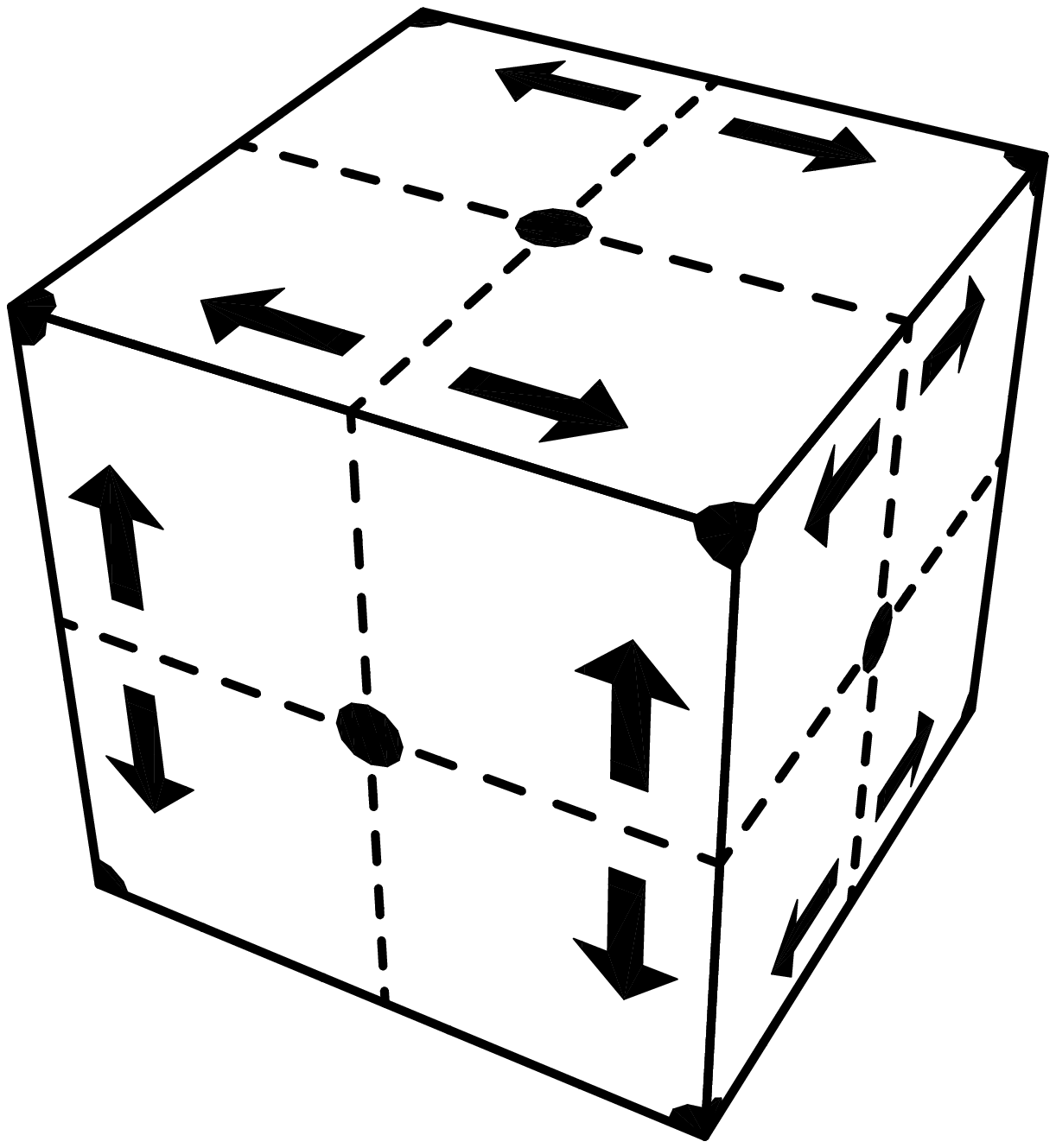}}&
\scalebox{\irredScale}{\includegraphics{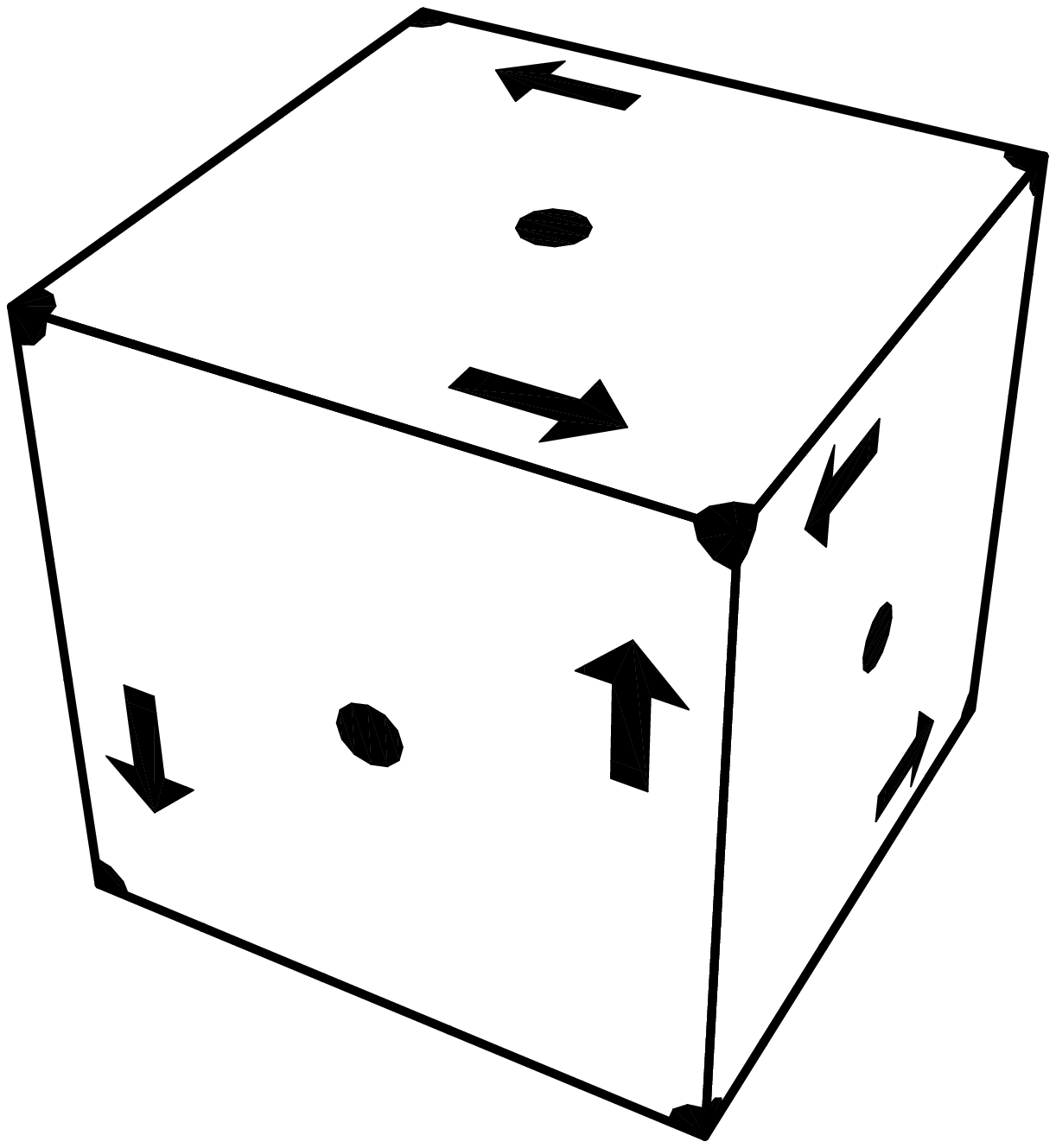}}
\\
$ \Oh \cong \s_4 \times \Z_2 $&
$ \oct \cong \s_4$&
$ \tet_d \cong \s_4$&
$ \tet_h \cong \A_4\times \Z_2$&
$ \tet \cong \A_4$
\end{tabular}
\caption{
Visualization of $\Oh$
and some of its subgroups.
The group orbit of a single arrow is drawn, suggesting a symmetric vector field.
These figures show the symmetry of the Lyapunov-Schmidt reduced gradient $\tilde g$ on 3-dimensional
critical eigenspaces in bifurcations with symmetry
occurring in PDE~(\ref{pde}) on the cube.
The dotted lines show the intersections of reflection planes with the cube.
The dots indicate 1-dimensional fixed-point subspaces of the
action, which intersect the cube at a vertex, edge, or face.
The EBL states that solution branches bifurcate in the direction of
these 1-dimensional fixed-point subspaces in $\tilde E$.
}
\label{irredSpaces3}
\end{figure}

In this section we describe the various matrix groups related to the symmetry of a cube.
We start with the symmetry group of
the cube $(-1,1)^3$.
The matrix form of this symmetry group is the 48-element
$\oct_h := \langle R_{90},R_{120},R_{180},-I_3 \rangle \le {\rm GL}_3(\R)$ where
\begin{equation*}
R_{90} = \left [ \begin{matrix}
0 & -1 & 0 \\
1 & 0 & 0 \\
0 & 0 & 1 \\
\end{matrix}
\right ], \quad
R_{120} = \left [ \begin{matrix}
0 & 0 & 1 \\
1 & 0 & 0 \\
0 & 1 & 0 \\
\end{matrix}
\right ], \quad
R_{180} = \left [ \begin{matrix}
0 & 1 & 0 \\
1 & 0 & 0 \\
0 & 0 & -1 \\
\end{matrix}
\right ]
\end{equation*}
as shown in Figure~\ref{R90}, and $I_3$ is the $3 \times 3$ identity matrix.

There is a slight complication in describing the symmetry group of the cube
$\Omega_3 := (0, \pi)^3$ that is the domain of our PDE, since $\Omega_3$ is not centered
at the origin.
The action of $\aut(\Omega_3) \cong \Oh $ on $\Omega_3$
is given by matrix multiplication about
the center $d = \pi/2 (1, 1, 1)$ of the cube.
That is,
for $\gamma \in \Oh$ and $x \in \Omega_3$,
the action is defined as
$$
\gamma \cdot x = d + \gamma(x - d) .
$$

The action of $\Oh \times \Z_2$
on the vector space of functions $u: \Omega_3 \rightarrow \R$
is given by
$$
((\gamma, \beta) \cdot u) (x) = \beta u(\gamma^{-1}\cdot x), \mbox{ for }
\gamma \in \Oh, \ \beta \in \Z_2 = \{ 1, -1 \}.
$$
It follows that the action of the generators of $\Oh\times \Z_2$ on eigenfunctions is
\begin{align*}
( R_{90}, 1) \cdot \psi_{i, j, k} & = (-1)^{j-1} \psi_{j,i,k} \\
(R_{120}, 1) \cdot \psi_{i, j, k} & = \psi_{k,i,j} \\
(R_{180}, 1) \cdot \psi_{i, j, k} & = (-1)^{k-1} \psi_{j,i,k} \\
(-I_3, 1) \cdot \psi_{i, j, k} & = (-1)^{i+j+k-1} \psi_{i,j,k} \\
(I_3, -1) \cdot \psi_{i, j, k} & = -\psi_{i,j,k}.
\end{align*}

There are three subgroups of $\Oh$ with 24 elements.  They are
$$
\oct = \langle R_{90}, R_{120}, R_{180} \rangle, \quad
\tet_d = \langle -R_{90}, R_{120}, -R_{180} \rangle, \quad  \mbox{and} \
\tet_h = \langle R_{90}^2, R_{120}, R_{180}, -I_3 \rangle
$$
shown in Figure~\ref{irredSpaces3}.
The group $\oct$ contains the rotational symmetries of the cube (or octahedron),
and is called the \emph{octahedral group}.
Note that $\Oh$ is the internal direct product $\oct \times \langle -I_3 \rangle$.

The groups $\tet_d$ and $\tet_h$ are related to the 12-element {\em tetrahedral group}
$$
\tet = \langle R_{90}^2, R_{120}, R_{180} \rangle,
$$
that contains all of the rotational symmetries of the tetrahedron.
Note that $\tet_h$ is the internal direct product $\tet \times \langle -I_3 \rangle$,
whereas $\tet_d$ does {\em not} contain $-I_3$.
The groups $\oct$ and $\tet_d$ are both isomorphic to the symmetric group $\s_4$.
The isomorphism can be proved by considering how the groups permute the 4 diagonals of the cube.
Similarly,
$\tet$ is isomorphic to $\A_4$, the alternating group.
The group names involving $\s_4$ and $\A_4$ are used by
the GAP program. In our GAP programs $\oct$ and $\tet_d$
are computed as irreducible representations of $\s_4$.
Our website, which is written automatically using the GAP output, also uses these names.

\end{section}

\begin{section}{Algorithms}
\label{algorithms}

The main mathematical algorithms used in the code for the results found in this paper
are the tGNGA, cGNGA, and the secant method with recursive bisection, all of which are developed and described in detail in~\cite{NSS3}.  We give a brief overview of these algorithms in Section~\ref{gnga}.
The current implementation of C++ code which supervises the execution of these algorithms has two substantial modifications not found in~\cite{NSS3}.

The first major improvement concerns the efficient way we now use our symmetry information to 
reduce the matrix dimension when setting up the system for the search direction $\chi$ used in 
the tGNGA.  In brief, each system uses only the rows and columns corresponding to the 
eigenfunctions having the symmetry of points on the branch.  The time savings can be substantial 
when seeking solutions with a lot of symmetry, since the numerical integrations required to form 
the systems are generally the most time-intensive computations we make.  We present the details 
in Subsection~\ref{blockhessian}.

Secondly, the current implementation is in parallel.  A serial implementation could not reproduce our results in a reasonable time.
To that end, in~\cite{NSS4} we developed a simple and easy to apply methodology for using high-level, self-submitting
parallel job queues in an MPI (Message Passing Interface \cite{MPI}) environment.
In that paper, we apply our parallel job queue techniques toward solving computational combinatorics problems,
as well as provide the necessary details for implementing our PDE algorithms in parallel C++ code in order to obtain the results found in the current article.
We include a high-level description in Section~\ref{mpq}.

\subsection{GNGA}
\label{gnga}

To follow branches and find bifurcations we take the parameter $s$ to be the $(M+1)^{\rm st}$ unknown.
When we say that $p=(a,s)\in\R^{M+1}$ is a solution we mean that
$\bu=\sum_{m=1}^M a_m\bpsi_m$ solves Equation~(\ref{dpde}) with parameter $s$,
that is $g=0$.
The $(M+1)^{\rm st}$ equation, $\kappa(a,s)=0$,
is chosen in two different ways, depending on whether we are implementing the tangent-augmented Newton's method (tGNGA) to force the following of a tangent of a bifurcation curve or
we are implementing the cylinder-augmented Newton's method (cGNGA) to force the switching to a new branch at a bifurcation point.
In either case, the iteration we use is:
\begin{itemize}
\item compute the constraint $\kappa$, gradient vector $g:=g_s(\bu)$, and Hessian matrix $h:=h_s(\bu)$
\item solve $\left[\begin{matrix}
h & \frac{\partial g}{\partial s} \\
(\nabla_a\kappa)^T & \frac{\partial\kappa}{\partial s}
\end{matrix}\right]
\left[\begin{matrix} \chi_a \\ \chi_s \end{matrix}\right]
=
\left[\begin{matrix} g \\ \kappa \end{matrix}\right]
$
\item $(a,s)\leftarrow (a,s) - \chi$, \ \ $u=\sum a_j \psi_j$.
\end{itemize}
\noindent
Equations~(\ref{grad}) and (\ref{hess}) are used to compute $g$ and $h$.
The $(M+1)^{\rm st}$ row of the matrix is defined by
$(\nabla_a\kappa, \frac{\partial\kappa}{\partial s})=\nabla\kappa\in\R^{M+1}$;
the search direction is
$\chi=(\chi_a,\chi_s)\in\R^{M+1}$.
Since this is Newton's method on $(g,\kappa)\in\R^{M+1}$ instead of just $g\in\R^M$,
when the process converges we have not only that $g=0$
(hence $p=(a,s)$ is an approximate solution to Equation~(\ref{pde})),
but also that $\kappa=0$.

\incmargin{1em}
\begin{algorithm2e}
\dontprintsemicolon
\linesnumbered
\hrule
\medskip
\SetKwFunction{KwPB}{pushBack}
\SetKwFunction{KwInsert}{insert}
\SetKwFunction{KwRemove}{remove}
\SetKw{KwBreak}{break}
\SetKw{KwWhile}{while}
\SetKwFor{Rep}{repeat}{}{}
\SetKwSwitch{Switch}{Case}{Other}{switch}{}{case}{otherwise}{}

\smallskip
\Rep{}{
  wait for a message from the boss \;  
    \Switch{the message is a}{
    \Case{follow branch job}{		
    		\While {branch is in window}{
				compute next point on branch with tGNGA \;
				\If {change in MI}{			
					call secant-bisection to find intervening bifurcation points \;
					\For {each intervening bifurcation point}{
					   put {\it find daughters} job on queue \;
					}
			   }
	   	}
	   	use interpolated guess and tGNGA to get last point on window boundary \;
    }
    \Case{find daughters job}{
      \For {each possible bifurcation subspace of critical eigenspace}{
			\While {new solutions still being found}{
				make a random guess in subspace and call cGNGA \;
				\If {new solution is nonconjugate to previously found solutions}{
			      put {\it follow branch} job on queue \;
      		}
      	}
      }
    }
    \Case{stop command}{
       stop \;
    }    
    }
}
\medskip
\hrule
\caption{\label{alg2}
Pseudo code for the main loop of the workers.  
This loop is entered after loading basis and symmetry files.
Whenever idle,  each worker accepts and runs jobs whenever such jobs exist.
The boss puts the trivial solution branch on the job queue as the first job.  
It manages the queue while the workers do their jobs, until the queue is empty, 
and then sends all workers a stop job.
}

\end{algorithm2e}


We use the tGNGA to follow branches.
The details are given in Algorithm~1 of~\cite{NSS3}.
In brief, given consecutive {\em old} and {\em current}
solutions  $p_\text{old}$ and $p_\text{cur}$ along a symmetry invariant branch, we compute the
(approximate) tangent vector $v=(p_\text{cur} - p_\text{old})/ \|p_\text{cur} - p_\text{old}\|\in\R^{M+1}$.
The initial \emph{guess} is then $p_\text{gs}=p_\text{cur} +cv$.
The speed $c$ has a minimum and maximum range, for example from 0.01 to 0.4,
and is modified dynamically according to various heuristics.
For example, this speed is decreased when the previous tGNGA call failed or the curvature of the branch is large,
and is increased toward an allowed maximum otherwise.
For the tGNGA, the constraint is that each iterate $p=(a,s)$ must lie on
the hyperplane passing through the initial guess $p_\text{gs}$, perpendicular to $v$.
That is, $\kappa(a,s) := (p-p_\text{gs})\cdot v$.
Easily, one sees that $(\nabla_a\kappa(a,s), \frac{\partial\kappa}{\partial s}(a,s)) = v$.
In general, if $f_s$ has the form $f_s(u) = s u + H(u)$, then $\frac{\partial g}{\partial s} = -a$.
Our function {\tt tGNGA}$(p_\text{gs}, v)$
returns, if successful,
a new solution $p_\text{new}$ satisfying the constraint.
Figure~\ref{alg2} shows how repeated tGNGA calls are made when a worker executes a branch following job.

The constraint used by the {\tt cGNGA} (see Algorithm~3 of~\cite{NSS3}) at a bifurcation point $p^*$
instead forces the new solution
$p_\text{new}$ to have a non-zero projection onto a subspace $E$ of the critical eigenspace $\tilde E$.
To ensure that we find the mother solution rather than a daughter,
we insist that the Newton iterates belong to the cylinder
$C :=\{(a, s) \in\R^{M+1}:\|P_E (a - a^*)\|=\eps\}$,
where $P_E$ is the orthogonal projection onto $E$
and the radius $\eps$ is a small fixed parameter.
At a symmetry breaking bifurcation the critical eigenspace is orthogonal to
the fixed-point subspace of the mother,
so the mother branch does not intersect the cylinder.
The constraint we use to put each Newton iterate on the cylinder is
$\kappa(a,s)=\frac12(\|P_E (a - a^*)\|^2-\eps^2)=0$.
The initial guess we use is $p_\text{gs}:= (a^*, s^*) +\eps (e,0)$, where $e$ is a
randomly chosen unit vector in $E$.  Clearly, $p_\text{gs}$ lies on the cylinder $C$.
A computation shows that
$\nabla_a \kappa (a,s) = P_E(a-a^*)$, and $\frac{\partial\kappa}{\partial s}(a,s) = 0$.
When successful, {\tt cGNGA($p^*, p_\text{gs}, E$)}
returns a new solution $p_\text{new}$ that lies on the cylinder $C$.

In the above paragraph, we take $E$ to be various low-dimensional subspaces of the critical eigenspace,
corresponding to the symmetries of solutions that are predicted by bifurcation theory.
For example, at an EBL bifurcation $E$ is spanned by a single eigenvector.
When the dimension of $E$ is greater than one, we call {\tt cGNGA} repeatedly with
several random choices of the critical eigenvector $e$.
The details are given in Equation~(7) and Algorithm~3 of~\cite{NSS3}.
The theory we apply does not guarantee a complete prediction of all daughter solutions.
Therefore we also call {\tt cGNGA} with $E$ equal to the full critical eigenspace.
In this way, if the dimension of the critical eigenspace is not too big
we have a high degree of confidence that we are capturing all relevant solutions,
including those that arise due to accidental degeneracy and
that are neither predicted nor ruled out by understood bifurcation theory.
The number of guesses in each subspace is heuristically dependent on the dimension of the subspace.
Too many guesses wastes time, and too few will cause bifurcating branches to be missed.
Figure~\ref{alg2} shows how repeated cGNGA calls are made when a worker executes a {\it find daughters} job.

We use the secant method to find bifurcation points.
In brief, when using the tGNGA to follow a solution branch and the
MI changes at consecutively found solutions,
say from $k$ at the solution $p_\text{old}$
to $k+\delta$ at the solution $p_\text{cur}$,
we know by the continuity of $D^2J_s$ that there exists a third, nearby solution $p^*$ where
$h$ is not invertible and the $r^{\rm th}$ eigenvalue of $h$ is zero,
where
$r=k+ \lceil\frac{\delta}2\rceil$.
Let $p_0 = p_\text{old}$, $p_1 = p_\text{cur}$, with $\beta_0$ and $\beta_1$
the $r^{\rm th}$ eigenvalues of $h$ at the points $p_0$ and $p_1$, respectively.

We effectively employ the vector secant method by iterating
\begin{itemize}
\item
$\displaystyle{ p_\text{gs} = p_i - \frac{(p_i-p_{i-1})\beta_i}{(\beta_{i}-\beta_{i-1})} }$
\item
$ p_{i+1} = {\tt tGNGA}(p_\text{gs}, v) $
\end{itemize}
\noindent
until the sequence $(p_i)$ converges.
The vector $v=(p_\text{cur} - p_\text{old})/\|p_\text{cur} - p_\text{old}\|$ is held fixed throughout,
while the value $\beta_i$ is the newly computed $r^{\rm th}$ eigenvalue
of $h$ at $p_i$.
If our function {\tt secant}$(p_\text{old}, p_\text{cur})$ is successful, it returns
a solution point $p^*=(a^*,s^*)$,
lying between $p_\text{old}$ and $p_\text{cur}$,
where $h$ has $\delta$ zero eigenvalues within some tolerance.
We take the critical eigenspace $\tilde E$ to be the span of the corresponding eigenvectors.
If $p^*$ is not a turning point, then it is a bifurcation point.

In fact, it is possible that several intervening bifurcation points exist.  
If the secant method finds a bifurcation point that has fewer than $d$ zero Hessian eigenvalues, 
there must be another bifurcation point in the interval.
In Figure~\ref{alg2}, the {\it secant-bisection} call refers to an implementation of Algorithm~2 
from~\cite{NSS3} entitled {\tt find$\_$bifpoints}, 
whereby such an occurrence triggers a bisection and a pair of recursive calls back to itself.  
Upon returning, each of the one or more found bifurcation points spawns its own {\it find daughters} job.
In turn, each time a (non-conjugate) daughter is found, a new {\it follow branch} job is put on the queue.

\subsection{The Block Diagonal Structure of the Hessian}
\label{blockhessian}

The majority of the computational effort for solving PDE~(\ref{pde})
using Newton's method comes from the computation of the entries of the Hessian matrix.
The time required can be drastically reduced by taking advantage of the block diagonal structure
of the Hessian that follows from the isotypic decomposition of $V = \R^M$,
the Galerkin space.

If the initial guess $u$ has symmetry $\Gam_i$,
then the isotypic decomposition (\ref{iso_decomp}) of the $\Gam_i$ action on $V$ is
$V = \bigoplus_k V_{\Gam_i}^{(k)}$,
where $k$ labels the irreducible representations of $\Gam_i$.
We assume that $k=0$ denotes the trivial representation, so
$\fix(\Gam_i) = V_{\Gam_i}^{(0)}$.

For any $u \in \fix(\Gam_i)$, the symmetry of the PDE implies that
the gradient $g_s(u)$ is also in $\fix(\Gam_i)$, and the Hessian,
evaluated at $u$, maps each of the isotypic components to itself.
That is,
$$
u \in \fix(\Gam_i) \implies g_s(u) \in \fix(\Gam_i) \ \mbox{and} \
h_s(u) \left (V_{\Gam_i}^{(k)} \right ) \subseteq V_{\Gam_i}^{(k)}.
$$
Thus, the Hessian is block diagonal in the basis $B_{\Gam_i}$ defined in Section~\ref{isoSub}.
A huge speedup of our program is obtained by only computing the Hessian restricted
to $\fix(\Gam_i)$ when doing Newton's method. 
After a solution is found, the block diagonal structure of the Hessian allows 
its efficient computation by avoiding integration of zero terms.
The full Hessian is required for the calculation of the MI.

Actually, the way we achieve a speedup in our numerical algorithm is not quite this simple.
In our implementation of the GNGA we always use $B_{\Gam_0}$, a basis of eigenvectors of the Laplacian.
The basis vectors are partitioned into bases
$B_{\Gam_0}^{(k)}$ for each
of the isotypic components of the $\Gam_0$ action on $V$.
We do not change the basis depending on the symmetry of the solution we are approximating.
Hence, we do not simply compute the blocks of the block diagonal Hessian.

When doing Newton's method, we use a {\em reduced Hessian} $\bar h_s$ in place of the
full Hessian $h_s$.
Define $P^{(k)}v$ to be the projection of $v$ onto $ V_{\Gam_i}^{(k)}$.
For $u \in \fix(\Gam_i)$,  the reduced Hessian is defined to be 
$$
\bar h_s(u)_{j,k} = 
\begin{cases}
h_s(u)_{j,k} & \text{ if }  P^{(0)}\psi_j \neq 0 \text{ and } P^{(0)}\psi_k \neq 0 \\
\lam_j - s  & \text{ if } j=k  \text { and } P^{(0)}\psi_j = 0 \\ 
0 & \text{ if } j\neq k \text{ and } ( P^{(0)}\psi_j = 0 \text { or } P^{(0)} \psi_k = 0).  
\end{cases}
$$
For each $u \in \fix(\Gam_i)$, assuming $h_s$ and $\bar h_s$ are nonsingular,
the Newton search direction is the solution to either system
$h_s(u)\chi =  g_s(u) $ or $\bar h_s(u) \chi = g_s(u)$.
The terms in the reduced
Hessian of the form $(\lam_j - s)\delta_{j,k}$ are included to make $\bar h_s$ nonsingular.
They are not strictly necessary since we find the least square solution $\chi$ with the smallest norm,
but they improve the performance of the LAPACK solver (dgelss).

After the solution is found, we identify those elements of the full Hessian that are known to be zero,
and avoid doing numerical integration for those elements.
In particular, $h_s(u)_{p,q} = 0$ if there is no $k$ such that
$P^{(k)}\psi_p \neq 0$ and $P^{(k)} \psi_q \neq 0$.

As a simple example for demonstration purposes, suppose that there are $M = 5$ modes,
and we are using the basis vectors $\psi_1,\ldots,\psi_5$.
Suppose further that the isotypic decomposition of the $\Gam_i$ action is
$V_{\Gam_i}^{(0)} = \spn\{\psi_1, \psi_3\}$,
$V_{\Gam_i}^{(1)} = \spn\{\psi_2\}$,
$V_{\Gam_i}^{(2)} = \spn\{\psi_4 + \psi_5\}$, and
$V_{\Gam_i}^{(3)} = \spn\{\psi_4 - \psi_5\}$.
The only nonzero projections are
$P^{(0)}\psi_1$, 
$P^{(0)}\psi_3$, 
$P^{(1)}\psi_2$, 
$P^{(2)}\psi_4$, 
$P^{(2)}\psi_5$, 
$P^{(3)}\psi_4$, and
$P^{(3)}\psi_5$. 
Thus, if $u \in \fix(\Gam_i)$, the gradient and Hessian have the form
\begin{equation}
\label{gradient_and_Hessian}
g_s(u) = \left [
\begin{smallmatrix}
* \\ 0 \\ * \\ 0 \\ 0
\end{smallmatrix}
\right ]
, \quad
h_s(u) = \left [
\begin{smallmatrix}
* & 0 & * & 0 & 0 \\
0 & * & 0 & 0 & 0 \\
* & 0 & * & 0 & 0 \\
0 & 0 & 0 & * & * \\
0 & 0 & 0 & * & * \\
\end{smallmatrix}
\right ].
\end{equation}
Note that a change of basis could be done to diagonalize
the lower right $2 \times 2$ block, but our program does not do this.
When we solve for the search direction $\chi$ in Newton's method, we use the
restricted Hessian
$$
\bar{h}_s(u) = \left [
\begin{matrix}
* & 0 & * & 0 & 0 \\
0 & \lam_2 - s & 0 & 0 & 0 \\
* & 0 & * & 0 & 0 \\
0 & 0 & 0 & \lam_4 - s & 0 \\
0 & 0 & 0 & 0 & \lam_5 - s \\
\end{matrix}
\right ],
$$
where only the $*$ terms are computed using numerical integration. 
Thus, in this small example, four numerical integrations are needed to compute the 
reduced Hessian for each step of
Newton's method, and nine numerical integrations are needed
to compute the full Hessian.

The speedup obtained in this manner is quite dramatic for solutions with high symmetry when $M$ is large.
As an example, for the PDE on the cube
the primary branch that bifurcates at $s = 3$ has symmetry $\Gam_2 \cong \Oh$.
The $\Gam_2$ action on $V$ has 10 isotypic components.
Our program, with $\tilde M=15$ and hence $M = 1848$, processes these solutions 10 to 15 times faster 
than it processes solutions
with trivial symmetry, where there is just one isotypic component and all $M(M+1)/2 = 1,708,474$
upper-triangular elements of the
Hessian need to be computed for every step of Newton's method.
Each one of these Hessian elements requires a sum over $N=29,729$ grid points (see Equation~(\ref{hess})).
Considering the whole process of finding a solution with $\Gam_2$ symmetry
using 3 or 4 steps of Newton's method, and then
computing the MI of this solution,
the majority of the computational effort
comes from the numerical integrations needed to compute
the full Hessian once after Newton's method converges.
Even with the speedup indicated in (\ref{gradient_and_Hessian}), the single
computation of $h_s(u)$ takes longer than the total time to perform all the other computations.
These include
the computation of the reduced Hessian $\bar {h}_s(u)$ and gradient $g_s(u)$, and
the LAPACK calls to solve $\bar h_s(u) \chi = g_s(u)$ at each Newton step,
as well as the LAPACK computation of the eigenvalues of the full Hessian evaluated at the solution.

\subsection{Parallel Implementation of Branch Following.}
\label{mpq}

%
%
%
It was necessary to implement our code in parallel in order to get timely results.  We use a library of
functions (MPQueue) presented in~\cite{NSS4} which make it easy to create and manage a parallel job queue.
For the current application of creating a bifurcation diagram, we choose a natural way to decompose the task into
two types of jobs,
namely {\it branch following} and {\it find daughters} jobs.
The boss starts MPI and puts the trivial solution branch on the queue as the first job.
After initialization and whenever idle,  each worker accepts and runs jobs whenever they exist.
The boss manages the queue while the workers do their jobs, until the queue is empty,
and then sends all workers a stop command.
Figure~\ref{alg2} shows how a parallel job queue is used to supervise the running of the jobs.

Normal termination for branch following jobs occurs when the branch exits a parameter interval for $s$.
When a change in MI is detected during branch following, 
the secant method is used to find an intervening bifurcation point with a proscribed
zero eigenvalue of the Hessian of $J$.
Recursive bisection is used if the number of zero eigenvalues of a found bifurcation point does not equal the observed
change in MI over a given subinterval, ensuring that all intervening bifurcation points are found.
Each bifurcation point spawns a {\it find daughters} job.
These jobs invoke the cGNGA with multiple random guesses from all possible bifurcation subspaces of the critical eigenspace.  
The number of attempts depends heuristically on the dimension of each subspace, stopping after sufficiently many 
tries have failed to find a new solution not in the group orbit of any previously found solution.
Each new, bifurcating solution spawns a {\it follow branch} job.

\begin{figure}
\scalebox{1.0}{\input{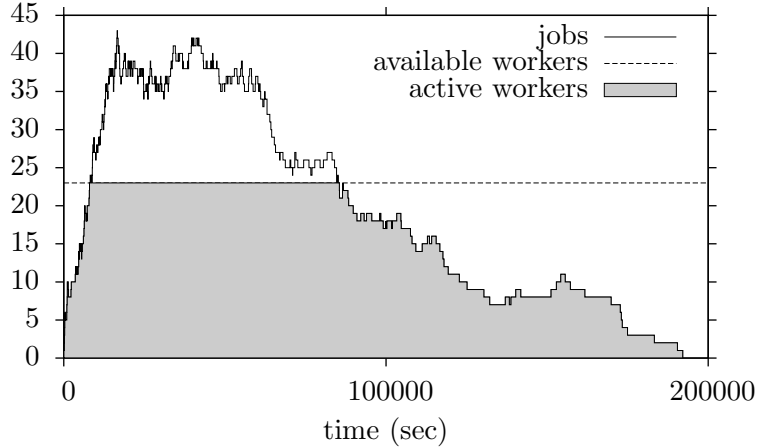}}
\caption{
The load diagram showing the number of jobs (branch following and bifurcation processing) and the
number of active workers.  There were 24 processors and thus 23 workers.  The jobs above the dotted line
are waiting in the job queue.
The total time for the run was 192,094 seconds $\approx$ 53.4 hours.
The average number of active workers is 15.6.
Thus, the run would have taken approximately 34 days with a single processor.
}
\label{load}
\end{figure}

We conclude this subsection with an example demonstrating the parallel run times encountered when generating our results
for the cube region.  In particular, Figure~\ref{load} shows the load diagram of a run 
for PDE~(\ref{pde}) on the cube with
$\tilde M = 15$, which gives $M = 1848$ modes and uses $N = 31^3$ gridpoints.
We followed all branches connected to the trivial branch with $0 \leq s \leq 13$, up to 4 levels of bifurcation.
The code found 3168 solutions lying on 111 non-conjugate branches, with 126 bifurcation (or fold) points.
This 2-day run used 24 processors and generated
all the data necessary for Figures~\ref{s0.10}, \ref{s0.10c}, \ref{bifdiagTd}, \ref{T_dBifContours},
and \ref{trivialSymContour}.
In Table 6.1 of \cite{NSS4} we provide runtime data for a similar problem using
a varying number of processors, in order to demonstrate scalability.

\subsection{Creating contour plots}

\label{contour}
\begin{figure}
\begin{tabular}{cccccccc}
\includegraphics{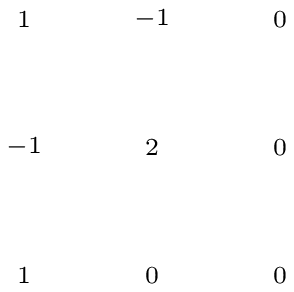} &  & \includegraphics{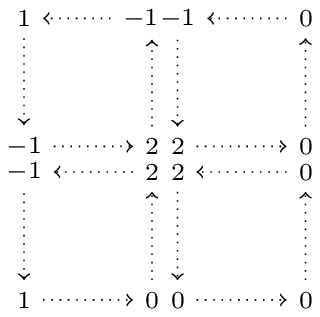} &  & \includegraphics{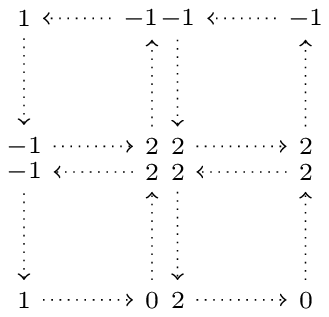} &  & \includegraphics{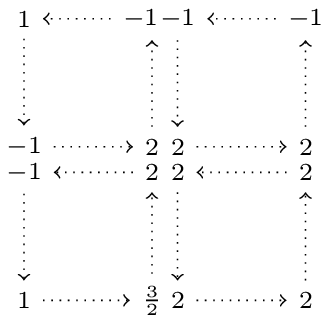} & \tabularnewline
 &  &  &  &  &  &  & \tabularnewline
(a) & $\rightsquigarrow$ & (b) & $\rightsquigarrow$ & (c) & $\rightsquigarrow$ & (d) & $\rightsquigarrow$\tabularnewline
\end{tabular}

\caption{
Square creation, followed by zero localization and zero purge.}
\label{fig:contour1}\end{figure}

\begin{figure}
\begin{tabular}{ccccc}
\includegraphics{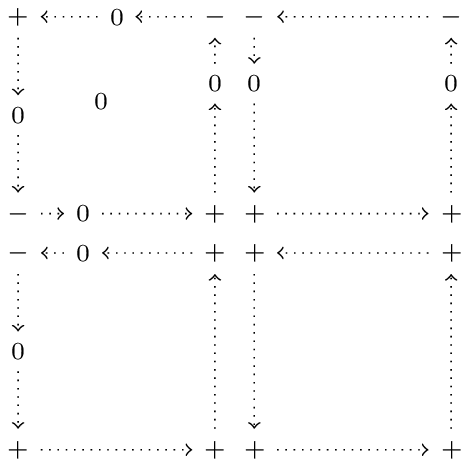} &  & \includegraphics{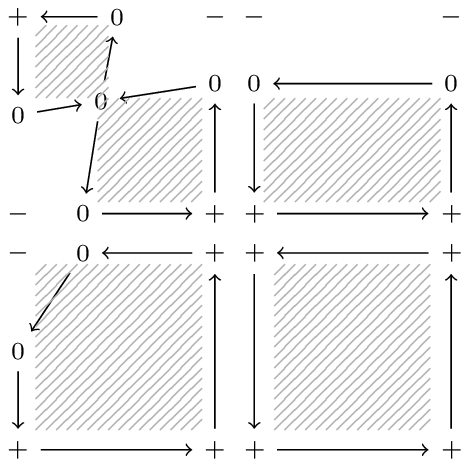} &  & \includegraphics{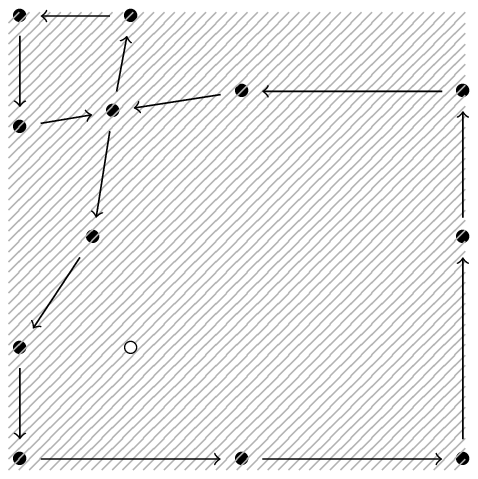}\tabularnewline
 &  &  &  & \tabularnewline
(e) & $\rightsquigarrow$ & (f) & $\rightsquigarrow$ & (g)\tabularnewline
\end{tabular}

\caption{
Boundary detection, followed by polygon creation and polygon merging.}
\label{fig:contour2}
\end{figure}


We now describe the process of creating a
contour plot for visually depicting an approximate solution on the square and cube.
For the square contour plot graphics found in Figures~\ref{square_plots} and \ref{seq4},
we use $\tilde M = 30$ leading to $N=61^2$ grid points.
This is a sufficient number of grid points to generate good contour plots.
The cube solutions visualized in the contour plot graphics found in
Figures~\ref{s0.10c}, \ref{bifdiag4c}, \ref{trivialSymContour}, \ref{T_dBifContours}, and~\ref{s14}
were all obtained via GNGA using
$\tilde M = 15$ leading to $M=1848$ modes and $N=31^3$ grid points.
This resolution is not quite sufficient for generating clear contour plots.
Using the $M$ coefficients and the exact known basis functions,
we reconstruct $u$ on a grid with $56^3$ points before applying the techniques described below.

When $\Omega$ is the cube,
since $u=0$ on the boundary we plot $u$ on the boundary of a smaller concentric cube.
The faces of this inner cube are $5$ gridpoints deep from the original boundary.
Thus each of the $6$ faces we draw has $u$ values on $(56-2\cdot 5)^2$ grid points.
The side length of the cube seen in the figures is about
$46/56 = 82\%$ the length of the original cube.
At the end of the process,
the contour plots on the six faces of the inner cube are assembled into a front and back view.
We use a standard perspective mapping from $\R^3$ to $\R^2$ to join three of the six faces
for each view.

In the remainder of this sub-section, we describe how to make a contour map for a function
$u: \Omega \rightarrow \R$ on the square or cube.
First of all, an algorithm chooses one function in the group orbit $\Gamma_0 \cdot u$
that makes the symmetry of the contour map agree with the flag diagram. 
It also replaces $u$ by $-u$ if this saves ink. For example,
if the function has a 3-fold symmetry about a diagonal of the cube, our algorithm chooses
a function that satisfies $(\R_{120}, 1)\cdot u  = u$ since the axis of rotation of $R_{120}$
points toward the viewer.
Then we have the values of $u$ on a square grid
(from the square or a face of the cube) as
shown in Figure~\ref{fig:contour1}(a).
We do some data pre-processing, depicted in Figures~\ref{fig:contour1}(b--d),
in order to get acceptable plots when there are many data points
with $u = 0$, which is common because of the symmetry of solutions.
After interpolation, Figure~\ref{fig:contour2}(e),
the positive region is the union of many polygons, mostly squares, Figure~\ref{fig:contour2}(f).
Redundant edges are eliminated to obtain a few many-sided polygons that are shaded,
Figure~\ref{fig:contour2}(g).
The contours are obtained by the same algorithm: It finds the many-sided polygons
enclosing the regions where $u \ge c$, but just the boundary is drawn.
Our implementation directly creates postscript files.
Without the elimination of redundant edges, the files would be of an unreasonably large size, and
the contours could not be drawn with the same algorithm.

We now provide the details.
Let $(\mathbf{x},a)$
represent a point where $\mathbf{x}\in\mathbb{R}^{n}$, $n = 2$ or $3$, is a grid
point and $a\in\mathbb{R}$ such that $u(\mathbf{x})\approx a$ for
a given solution $u$. We use the following steps as visualized by
Figures~$\ref{fig:contour1}$ and \ref{fig:contour2}.
\begin{enumerate}
\item Square creation (a)$\rightsquigarrow$(b): In the first step, we break
the square grid into individual squares to be handled separately.
We represent one of these squares with the cycle $((\mathbf{x}_{1},a_{1})\cdots(\mathbf{x}_{4},a_{4}))$
containing the corner points.
\item Zero localization (b)$\rightsquigarrow$(c): We eliminate several
zero values in a row by applying the rule
\[
\cdots(\mathbf{x},a)(\mathbf{y}_{1},0)\cdots(\mathbf{y}_{n},0)(\mathbf{z},b)\cdots\rightsquigarrow\cdots(\mathbf{x},a)(\mathbf{y}_{1},a)\cdots(\mathbf{y}_{n},b)(\mathbf{z},b)\cdots
\]
to each cycle where the variables $a$ and $b$ represent non-zero
values.
\item Zero purge (c)$\rightsquigarrow$(d): We eliminate the zero values
without a sign change by applying the rule
\[
\cdots(\mathbf{x},a)(\mathbf{y},0)(\mathbf{z},b)\cdots\rightsquigarrow\cdots(\mathbf{x},a)(\mathbf{y},(a+b)/2)(\mathbf{z},b)\cdots
\]
where $ab>0$.
\item Boundary detection (d)$\rightsquigarrow$(e): We insert zero points
between sign changes using the rule
\[
\cdots(\mathbf{x},a)(\mathbf{y},b)\cdots\rightsquigarrow\cdots(\mathbf{x},a)\left(\frac{|b|\mathbf{x}+|a|\mathbf{y}}{|a|+|b|},0\right)(\mathbf{y},b)\cdots
\]
where $ab<0$. The location of the new zero point is determined using
linear interpolation. The number $k$ of zeros in the cycle after
this step must be $0$, $2$ or $4$.
\item Polygon creation (e)$\rightsquigarrow$(f): Now we create polygons
with shaded inside to indicate positive values.

\begin{enumerate}
\item If $k=0$ and $a_{i}>0$ for all $i$ in the cycle $((\mathbf{x}_{1},a_{1})\cdots(\mathbf{x}_{4},a_{4}))$,
then we create the polygon with these corner points. This occurs in
the lower-right square of Figure~\ref{fig:contour2}(e). Note that
if $a_{i}<0$ then no polygon is created.
\item If $k=2$, then we create a polygon for each cyclic pattern of the
form
\[
(\mathbf{x},0)(\mathbf{y}_{1},a_{1})\cdots(\mathbf{y}_{n},a_{n})(\mathbf{z},0)
\]
with $a_{1},\ldots,a_{n}$ all positive. There is an example of
this with $n=2$ in the upper-right square and an example with $n=3$
in the lower-left square of Figure~\ref{fig:contour2}(e). It is
also possible to have $n=1$, but this is not shown in the example.
\item If $k=4$, then we find the intersection $(\mathbf{c},0)$ of the
line segments joining opposite zeros. Then we look for cyclic patterns
of the form
\[
(\mathbf{x},0)(\mathbf{y},a)(\mathbf{z},0)
\]
with $a$ positive. For each such pattern we create the polygon with
corner points
\[
(\mathbf{x},0),(\mathbf{y},a),(\mathbf{z},0),(\mathbf{c},0).
\]
This happens in the upper-left square of Figure~\ref{fig:contour2}(e)
\end{enumerate}
\item Polygon merging (f)$\rightsquigarrow$(g): We combine polygons sharing
a side into a single polygon. This gives the zero set which is the
boundary of the shaded positive region. This single polygon is written
directly to a postscript file. Polygon merging is a time consuming
operation, but writing the merged polygons instead of the individual
ones reduces the size of the postscript file significantly.
\item Level curve creation: Level curves are created using the previous
steps. We use the zero set of $u-c$ as the level curve of $u=c$.
The original grid values are changed by the shift value $c$ and only
the final merged polygon is written to the postscript file without
shading. Note that polygon merging is essential for finding level
curves.
\item Local extremum dots (g):  White dots are drawn at the estimated position
of local extrema in the $u > 0$ region, as indicated in Figure~\ref{fig:contour2}(g).
Every interior data point $(\mathbf{x}, a)$, is checked to see if $a$ is at least as large
(or at least as small) as $b_i$ for the eight neighbors $(\mathbf{x}_i, b_i)$.  For the
data in Figure~\ref{fig:contour1}(a), only the center point is checked, and it is indeed
a local maximum grid point.  For each local extremum grid point, a more precise estimate
of the extreme point is found by fitting a quadratic function to the $9$ data points
with the extreme grid point at the center.  Since a quadratic function $f: \R^2 \rightarrow \R$
has $6$ constants, the least squares solution for the $9$ equations in $6$ unknowns is found.
Then standard calculus gives the position of the extreme point of $f$, and a dot is drawn there.
For the data in \ref{fig:contour1}(a), the maximum point of the fitted $f$ is located half way to the grid
point to its lower left.  Similarly, black dots are drawn at the estimated positions of the
local extrema in the white ($u < 0$) region.
\end{enumerate}

\end{section}


\begin{section}{The PDE on the Square and $\D_4$ Symmetry}
\label{square}

In this section we re-visit the case where $\Omega$ is the unit square.
We first studied the square case in \cite{NS}, without the benefit of automation
or our recent improvements in branch following.
In that paper, all of the symmetry analysis was done by hand.
Each of the few bifurcation points we analyzed were essentially small projects in themselves, as were monotonic
segments of the branches connecting them.
In the present article, we use our automatically generated bifurcation digraph
and new Newton algorithms to quickly reproduce the old results and then go
much further ($s \gg \lambda_1$) in finding solutions of every possible symmetry type.
We also show that we can handle some accidental degeneracies that occur in the square case.

\def\psize{.13}
\def\ya{4}
\def\yb{9}
\def\yc{14}
\def\yd{18}
\begin{figure}
\scalebox{1}{
\xygraph{ !{0;<.9cm,0cm>:}
[]="O"
"O"
[r(0)] [u(0)]{
               \scalebox{\psize}{
               \includegraphics{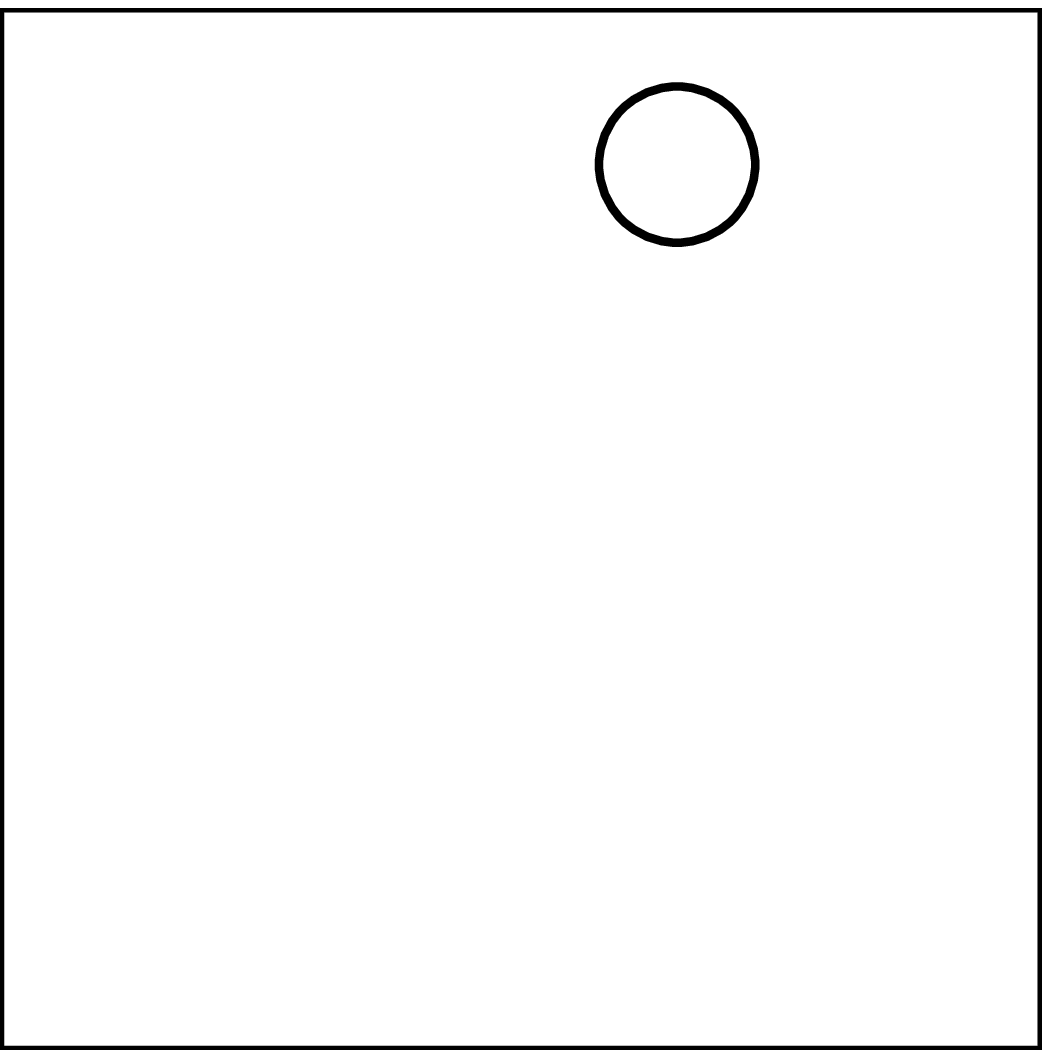}
                } }="v19"
"O"
[r(-5)] [u(\ya)]{
               \scalebox{\psize}{
               \includegraphics{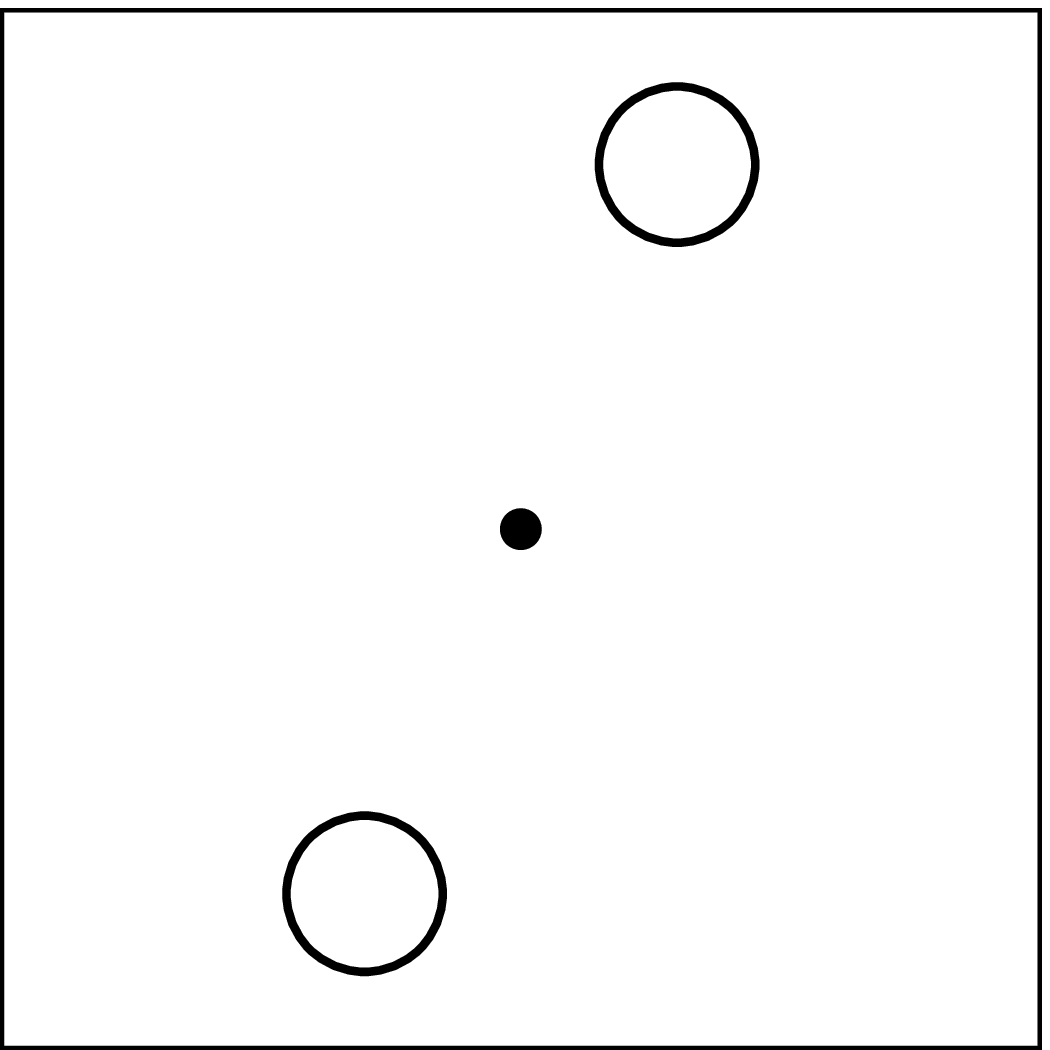}
                } }="v17"
"O"
[r(0)] [u(\ya)]{
               \scalebox{\psize}{
               \begin{tabular}{ccc}
               \includegraphics{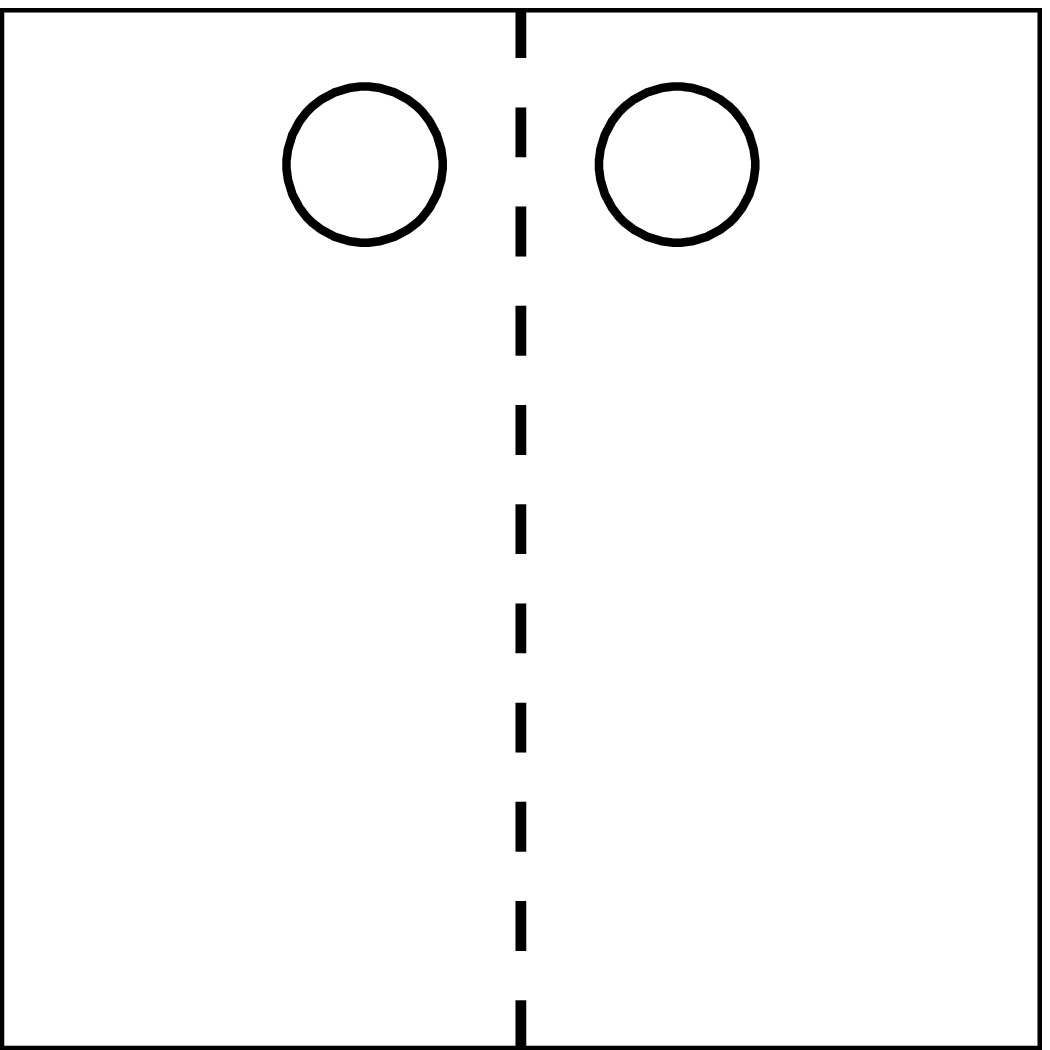} & \hspace{0.3in} &
               \includegraphics{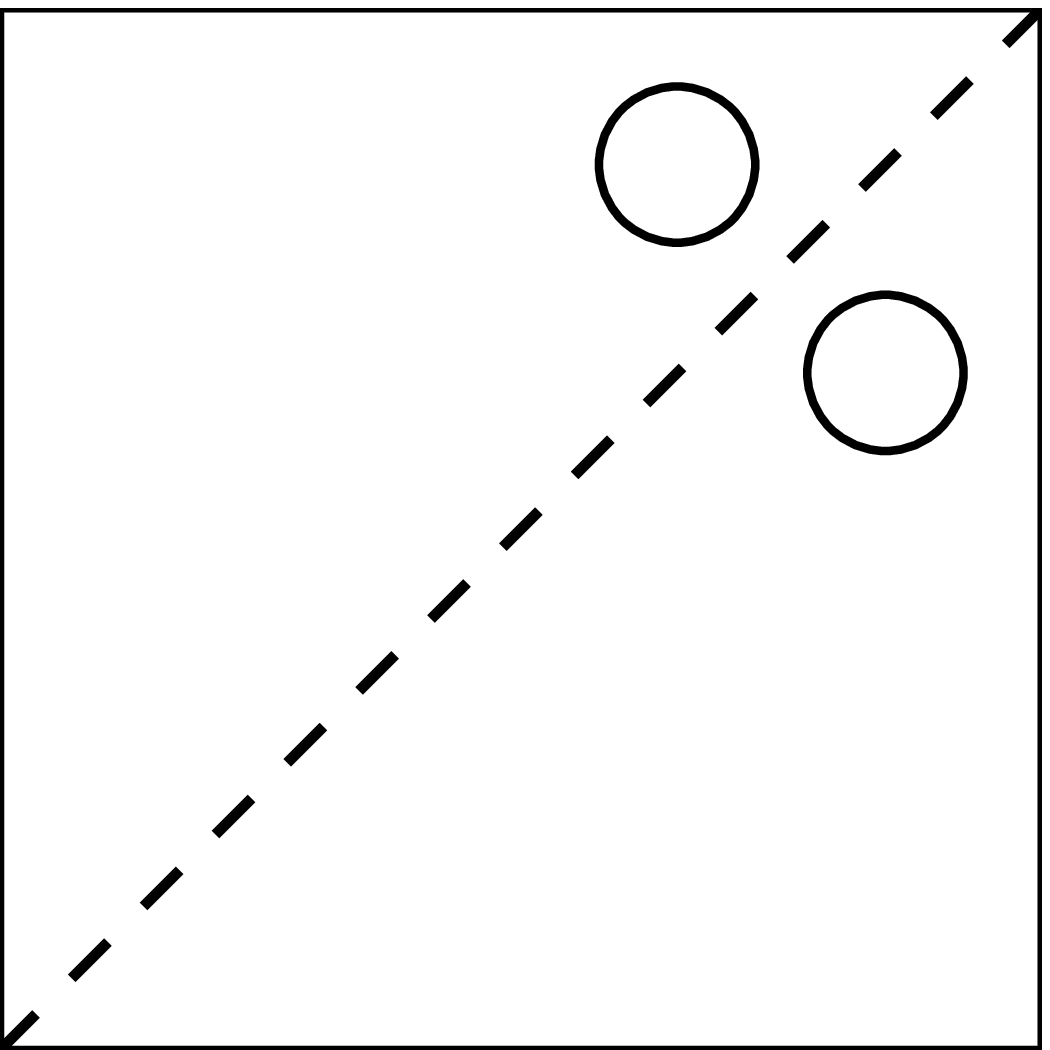}\\ [0.5in]
               \includegraphics{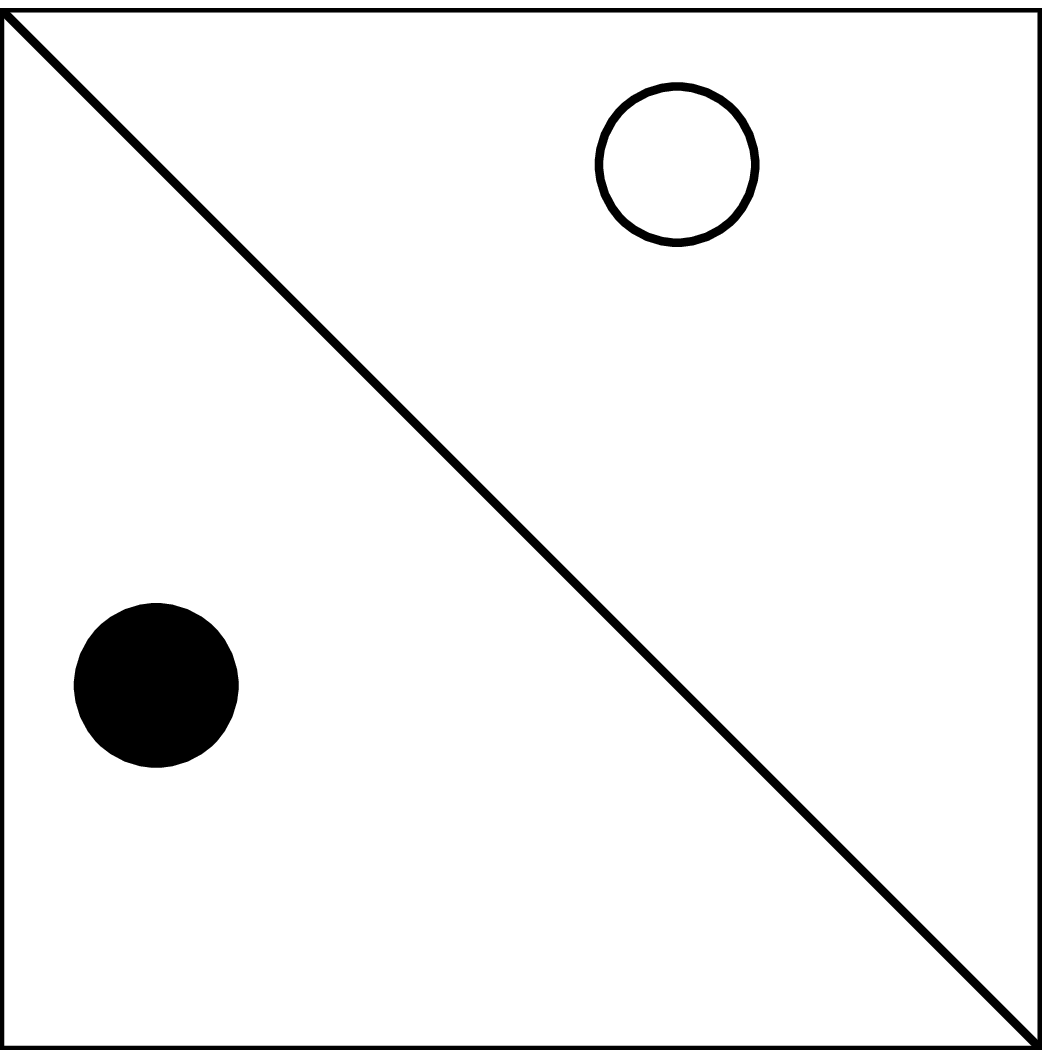} & \hspace{0.3in} &
               \includegraphics{figures/s13.eps}\\
               \end{tabular}
               } }="v13"
"O"
[r(5)] [u(\ya)]{
               \scalebox{\psize}{
               \includegraphics{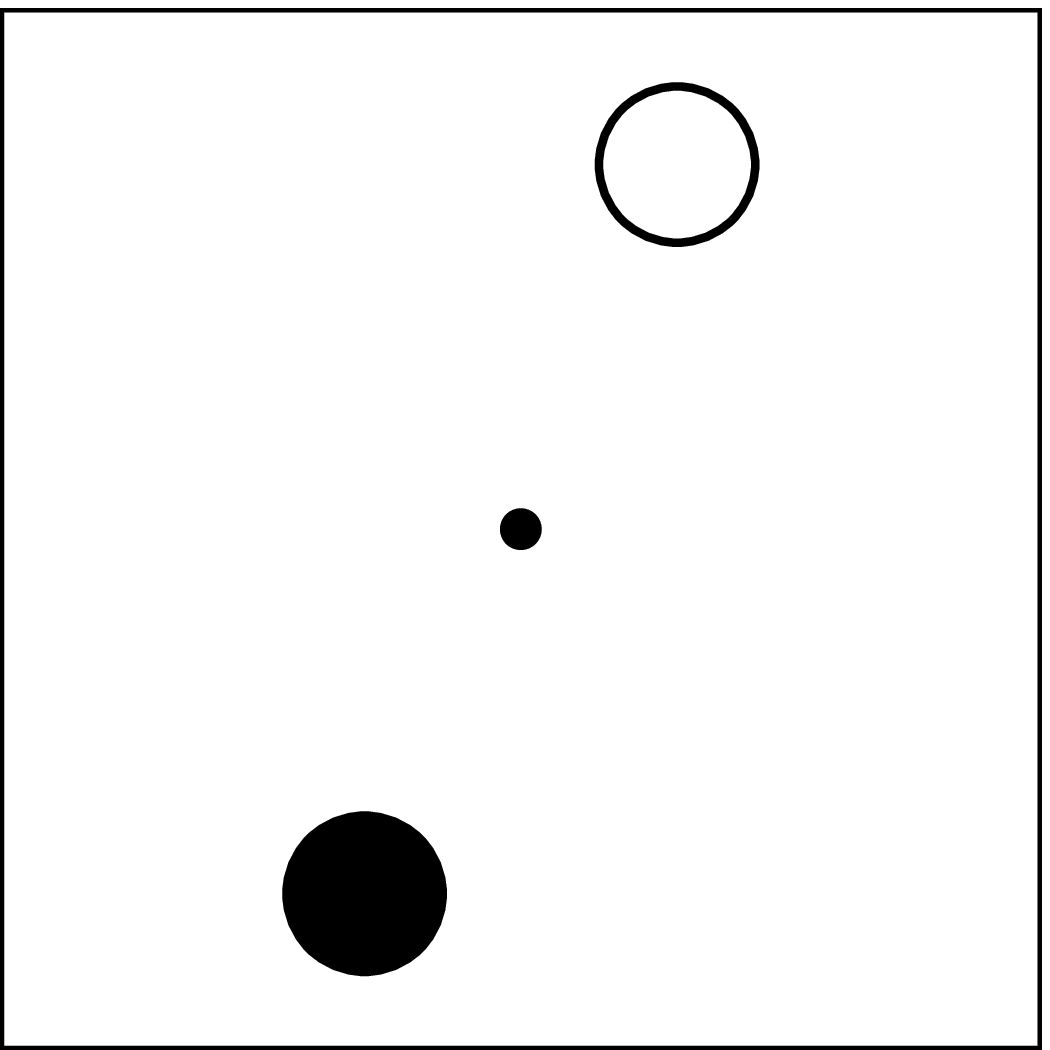}
                } }="v15"
"O"
[r(0)] [u(\yb)]{
               \scalebox{\psize}{
               \begin{tabular}{ccc}
               \includegraphics{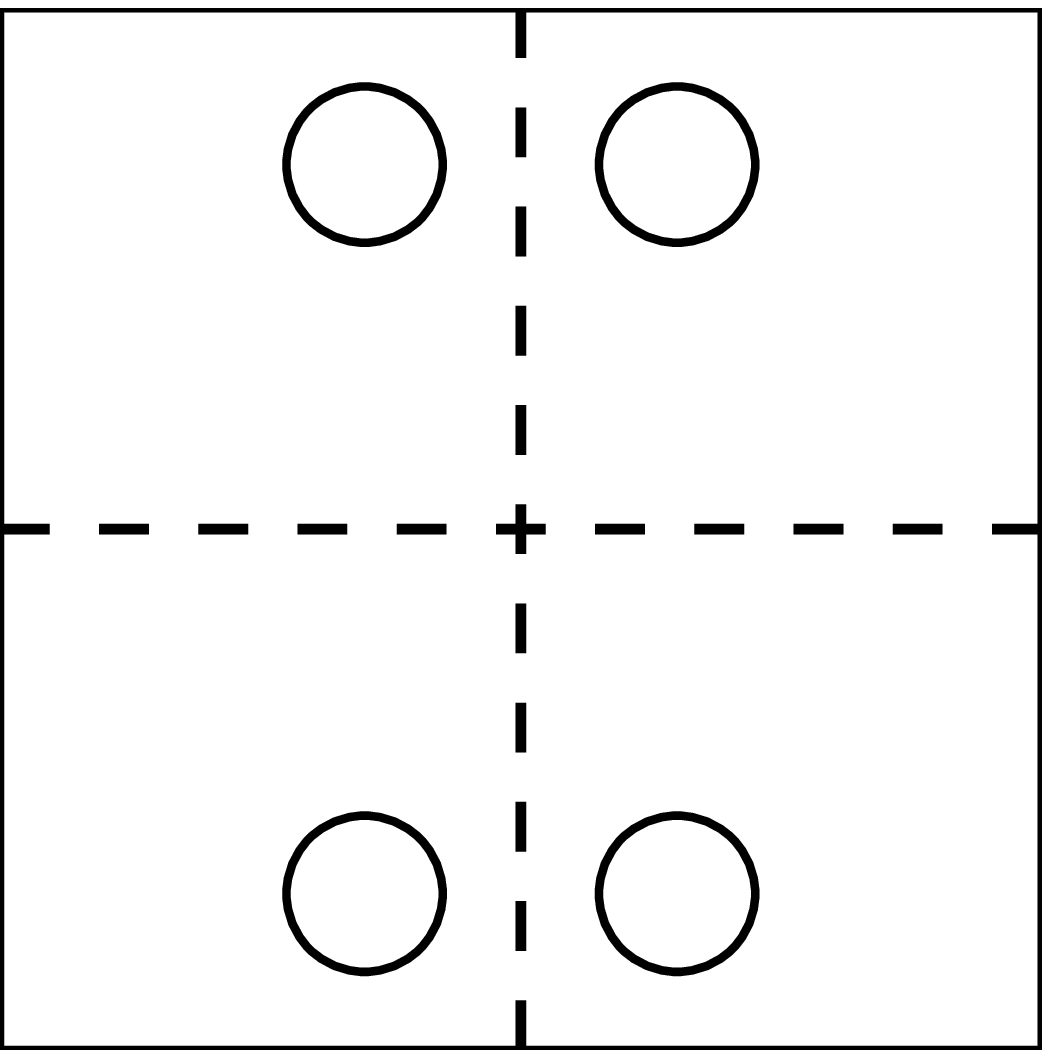}&  \hspace{0.3in} &
               \includegraphics{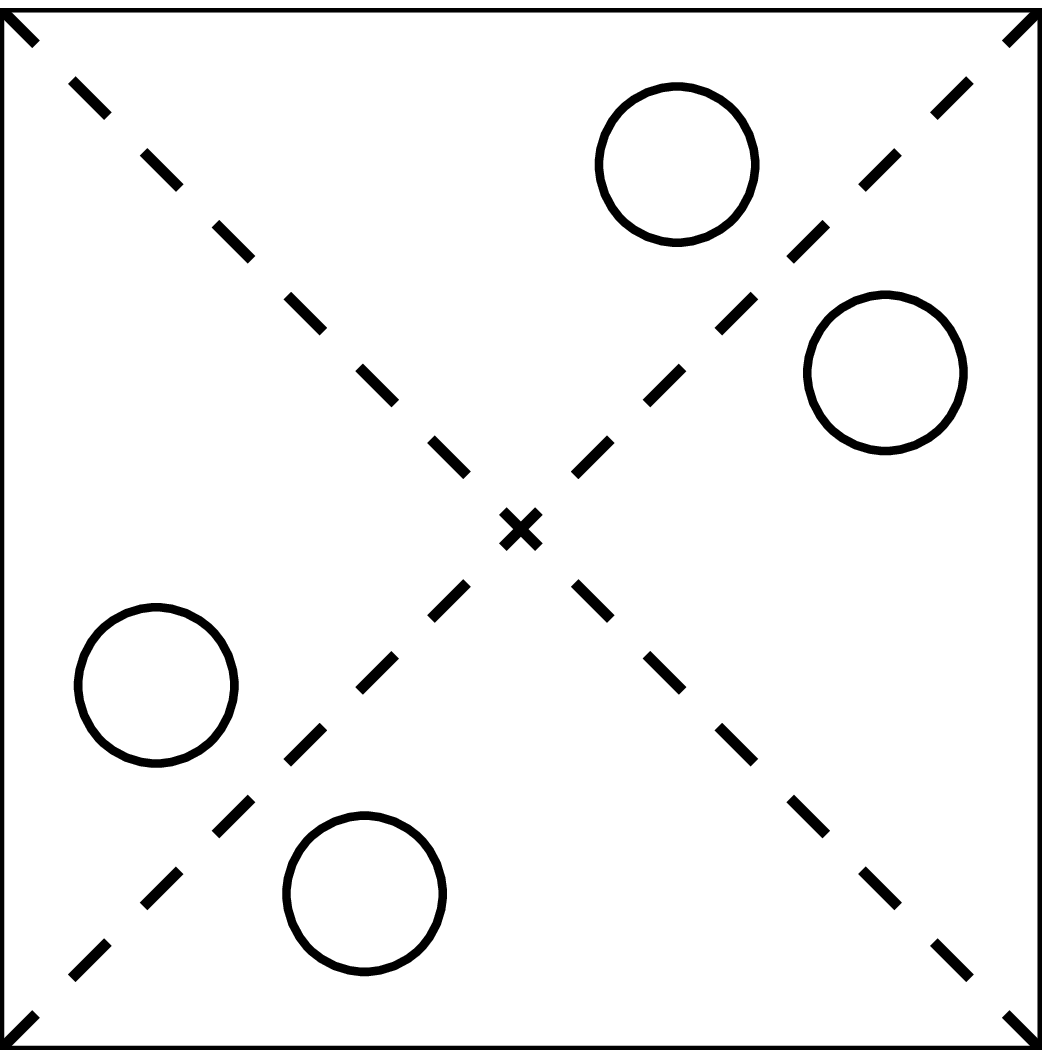}\\ [.5in]
               \includegraphics{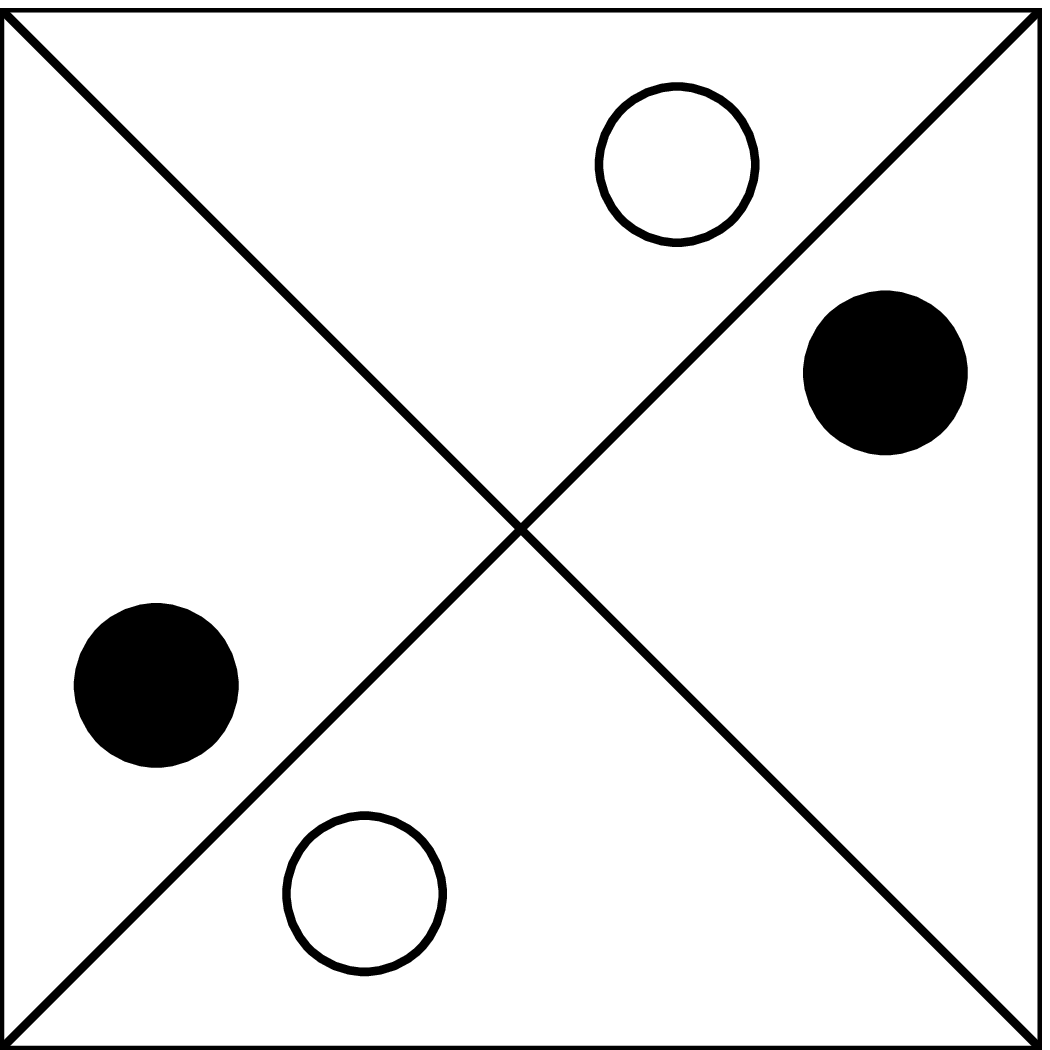}&  \hspace{0.3in} &
               \includegraphics{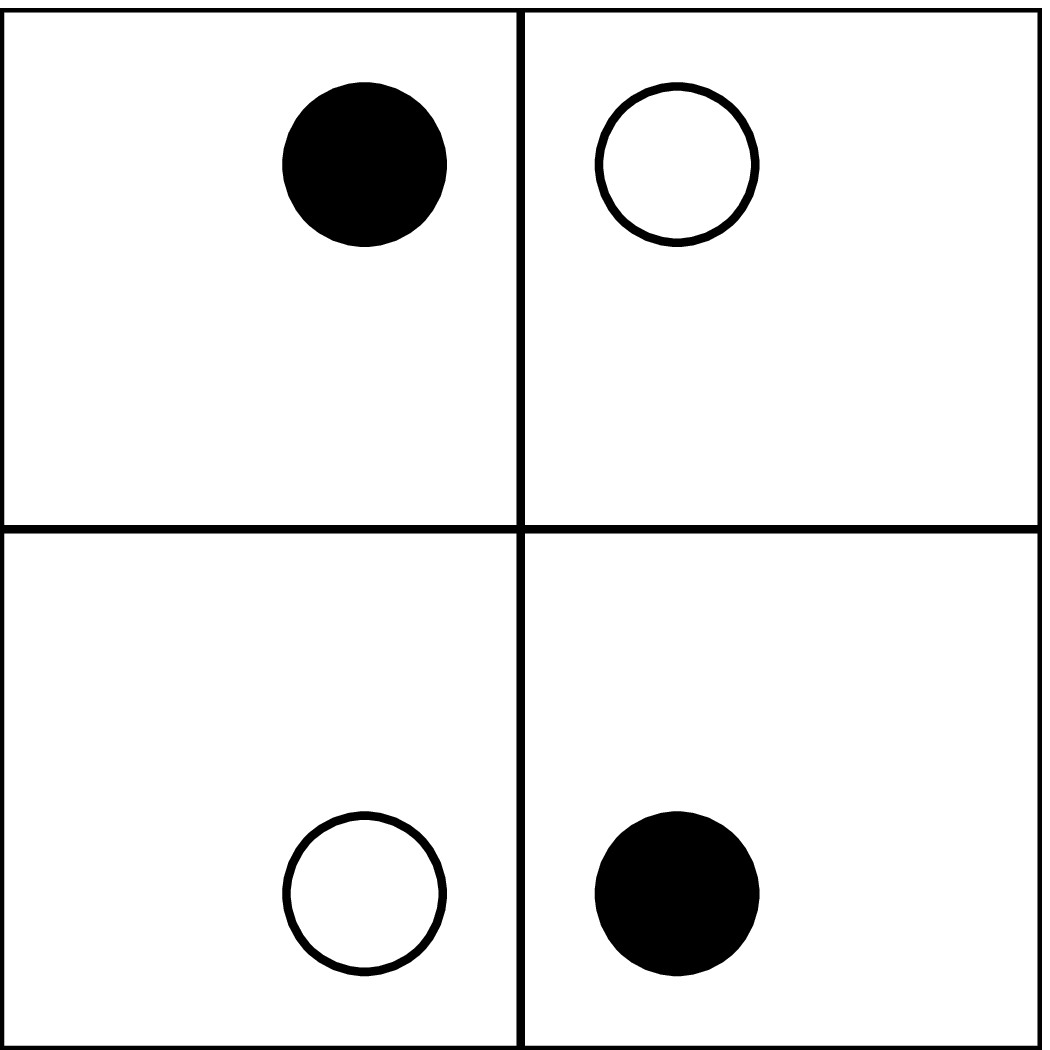}\\
               \end{tabular}
                } }="v7"
"O"
[r(-8)] [u(\yb)]{
               \scalebox{\psize}{
               \begin{tabular}{c}
               \includegraphics{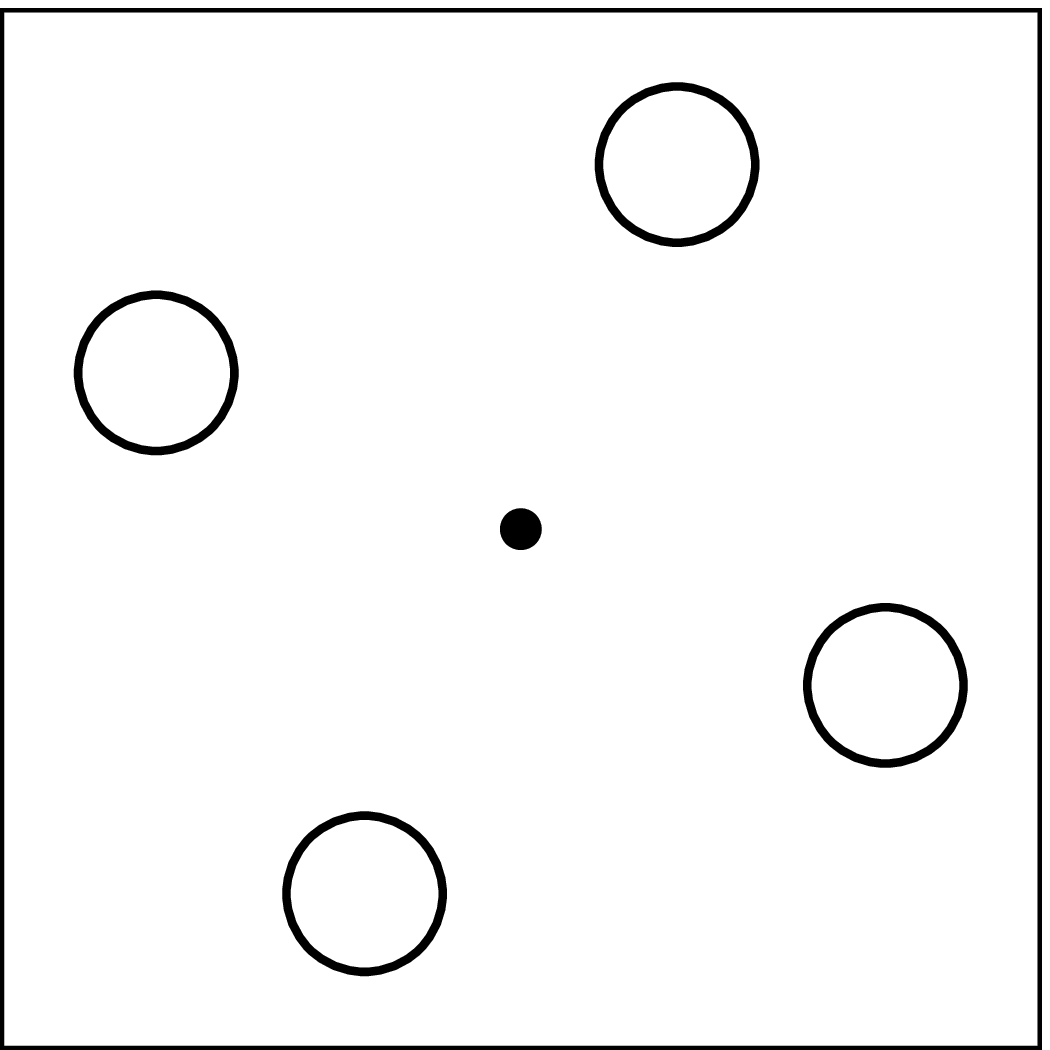}\\[.5in]
               \includegraphics{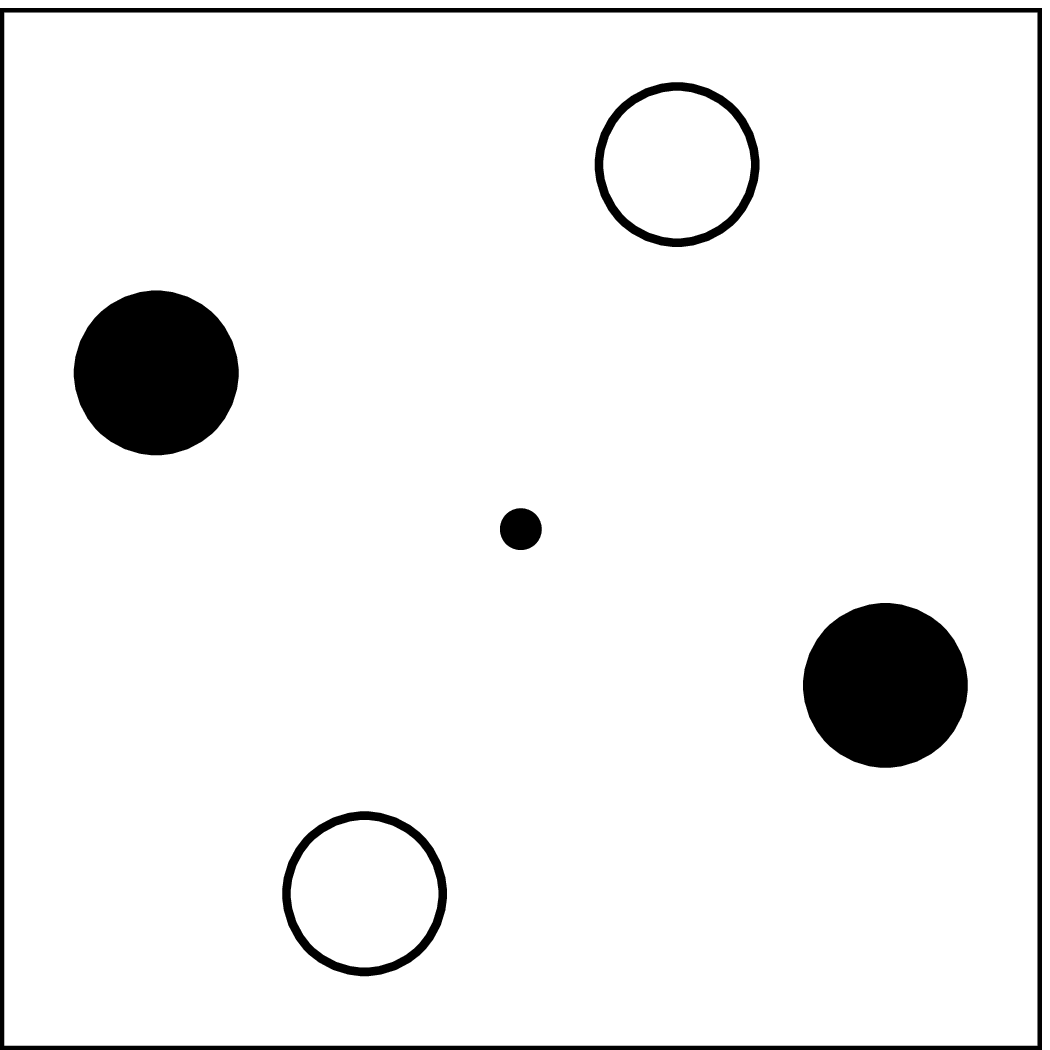}\\
               \end{tabular}
                } }="v6"
"O"
[r(8)] [u(\yb)]{
               \scalebox{\psize}{
               \begin{tabular}{c}
               \includegraphics{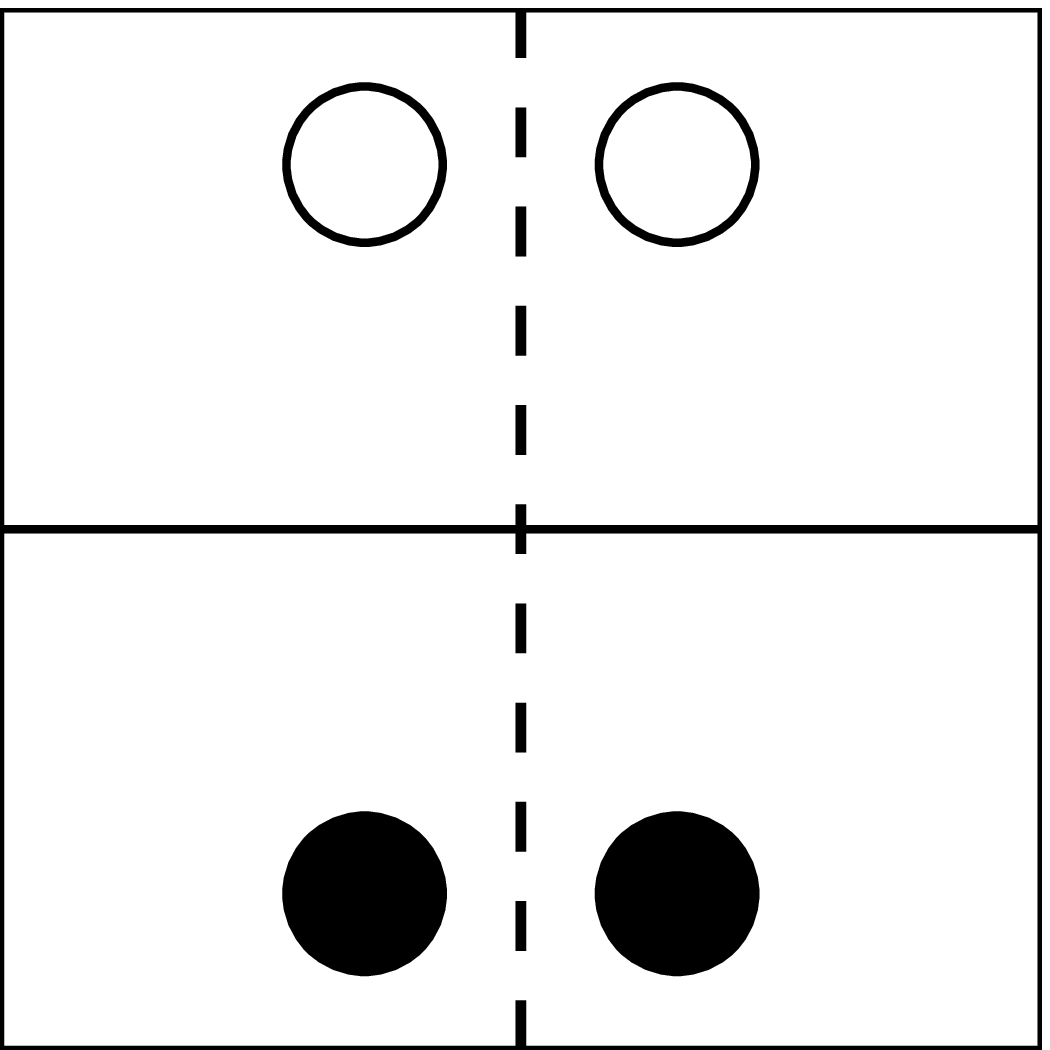} \\ [.5in]
               \includegraphics{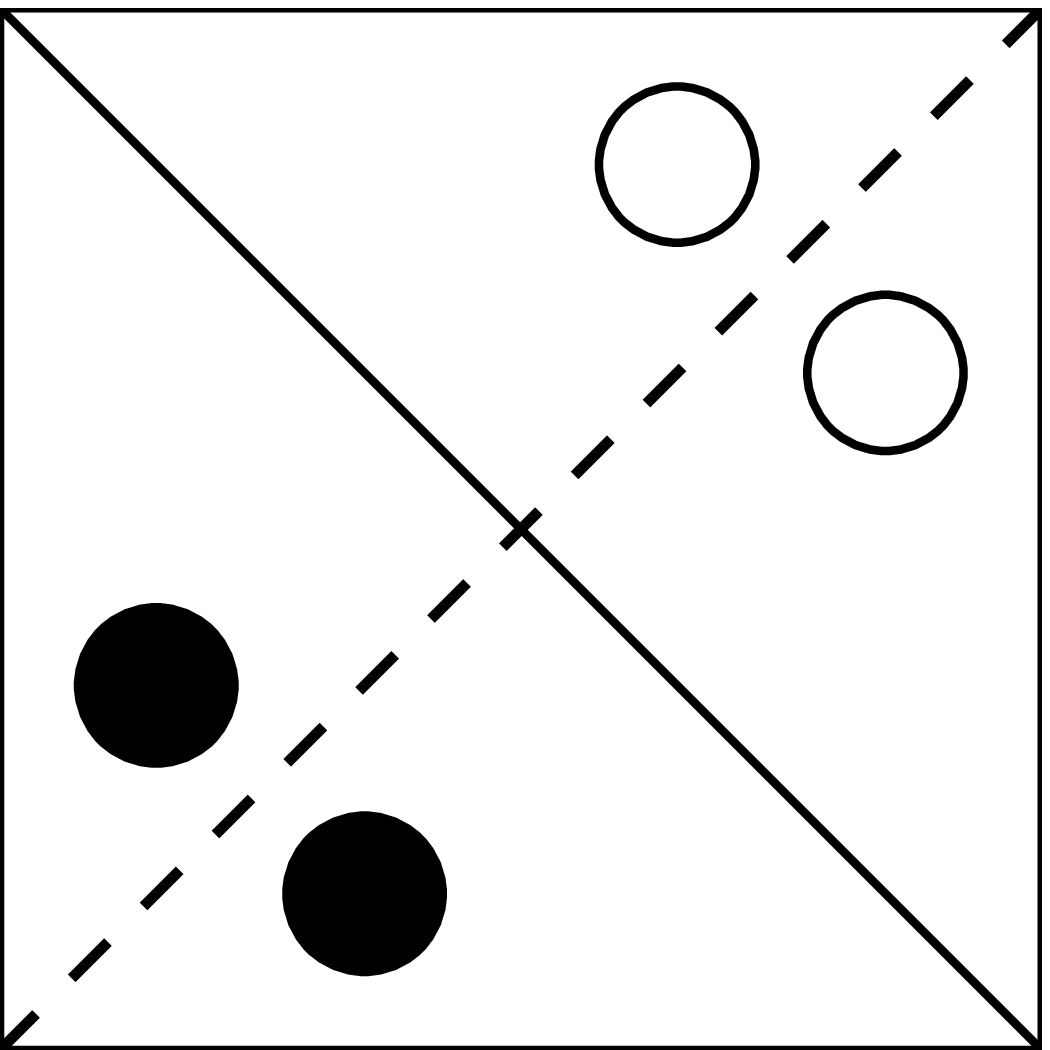}\\
               \end{tabular}
                } }="v5"
"O"
[r(0)] [u(\yc)]{
               \scalebox{\psize}{
               \begin{tabular}{ccc}
               \includegraphics{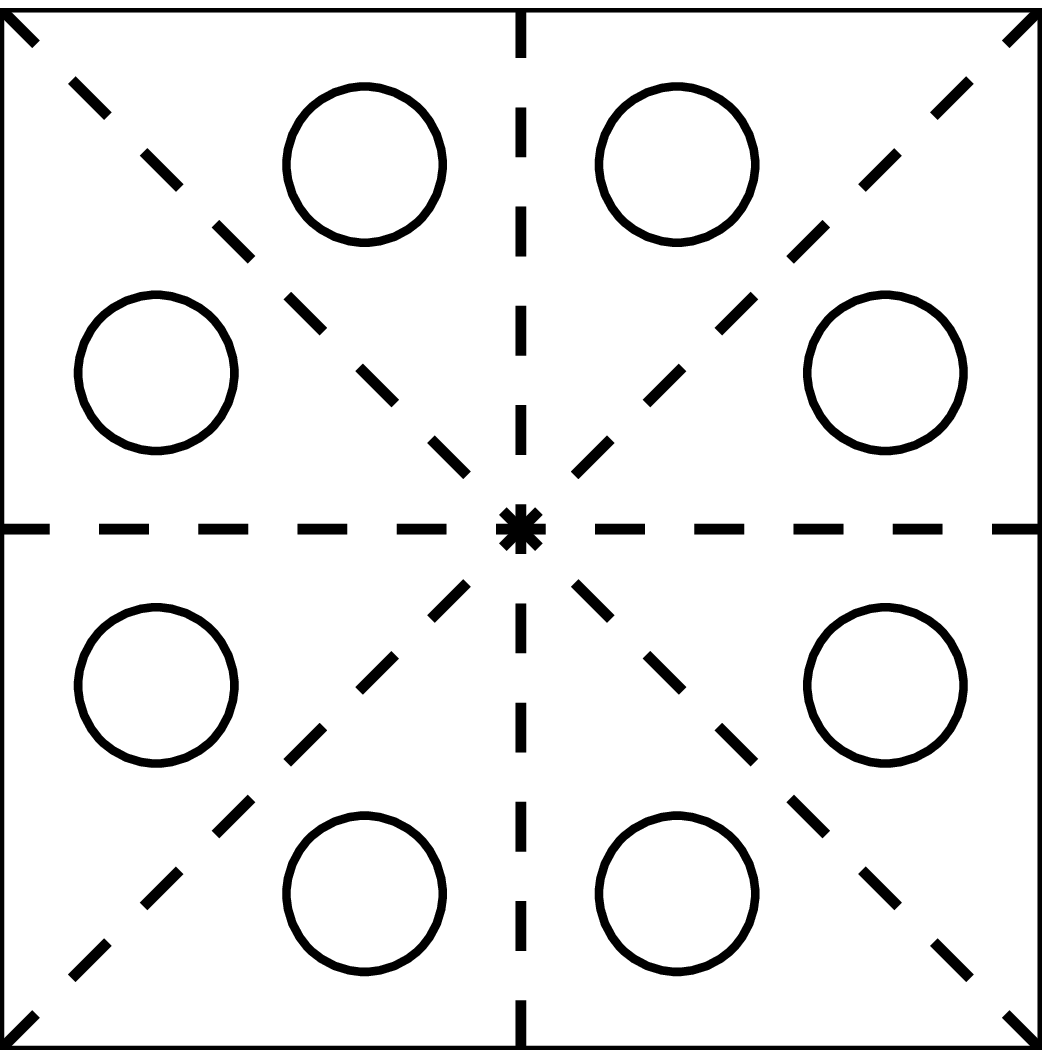}&  \hspace{0.3in} &
               \includegraphics{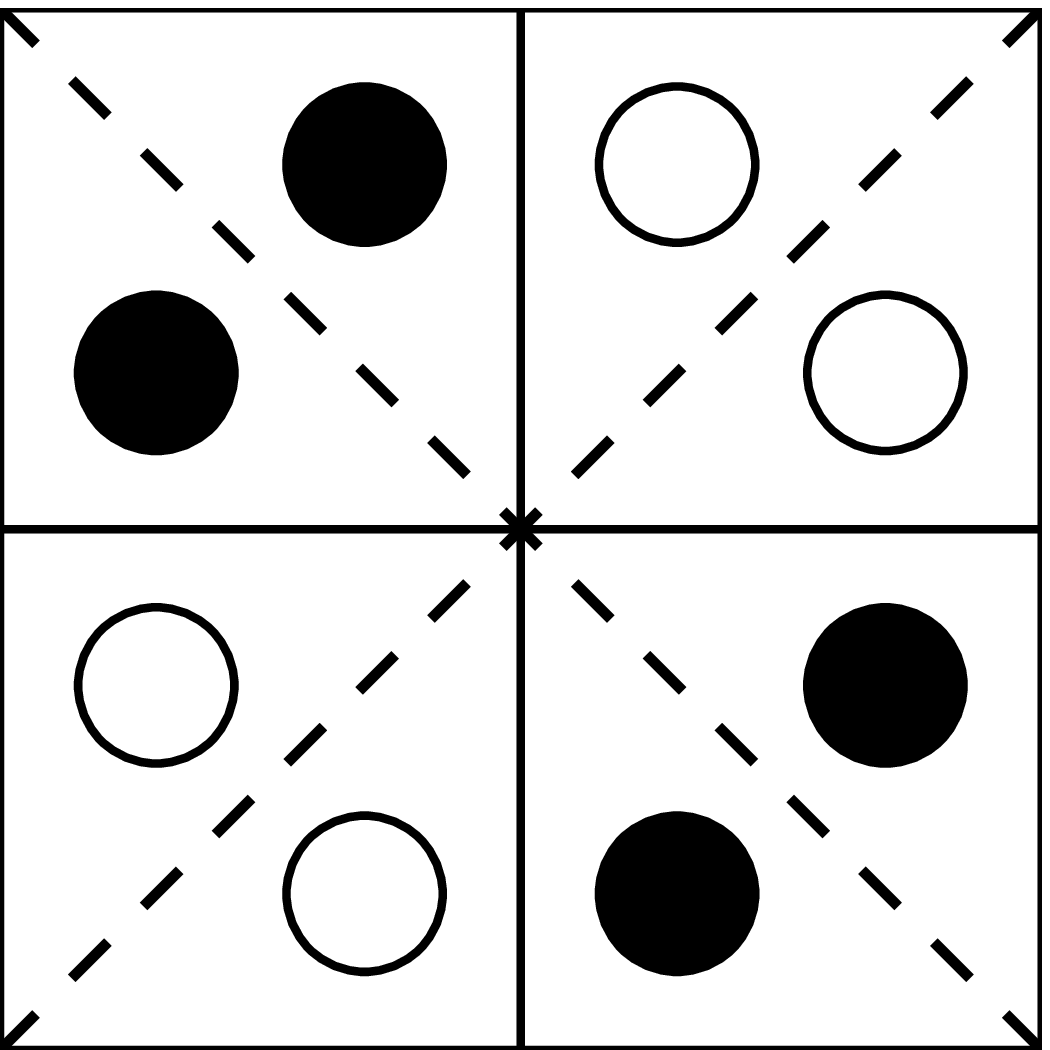}\\ [.5in]
               \includegraphics{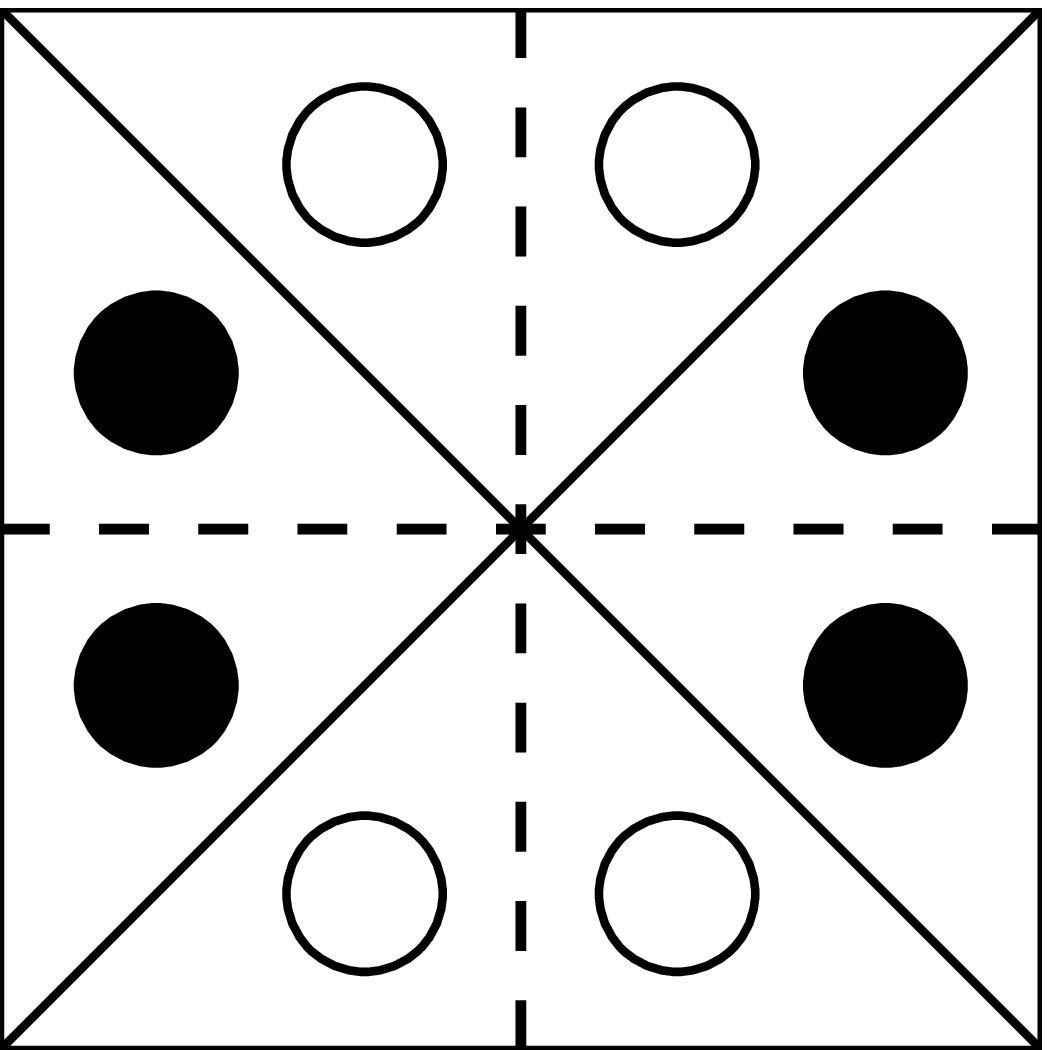}&  \hspace{0.3in} &
               \includegraphics{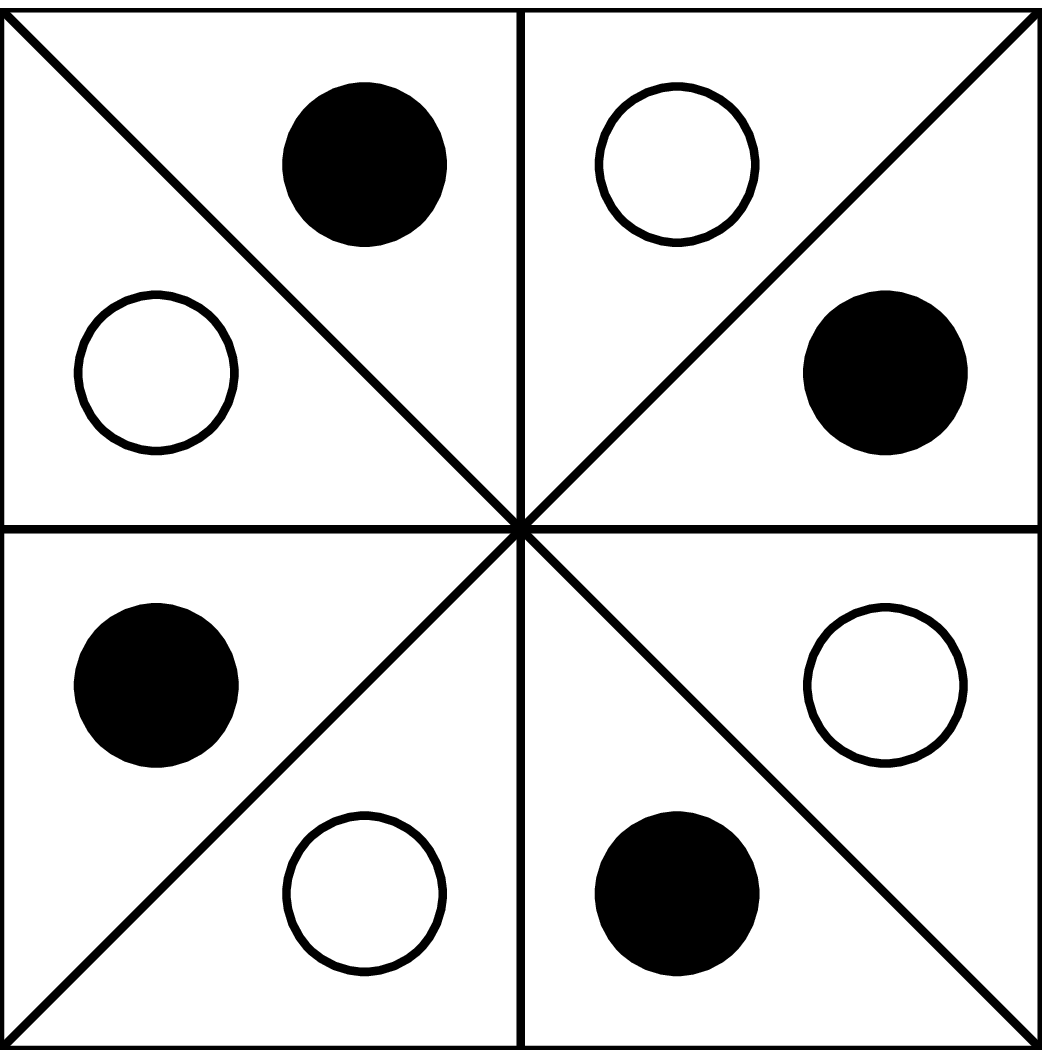}\\
               \end{tabular}
                } }="v1"
"O"
[r(0)] [u(\yd)]{
               \scalebox{\psize}{
               \includegraphics{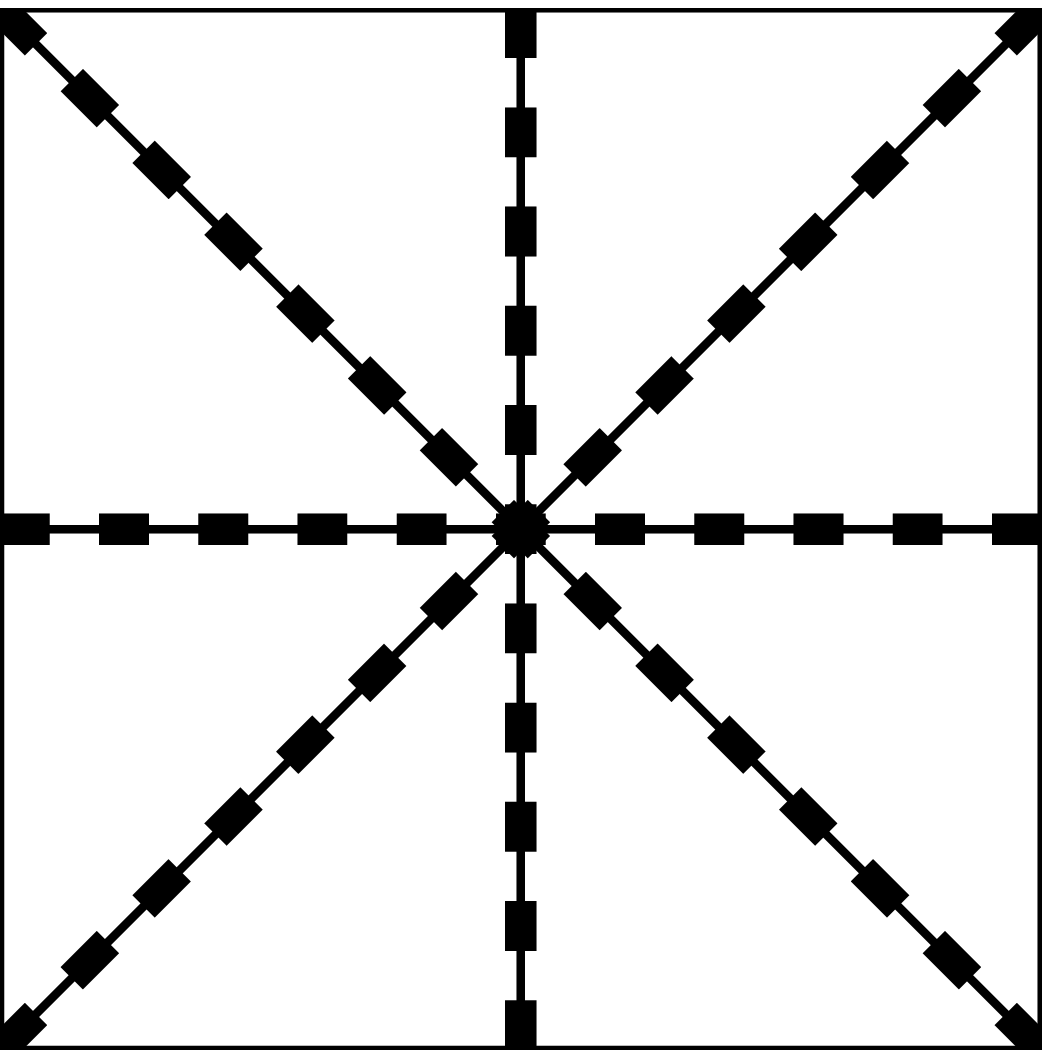}
                } }="v0"
"v0":_(.4){\Z_2}^<>(0.2)4"v1"
"v0":^(.5){\D_4}@/^1cm/^<>(0.05)2"v5"
"v1":_(0.55){\Z_2}^<>(.9)2"v6"
"v1":_(.5){\Z_2}^<>(0.1)2^<>(0.8)2"v7"
"v1":^(.5){\D_4}@/^3cm/^(.2)2_(.8)2"v13"
"v5":^(.45){\Z_2}_(.175)2"v13"
"v5":^(.55){\Z_2}_(.75)2"v15"
"v6":_(.6){\Z_2}^(.75)2"v17"
"v6":@{.>}_{\Z_4}@/_3cm/_<>(0.95)2"v19"
"v7":_(.5){\Z_2}"v13"
"v7":_(.6){\Z_2}^(.75)4"v17"
"v13":_(.6){\Z_2}^<>(0.8)4"v19"
"v15":^{\Z_2}"v19"
"v17":_{\Z_2}"v19"
}
}
\begin{center}
\begin{tabular}{ccc}
\\
\hline
\\
$\quad$ \scalebox{.15}{\includegraphics{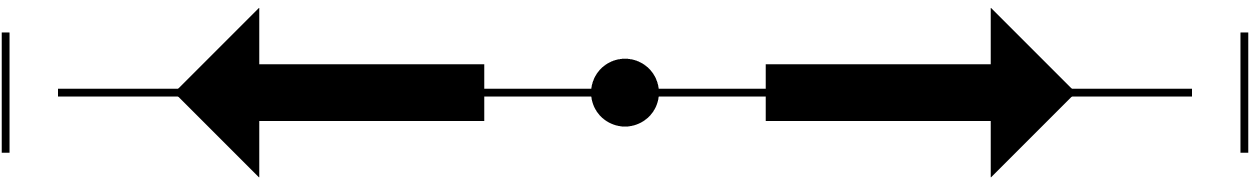}} $\quad$ &
$\quad$ \scalebox{.15}{\includegraphics{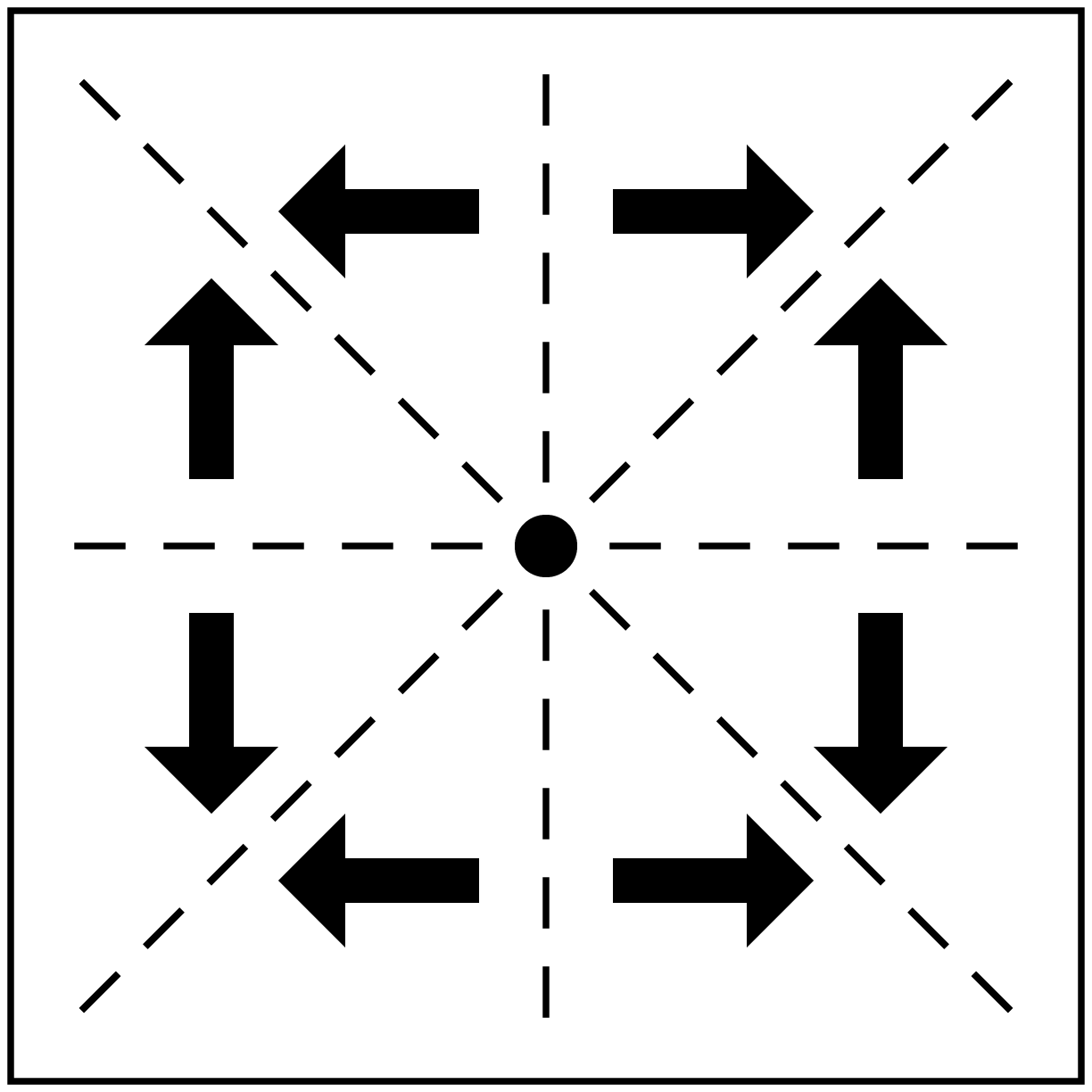}} $\quad$ &
$\quad$ \scalebox{.15}{\includegraphics{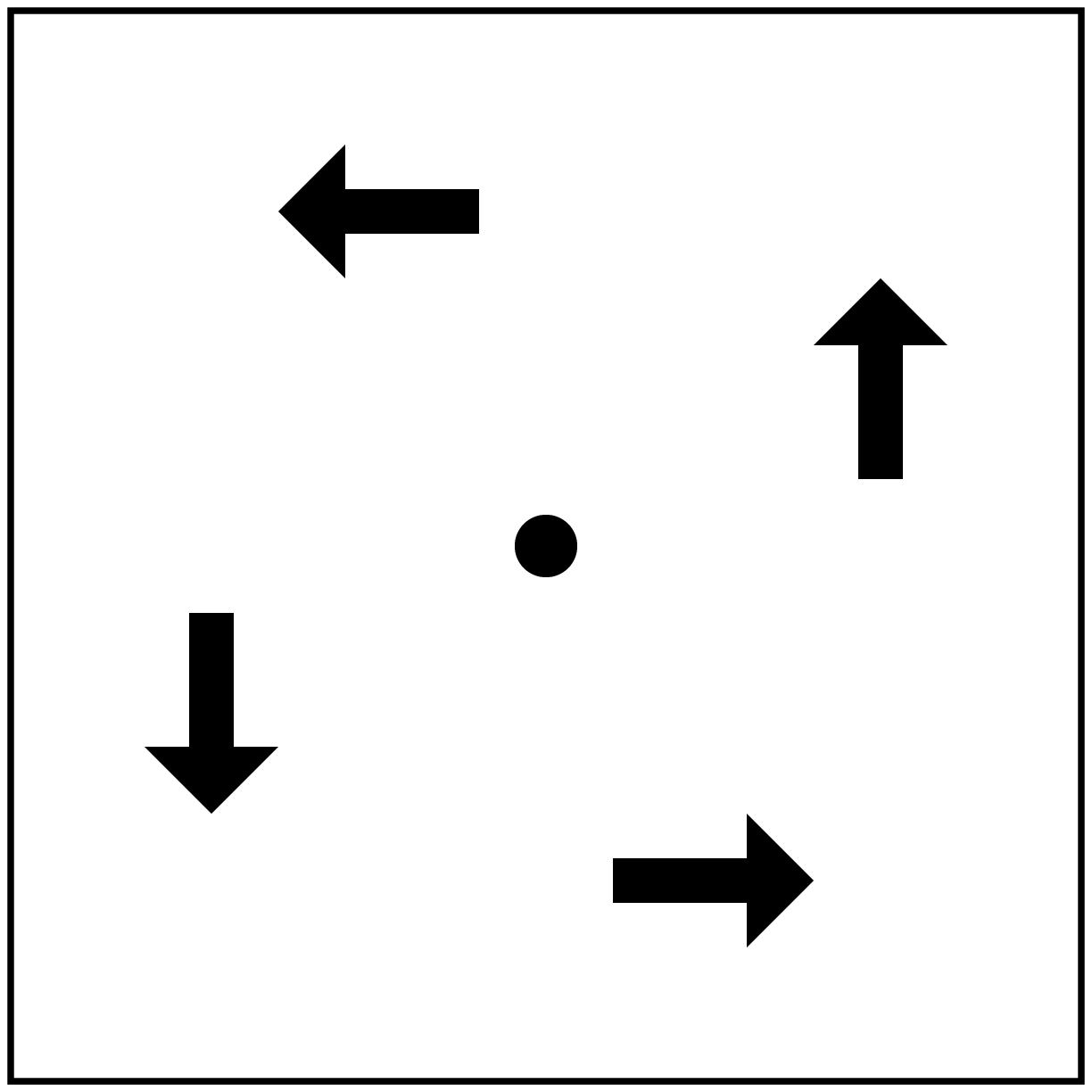}} $\quad$ \\
$\Z_2$&
$\D_4$&
$\Z_4$
\end{tabular}
\end{center}
\caption
{\label{D4_digraph}
Condensed bifurcation digraph for PDE on the square, and the irreducible spaces for the
generic bifurcations in the digraph. 
}

\end{figure}

\def\sbsize{0.2}
\def\sss{0.1}
\def\figheight{3.1cm}

\begin{figure}
\begin{center}
\begin{tabular}{cccc}

\vphantom{{\rule{0cm}{\figheight}}}
   {\scalebox{\sbsize}{\includegraphics{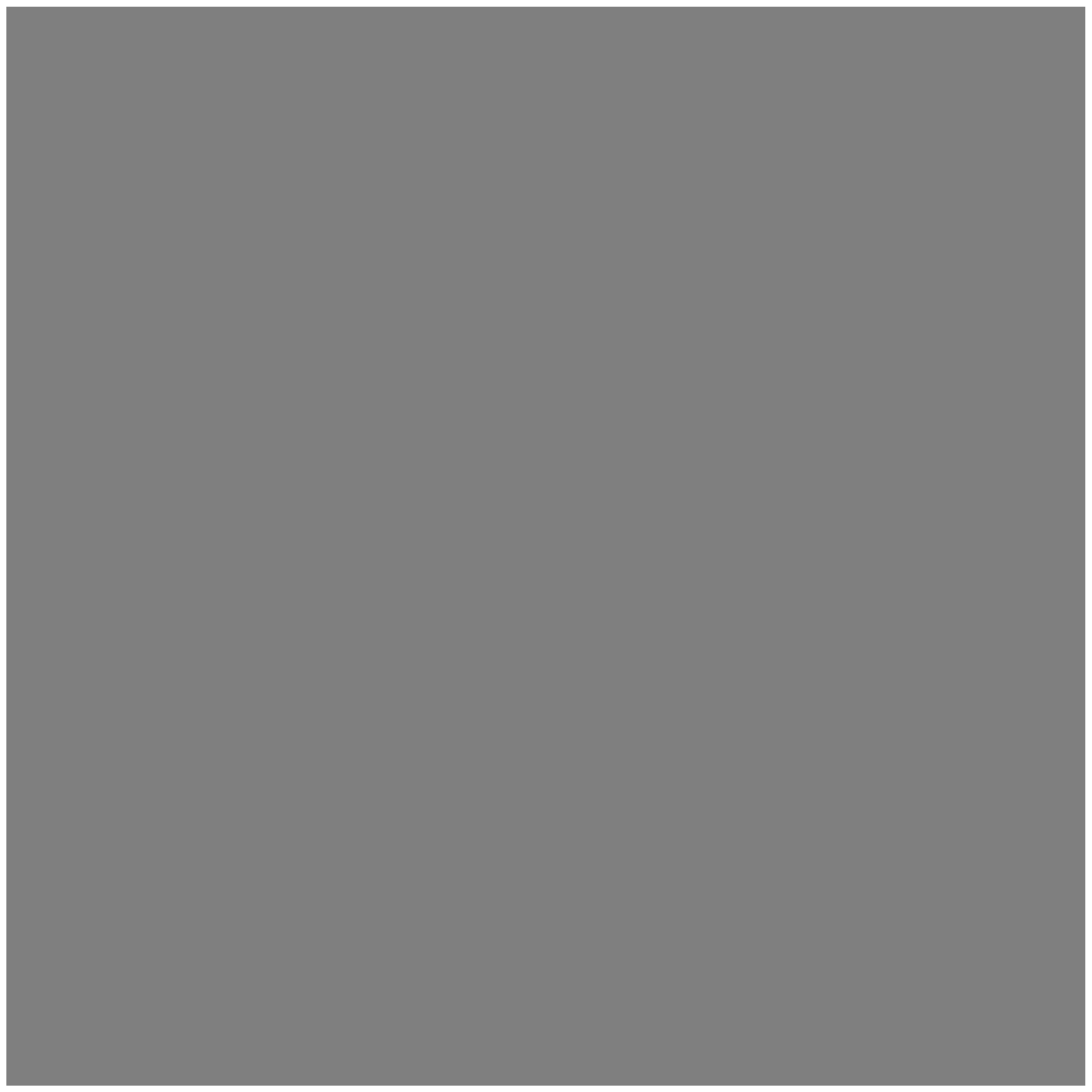}}}
&
   {\scalebox{\sbsize}{\includegraphics{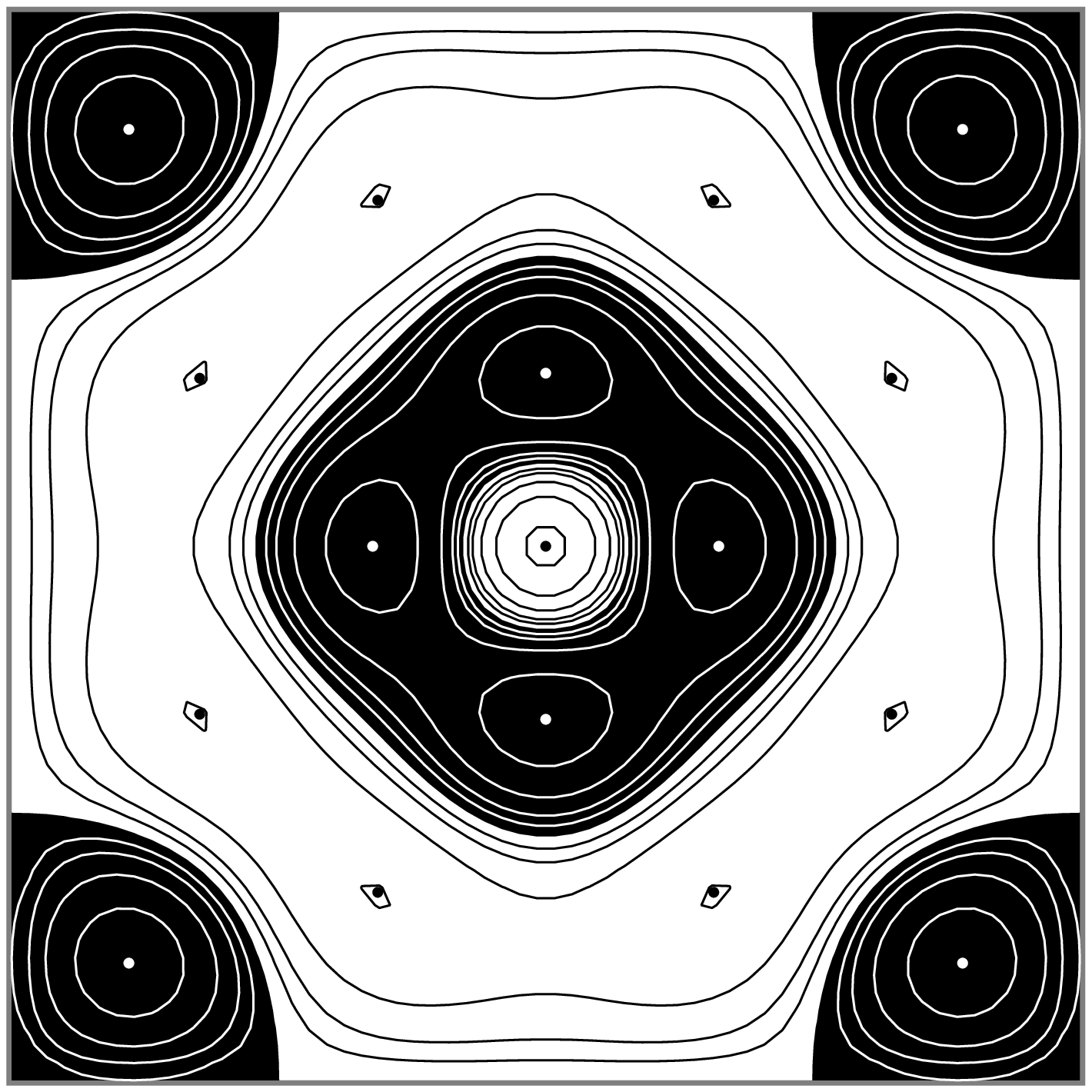}}}
&
   {\scalebox{\sbsize}{\includegraphics{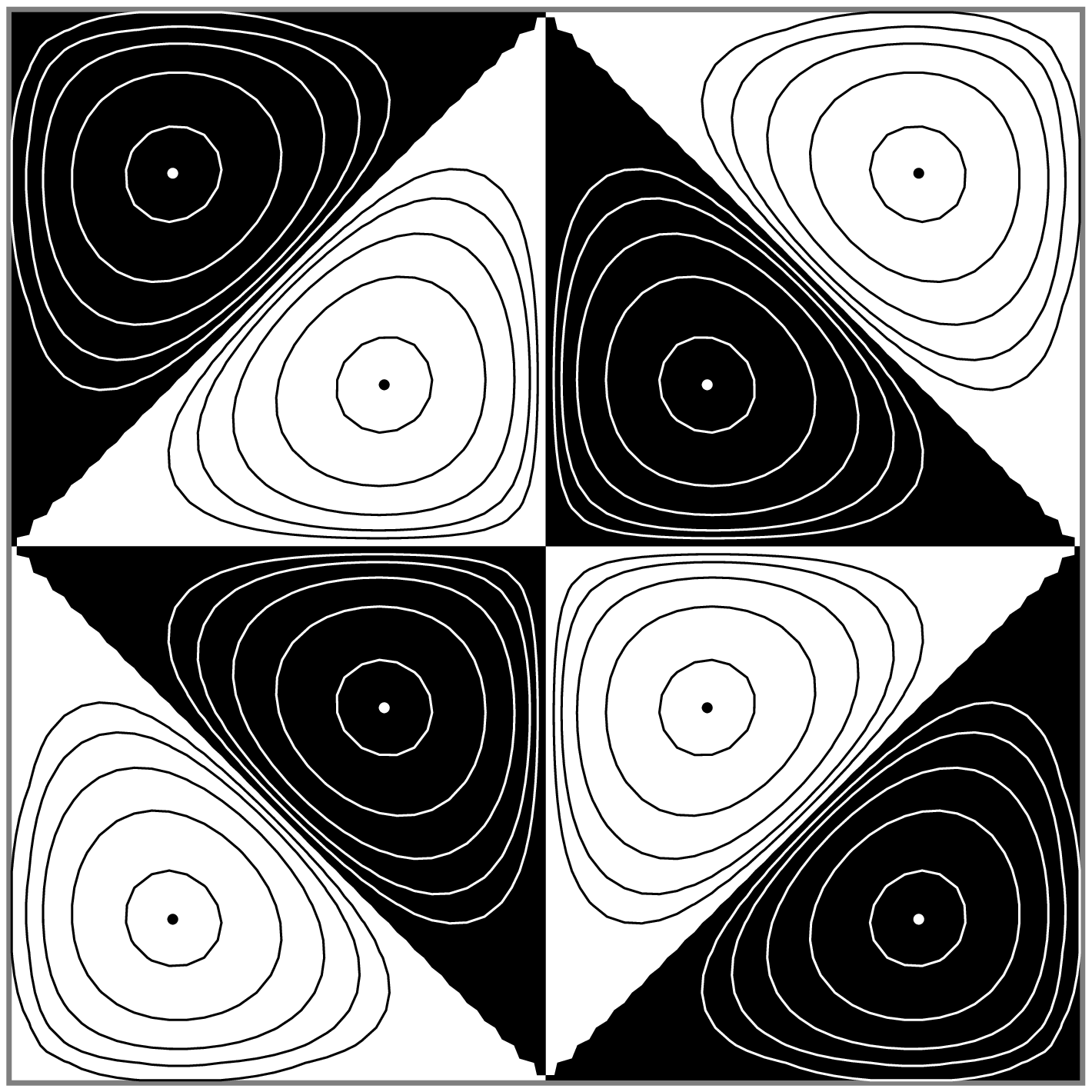}}}
&
   {\scalebox{\sbsize}{\includegraphics{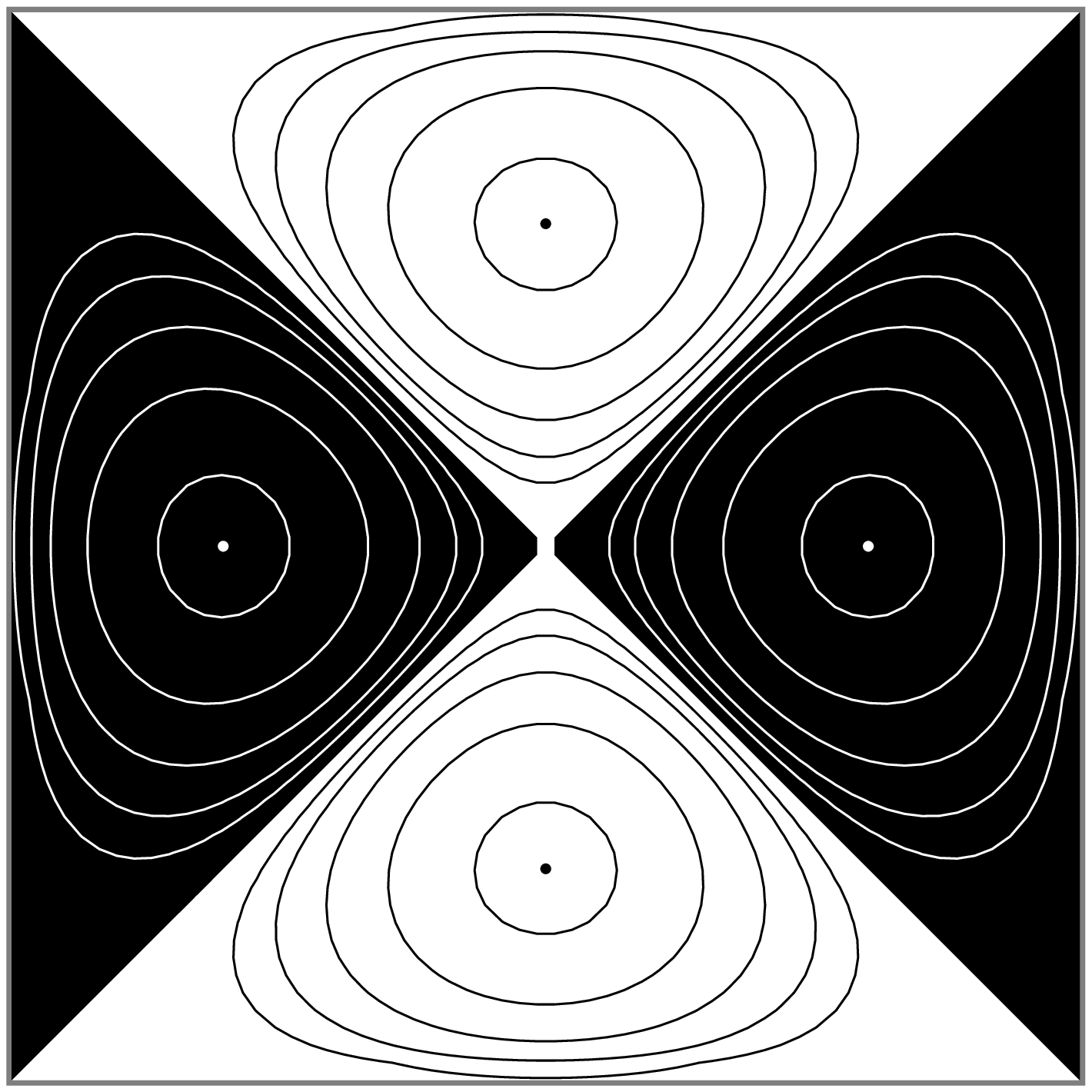}}}
 \\
%
\scalebox{\sss}{\includegraphics{figures/s0.eps}} &
\scalebox{\sss}{\includegraphics{figures/s4.eps}} &
\scalebox{\sss}{\includegraphics{figures/s3.eps}} &
\scalebox{\sss}{\includegraphics{figures/s2.eps}}
\\[3pt]

\hline

\vphantom{{\rule{0cm}{\figheight}}}
   {\scalebox{\sbsize}{\includegraphics{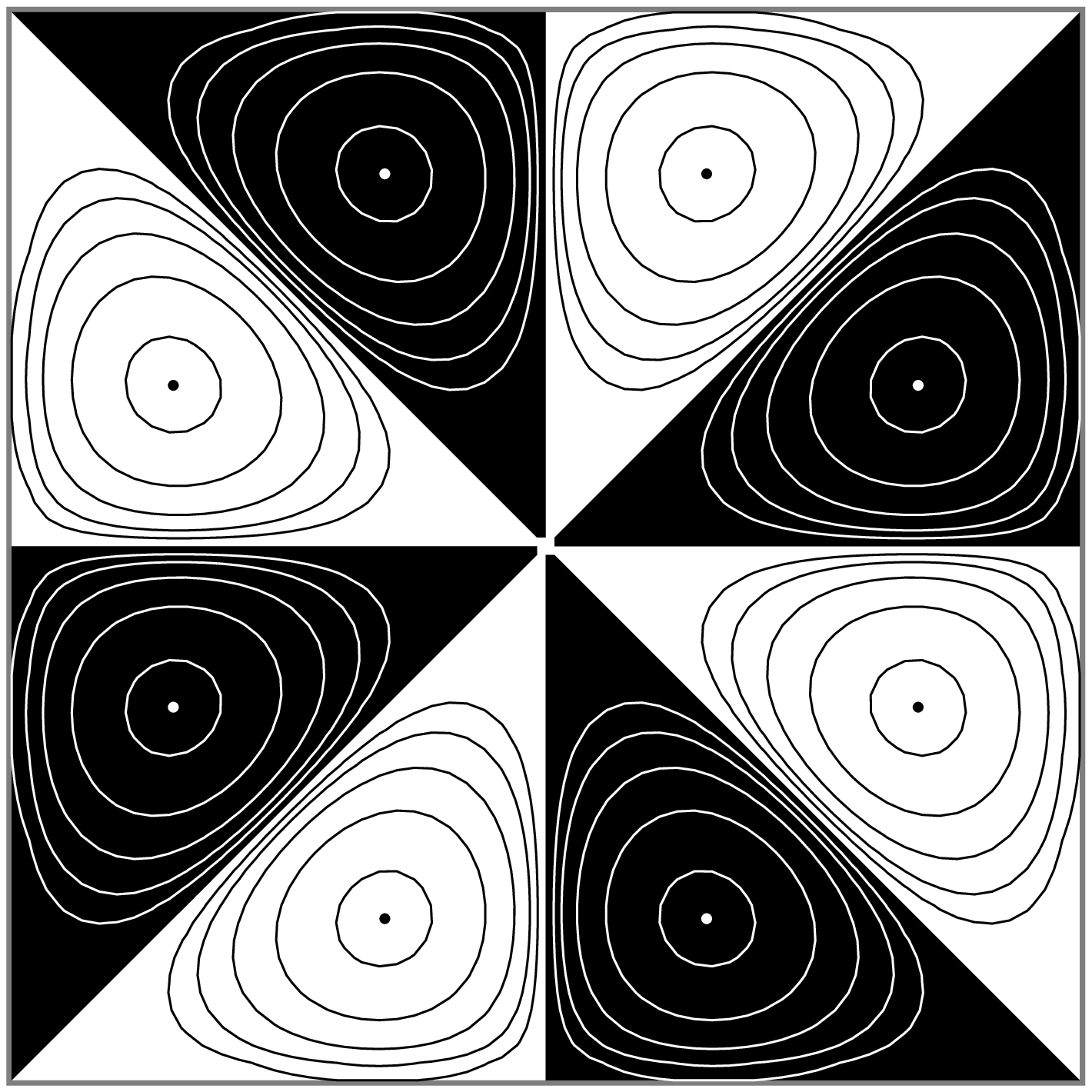}}}
&
   {\scalebox{\sbsize}{\includegraphics{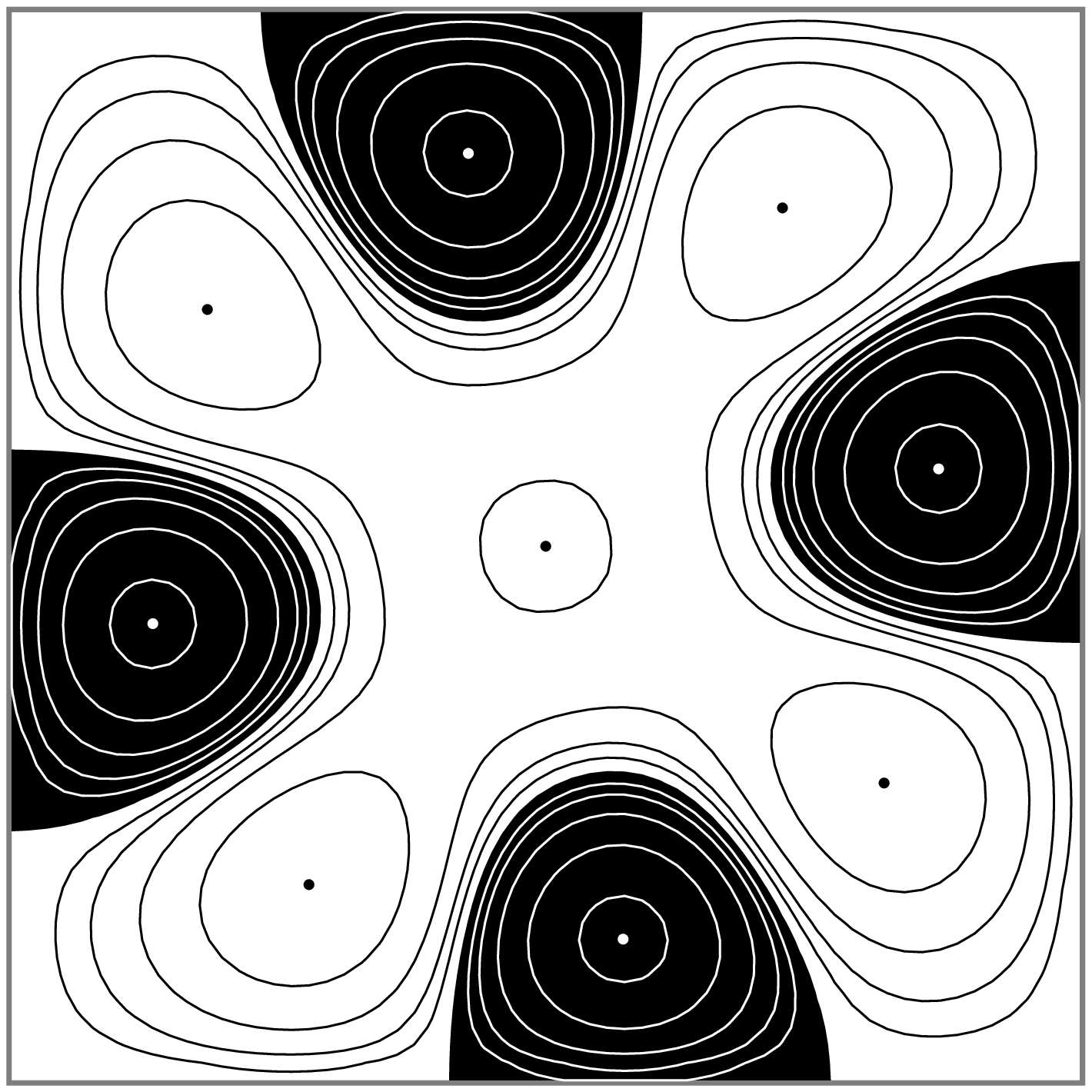}}}
&
   {\scalebox{\sbsize}{\includegraphics{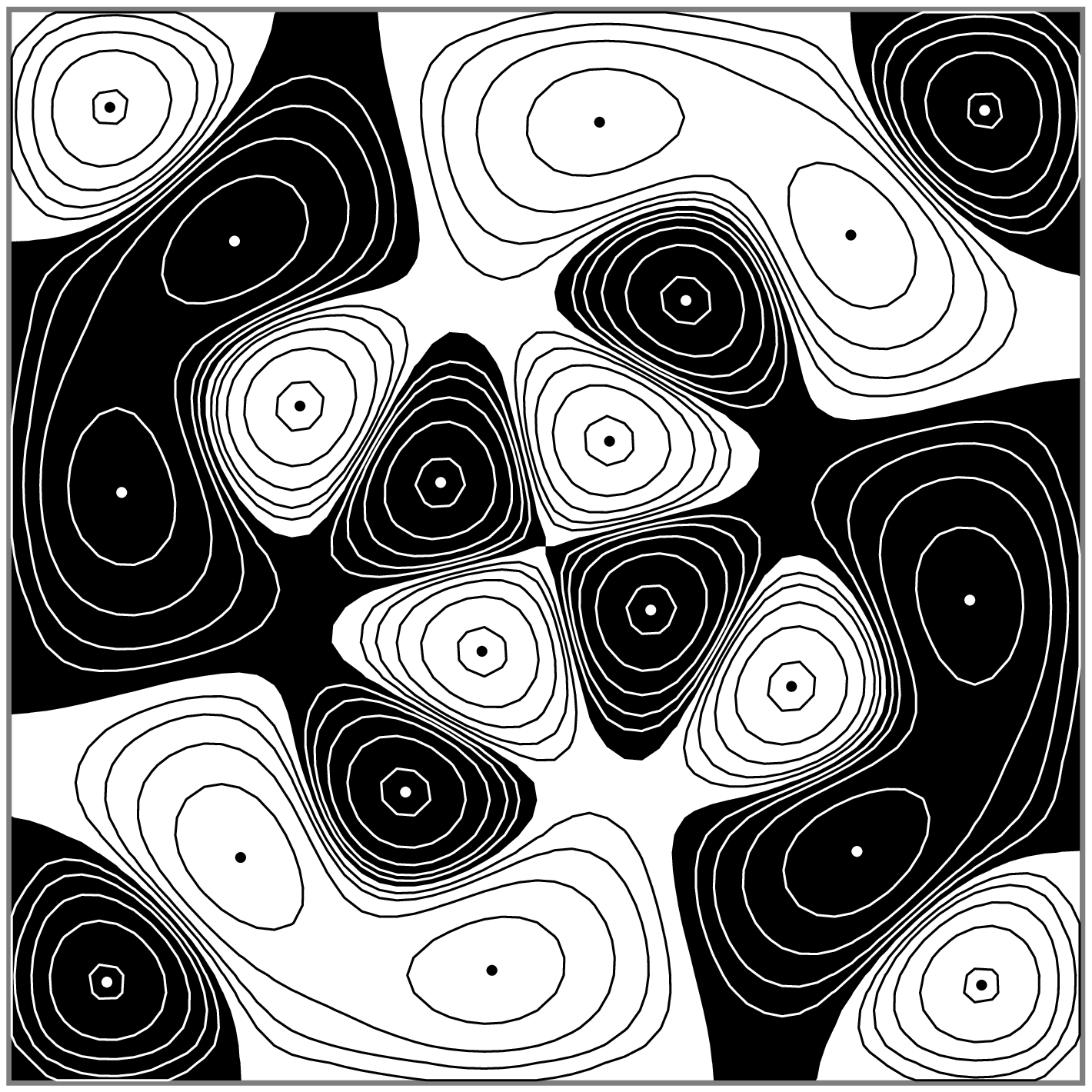}}}
&
   {\scalebox{\sbsize}{\includegraphics{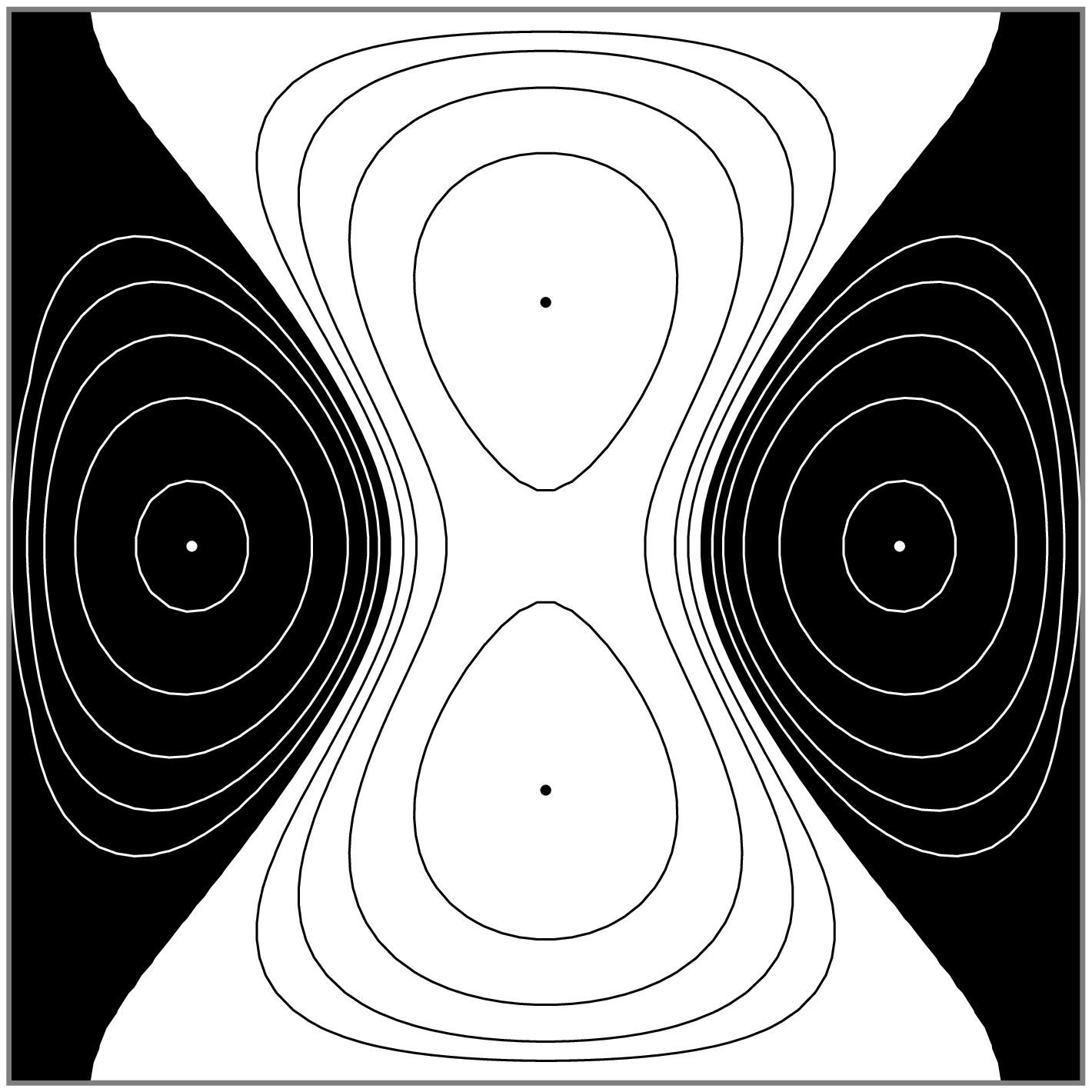}}}
 \\
\scalebox{\sss}{\includegraphics{figures/s1.eps}}&
\scalebox{\sss}{\includegraphics{figures/s9.eps}}&
\scalebox{\sss}{\includegraphics{figures/s6.eps}}&
\scalebox{\sss}{\includegraphics{figures/s10.eps}}
\\[3pt]

\hline
\vphantom{{\rule{0cm}{\figheight}}}
   {\scalebox{\sbsize}{\includegraphics{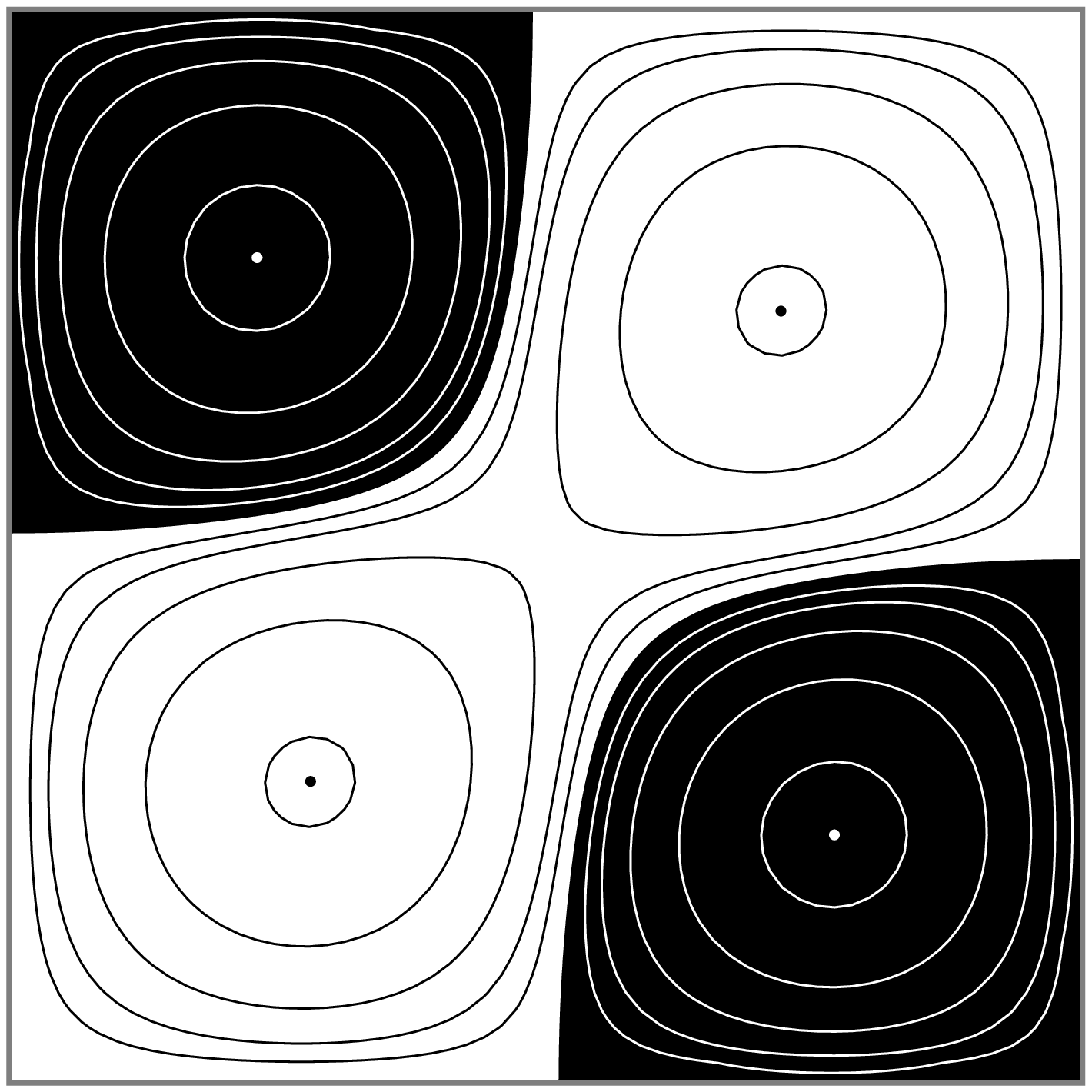}}}
&
   {\scalebox{\sbsize}{\includegraphics{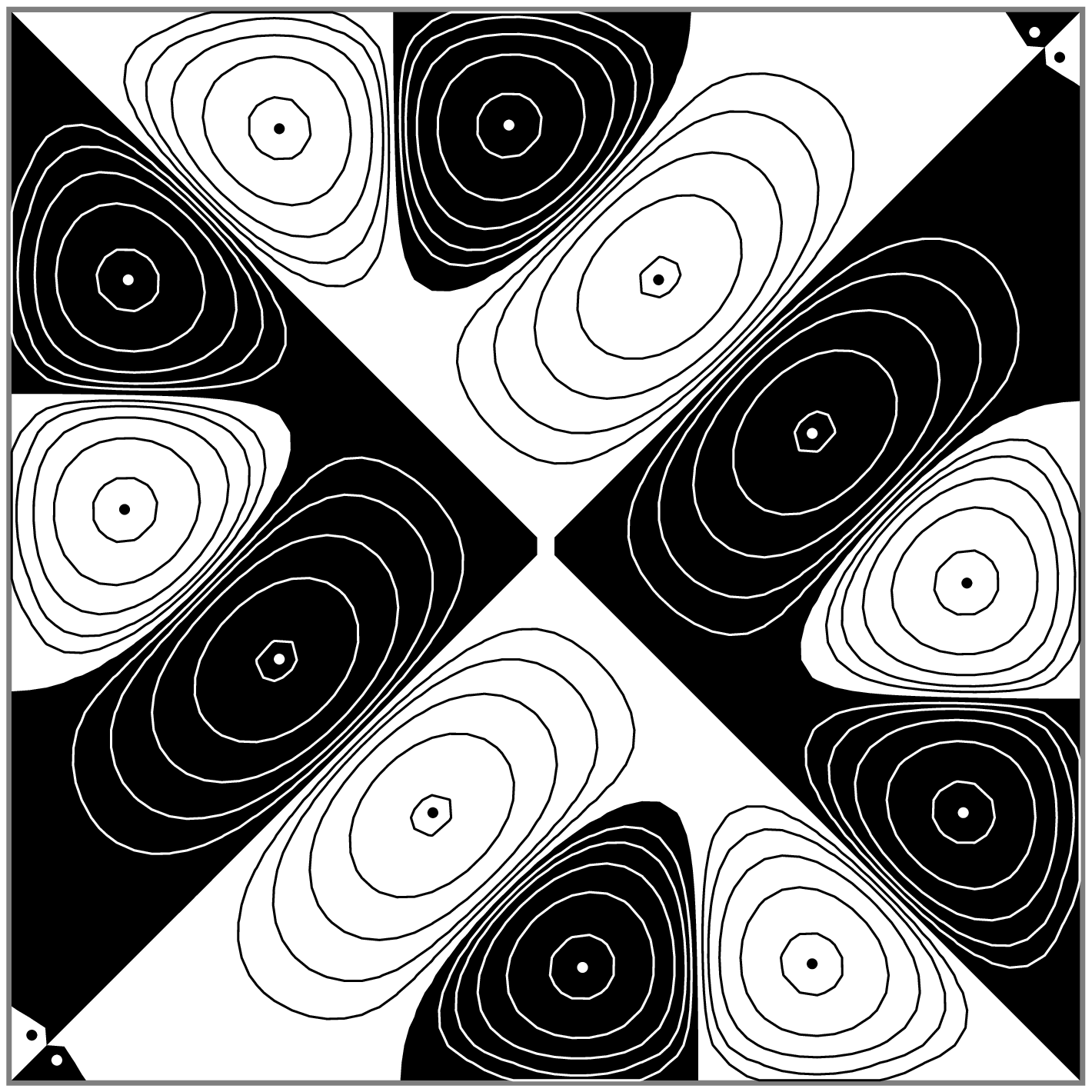}}}
&
   {\scalebox{\sbsize}{\includegraphics{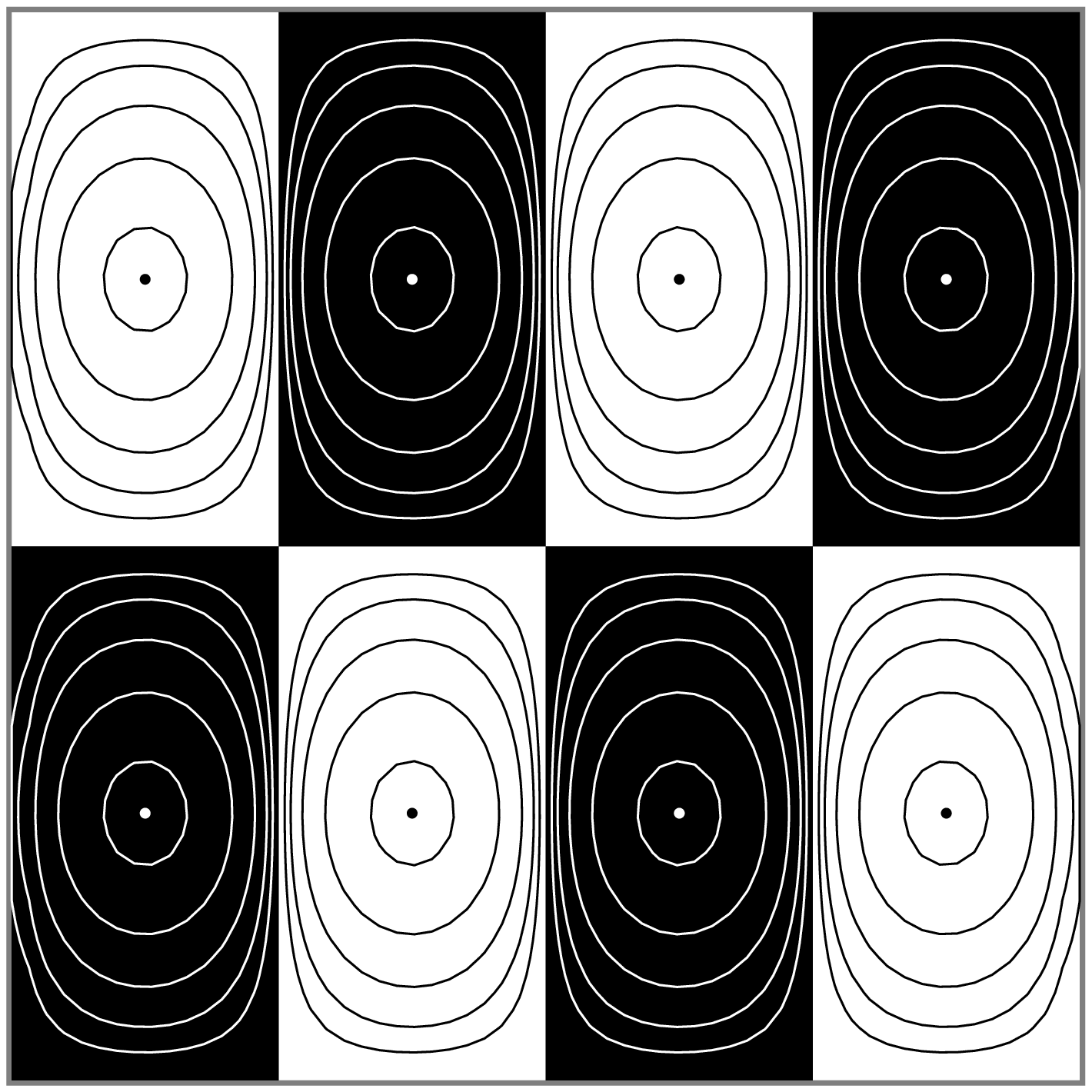}}}
&
   {\scalebox{\sbsize}{\includegraphics{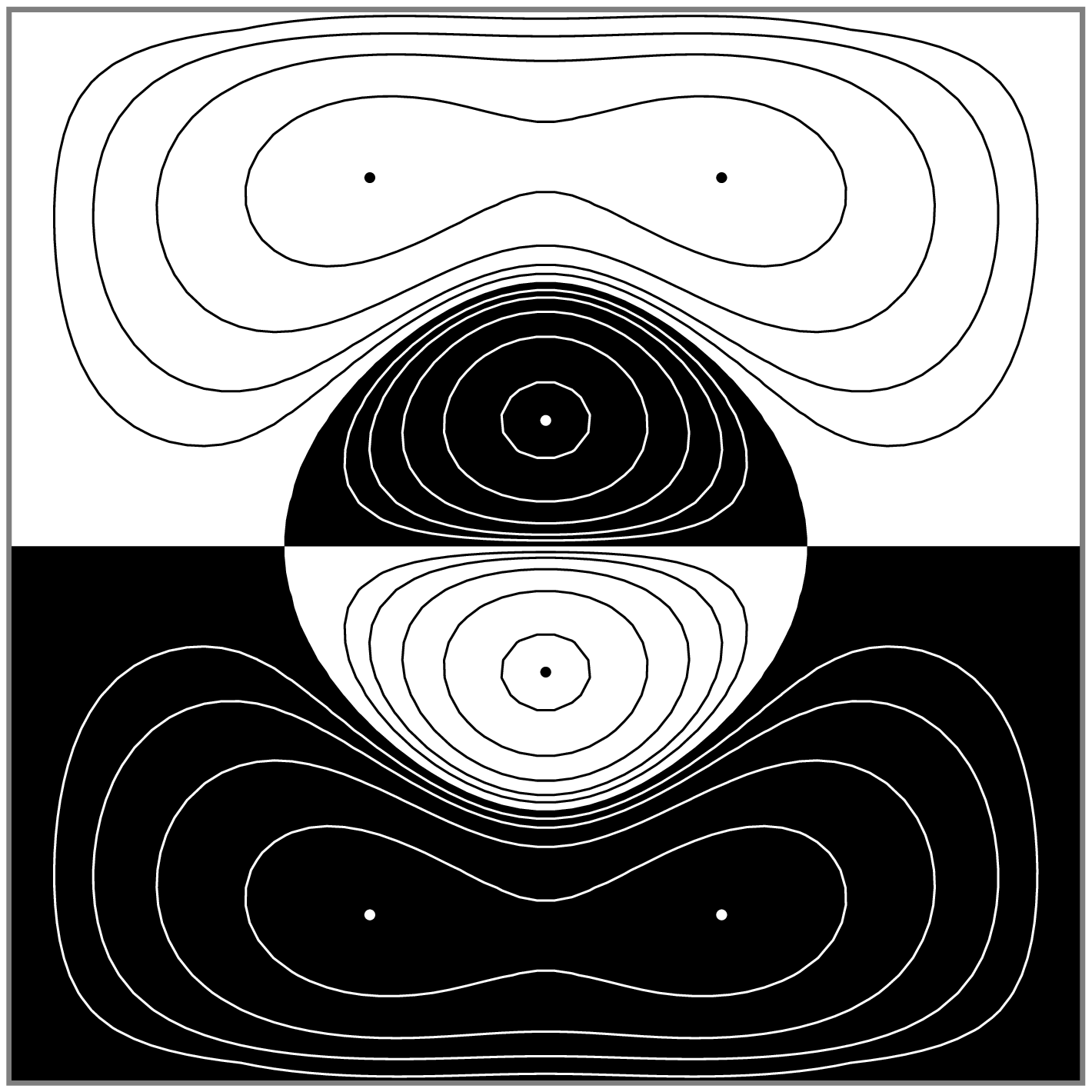}}}
 \\
\scalebox{\sss}{\includegraphics{figures/s12.eps}} &
\scalebox{\sss}{\includegraphics{figures/s11.eps}} &
\scalebox{\sss}{\includegraphics{figures/s7.eps}} &
\scalebox{\sss}{\includegraphics{figures/s5.eps}}
\\[3pt]

\hline

\vphantom{{\rule{0cm}{\figheight}}}
    {\scalebox{\sbsize}{\includegraphics{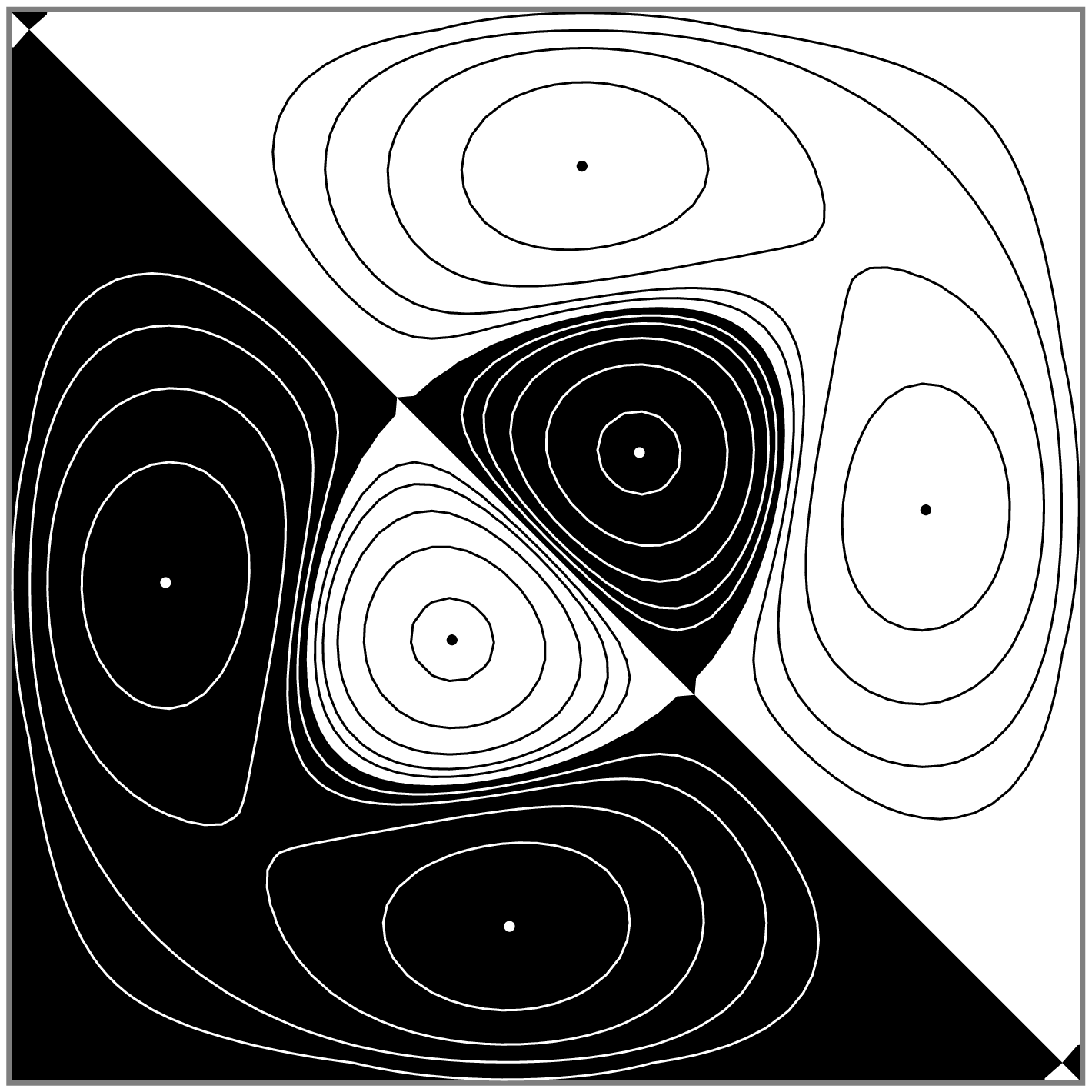}}}
&
   {\scalebox{\sbsize}{\includegraphics{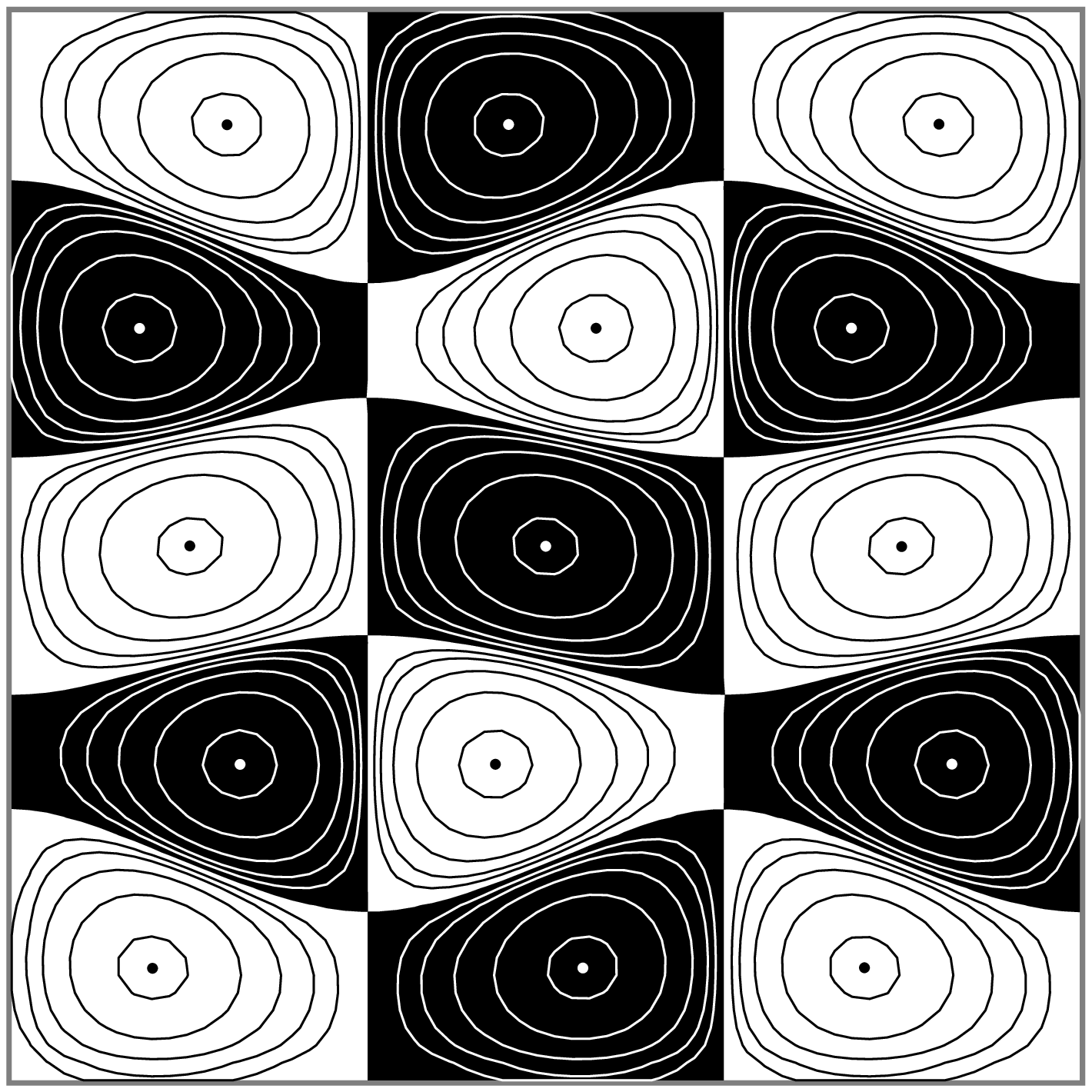}}}
&
   {\scalebox{\sbsize}{\includegraphics{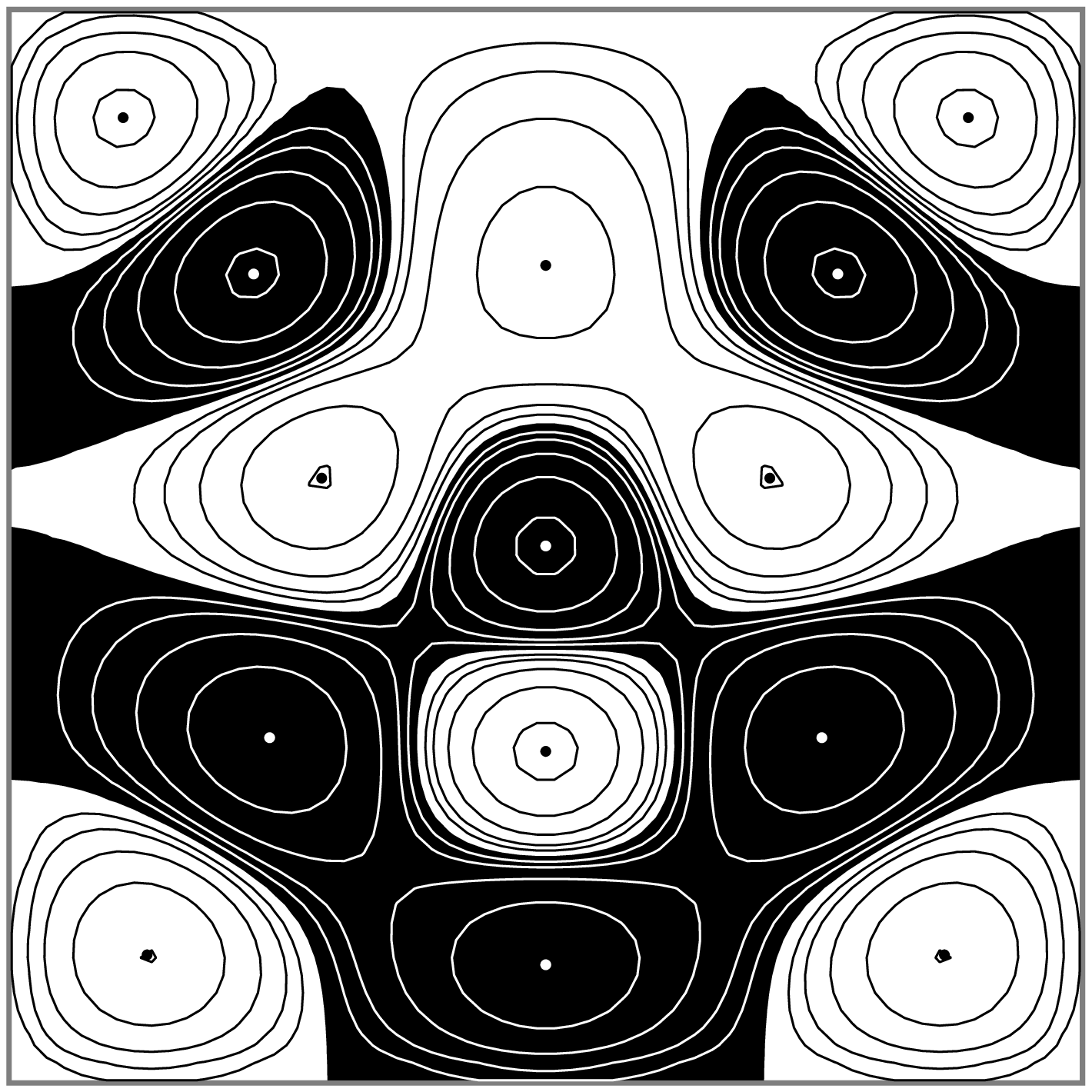}}}
&
   {\scalebox{\sbsize}{\includegraphics{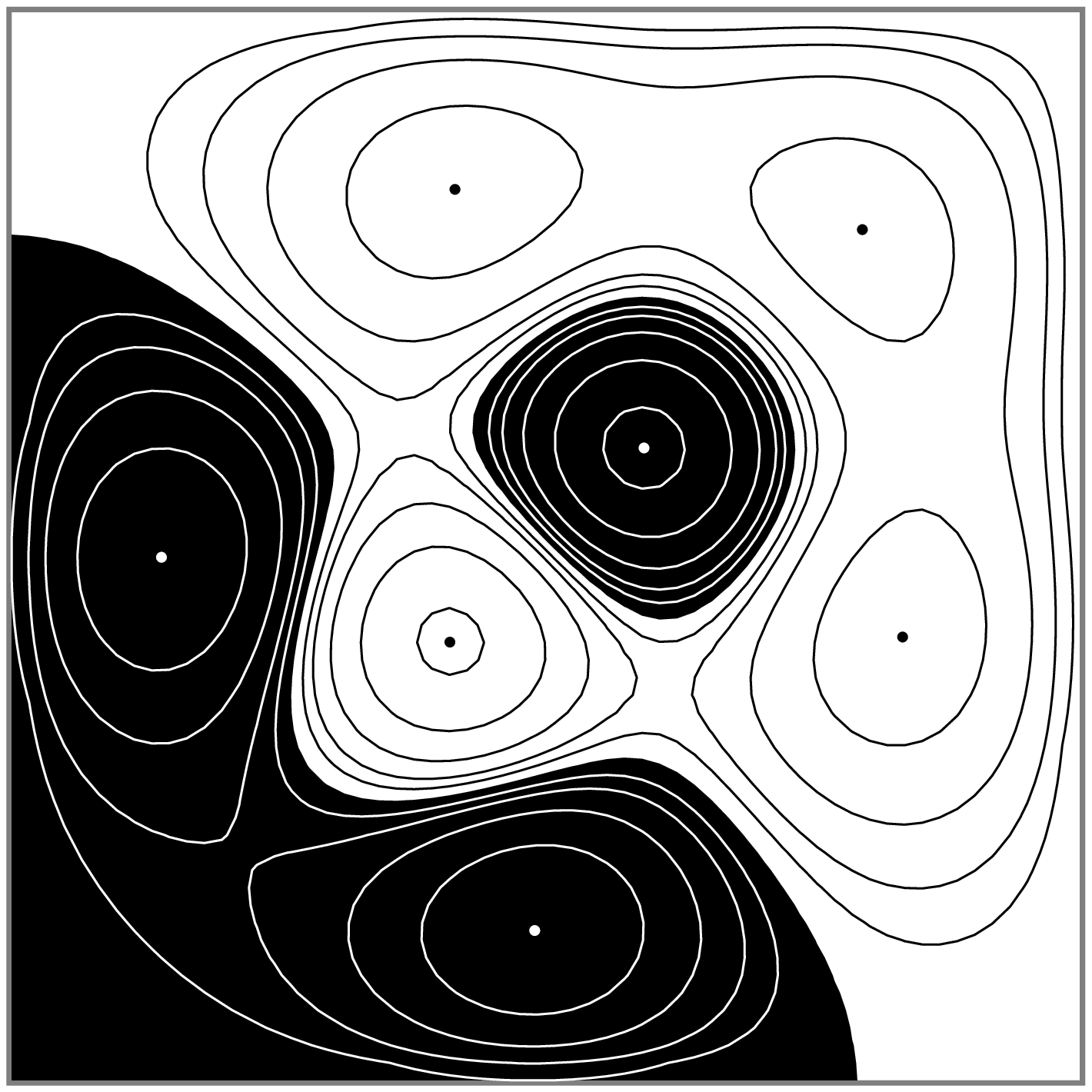}}}
 \\
\scalebox{\sss}{\includegraphics{figures/s8.eps}}  &
\scalebox{\sss}{\includegraphics{figures/s17.eps}} &
\scalebox{\sss}{\includegraphics{figures/s14.eps}} &
\scalebox{\sss}{\includegraphics{figures/s18.eps}}
\\[3pt]

\hline

\vphantom{{\rule{0cm}{\figheight}}}
   {\scalebox{\sbsize}{\includegraphics{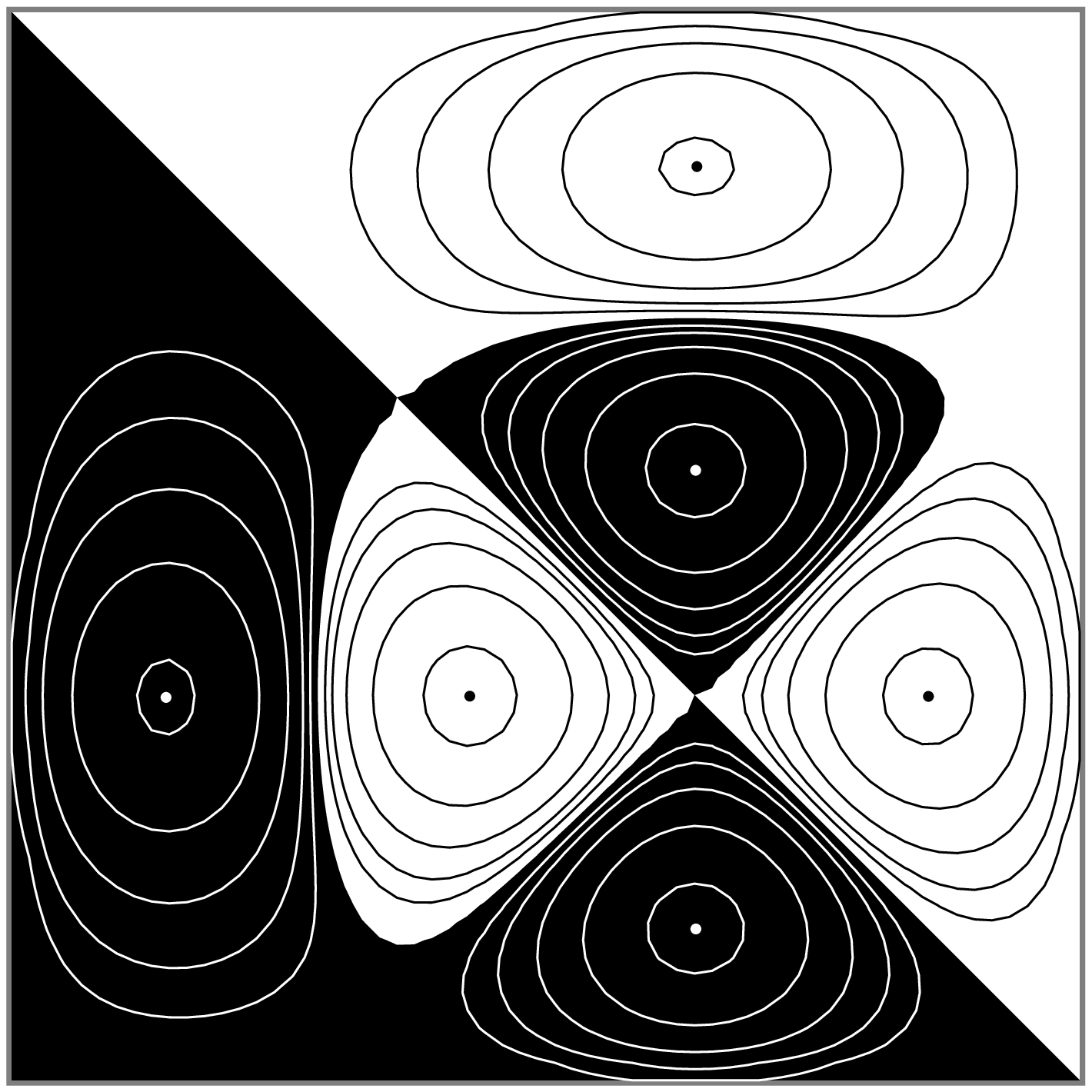}}}
&
   {\scalebox{\sbsize}{\includegraphics{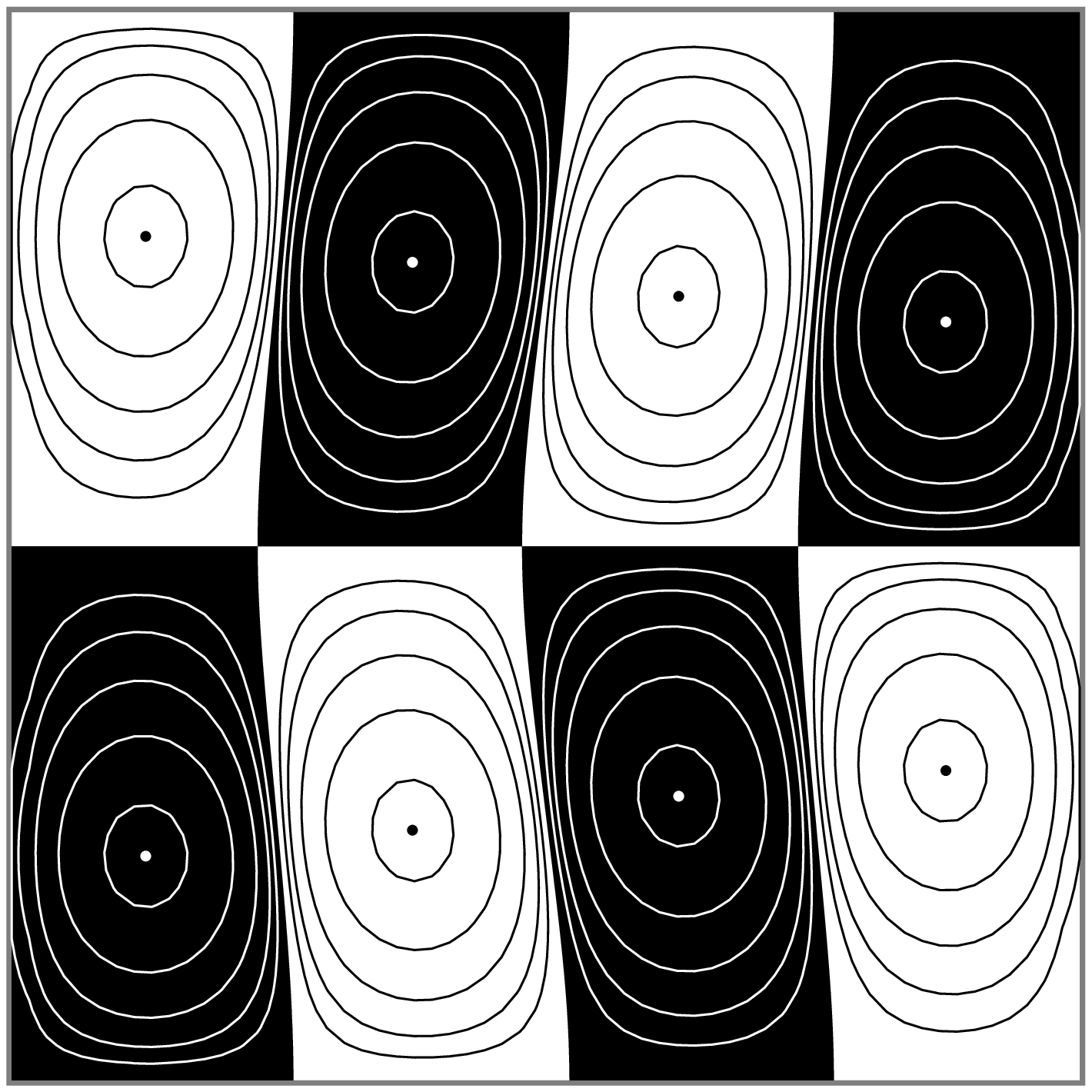}}}
&
   {\scalebox{\sbsize}{\includegraphics{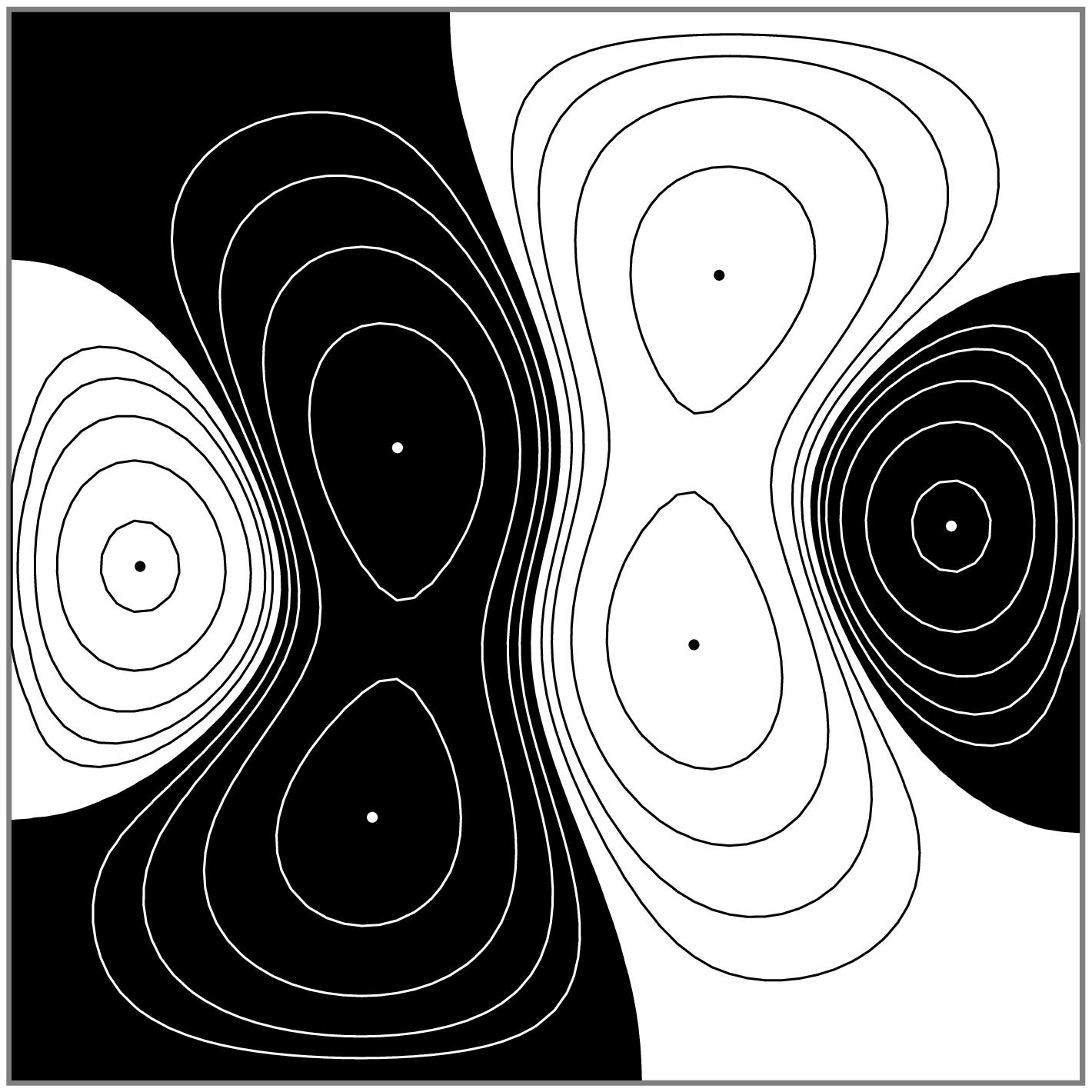}}}
&
   {\scalebox{\sbsize}{\includegraphics{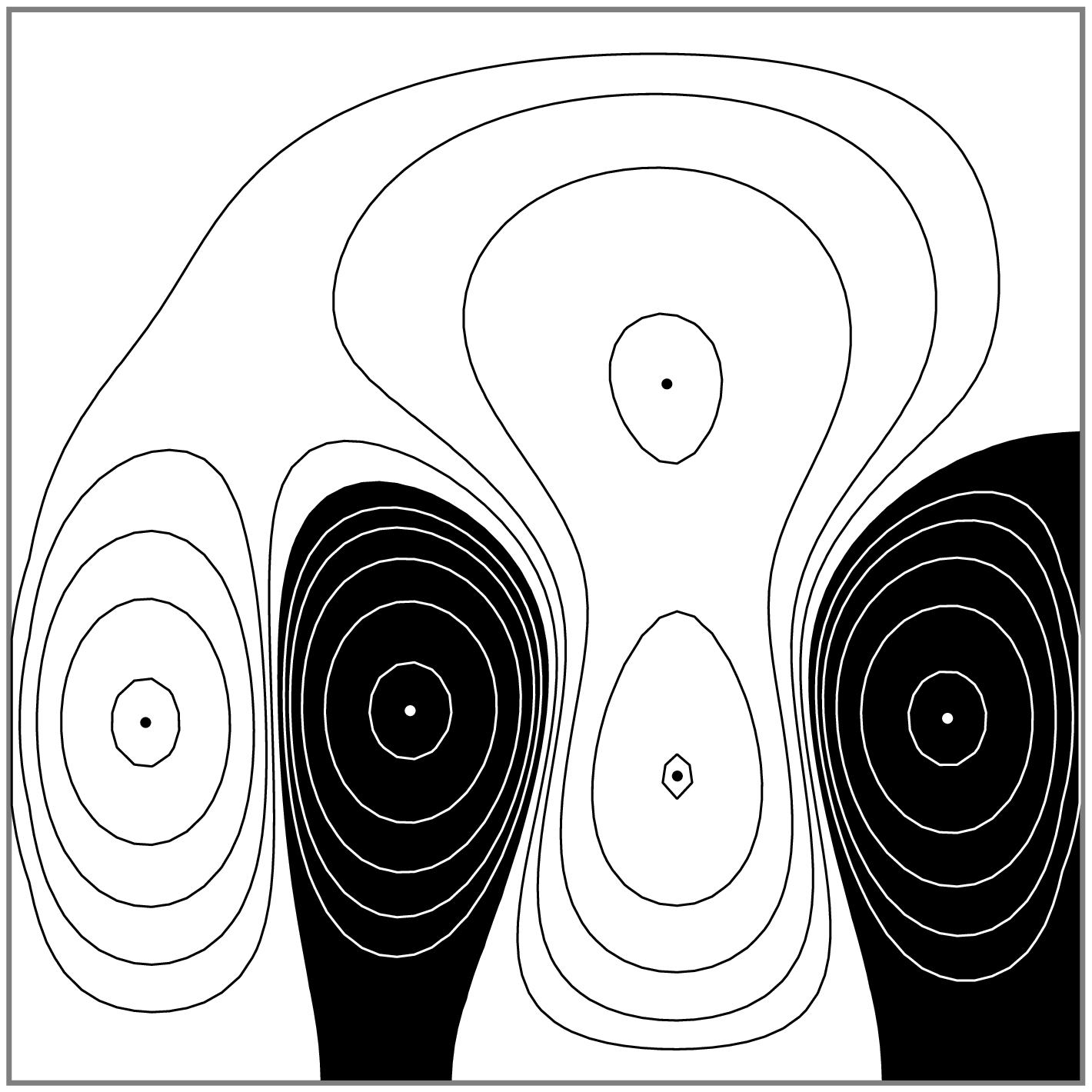}}}
 \\
\scalebox{\sss}{\includegraphics{figures/s16.eps}} &
\scalebox{\sss}{\includegraphics{figures/s13.eps}} &
\scalebox{\sss}{\includegraphics{figures/s15.eps}} &
\scalebox{\sss}{\includegraphics{figures/s19.eps}}
\\
%
\end{tabular}
\caption{\label{square_plots}
Contour plots of solutions on the square of each symmetry type, with schematic diagrams
(see Figure~\ref{D4_digraph}).
}
\end{center}
\end{figure}

Figure~\ref{D4_digraph} contains the condensed bifurcation digraph  for PDE~(\ref{pde}) 
on a region with $\D_4$ symmetry.
This digraph,
along with the files required by our GNGA code for following bifurcations,
was automatically generated by our code.
For clarity, we have chosen here to annotate the vertices of the digraph with schematic diagrams, i.e.,
contour plots of step functions on the
square which display the proper symmetries in a visually obvious way.
A solid line in the schematic diagrams represents
a nodal line, whereas a dashed line is a line of reflectional symmetry and a dot is a center
of rotational symmetry.
Contour plots of actual solutions can be found in Figure~\ref{square_plots}.
The bifurcation digraph was described in Subsection \ref{functSym} and a more thorough discussion 
is in \cite{NSS3}.

The bifurcation digraph in Figure~\ref{D4_digraph} is condensed,
as described in \cite{NSS3}.
The symmetry types are grouped into condensation classes, and
not all of the arrows are drawn.
For example, one condensation class is the block of four symmetry types near the top.
The condensation class has three arrows emanating from it, but in the un-condensed bifurcation digraph
there are 5 arrows emanating from {\em each} of the 4 solution types in the block.
The little numbers near the arrow tails count the number of arrows emanating from each solution.
Similarly, the little numbers near the arrow heads count the number of arrows ending at each of the
solution types in the condensation class.

At the top of Figure~\ref{D4_digraph}
is the trivial function $u\equiv0$ whose symmetry is all of $\D_4 \times \Z_2$.
At the bottom is a function with trivial symmetry, i.e.,
whose symmetry is the group containing only the identity.
There are four generic bifurcations with $\Z_2$ symmetry from the trivial branch.
For these bifurcations the critical eigenspace $\tilde E$
is the one-dimensional irreducible subspace for $\Z_2$ and there is a pitchfork bifurcation creating two
solution branches, $(u_s, s)$ and $(-u_s, s)$ at some point $(0, s^*)$.
The figure at the bottom left indicates the symmetry of a vector field in $\tilde E$.
We can think of this as the ODE on the one-dimensional center manifold,
or Lyapunov-Schmidt reduced bifurcation equation $\tilde g=0$.
There is one generic bifurcation with $\D_4$ symmetry from the trivial branch.
Here the two-dimensional critical eigenspace $\tilde E$ has lines of reflection symmetry across ``edges''
and non-conjugate lines of reflection symmetry across ``vertices'', as shown in the bottom middle part
of Figure~\ref{D4_digraph}.
At the bifurcation there are
two conjugacy classes of solution branches,
therefore the bifurcation digraph has two arrows labeled $\D_4$ coming out of the trivial solution.
On the condensed bifurcation digraph these two arrows are collapsed into one.
Note that there are several more bifurcations with $\Z_2$ symmetry
or with $\D_4$ symmetry in the bifurcation digraph.
For example, each of the 4 solution types in the second row can have a bifurcation with $\D_4$ symmetry.

There is only one more type of generic bifurcation that occurs in PDE~(\ref{pde}) on the square: a
bifurcation with $\Z_4$ symmetry.  This is a generic bifurcation in this {\em gradient} system~\cite{NSS3},
and the daughter solutions can be anywhere in the two-dimensional irreducible subspace (except the origin)
since there are no lines of reflection symmetry.
Note the use of a dotted arrow type for this bifurcation, and that there are no dashed arrows in this particular bifurcation digraph.
The reduced bifurcation equations  on the critical eigenspace at this bifurcation has the symmetry
indicated at the lower right part of Figure~\ref{D4_digraph}.

Figure~\ref{square_plots} contains the contour plot of
an example solution to Equation~(\ref{pde}) at $s = 0$
for each of the 20 possible symmetry types on the square.
The contour heights are $\pm c 2^{-h}$ with $h\in\{0,\ldots ,4\}$ and 
an appropriate $c$ near $\max(|u|)$, to give more contours near $u = 0$.
These figures were made with $\tilde M = 30$, meaning that the largest frequency in each direction
is 30. Thus the mode with the smallest eigenvalue that is left out of the basis is $\psi_{31,1}$.
This leaves $M = 719$ modes in our basis.
Unlike \cite{NS}, where an initial guess for each branch needed to be input by humans, the
solutions in Figure~\ref{square_plots} were found automatically by
following all of the primary branches that bifurcated from the trivial solution, and all of the secondary
branches, etc., recursively as described in \cite{NSS3}.

\begin{figure}
\scalebox{1.0}{\input{figures/bifdiag3Z2s.tex}}
\raise3cm\hbox{\scalebox{1}{\includegraphics{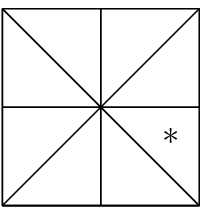}}}
\caption{
\label{bifdiag}
A partial bifurcation diagram for PDE (\ref{pde}) on the square, showing just one primary branch
and selected daughter branches.
The value of $u(x^*, y^*)$ vs. $s$ is plotted,
where $(x^*, y^*)$
is a {\em generic point} of the square,
as shown in the figure on the right.
A generic point is
not on any of the lines of reflection symmetry,
and we choose a point equidistant from those lines and the boundary.
The open dots show bifurcation
points, and the group of the bifurcation is shown for some.
The small numbers indicate the Morse Index of the solutions, and solid lines show an even MI whereas
dashed lines show an odd MI.
The solid dots at $s = 0$, from top to bottom, correspond to the contour plots in Figure~\ref{seq4},
from left to right.
}
\end{figure}

In addition to generic bifurcations, the PDE on the square has degenerate bifurcations due to the
``hidden symmetry'' of translation in the space of periodic functions~\cite{Gomes}.
For example, the bifurcation point at $s = \lam_{3,5} = 34$ on the trivial branch $a^* = 0$ has
an accidental degeneracy of Type 2 as defined in \cite{NSS3}.
Figure~\ref{bifdiag} shows a partial bifurcation diagram containing this point and
3 levels of branches bifurcating from it;
corresponding contour plots of solutions are found in Figure~\ref{seq4}.
A bifurcation diagram showing branches that bifurcate at $s \leq \lam_{2,3} = 13$ 
is shown in \cite{NS}.

The 2-dimensional critical eigenspace at this primary bifurcation point is
$\tilde E = \spn \{ \psi_{3,5}, \psi_{5, 3} \}$.
The trivial branch, whose symmetry is $\Gamma_0 \cong \D_4 \times \Z_2$,
undergoes a bifurcation with
$\Gamma_0 / \Gamma_0' \cong \Z_2 \times \Z_2 $ symmetry
at $s = 34$.  The action of $\Z_2 \times \Z_2$ on $\tilde E$ is generated by
$$
b \psi_{3,5} + c \psi_{5,3}
\mapsto
c \psi_{3,5} + b \psi_{5,3}
\quad
\mbox{and}
\quad
b \psi_{3,5} + c \psi_{5,3}
\mapsto
-b \psi_{3,5} - c \psi_{5,3}
.$$
Our code uses the ordered basis
$(\psi_{3,5}+\psi_{5,3}, \psi_{3,5}-\psi_{5,3})$.
In this basis, the action of $\Gamma_0/\Gamma_0'$
on $\tilde E$ is isomorphic to the natural action of 
$$\langle \left[
\begin{smallmatrix} 1&0 \\ 0 & -1 \end{smallmatrix}\right],
\left[
\begin{smallmatrix} -1 & 0 \\ 0 & -1 \end{smallmatrix}
\right]\rangle\cong\Z_2\times \Z_2$$ 
on $[\tilde E]=\R^2 = \R \oplus \R$. 
Note that $\tilde E$ is not an irreducible space;
this is a degenerate bifurcation.
The basis vectors were chosen to span the two one-dimensional irreducible subspaces, which are also
fixed-point subspaces of the $\Gam_0/\Gam_0'$ action on $\tilde E$ and therefore $\tilde g$-invariant
subspaces.
Each of these subspaces is used as an $E \subseteq \tilde E$ in the
{\tt cGNGA} algorithm described in Section~\ref{gnga}.
A pitchfork bifurcation generically occurs in each of these one-dimensional
invariant spaces.
\begin{figure}
\[
\xymatrix{
\scalebox{\sbsize}{\includegraphics{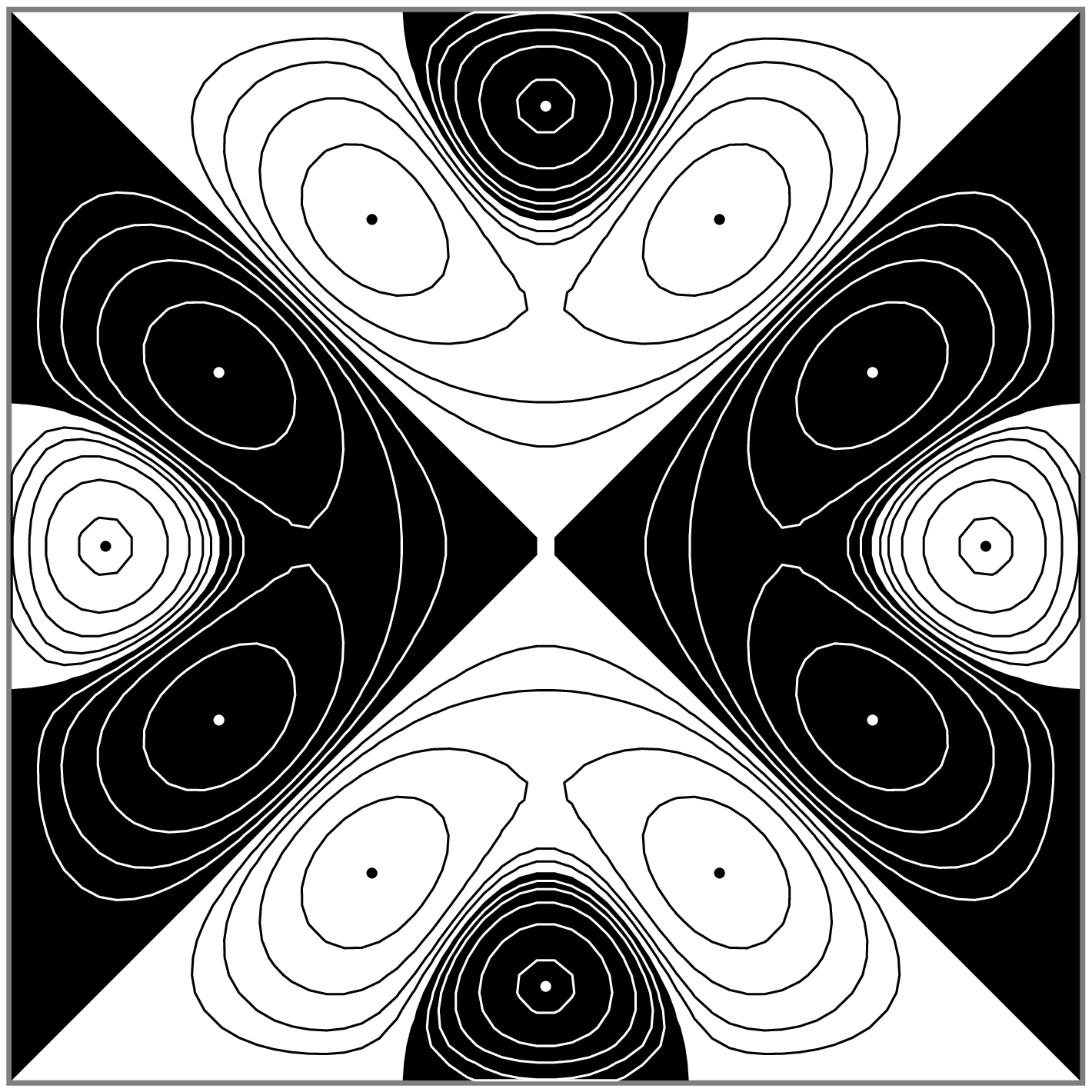}} &
\scalebox{\sbsize}{\includegraphics{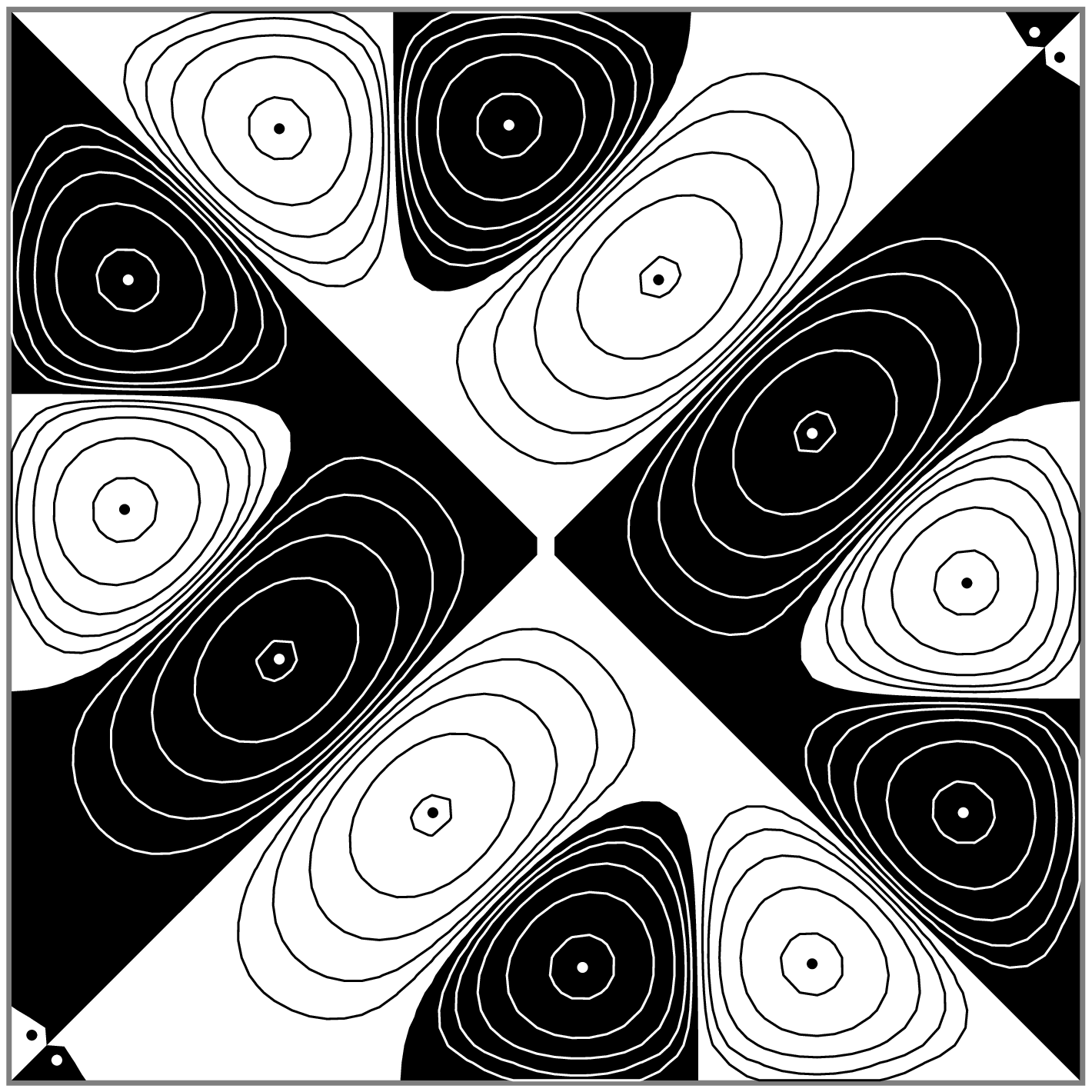}} &
{\scalebox{\sbsize}{\includegraphics{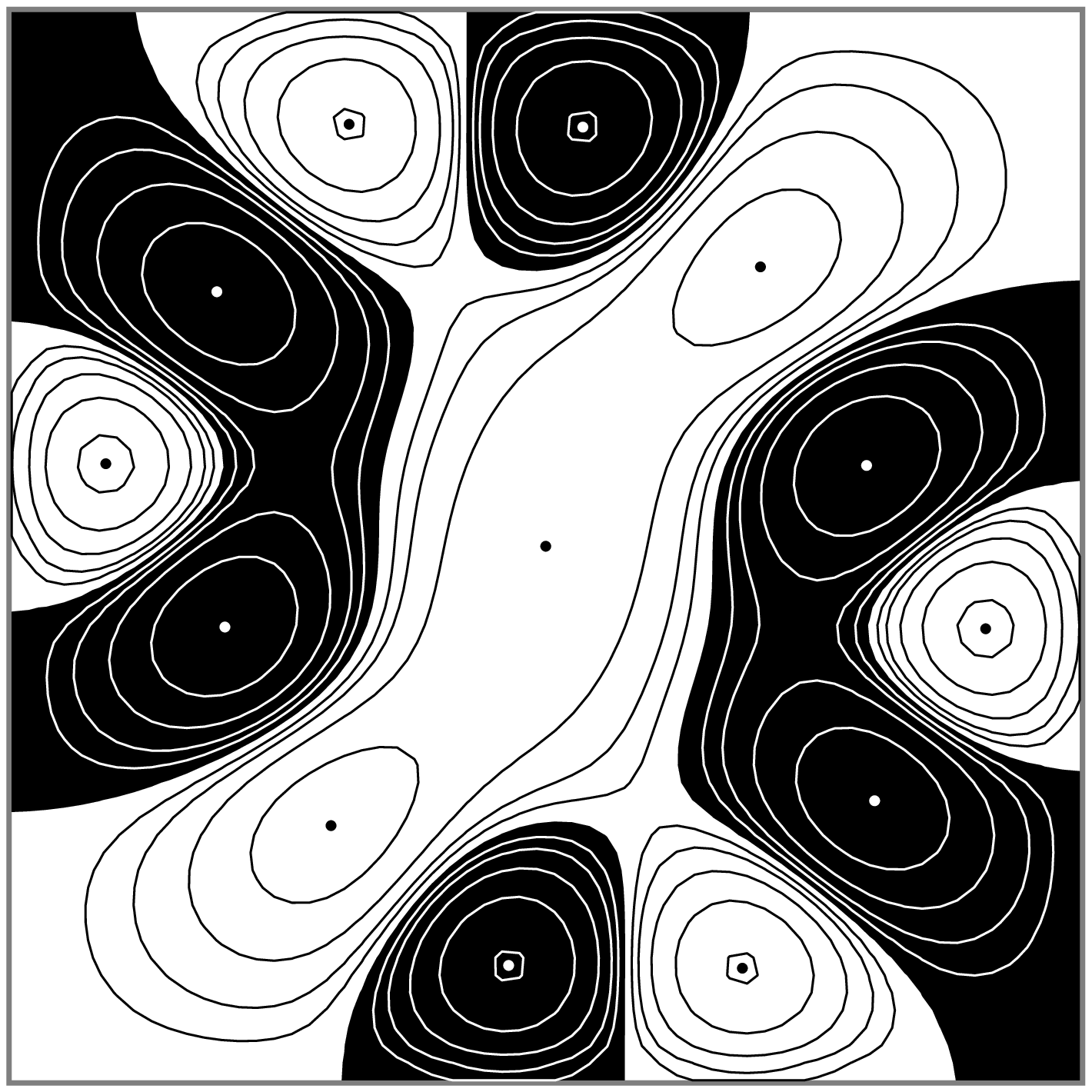}}} &
   {\scalebox{\sbsize}{\includegraphics{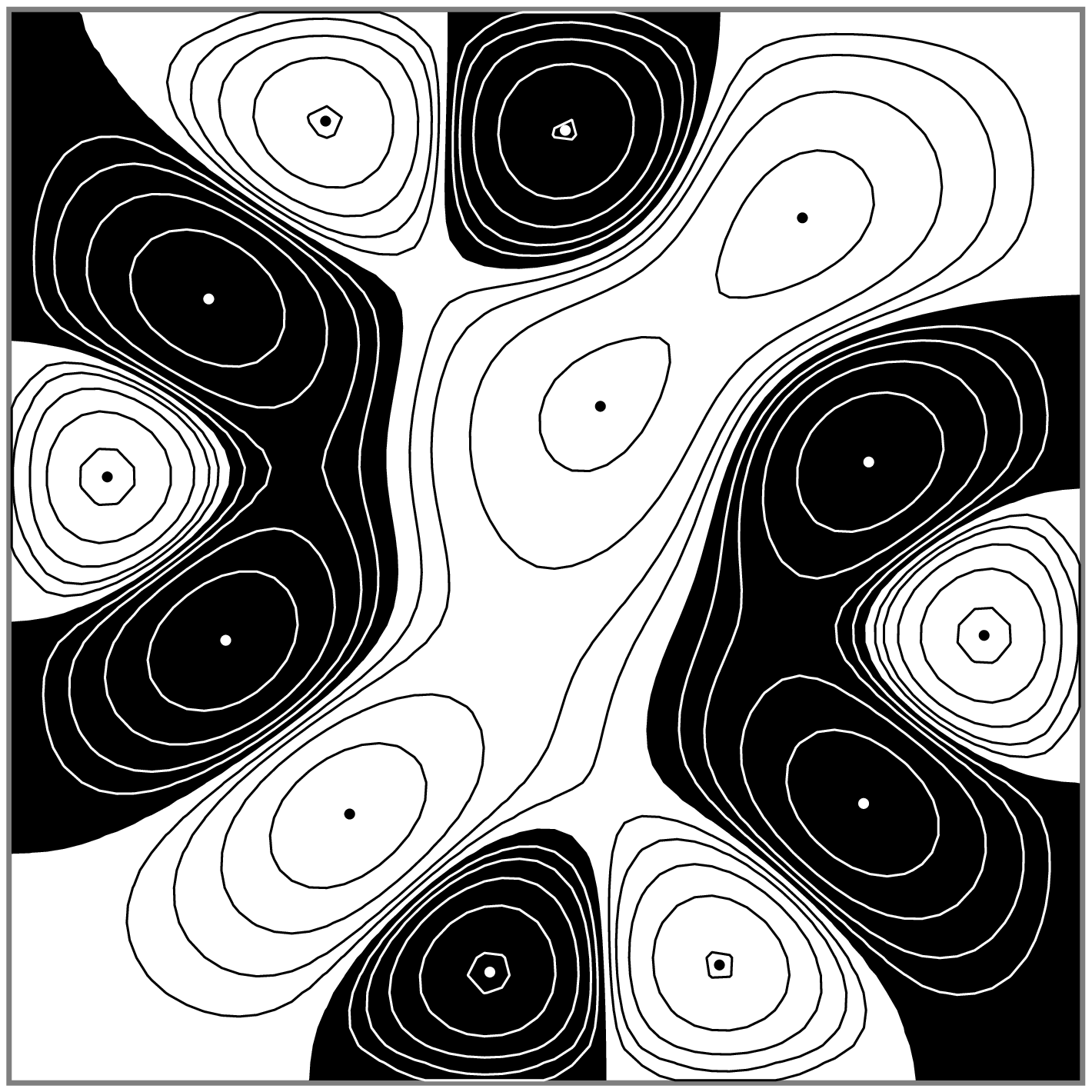}}}
\\
\scalebox{\sss}{\includegraphics{figures/s2.eps}}   \ar[r]^{\Z_2} &
\scalebox{\sss}{\includegraphics{figures/s11.eps}} \ar[r]^{\Z_2} &
\scalebox{\sss}{\includegraphics{figures/s17.eps}} \ar[r]^{\Z_2} &
\scalebox{\sss}{\includegraphics{figures/s19.eps}}
}
\]
\caption{A sequence of solutions obtained at $s = 0$ by following a chain of  $\Z_2$  bifurcations
from the trivial solution with full $\D_4 \times \Z_2$ symmetry, down to a solution
with trivial symmetry.
The sequence of bifurcations shown is a path in the bifurcation digraph in Figure~\ref{D4_digraph}.
The first solution shown is on a primary branch bifurcating at $s=\lambda_{3,5}=\lambda_{5,3}=34$,
and is approximately a multiple of $\psi_{3,5} - \psi_{5,3}$.
The four contour plots represent the solutions indicated by the four black dots in Figure~\ref{bifdiag}.
}
\label{seq4}

\end{figure}

The one-dimensional subspaces $\spn \{ \psi_{3,5}\}$
and
$\spn \{ \psi_{5,3} \} \subseteq \tilde E$, corresponding to
$\spn\{(1,1)\}$ and 
$\spn \{(-1, 1)\} \subseteq \R^2$, respectively, are not fixed-point subspaces. 
However, they are
AIS (see Section~\ref{functSym}) of $\tilde g: \tilde E \rightarrow \tilde E$,
since $\psi_{3,5}$ (and $\psi_{5,3}$) can be periodically extended to tile the plane with a solution
to the PDE (\ref{pde}).
There is a primary 
branch of
solutions
which is tangent to
$\{(u, s) = (a \psi_{3,5}, 34) \mid a \in \R\}$ at $(0, 34)$.
Thus, there are at least three (conjugacy classes of) solution branches bifurcating from this 
degenerate
bifurcation with $\Z_2 \times \Z_2$ symmetry.
Near the bifurcation, the nontrivial solutions are approximately multiples of
$\psi_{3,5} + \psi_{5,3}$, $\psi_{3,5} - \psi_{5,3}$, or $\psi_{3,5}$ (or its conjugate 
$\psi_{5,3}$).
Figure~\ref{bifdiag} shows a partial bifurcation diagram which follows the primary
branch which is approximately a multiple of $\psi_{3,5} - \psi_{5,3}$ near the bifurcation.

Figure~\ref{seq4} shows contour plots of example solutions along a particular path in the bifurcation digraph shown in Figure~\ref{D4_digraph}.
The primary branch is created at the degenerate bifurcation with $\Z_2 \times \Z_2$ symmetry at $s = 34$ discussed above.
The critical eigenspace is two-dimensional at $s = \lam_{3, 5} = 34$, and this bifurcation is not on the bifurcation digraph, Figure~\ref{D4_digraph}, which only shows {\em generic} bifurcations.
There are two primary branches conjugate to the one shown, four secondary branches, eight tertiary, and
16 branches conjugate to the solution with trivial symmetry shown in Figure~\ref{seq4}.
At each bifurcation our GNGA code follows exactly one of the conjugate branches that bifurcate.

\begin{figure}
\scalebox{0.7}{\input{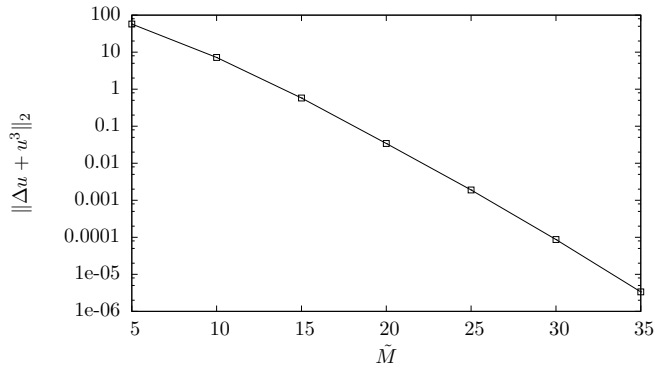}}
%
\caption{
A measure of the error of our approximation to the CCN solution (minimal energy sign-changing
solution \cite{CCN}) of PDE~(\ref{pde}) 
at $s = 0$,
as a function of $\tilde M$.  The region used was $\Omega = (0, 1)^2$ to
facilitate comparison with Figure~7(b) of \cite{NS}.
}
\label{L2pde}

\end{figure}
Figure~\ref{L2pde} contains a numerical demonstration of the convergence of the GNGA as the number of modes increases.
In Figure~7(b) of \cite{NS}, we previously provided a portion of a similar graph of the $L^2$ norm of $\Delta u + u^3$ vs. $\tilde M$,
for a particular solution $u$.
That graphic was not entirely convincing in showing convergence to 0.
Using our current code
with larger values of $\tilde M$ and a smaller convergence tolerance in Newton's Method,
we re-computed the data for the same problem and obtained the more accurate numerical result displayed here.
\end{section}

\begin{section}{The PDE on the Cube}
\label{cube}

The PDE~(\ref{pde}) on the  cube has a rich array of bifurcations with symmetry.
If $\Omega$ is a planar region, only bifurcations with $\D_n$ or $\Z_n$ symmetry are present in
the bifurcation digraph.
The bifurcation digraph of the PDE on the cube includes bifurcations with $\Z_2$ symmetry, 
which have 1-dimensional critical eigenspaces.
The digraph also includes bifurcations with $\Z_n$ or $\D_n$ symmetry,
$n \in \{3, 4, 6\}$, for which the critical eigenspace is 2-dimensional.
A novel feature of the PDE on the cube is that there are
5 bifurcations with symmetry that have 3-dimensional critical eigenspaces, shown in 
Figure~\ref{irredSpaces3}.

Recall \cite{NSS3} that a generic {\em bifurcation with $\Gam$ symmetry} has a critical
eigenspace $\tilde E$ that is a faithful, irreducible representation space of $\Gam$.
Faithful means that only the identity in $\Gam$ acts trivially on $\tilde E$, and irreducible
means that no proper subspace or $\tilde E$ is $\Gam$-invariant.

Faithful, irreducible representation spaces for $\D_4$ and $\Z_4$ were shown in Figure~\ref{D4_digraph},
and the generalization to $\D_n$ and $\Z_n$ is obvious.
Note that there are no such representation spaces for $\Z_2 \times \Z_2$, since no 1-dimensional
representation space is faithful,
and every 2 or higher-dimensional representation space is reducible.
Recall that the bifurcation with
$\Z_2\times\Z_2$ symmetry that occurs at $s = 34$ in Figure~\ref{bifdiag}
is not {\em generic}.  That bifurcation point has a Type-2 degeneracy \cite{NSS3}.

This section contains our main new numerical results,
namely approximate solutions to Equation~(\ref{pde}) on the cube.
For convenience,
we denote the $99$ symmetry types of solutions to this PDE on this region by $S_0,\ldots,S_{98}$.
Obtaining accurate approximations on this 3-dimensional region requires a large number of gridpoints.
We use a parallel implementation as described in Section~\ref{mpq}.

In Subsection~\ref{subs:bifDig}, we give an overview of features of the bifurcation digraph, 
which is too 
big to include in its entirety in a single document. 
We describe how our companion website \cite{cubeWEB}  can 
be used to navigate the digraph in order to view graphics and understand various symmetry 
information across the spectrum of solutions.
In the remaining subsections, we include 
a survey of our numerical results which showcase our analysis
of the bifurcations with most interesting symmetries.  

\subsection{The Bifurcation Digraph}
\label{subs:bifDig}

The bifurcation digraph of the $\Oh \times \Z_2$ action on $V = G_M$ or $H$ is far too 
complicated to display
as a figure in this paper.
Our web site \cite{cubeWEB} 
has a page for each symmetry type $S_i$, encoding the arrows
emanating from this symmetry type, along with additional information. 

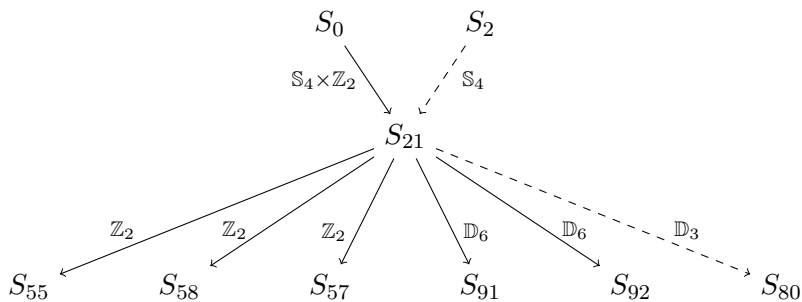
\begin{figure}
\begin{tabular}{|l|}
\hline
Symmetry types, partitioned into condensation classes. \\
Non-isomorphic types are separated with a `:' \\
:0: 1 2 3 4 : $\cdots$  : 19 : 20 21 22 23 : $\cdots$ : 87 : 88 . 89 90 91 92 . 93 94 . 95 . 96 97 : 98 \\
\hline
Bifurcates from: 0 2 \\
\hline
Symmetry type: 21, representatives are isomorphic to $\D_6$ \\
(Flag diagram and back and front contour plots appear here.)\\
\hline
Bifurcation with (symmetry) to (symmetry type): \\
\qquad$\mid
\Z_2  \rightarrow 55 \mid
\Z_2  \rightarrow 58 \mid
\Z_2  \rightarrow 57 \mid
\D_6  \rightarrow 91 \ 92 \mid
\D_3  - \rightarrow 80
$ \\
\hline
View: \ 55 \ 58 \ 57 \  91 \ 92 \  80 \\
\hline
\end{tabular}
\begin{center}
\begin{tikzpicture}
\node (s21) at (5,2) {$S_{21}$};
\node (s55) at (0,0) {$S_{55}$};
\draw[->] (s21) to node[pos=.8,above]{$\scriptstyle\mathbb{Z}_2$} (s55);
\node (s58) at (2,0) {$S_{58}$};
\draw[->] (s21) to node[pos=.85,above] {$\scriptstyle\mathbb{Z}_2$} (s58);
\node (s57) at (4,0) {$S_{57}$};
\draw[->] (s21) to node[pos=.7,left] {$\scriptstyle\mathbb{Z}_2$} (s57);
\node (s91) at (6,0) {$S_{91}$};
\draw[->] (s21) to  node[pos=.7,right] {$\scriptstyle\mathbb{D}_6$} (s91) ;
\node (s92) at (8,0) {$S_{92}$};
\draw[->] (s21) to node[pos=.85,above] {$\scriptstyle\mathbb{D}_6$} (s92);
\node (s80) at (10,0) {$S_{80}$};
\draw[dashed,->] (s21) to node[pos=.8,above] {$\scriptstyle\mathbb{D}_3$} (s80) ;
\node (s0) at (4,3.5) {$S_0$};
\node (s2) at (6,3.5) {$S_2$};
\draw[dashed,->] (s2) to  node[pos=.5,right] {$\ \scriptstyle\mathbb{S}_4$} (s21) ;
\draw[->] (s0) to  node[pos=.5,left] {$\scriptstyle\mathbb{S}_4\times\mathbb{Z}_2$} (s21) ;
\end{tikzpicture}
\end{center}
\caption{A schematic representation of the page for symmetry type $S_{21}$ 
from the companion web site \cite{cubeWEB}, along with the corresponding arrows in the
bifurcation digraph.
The labels on the top two arrows are found in the pages for symmetry types $S_0$ and $S_2$, respectively.   
}
\label{webPageFigure}
\end{figure}

An example from the companion web site for symmetry 
type $S_{21}$ is shown in the top half of Figure~\ref{webPageFigure}.
The first box contains links to all the symmetry type pages.
The 99 symmetry types are grouped into isomorphism classes by colons.
The isomorphism classes are further subdivided into condensation 
classes by periods.
The abbreviated list in 
Figure \ref{webPageFigure} shows, for example, that $S_0$, $S_{19}$, $S_{87}$ and $S_{99}$ are
singleton condensation classes, and
that $\{S_{20}, S_{21}, S_{22}, S_{23}\}$ 
is a condensation class.
The symmetry types $S_{88}$ through $S_{97}$ are all isomorphic. 
These 11 symmetry types are separated into 5 condensation classes.  
For example $\{S_{93}, S_{94} \}$
is a condensation class.

The second box indicates that there are arrows in the bifurcation digraph
pointing from $S_0$ and from $S_2$ to $S_{21}$.

The third box indicates that any $\Gam_i \in S_{21}$ is isomorphic to $\D_6$, and
contains the graphics on the web page.  There is a flag diagram 
for every symmetry type, and contour plots if our computer program found a solution 
with this symmetry type.  About half of the symmetry types feature contour plots.

The fourth box encodes the
6 arrows in the bifurcation digraph emanating from $S_{21}$, as shown in the bottom half
of Figure~\ref{webPageFigure}.
In addition, the arrows are separated into the 5 generic
bifurcations with symmetry coming from the 5 nontrivial
irreducible representations of $\D_6$.
The symmetry types in any condensation class
have an identical pattern of generic bifurcations, with different labels of the symmetry types.
Thus, there are 6 arrows emanating from each of the symmetry types in the class
$\{S_{20}, S_{21}, S_{22}, S_{23}\}$.
In the condensed bifurcation digraph, these 24 arrows are represented by just 6 arrows.

The final box contains buttons to view selected daughter flag diagrams.

\subsection{Bifurcations from the first three eigenvalues}

\begin{figure}
\scalebox{1.0}{\input{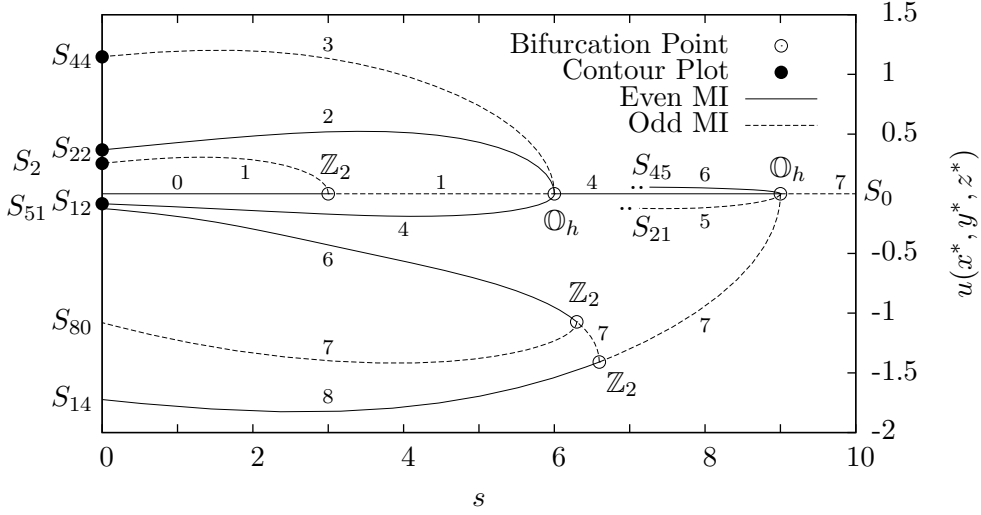}}
\caption{
Partial bifurcation diagram for PDE~(\ref{pde}) on the cube, showing the
first three bifurcation points of the trivial branch.
The two bifurcations with $\Oh$ symmetry and three bifurcations with $\Z_2$ symmetry are indicated
by open circles.  The symmetry type of the trivial branch is $S_0$, as indicated on the right.
The symmetry types
of the other branches are shown on the left.
The small numbers indicate the Morse Index of the solution.
The solution branches with symmetry type $S_{21}$ and $S_{45}$ are truncated to simplify the diagram.
The solid dots at the left correspond to the contour maps
shown in Figure~\ref{s0.10c}, and
contour maps for the other solution branches are shown in
Figure~\ref{secondOh}. 
}
\label{s0.10}
\end{figure}

Figure~\ref{s0.10} shows the bifurcation diagram for the solution branches
that are connected to the trivial solution branch with $s \leq 10$.
The value of $u$ at a generic point is plotted against the parameter $s$.
The trivial branch has a bifurcation with $\Z_2$ symmetry at $s = \lam_{1,1,1} = 3$ where
the MI changes from 0 to 1.
The trivial branch has two
bifurcations with $\Oh$ symmetry, at $s = \lam_{1,1,2} = 6$ and at $s = \lam_{1,2,2} = 9$.
The MI changes by 3 at each of these bifurcations, indicating a 3-dimensional critical eigenspace $\tilde E$.

At $s = 3$, the symmetry of the mother branch is $\Gam_0 = \Oh \times \Z_2$.
The critical eigenspace is $\tilde E = \spn \{ \psi_{1,1,1} \}$.  All of the reflections and rotations
in $\Oh$ act trivially on $\tilde E$.  That is,
$\Gam_0' = \langle (R_{90}, 1), (R_{120},1), (R_{180}, 1), (I_3, 1) \rangle = \Gam_2 \cong \Oh$.
Thus, the effective symmetry of the bifurcation is
$\Gam_0/\Gam_0' = \langle (I_3, -1) \Gam_0' \rangle \cong \Z_2 = \{-1, 1\}$.
The primary branch created at $s = 3$ has symmetry $\Gam_2$, and symmetry type $S_2 = \{ \Gam_2 \}$,
as shown in Figures~\ref{s0.10} and \ref{s0.10c}.

\newcommand{\flagScale}{.34}
\newcommand{\solScale}{.221}

\begin{figure}
\begin{tabular}{rccc}
\raise2cm\hbox{$S_{2}$}\hskip1cm
\scalebox{\flagScale}{\includegraphics{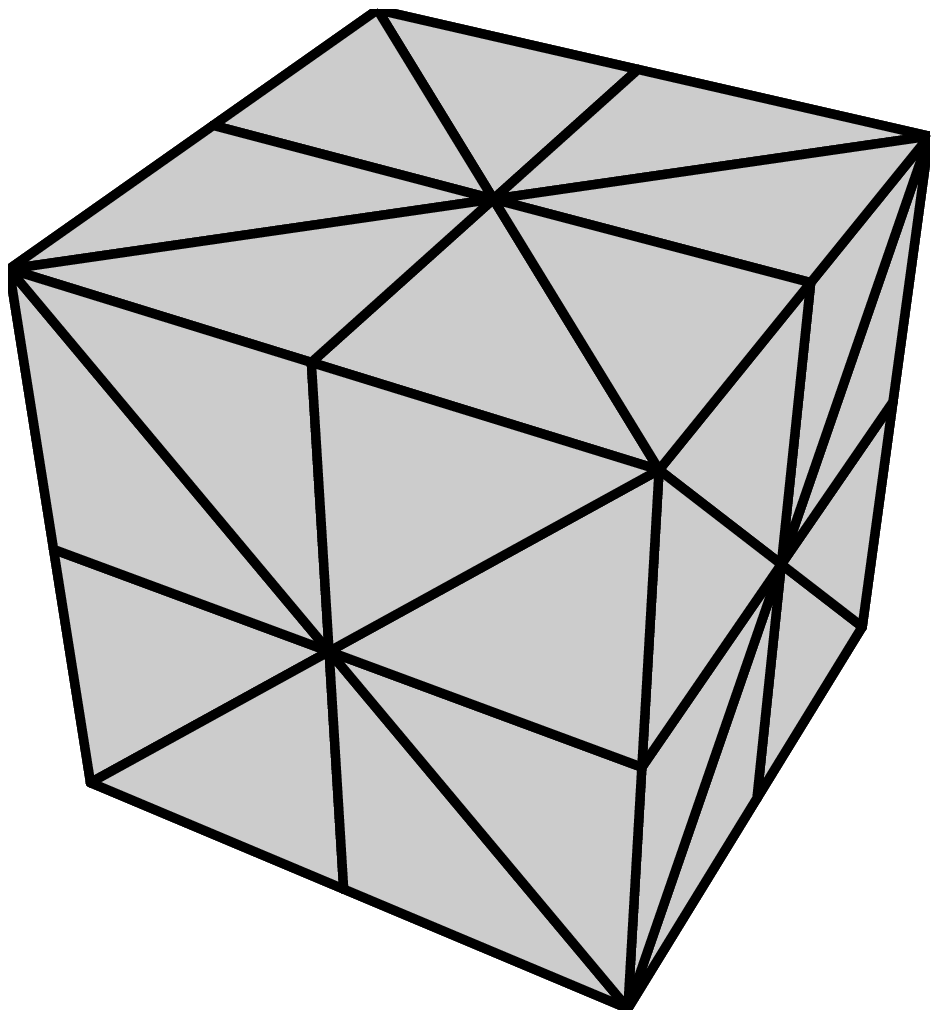}} &
\scalebox{\solScale}{\includegraphics{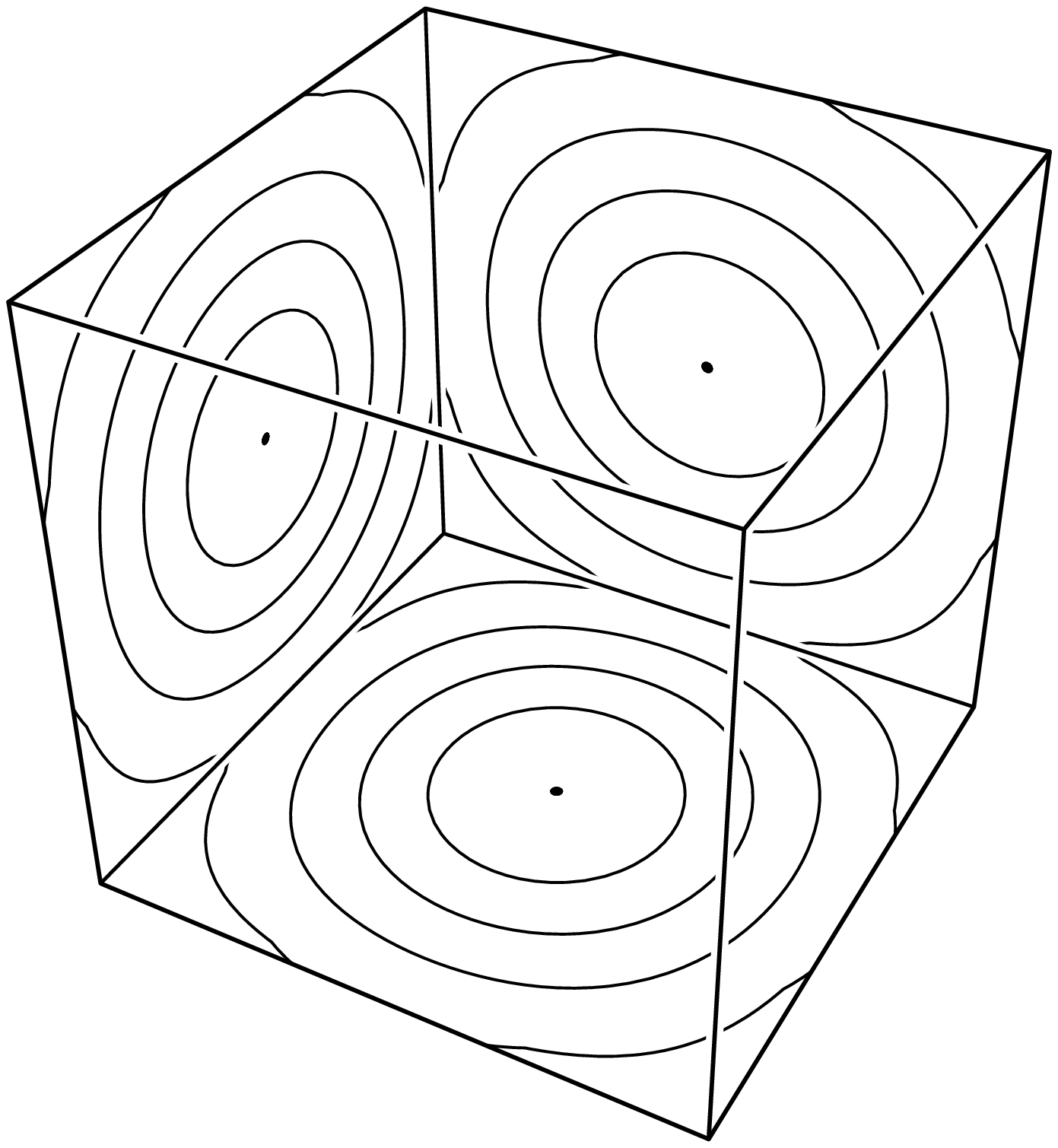}}  &
\scalebox{\solScale}{\includegraphics{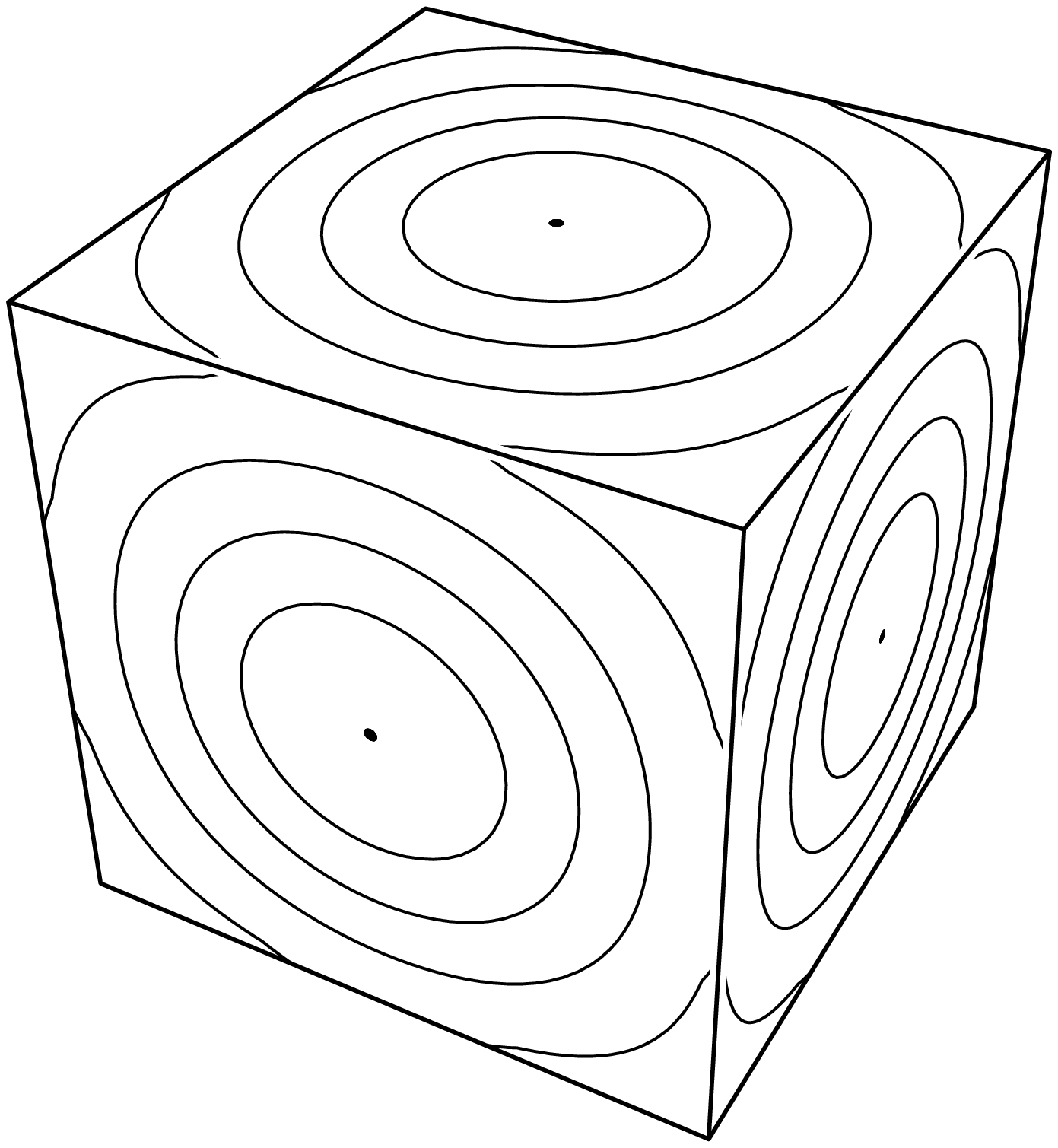}} \\
\raise2cm\hbox{$F$: $S_{12}$}\hskip1cm
\scalebox{\flagScale}{\includegraphics{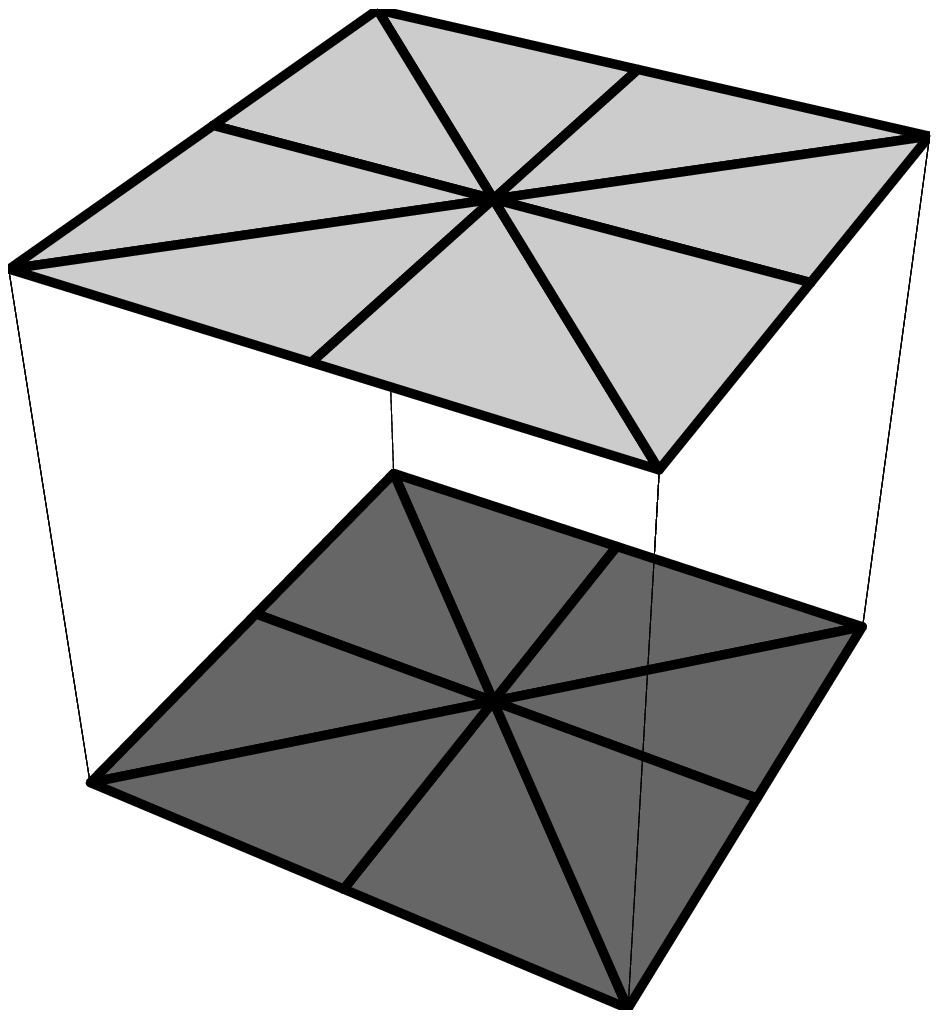}} &
\scalebox{\solScale}{\includegraphics{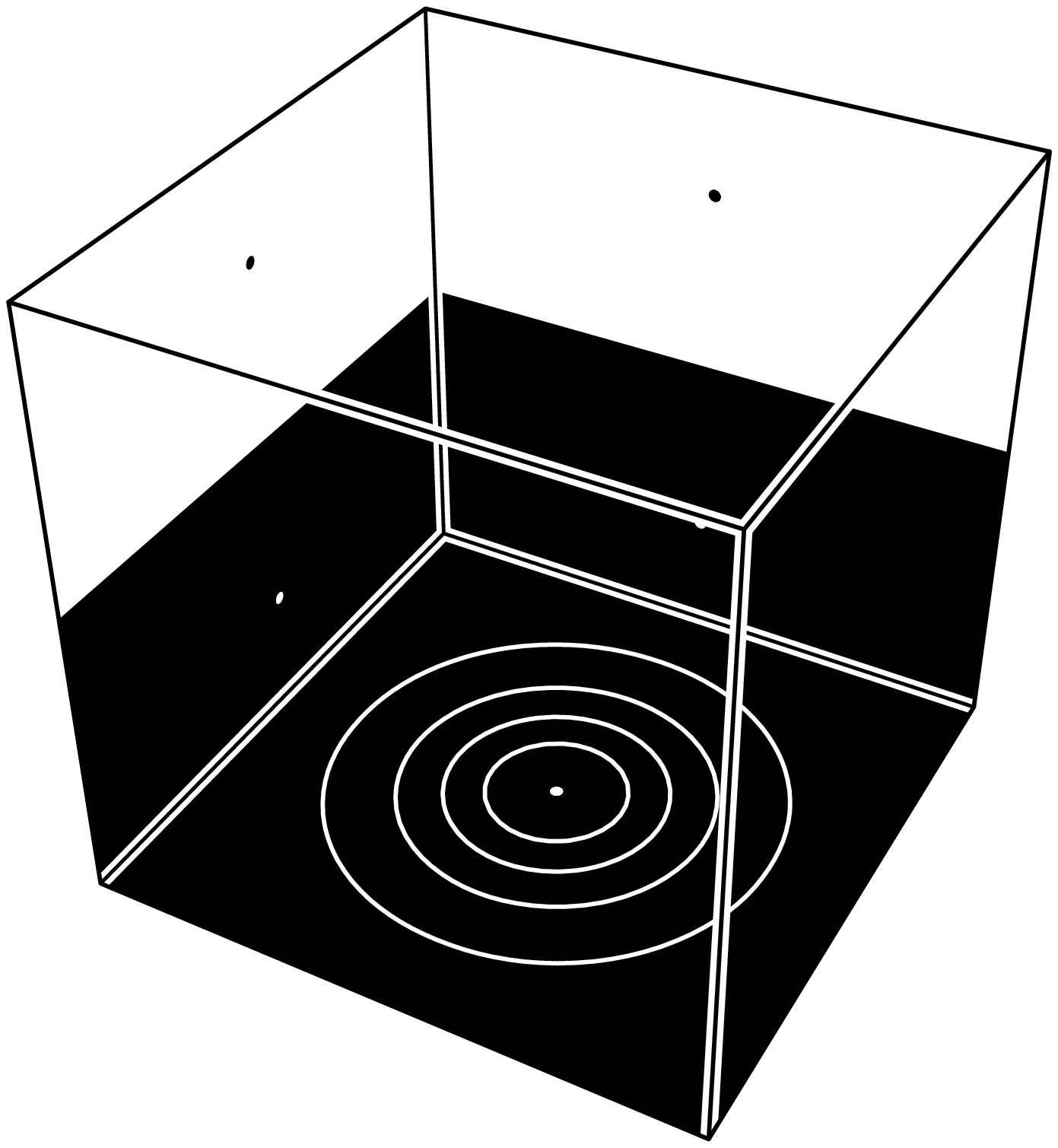}}  &
\scalebox{\solScale}{\includegraphics{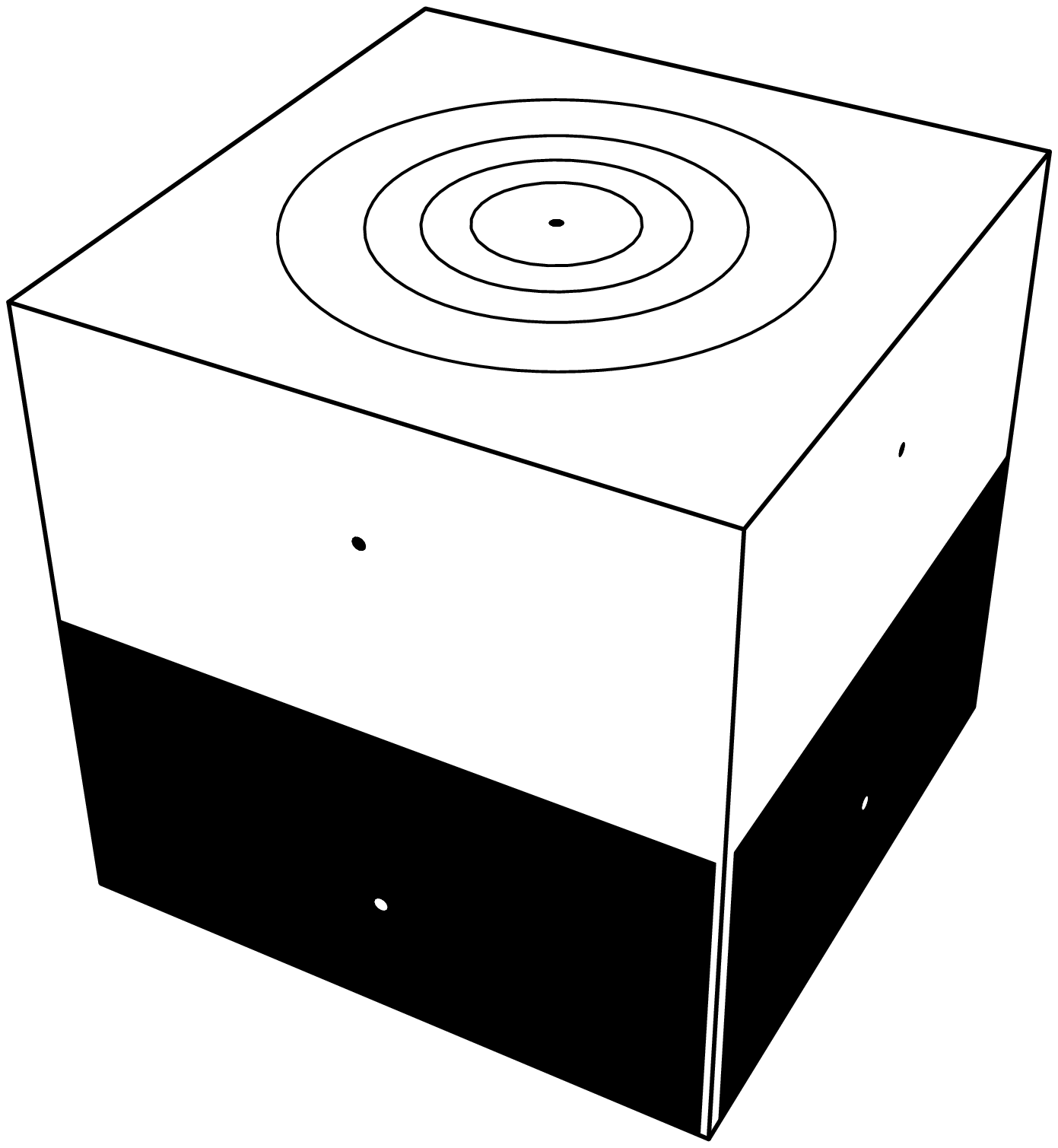}} \\
\raise2cm\hbox{$V$: $S_{22}$}\hskip1cm
\scalebox{\flagScale}{\includegraphics{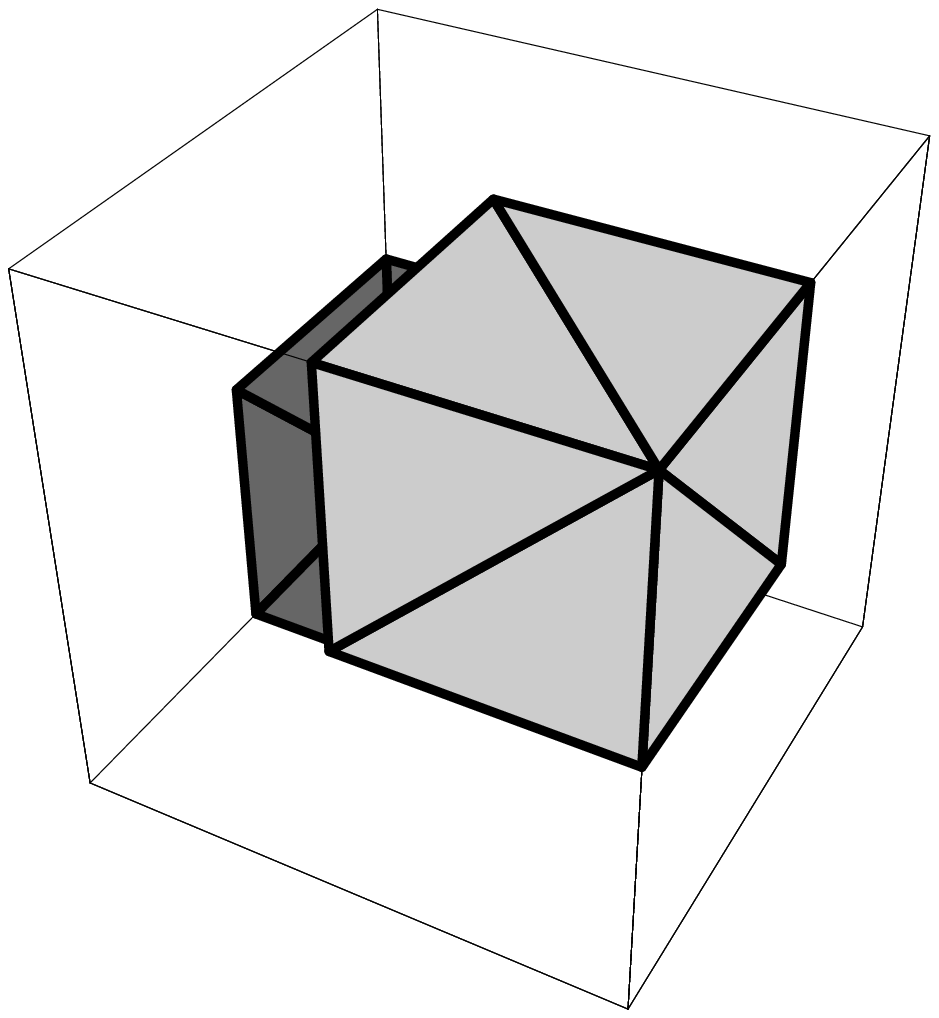}} &
\scalebox{\solScale}{\includegraphics{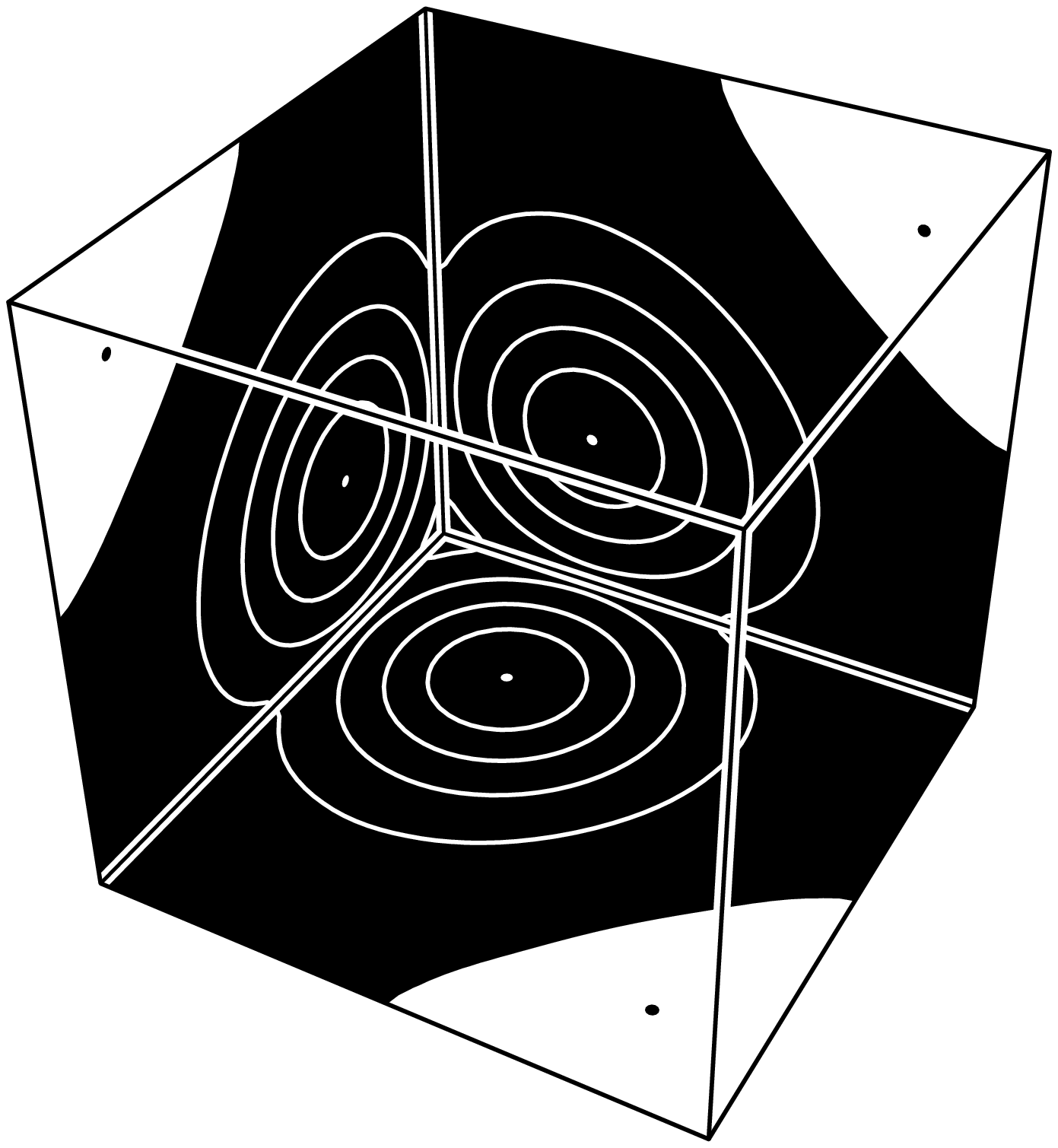}}  &
\scalebox{\solScale}{\includegraphics{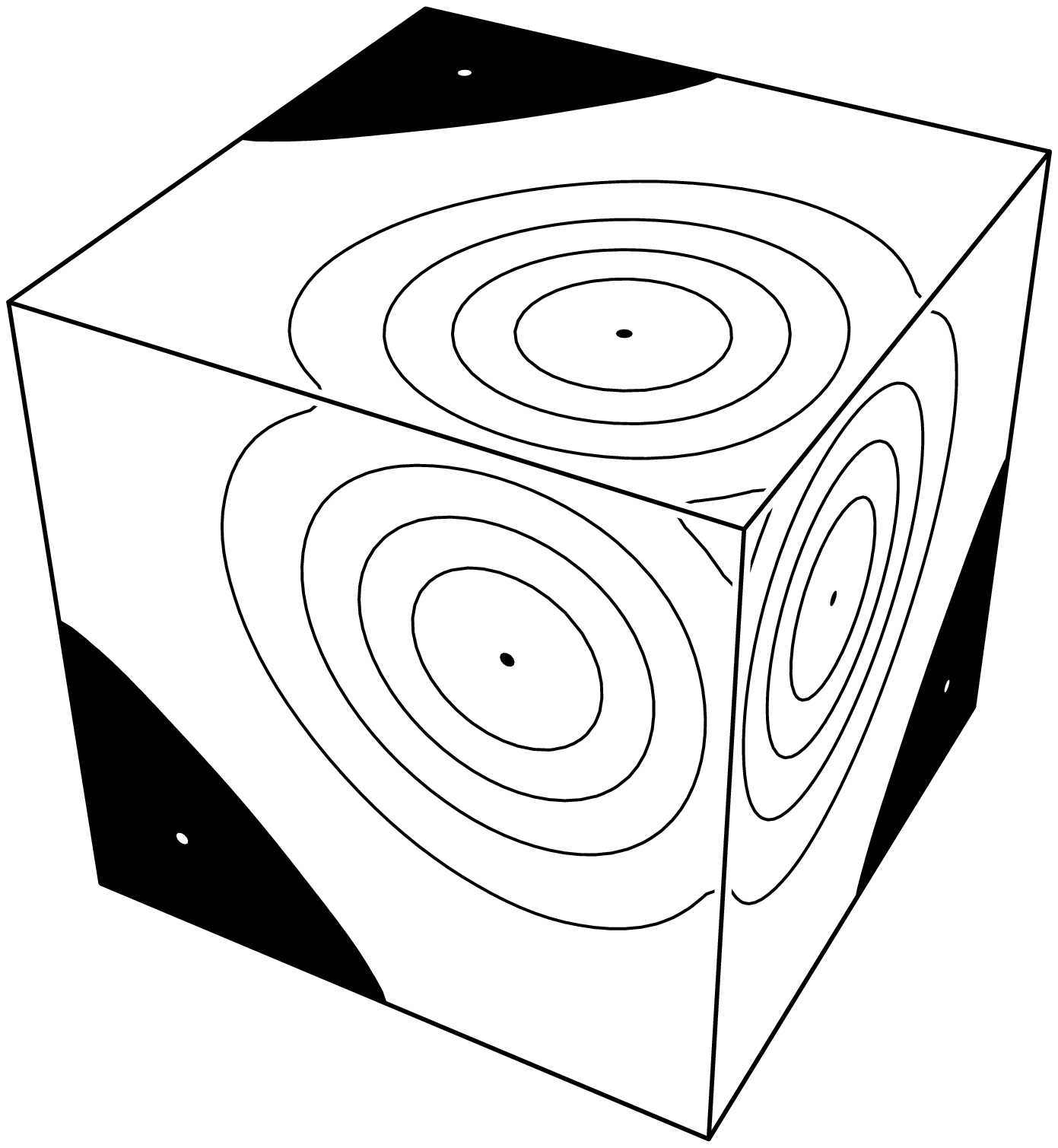}} \\
\raise2cm\hbox{$E$: $S_{44}$}\hskip1cm
\scalebox{\flagScale}{\includegraphics{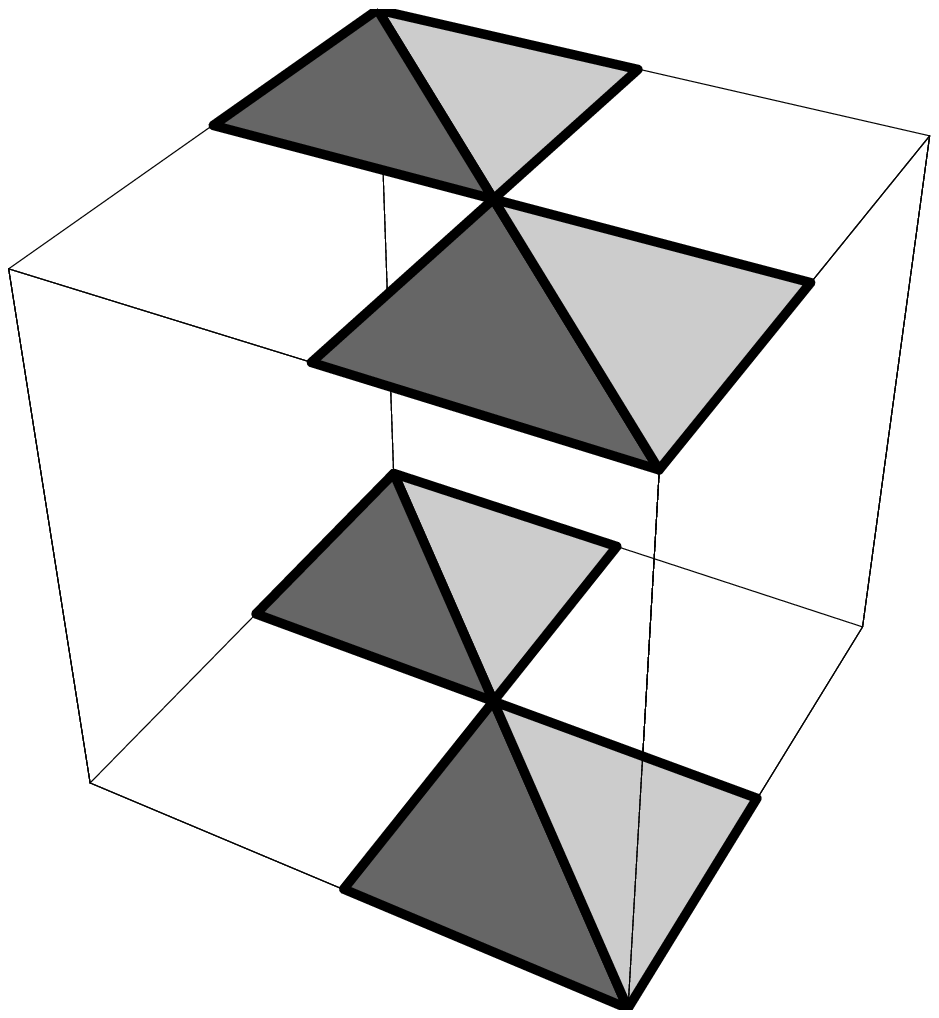}} &
\scalebox{\solScale}{\includegraphics{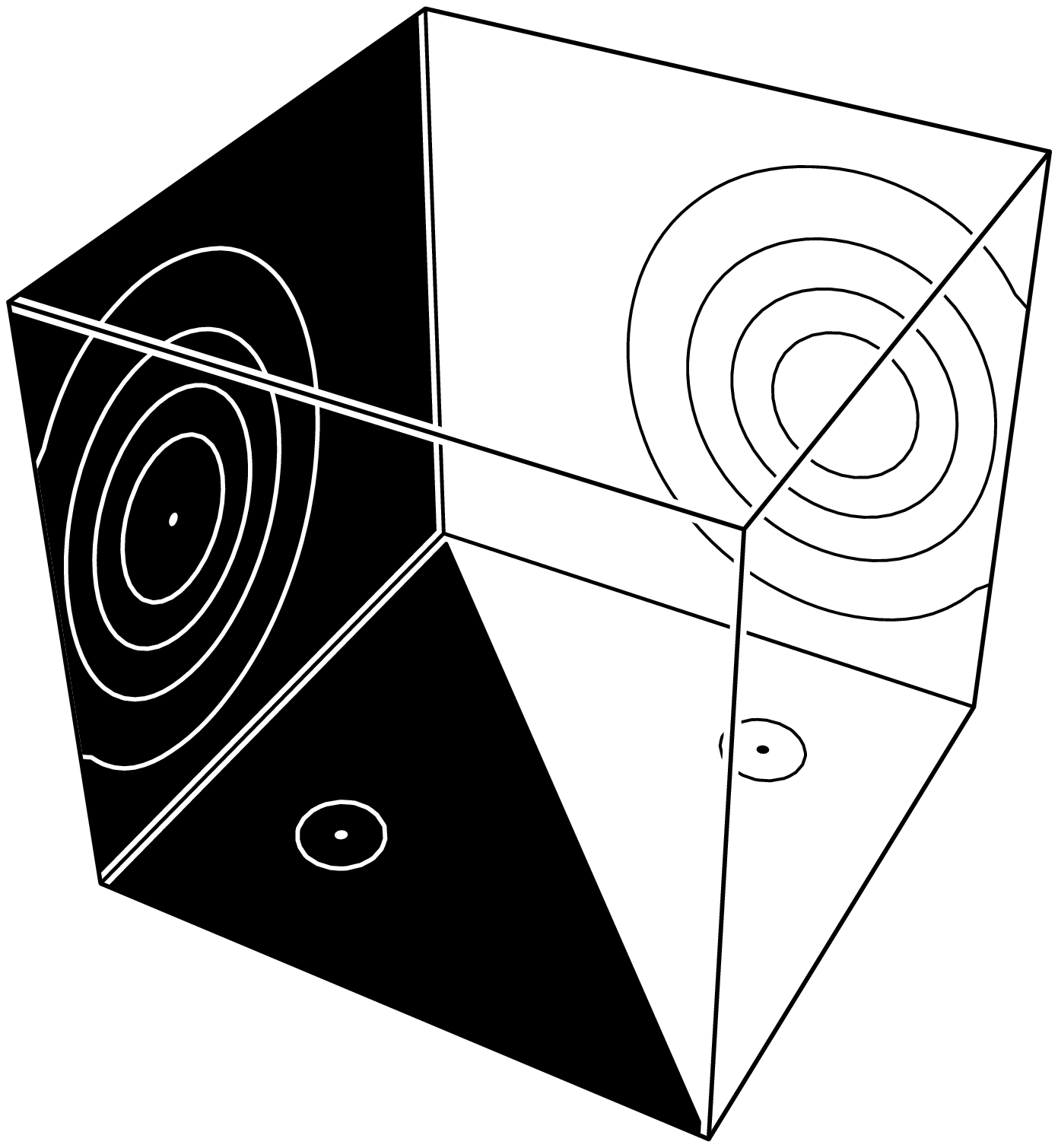}}  &
\scalebox{\solScale}{\includegraphics{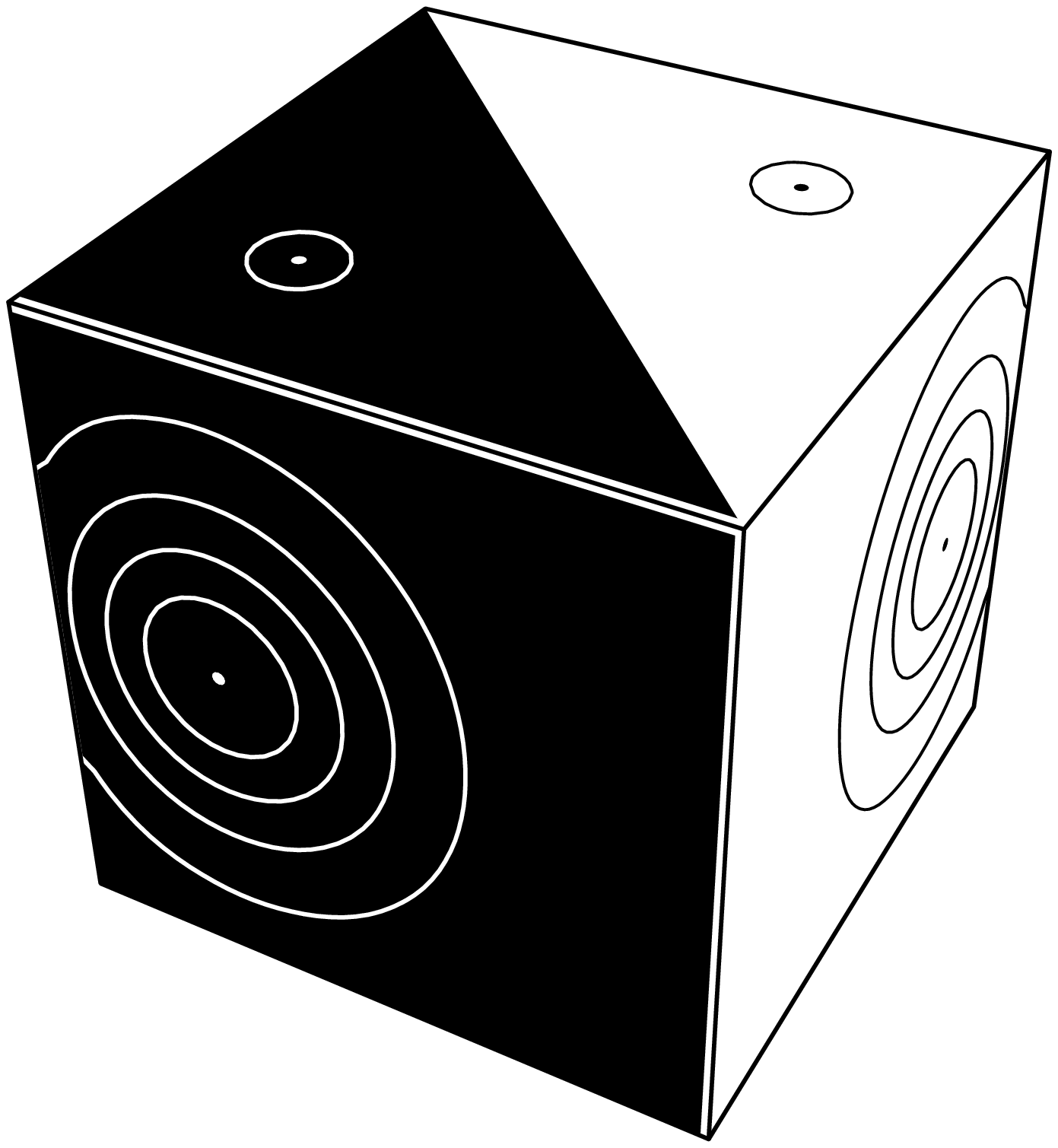}} \\
\end{tabular}
\caption{Contour plots for the solutions indicated with a solid dot in Figure~\ref{s0.10}.
The contour lines for functions on the cube are equally spaced.
The positive solution, with symmetry type $S_2$, bifurcates at $s = 3$.
Representatives of the face, vertex, and edge solutions 
(denoted by $F$, $V$ and $E$) that bifurcate at $s = 6$ are shown.
The vertex solution with symmetry type $S_{22}$ is the CCN solution.
The names derive for the position of the critical points of 
$\tilde g: \tilde E \rightarrow \tilde E$.
The equivariance of the vector field $\tilde g$ is shown in Figure~\ref{irredSpaces3}.
}
\label{s0.10c}
\end{figure}

\begin{figure}
\begin{tabular}{rccc}
\raise2cm\hbox{$V$: $S_{21}$}\hskip1cm
\scalebox{\flagScale}{\includegraphics{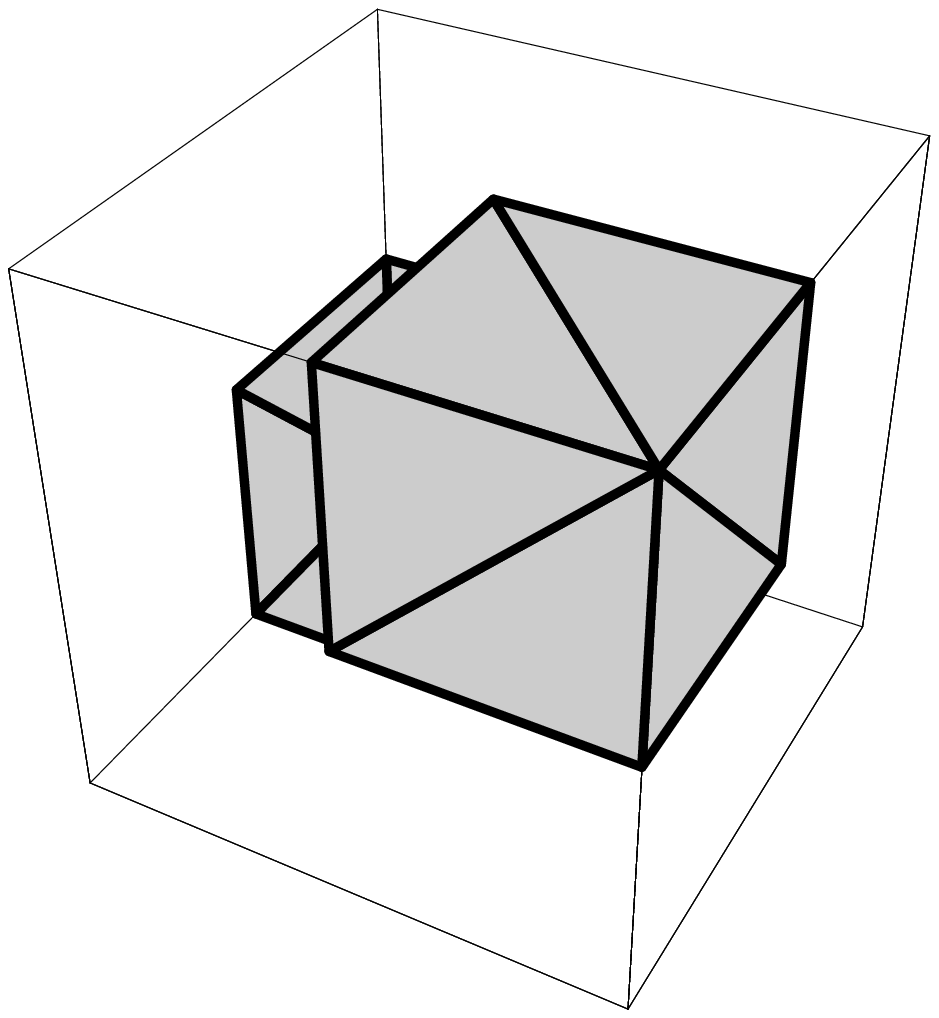}} &
\scalebox{\solScale}{\includegraphics{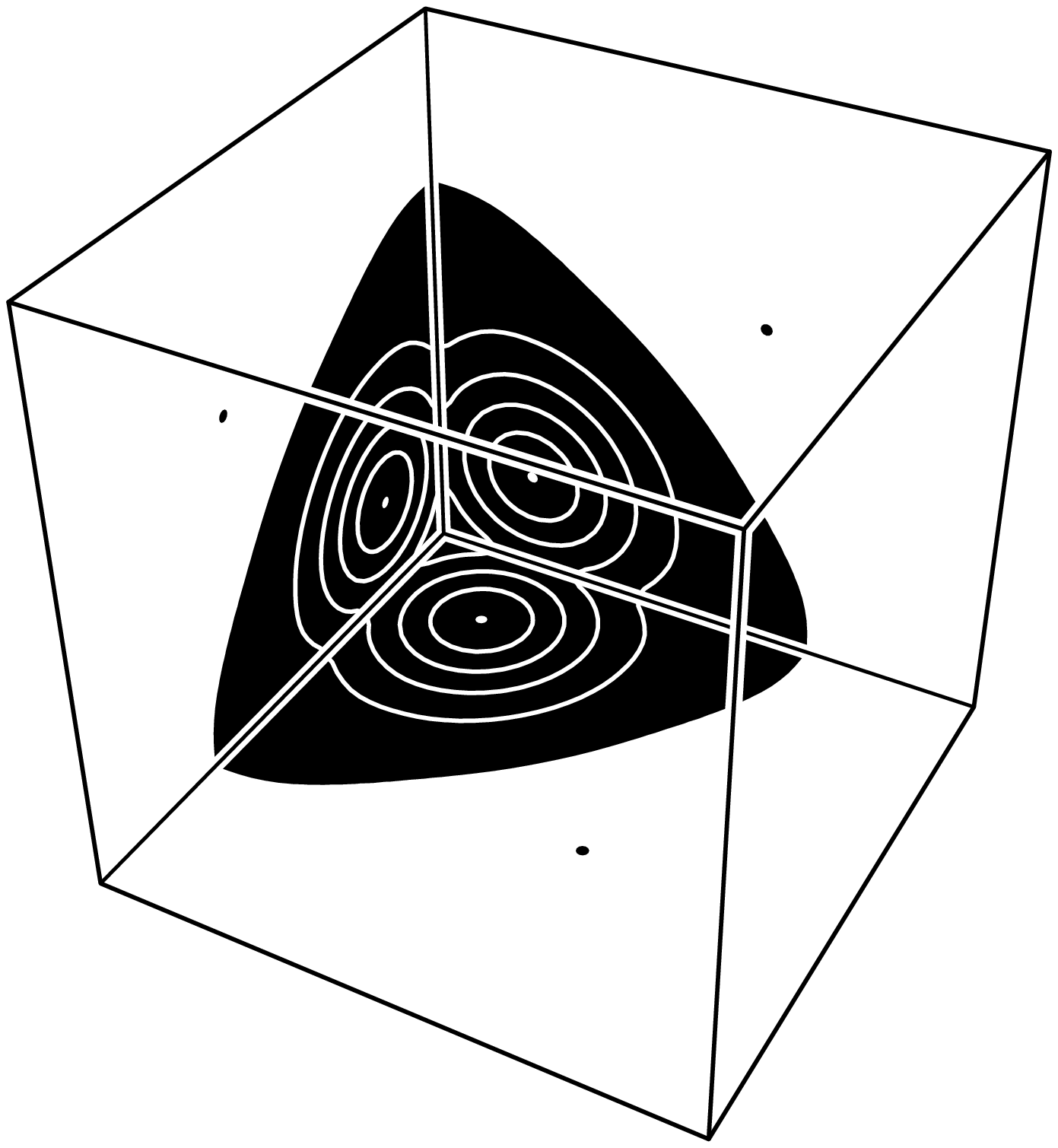}}  &
\scalebox{\solScale}{\includegraphics{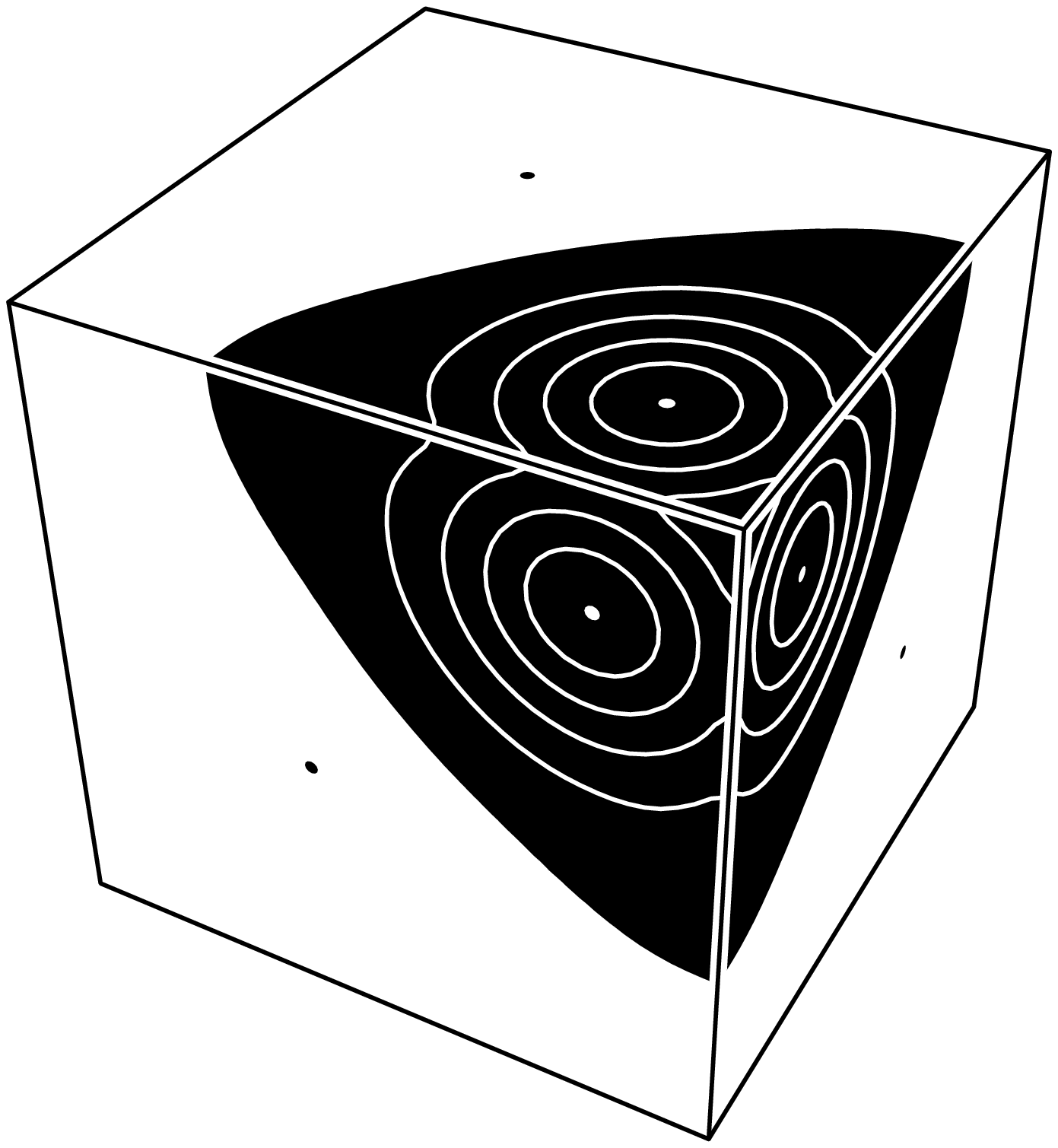}} \\
\raise2cm\hbox{$E$: $S_{45}$}\hskip1cm
\scalebox{\flagScale}{\includegraphics{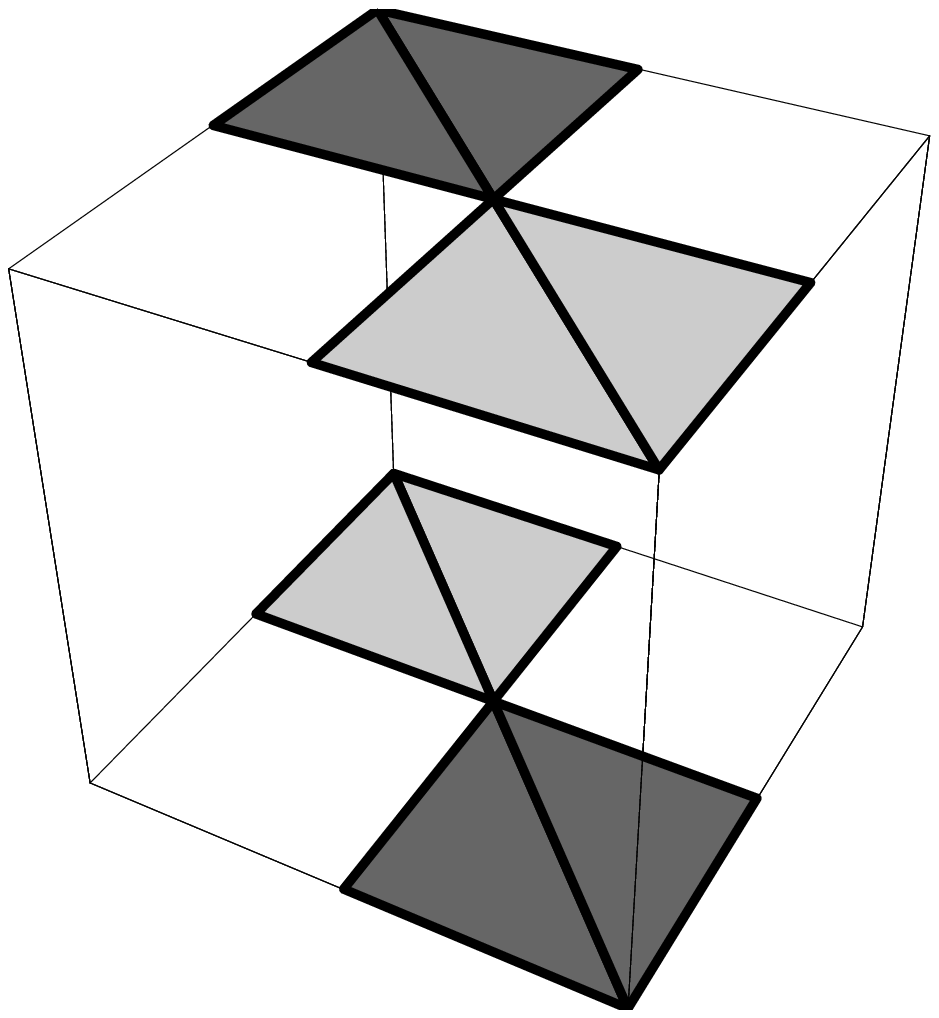}} &
\scalebox{\solScale}{\includegraphics{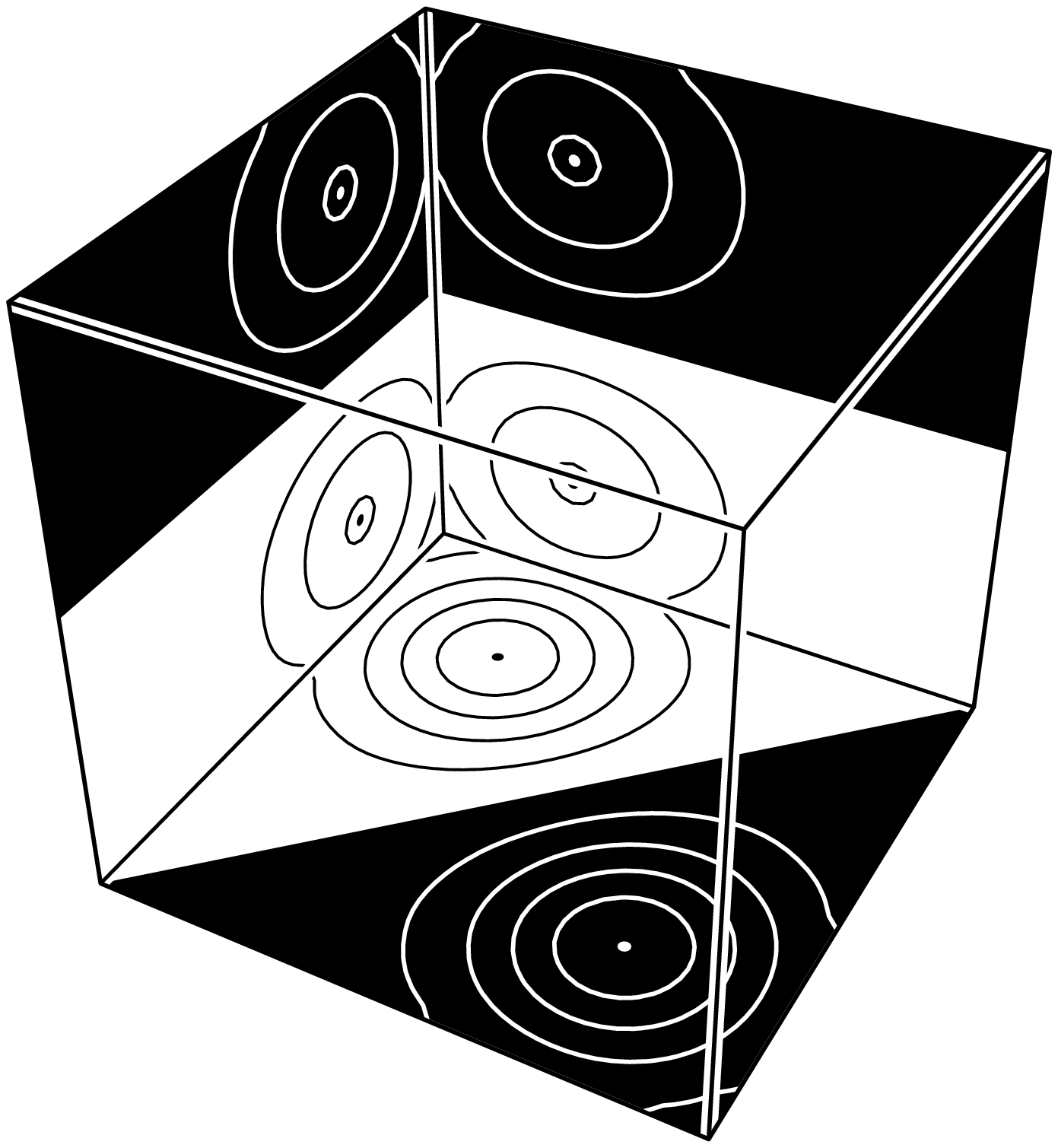}}  &
\scalebox{\solScale}{\includegraphics{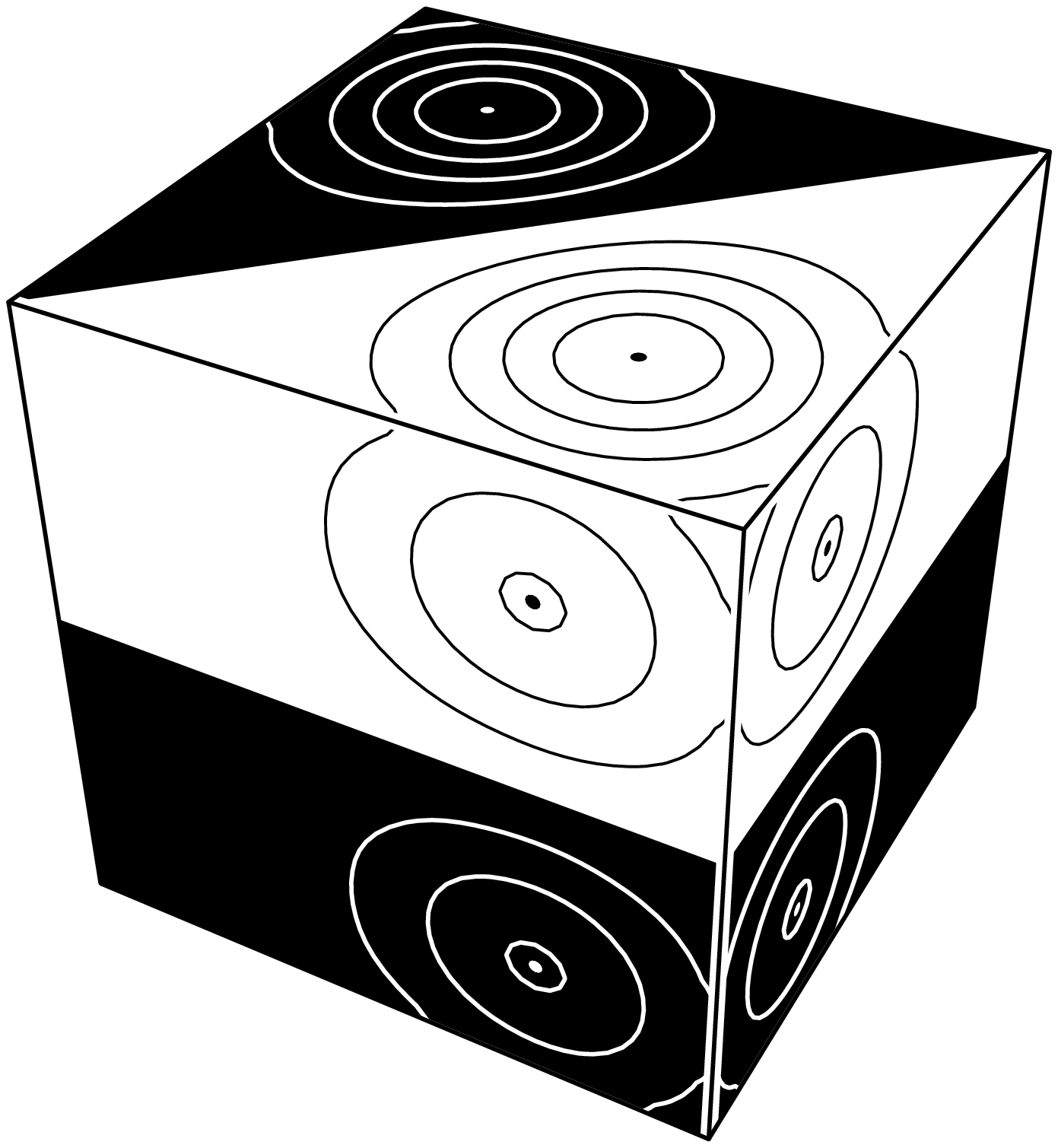}} \\
\raise2cm\hbox{$F$: $S_{14}$}\hskip1cm
\scalebox{\flagScale}{\includegraphics{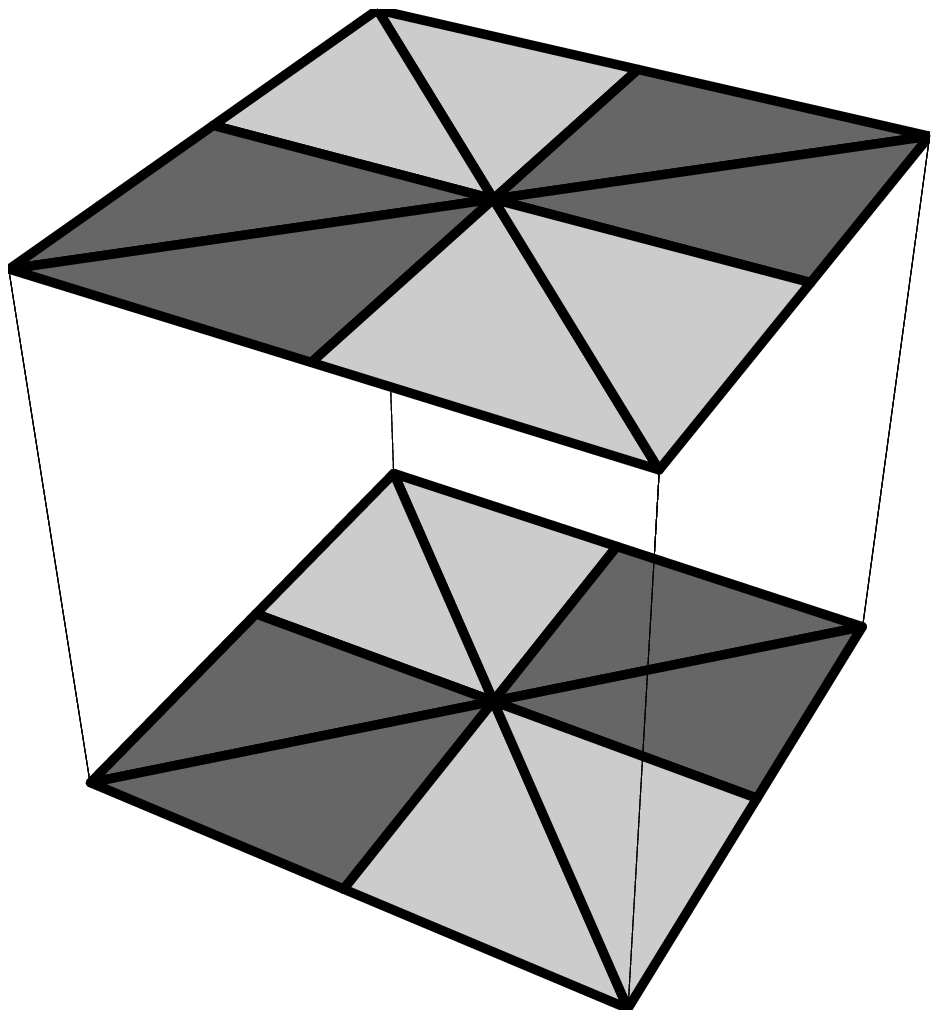}} &
\scalebox{\solScale}{\includegraphics{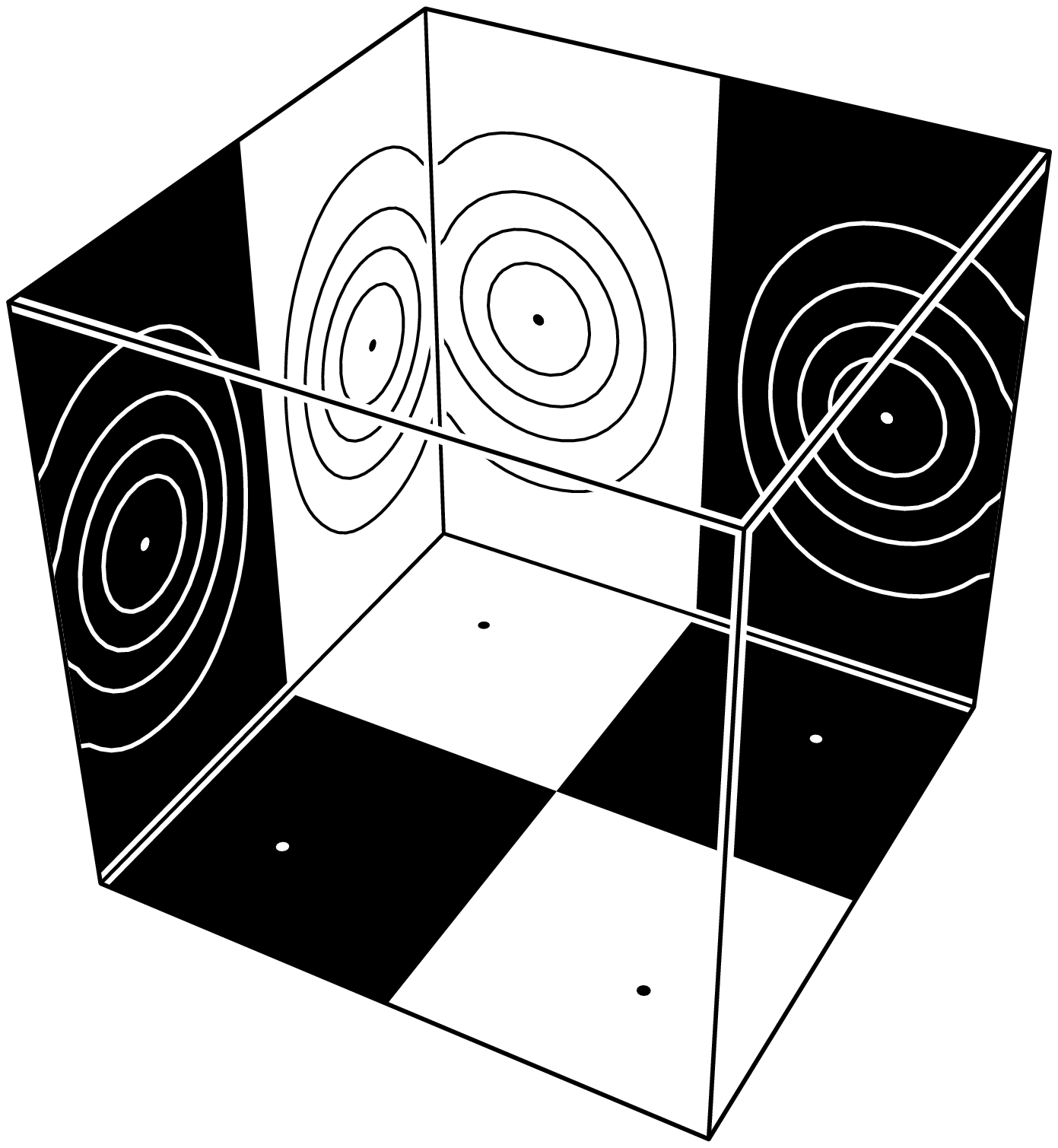}}  &
\scalebox{\solScale}{\includegraphics{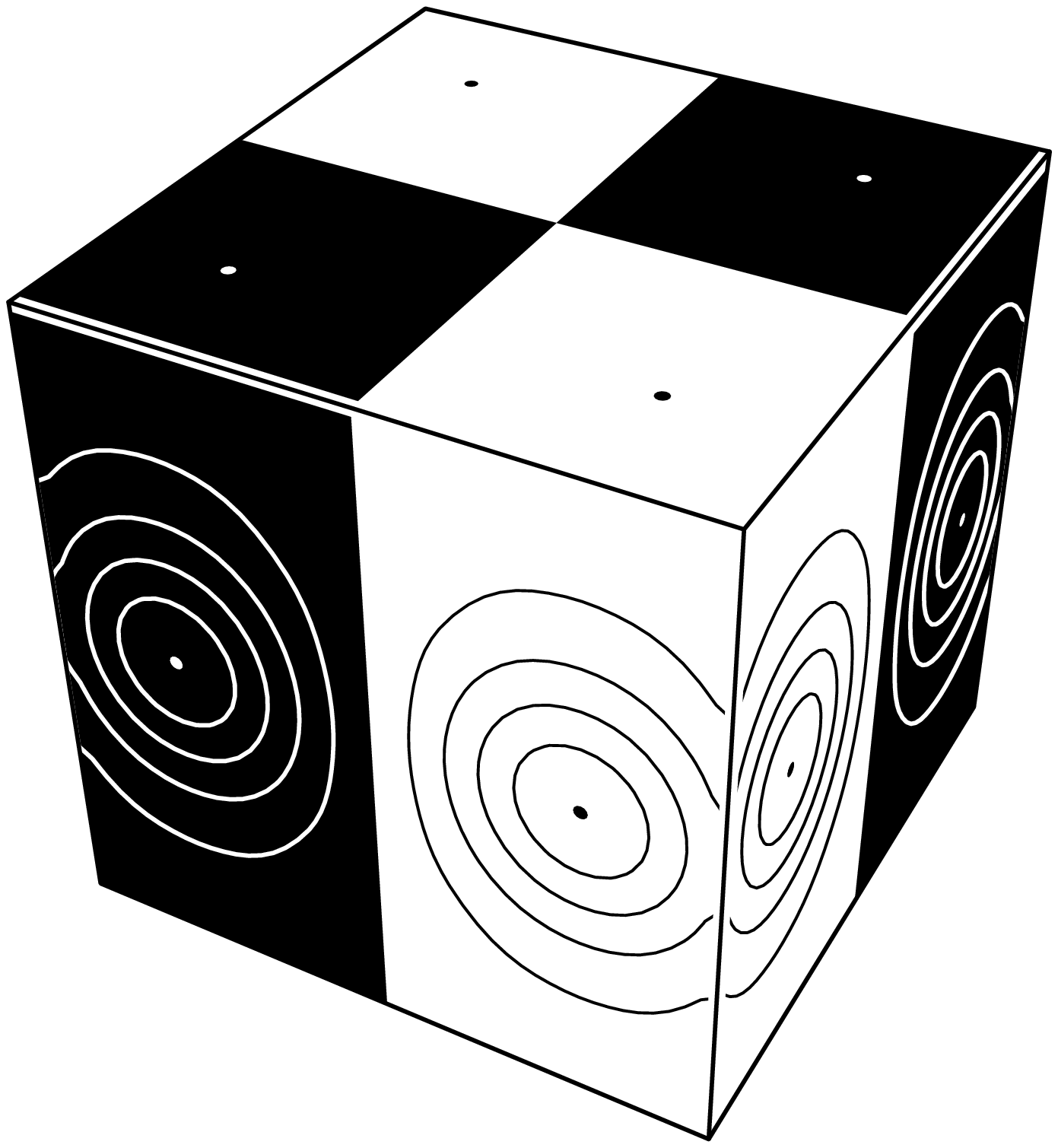}} \\
\raise2cm\hbox{Secondary: $S_{51}$}\hskip1cm
\scalebox{\flagScale}{\includegraphics{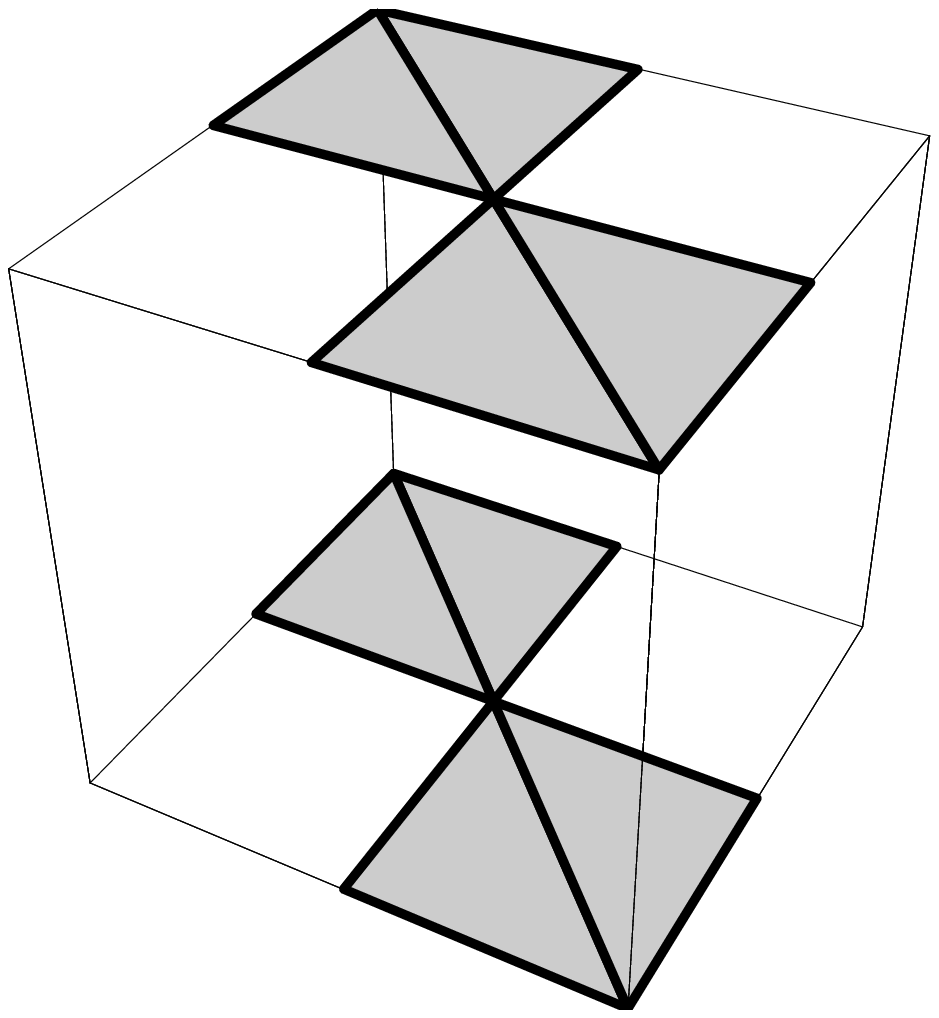}} &
\scalebox{\solScale}{\includegraphics{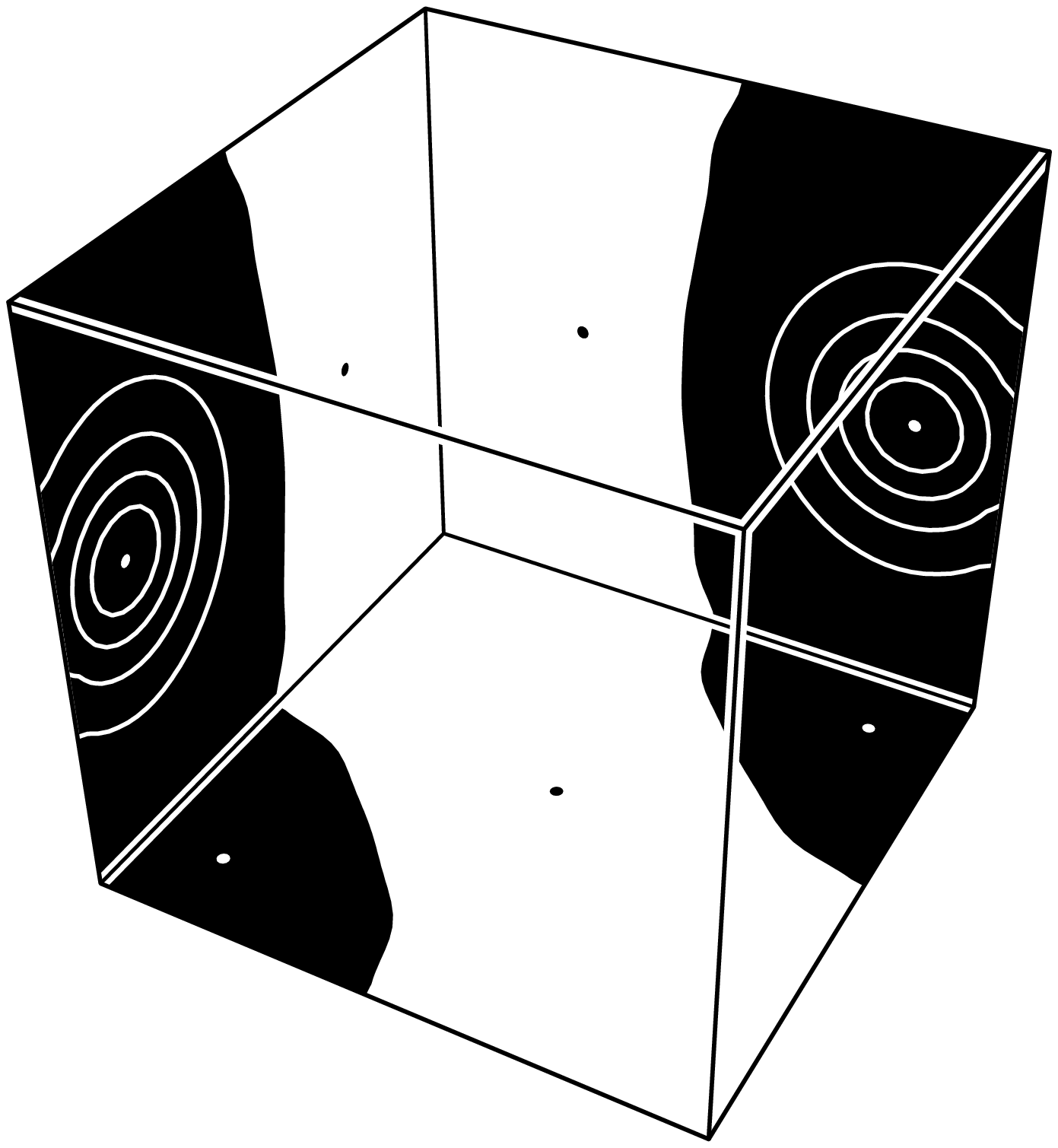}}  &
\scalebox{\solScale}{\includegraphics{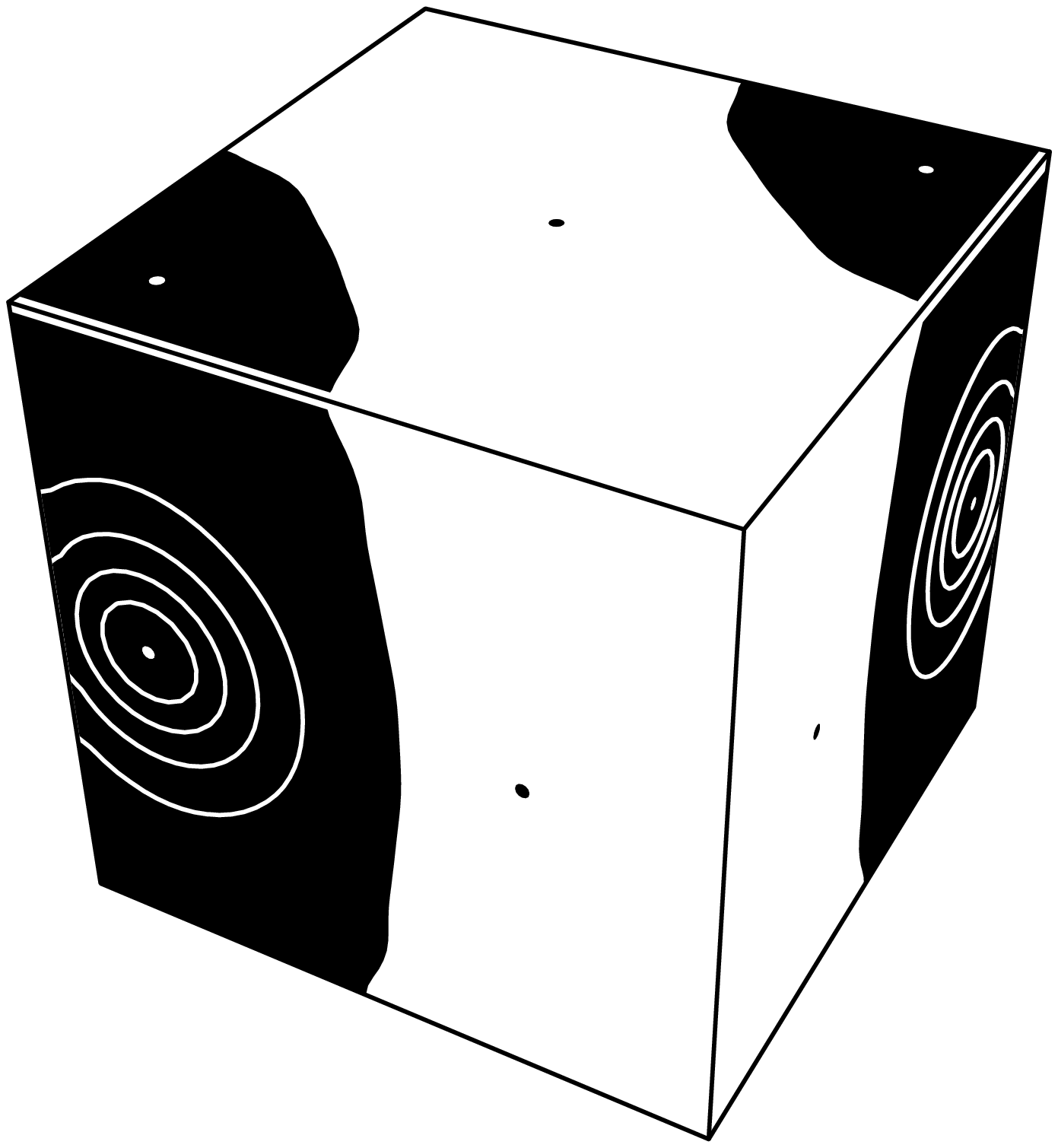}} \\
\raise2cm\hbox{Tertiary: $S_{80}$}\hskip1cm
\scalebox{\flagScale}{\includegraphics{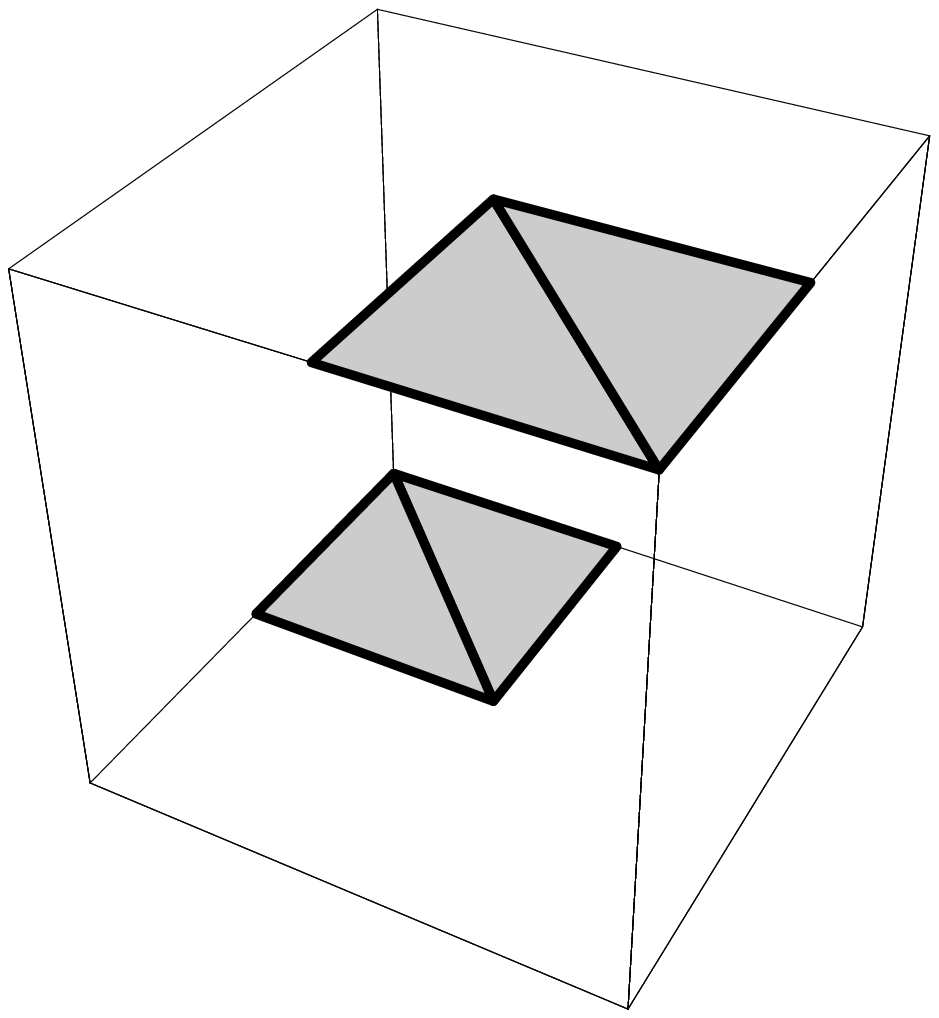}} &  
\scalebox{\solScale}{\includegraphics{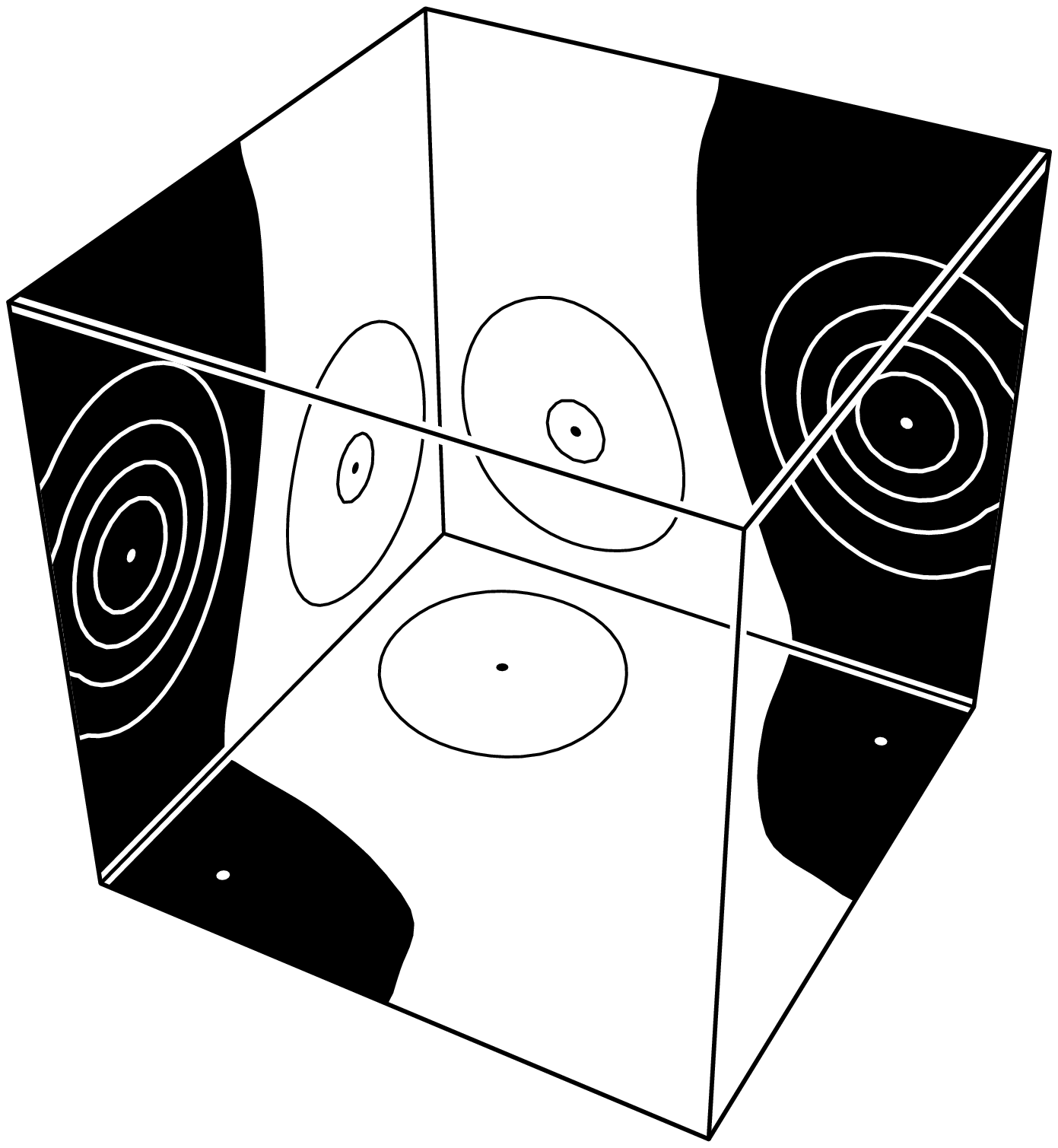}}  &  
\scalebox{\solScale}{\includegraphics{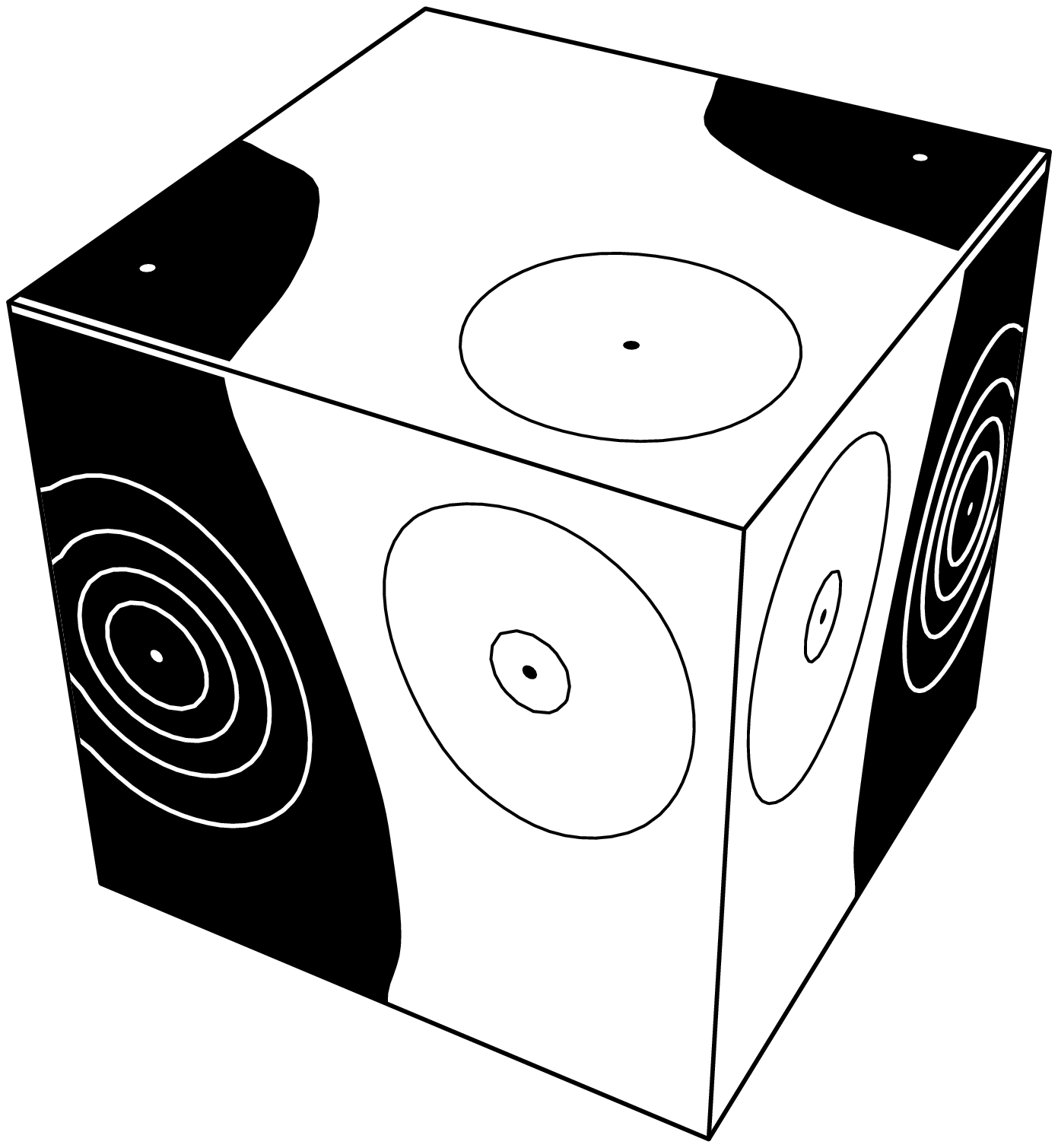}} \\
\end{tabular}
\caption{Contour plots for the vertex, edge, and face solutions that bifurcate at $s = 9$
in Figure~\ref{s0.10}, along with a daughter and granddaughter of the face solution.
The contour plots are for the solutions at $s = 0$.
}
\label{secondOh}
\end{figure}

The trivial solution undergoes a bifurcation with $\Oh$ symmetry at both $s = \lam_{1,1,2} = 6$ and
$s = \lam_{1,2,2} = 9$.
The bifurcations are very similar.
The critical eigenspaces for the two bifurcations are
$$
\tilde E_6 = \spn\{ \psi_{2,1,1} , \psi_{1,2,1} , \psi_{1,1,2} \}, \ \text{and} \
\tilde E_9 = \spn\{ \psi_{1,2,2} , \psi_{2,1,2} , \psi_{2,2,1} \},
$$
respectively.
The kernel of the action of $\Gamma_0 = \Oh \times \Z_2$ on $\tilde E_6$ is
$ \Gam_0' = \langle (I_3, -1) \rangle $
and the kernel of the action of $\Gamma_0$ on $\tilde E_9$ is $\Gam_0' = \langle (-I_3, -1) \rangle$.
In both cases,
$$
\Gam_0/\Gam_0' = \langle (R_{90}, 1) \Gam_0', (R_{120}, 1) \Gam_0', (R_{180}, 1) \Gam_0' \rangle \cong \Oh.
$$
The reduced gradient $\tilde g: \tilde E \rightarrow \tilde E$ in both cases has the equivariance indicated
in the first image of Figure~\ref{irredSpaces3}.
The Equivariant Branching Lemma (EBL) guarantees that under certain non-degeneracy conditions there
is a bifurcating branch tangent to each
one-dimensional fixed-point spaces in $\tilde E$.
These fixed-point subspaces intersect a cube in $\tilde E$ at the center of a
face, the center of an edge, or a vertex of the cube.
Thus, each EBL branch is made up of face, edge, or vertex solutions.
The standard choice of representative in each symmetry type is $\Gam_i \in S_i$, for $0 \leq i \leq 98$.
The contour plots of the bifurcating solutions in Figure~\ref{s0.10c} show the solution in the fixed-point
subspace indicated here:
$$
\begin{aligned}
\text{face }\ 
& [ \fix(\Gam_{12}, \tilde E_6) ] = [ \fix(\Gam_{14}, \tilde E_9) ] = 
  \{ (0, 0, a) \mid a \in \R\},\\
\text{vertex }\ 
& [ \fix(\Gam_{22}, \tilde E_6) ] = [ \fix(\Gam_{21}, \tilde E_9) ] = 
  \{ (a, a, a) \mid a \in \R\},\\
\text{edge }\ 
& [ \fix(\Gam_{44}, \tilde E_6) ] = \{ (-a, a, 0) \mid a \in \R\} \ \text{is conjugate to} \\
& [ \fix(\Gam_{45}, \tilde E_9) ] = \{ (a, a, 0) \mid a \in \R\}.
\end{aligned}
$$
The symmetry type containing $\Gam_{12}$, denoted $S_{12}$, has 3 elements, and the three conjugate face
directions in $[\tilde E]$ are the coordinate axes.
Similarly, $[ \Gam_{22} ] = S_{22}$ has 4 elements, corresponding to the four diagonals through vertices of
the cube centered in $\R^3$.
The edge solutions bifurcating at $s = 6$ have symmetry type $[\Gam_{44} ] = S_{44}$, which has 6 elements.

For the bifurcation at $s = 6$,
the geometry in $\tilde E_6$ is mirrored in the geometry of the solutions.
For example,
the maximum $u$ value for a vertex solution (type $S_{22}$) lies on the line from the origin to a vertex in $\Omega$.
Similarly, the maximum $u$ value for an edge or face solution is on the line from the origin to an edge
or face, respectively.
The three face directions in $\tilde E$ are $\psi_{2,1,1}$, $\psi_{1,2,1}$, and $\psi_{1,1,2}$, which
can be thought of as ``$x$'', ``$y$'', and ``$z$'' functions.
Note that the face solution in Figure~\ref{s0.10c} is approximately a multiple of $\psi_{1,1,2}$.

The geometry in $\tilde E_9$ for the bifurcation at $s = 9$ is the same.  The
geometry of the bifurcating solutions, shown in Figure~\ref{secondOh}, is more subtle though.
The face solutions have symmetry type $S_{14}$,
and the ``$z$'' eigenfunction is $\psi_{2,2,1}$.  
The face and edge solutions have a line where two nodal
planes intersect at right angles, and these lines intersect the midpoint of a face and edge of $\Omega$,
respectively.  However, the vertex solutions do not have an intersection of nodal planes.  Instead,
the vertex solutions have an axis of three-fold symmetry that intersects a vertex in $\Omega$.

The face solutions (type $S_{14}$) that bifurcate at $s = 9$ have a bifurcation at $s \approx 6.60$,
as seen in Figure~\ref{s0.10}.  The secondary branch (type $S_{51}$) itself has a bifurcation
that creates a tertiary branch with type $S_{80}$.  Solutions from these new branches are shown in 
Figure~\ref{secondOh}.

\subsection{A degenerate bifurcation with $\D_6$ symmetry}

\begin{figure}
\scalebox{1.0}{\input{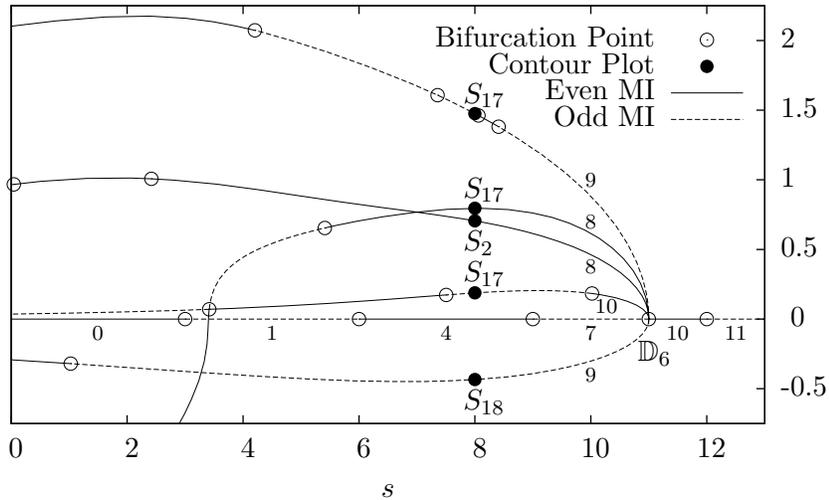}}
\caption{
Bifurcation with $\D_6$ symmetry in PDE~(\ref{pde}) on the cube, at $s = 11$.
The symmetry of $\tilde g$ on the critical eigenspace $\tilde E$
is shown in Figure~\ref{D6bif}.
There are 5 (conjugacy classes of) branches that bifurcate at $s = 11$.  One branch in each class
is shown in this bifurcation diagram,
and the solid dots indicate the solutions with the contour plots in Figure~\ref{bifdiag4c}.
(The upper solid dot is very close to a bifurcation point.)
There is an anomaly-breaking bifurcation at $s \approx 3.417$, where mother and daughter both have
symmetry type $S_{17}$.
}
\label{bifdiag4}
\end{figure}

\begin{figure}
\begin{tabular}{ccc}
\scalebox{.36}{\includegraphics{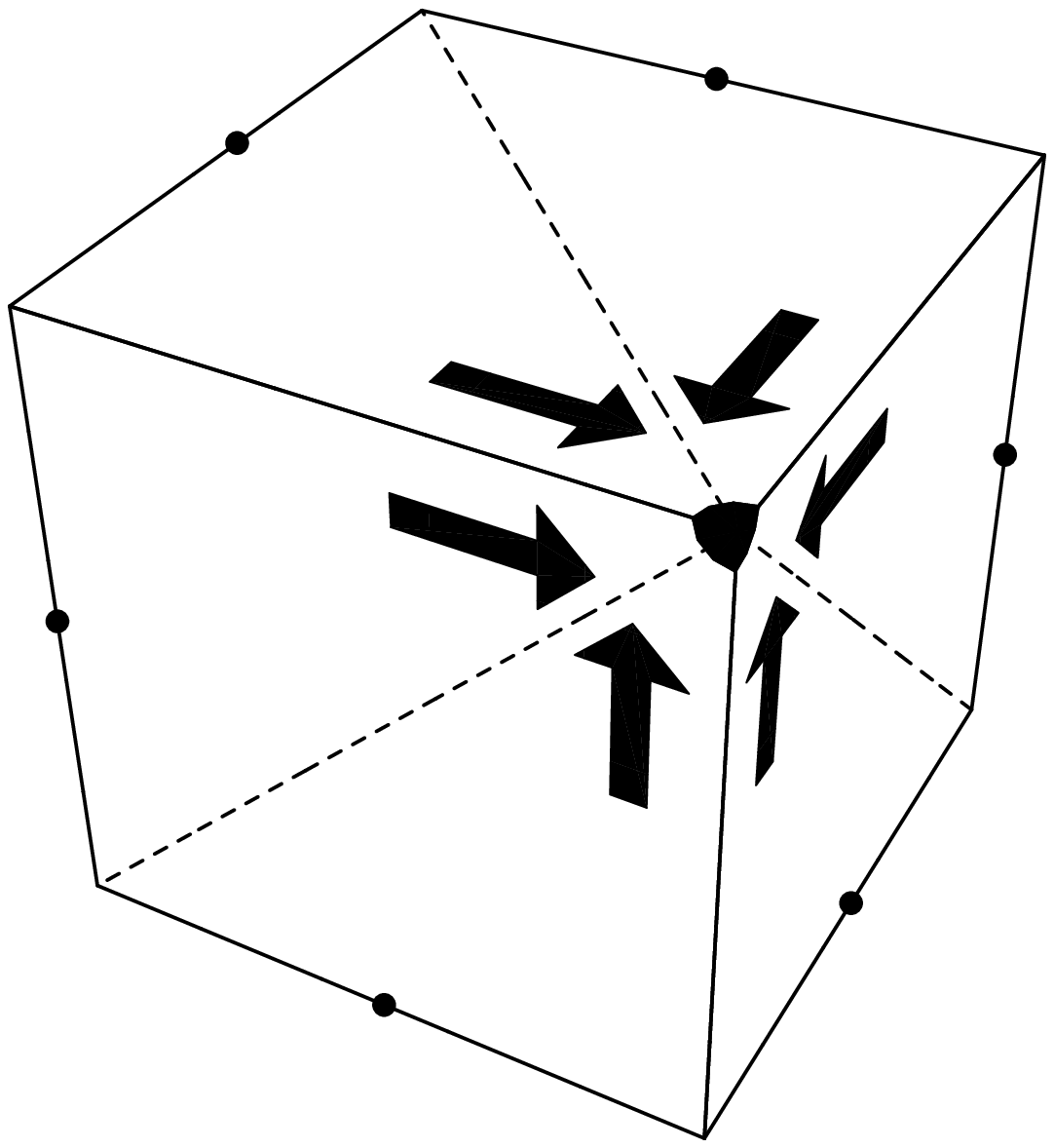}} &
\scalebox{.36}{\includegraphics{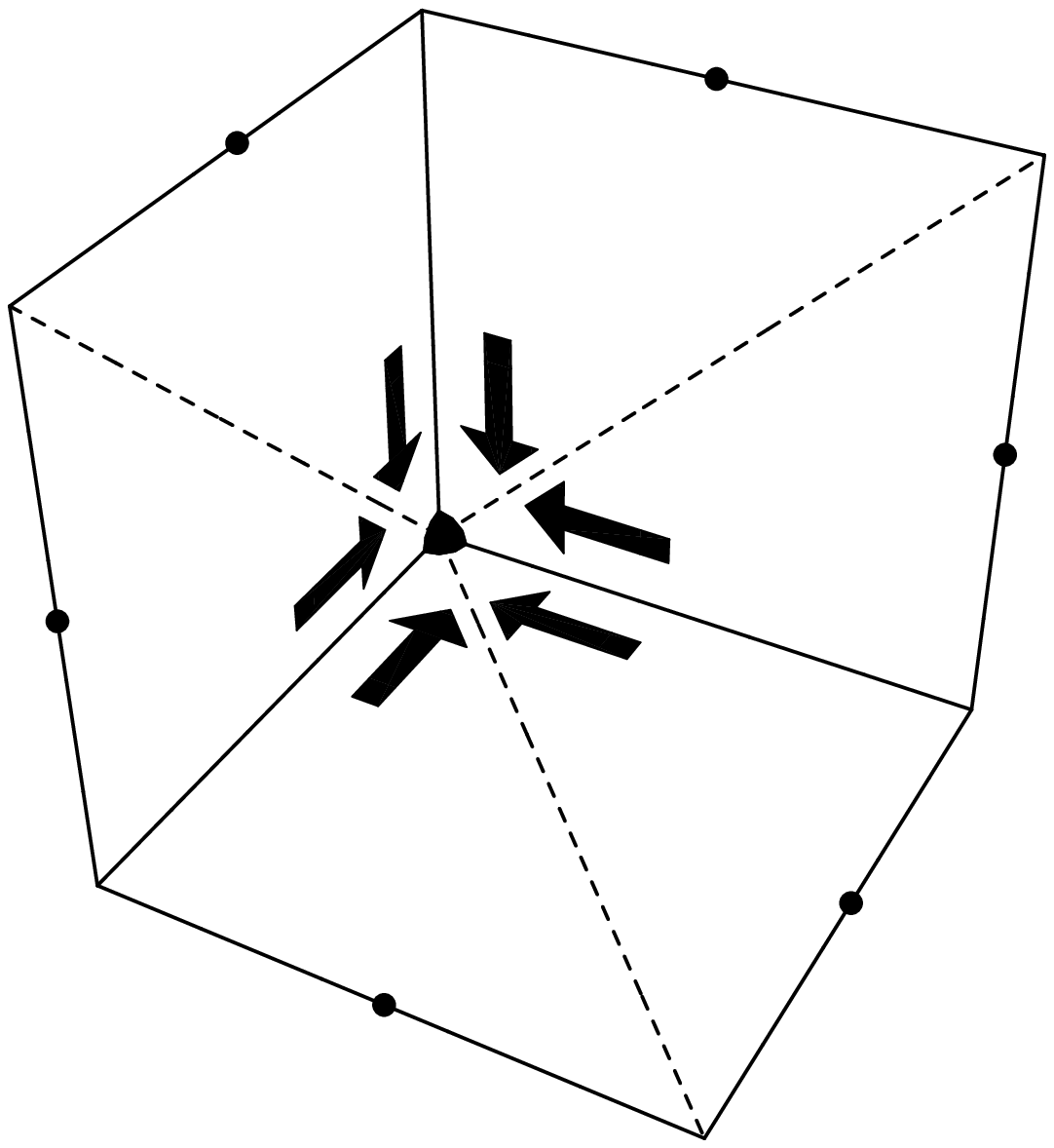}} &
\scalebox{.36}{\includegraphics{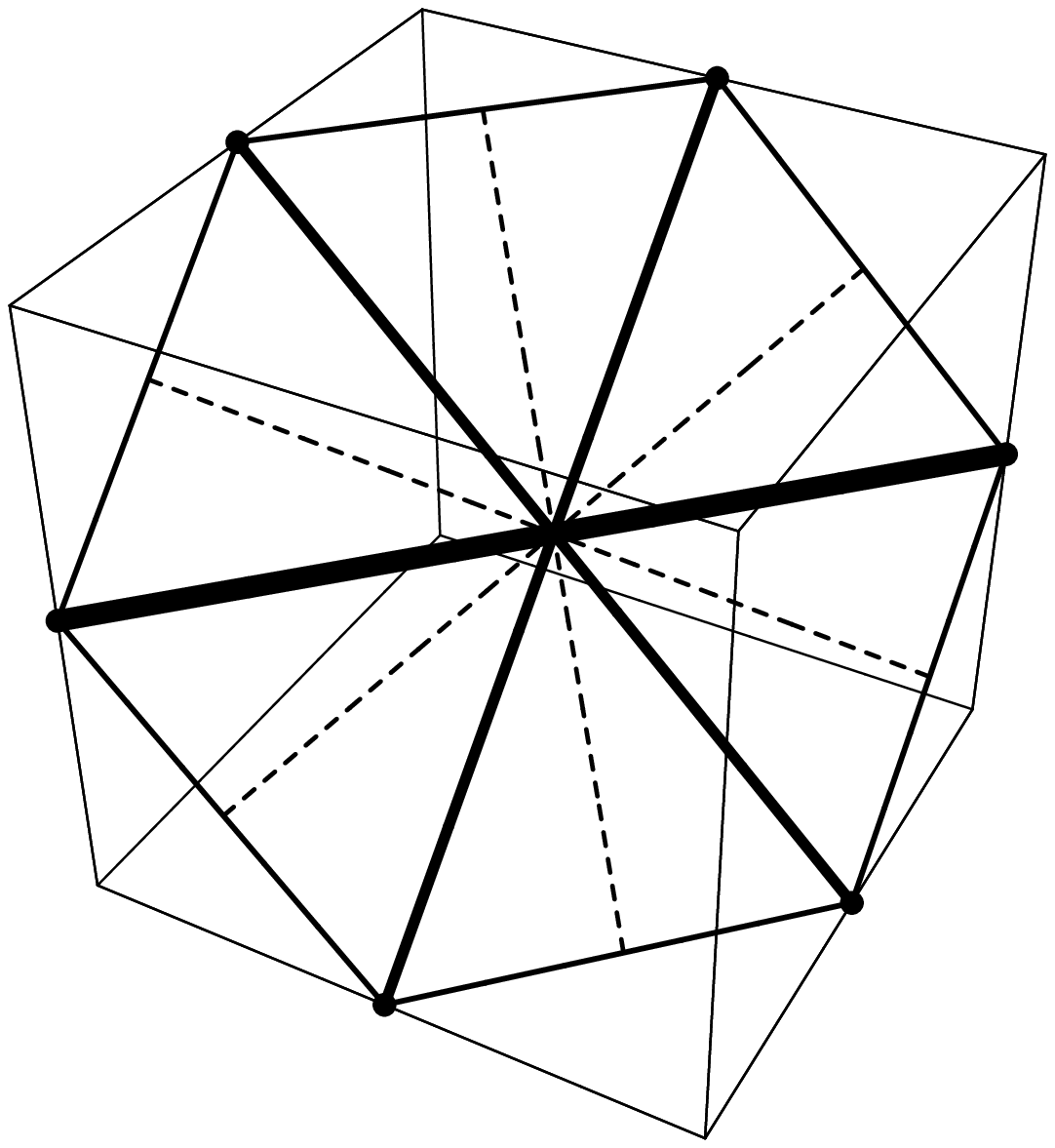}}
\end{tabular}
\caption{
The critical eigenspace $\tilde E$ at the bifurcation with $\D_6$ symmetry
of the trivial solution at $s = 11$.
The first two figures are similar to those of Figure~\ref{irredSpaces3}.
The arrows show the symmetry of the reduced gradient in $\tilde E$.
This is a degenerate bifurcation since $\tilde E$ is not an irreducible
representation space.  One of the diagonals of the cube in $\tilde E$ is
an irreducible subspace. The orthogonal subspace, which intersects the cube in a
hexagon as shown in the third figure, is another irreducible subspace.
}
\label{D6bif}
\end{figure}

\begin{figure}
\begin{tabular}{rccc}
\raise2cm\hbox{$S_{17}$}\hskip1cm
\scalebox{\flagScale}{\includegraphics{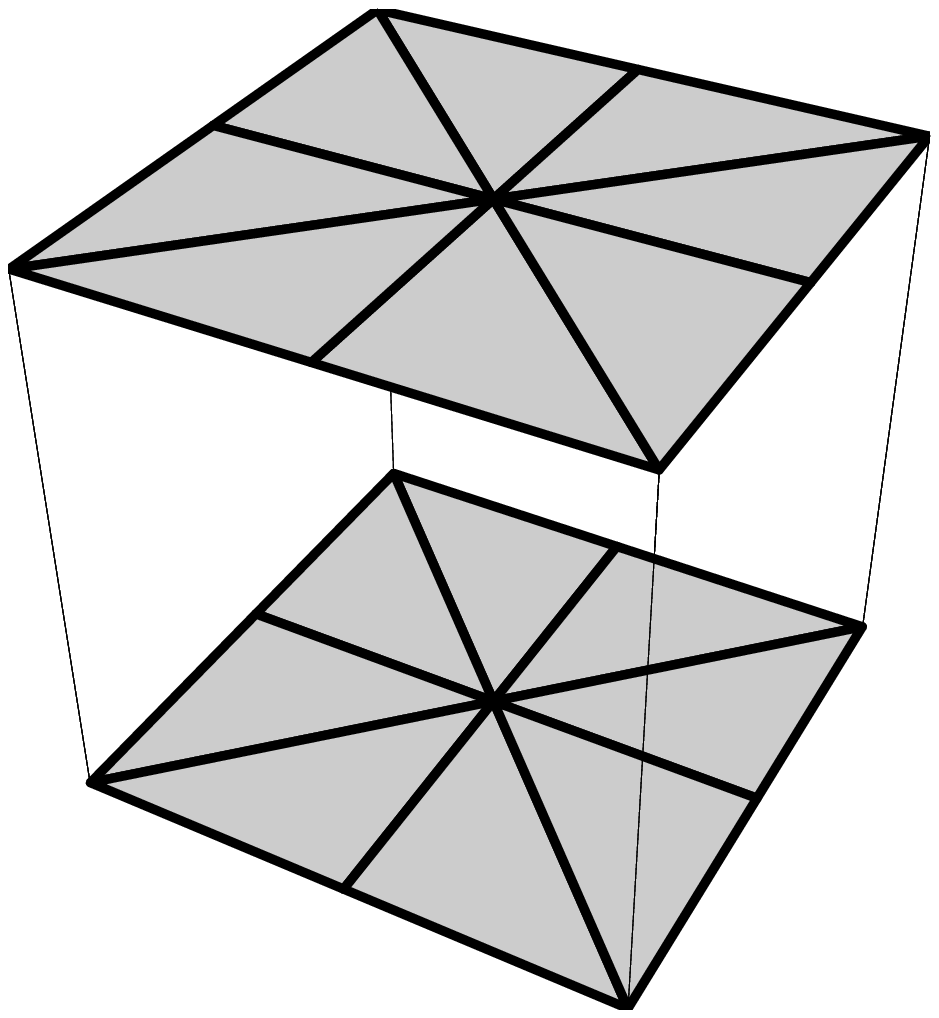}} &
\scalebox{\solScale}{\includegraphics{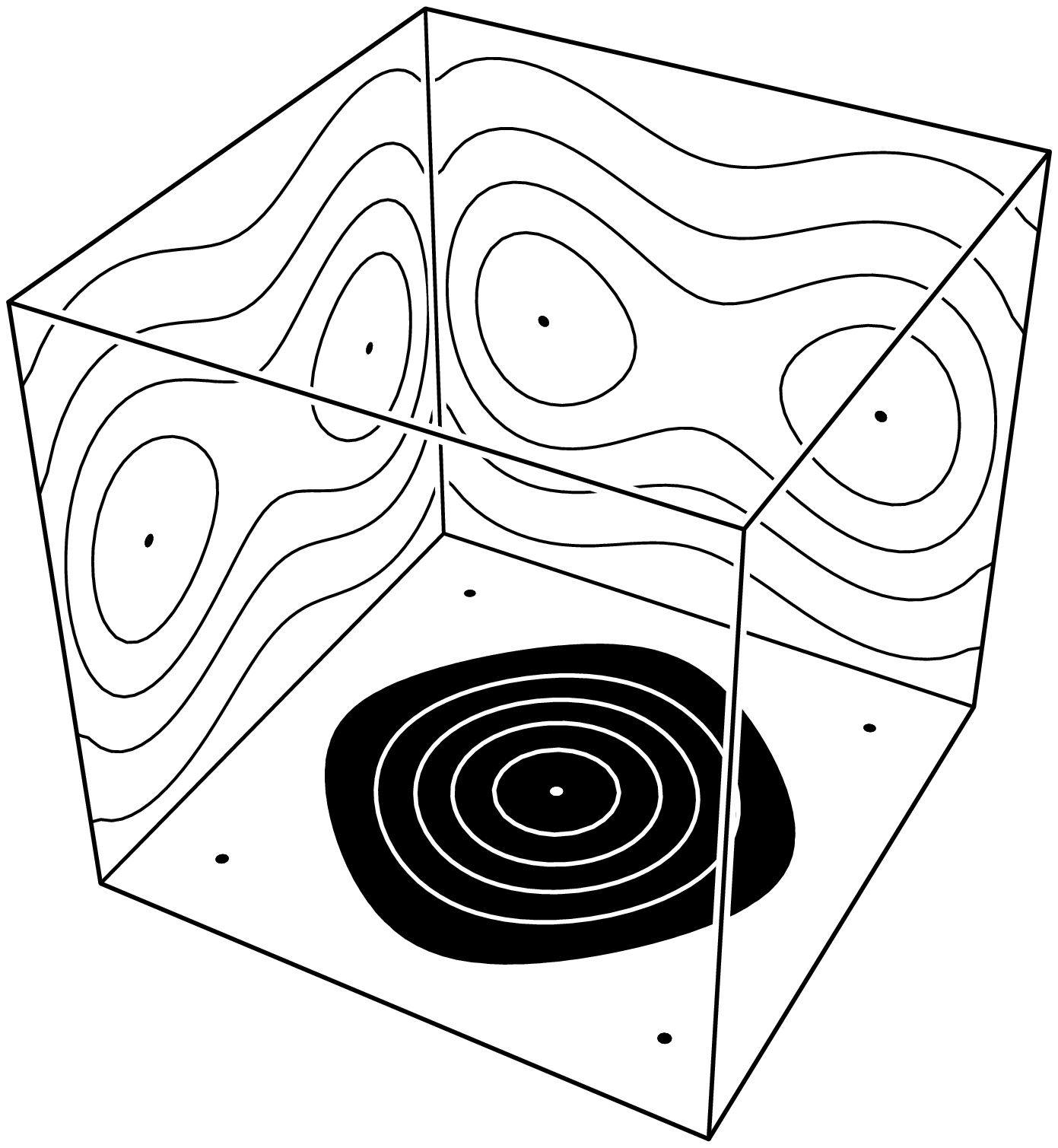}}  &
\scalebox{\solScale}{\includegraphics{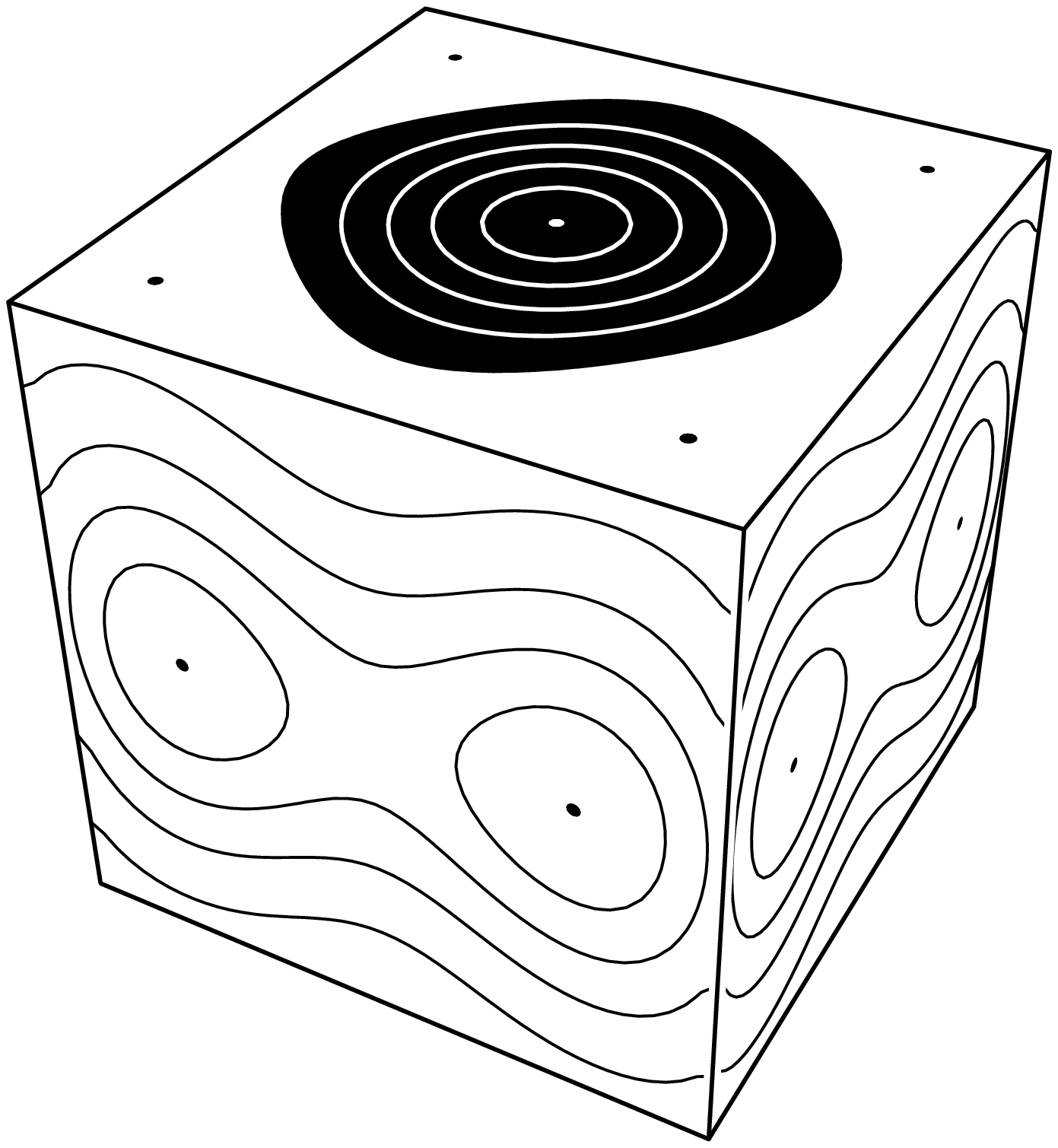}} \\
\raise2cm\hbox{$S_{17}$}\hskip1cm
\scalebox{\flagScale}{\includegraphics{figures/cube/fig.17.eps}} &
\scalebox{\solScale}{\includegraphics{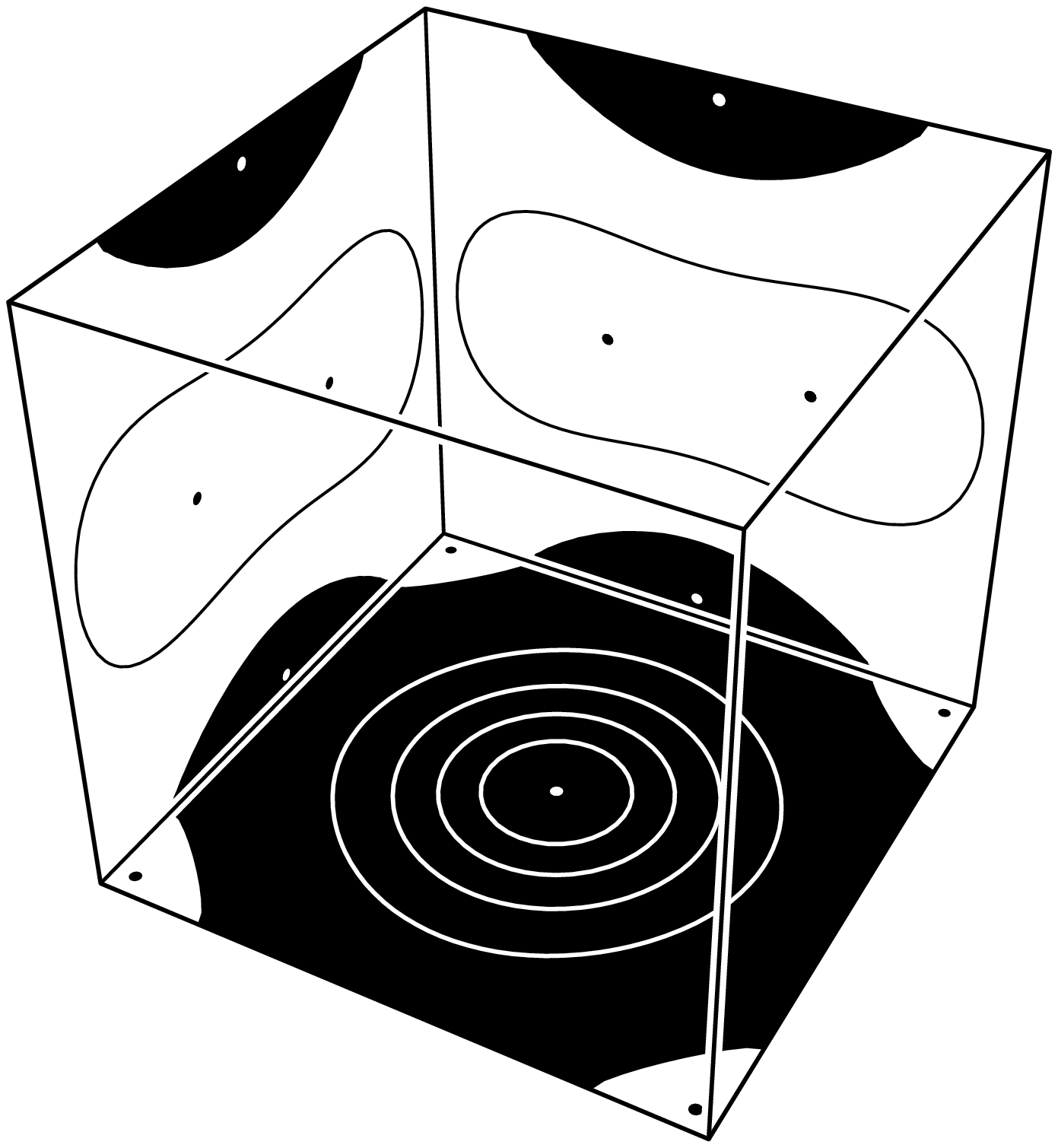}}  &
\scalebox{\solScale}{\includegraphics{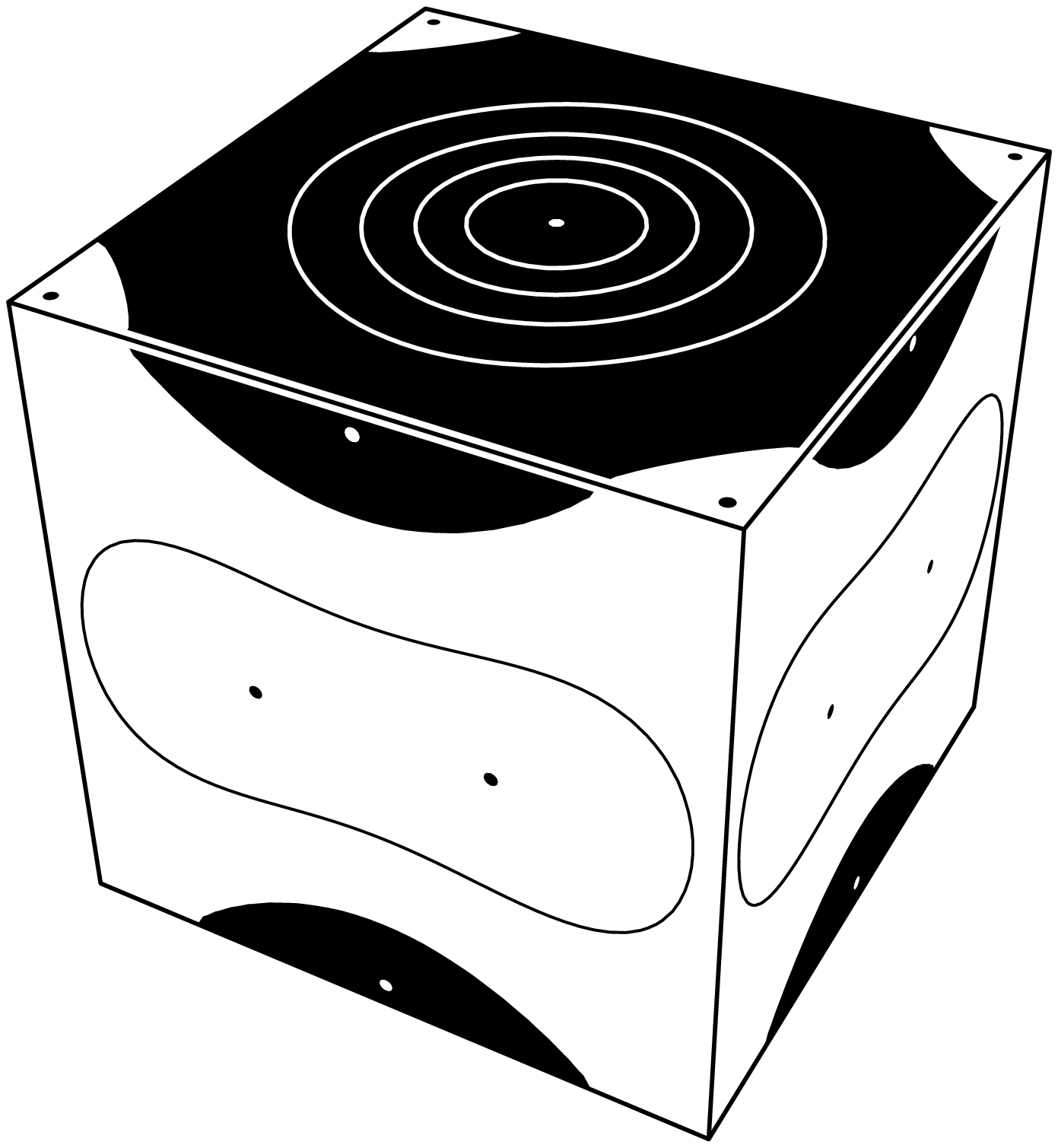}} \\
\raise2cm\hbox{$S_{2}$}\hskip1cm
\scalebox{\flagScale}{\includegraphics{figures/cube/fig.2.eps}} &
\scalebox{\solScale}{\includegraphics{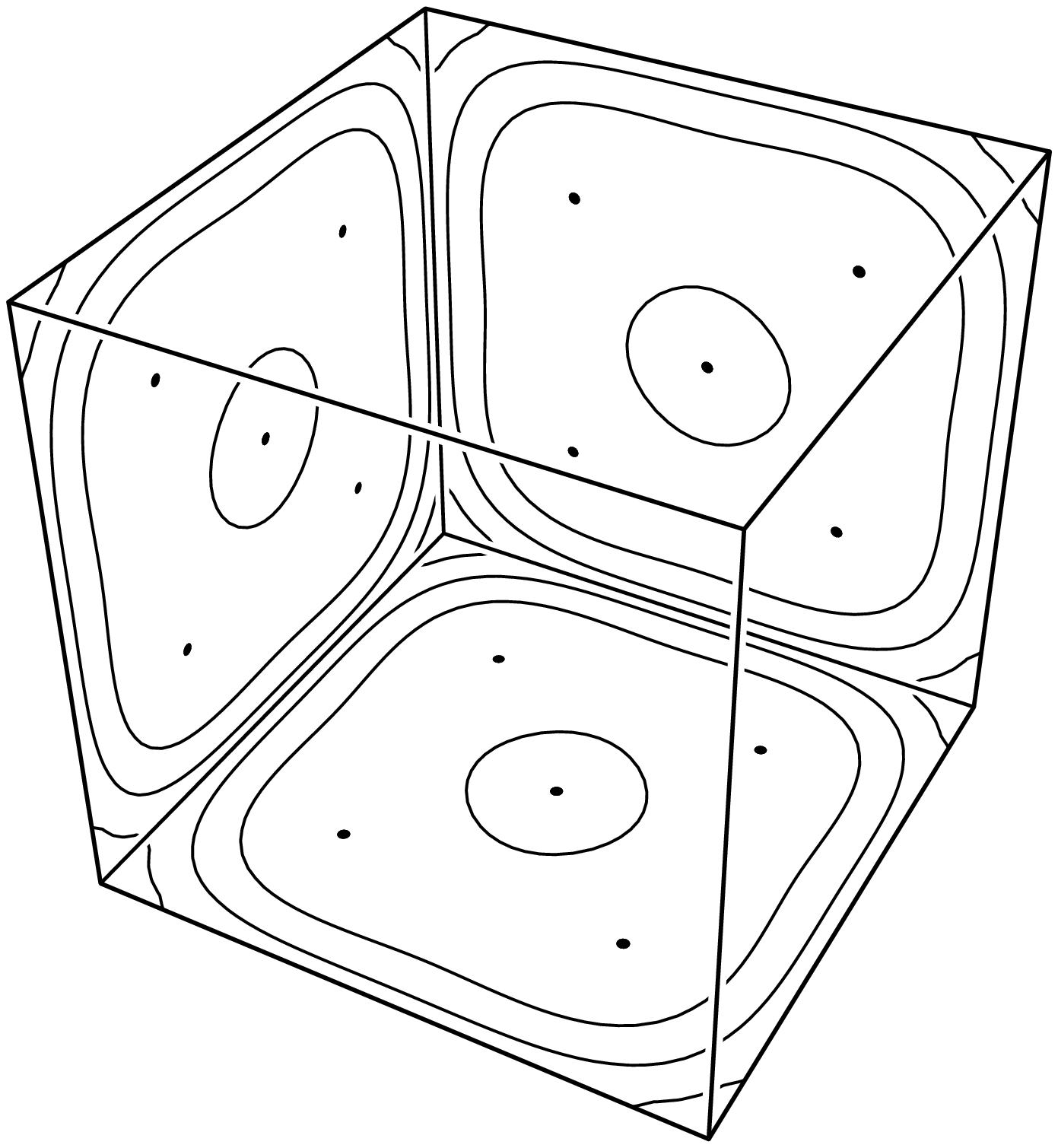}}  &
\scalebox{\solScale}{\includegraphics{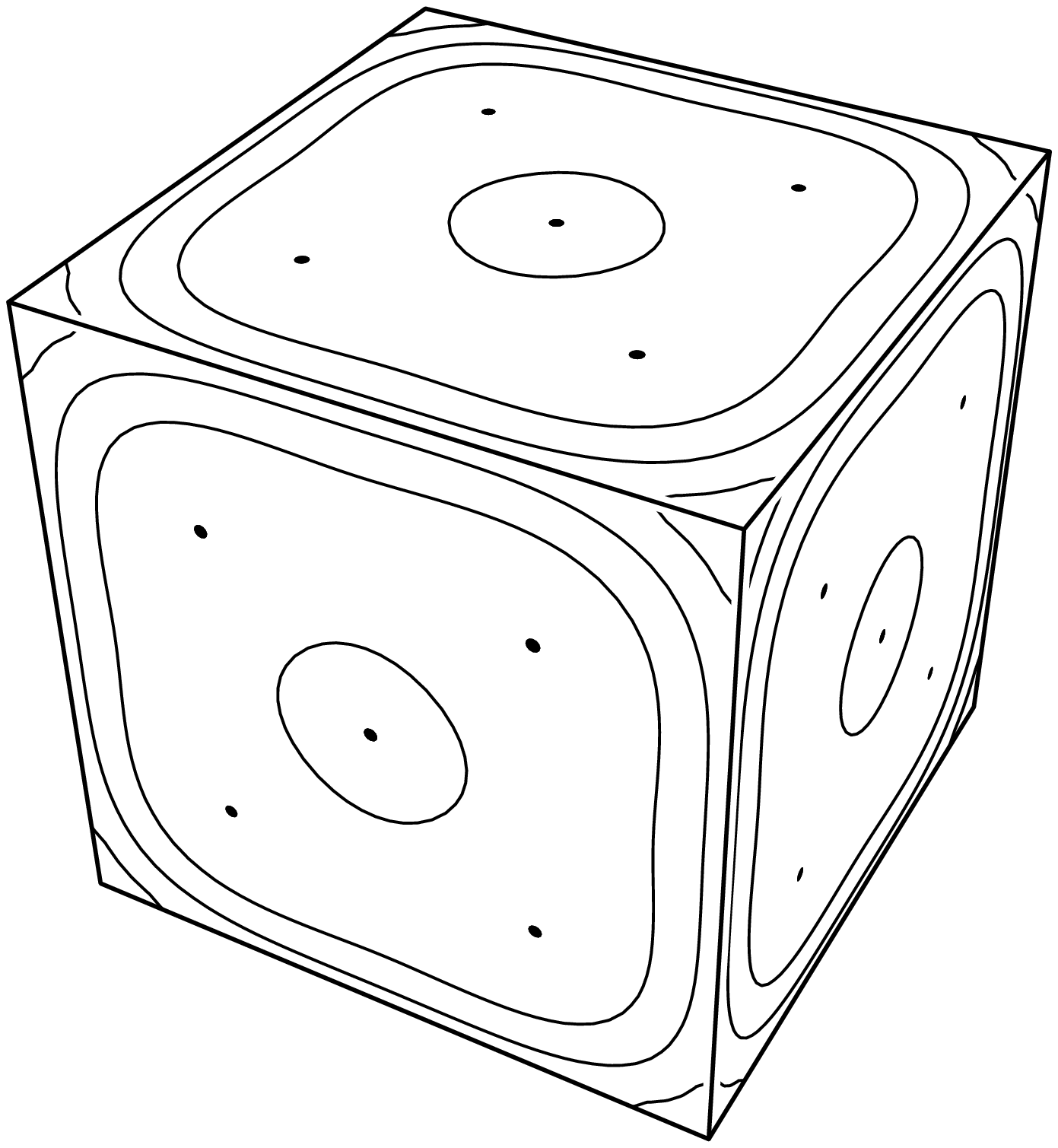}} \\
\raise2cm\hbox{$S_{17}$}\hskip1cm
\scalebox{\flagScale}{\includegraphics{figures/cube/fig.17.eps}} &
\scalebox{\solScale}{\includegraphics{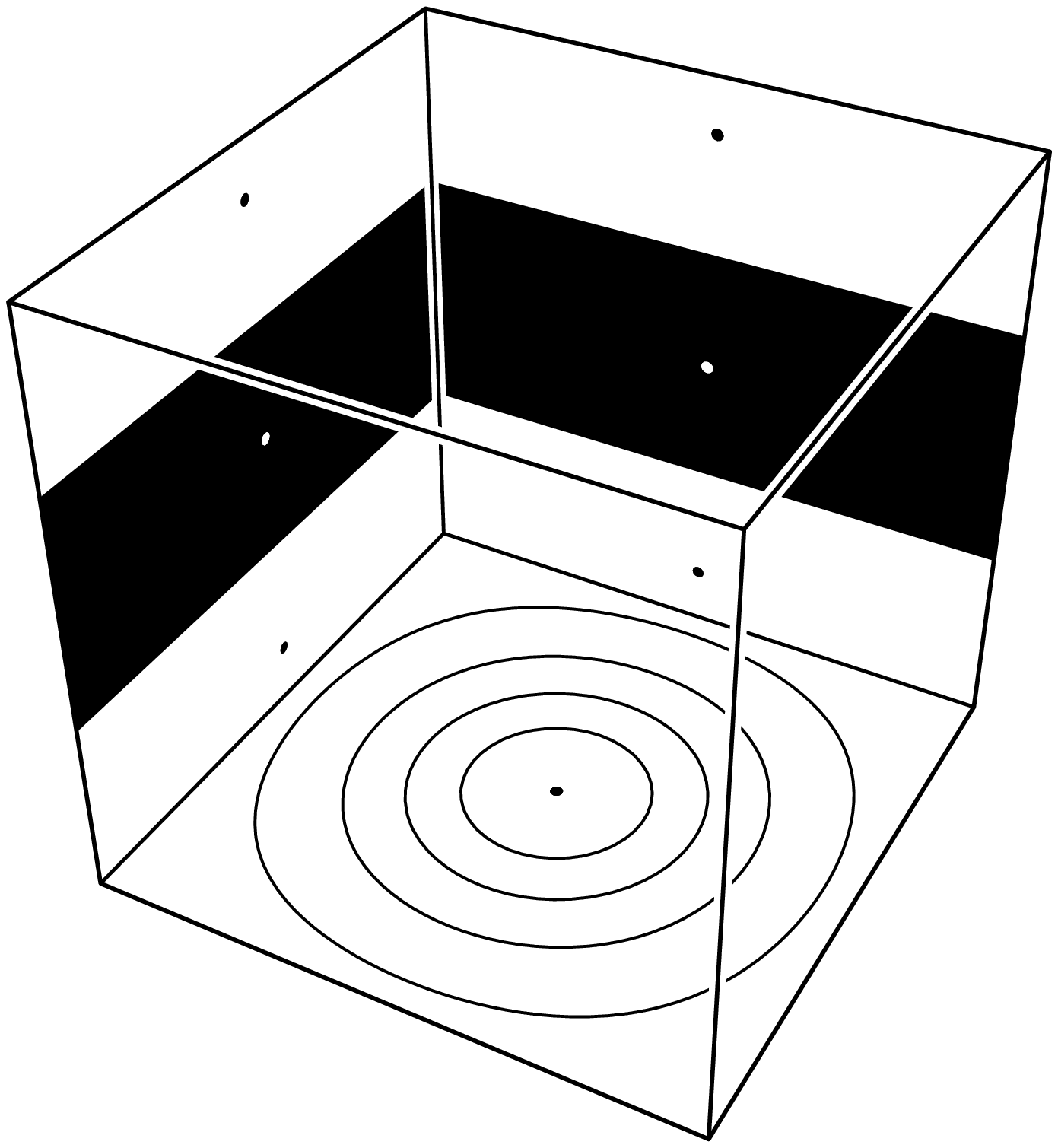}}  &
\scalebox{\solScale}{\includegraphics{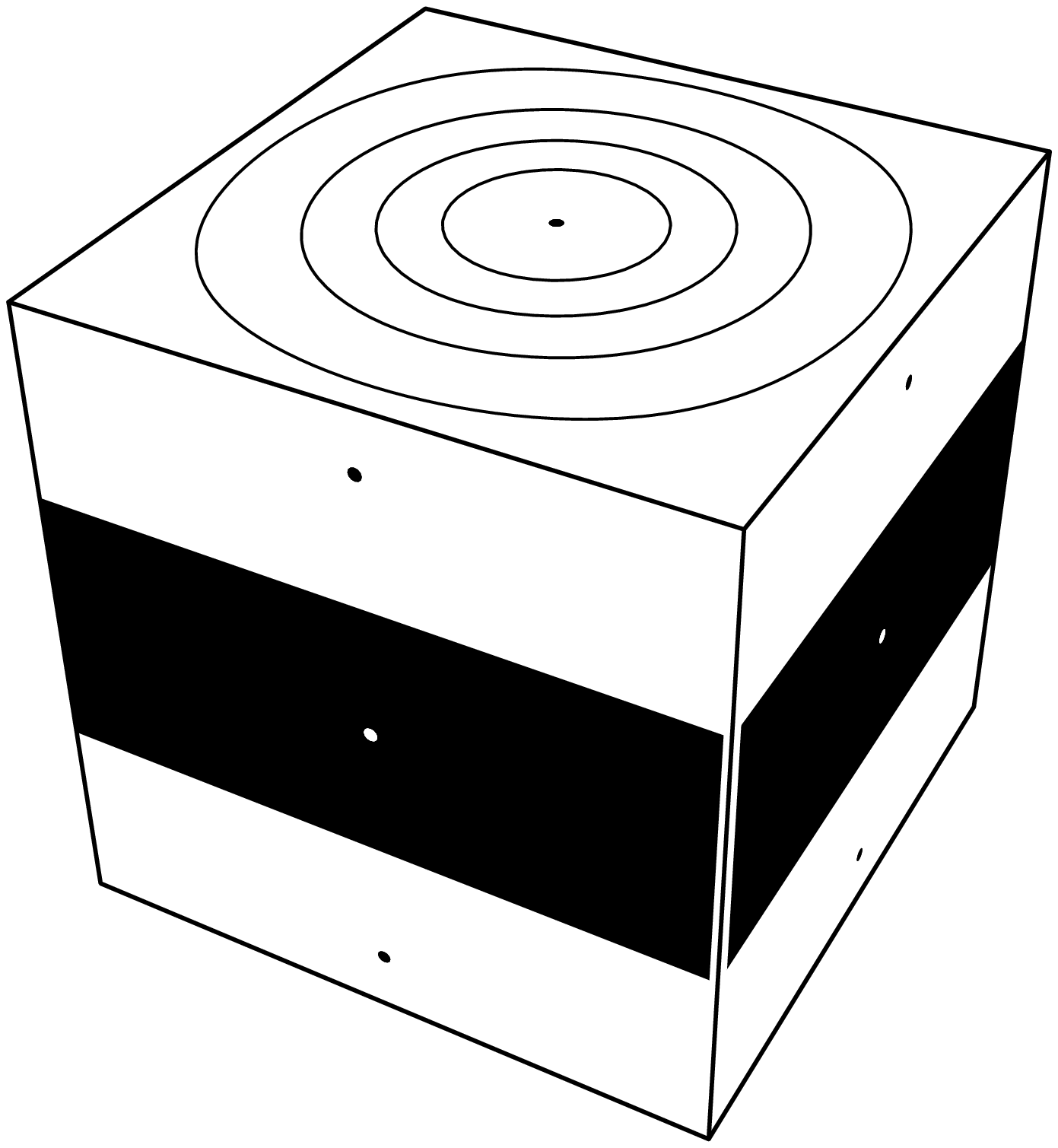}} \\
\raise2cm\hbox{$S_{18}$}\hskip1cm
\scalebox{\flagScale}{\includegraphics{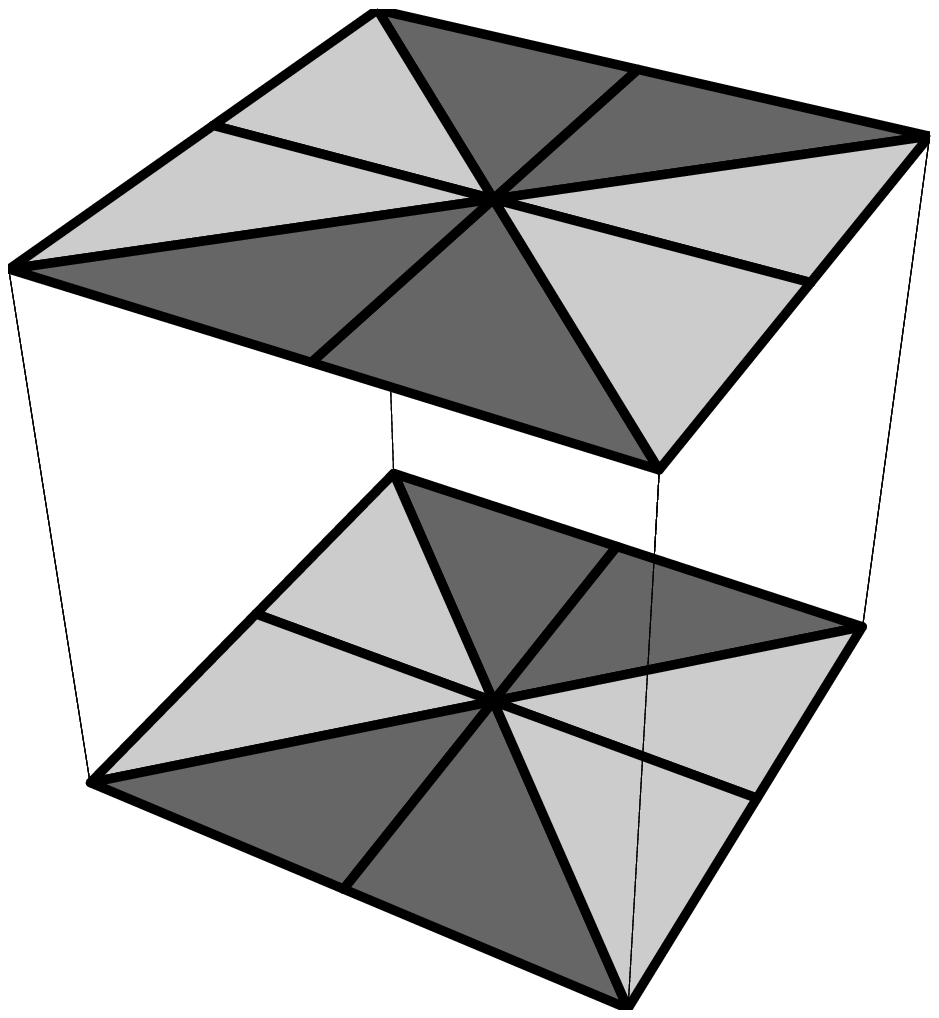}} &
\scalebox{\solScale}{\includegraphics{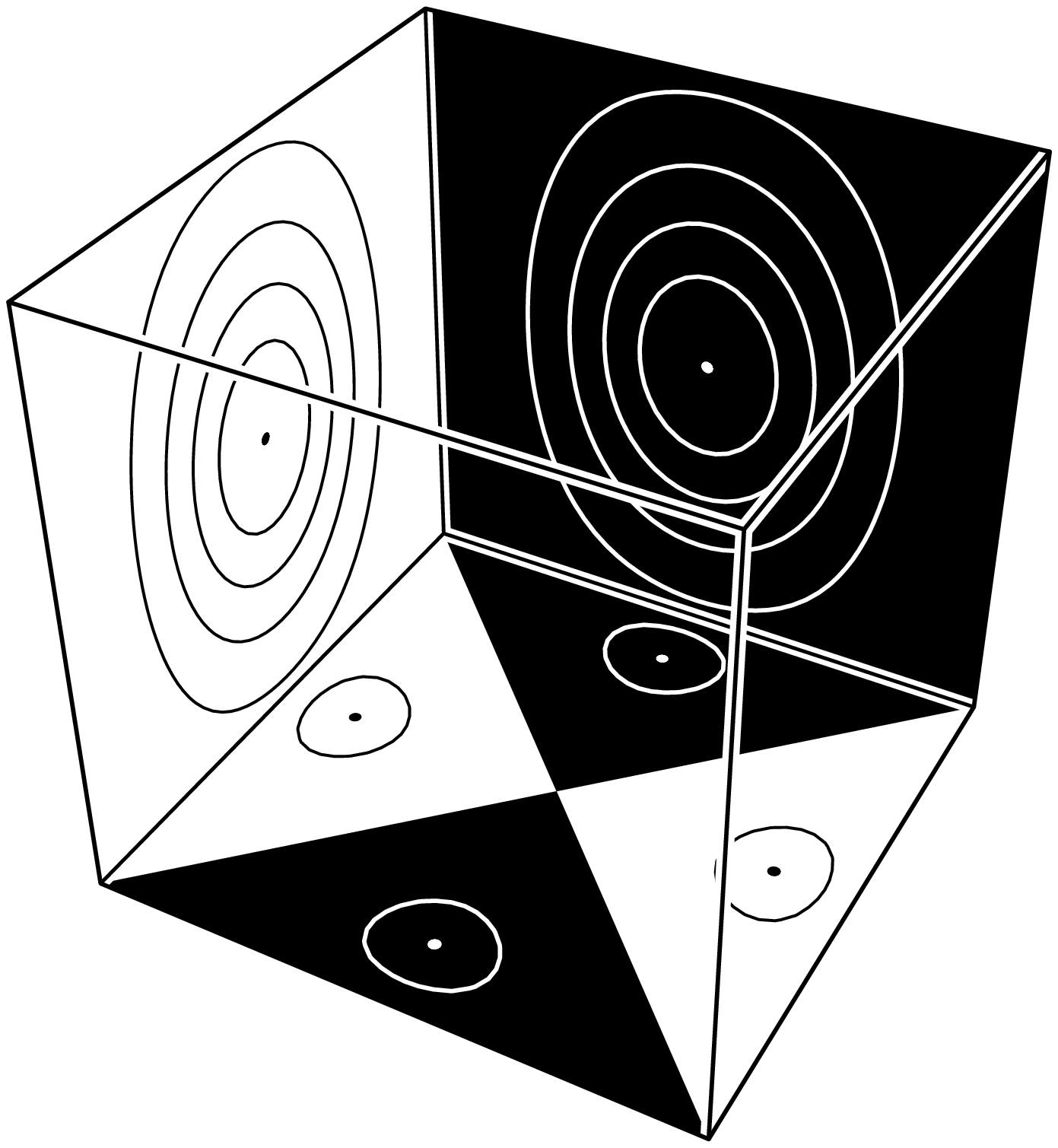}}  &
\scalebox{\solScale}{\includegraphics{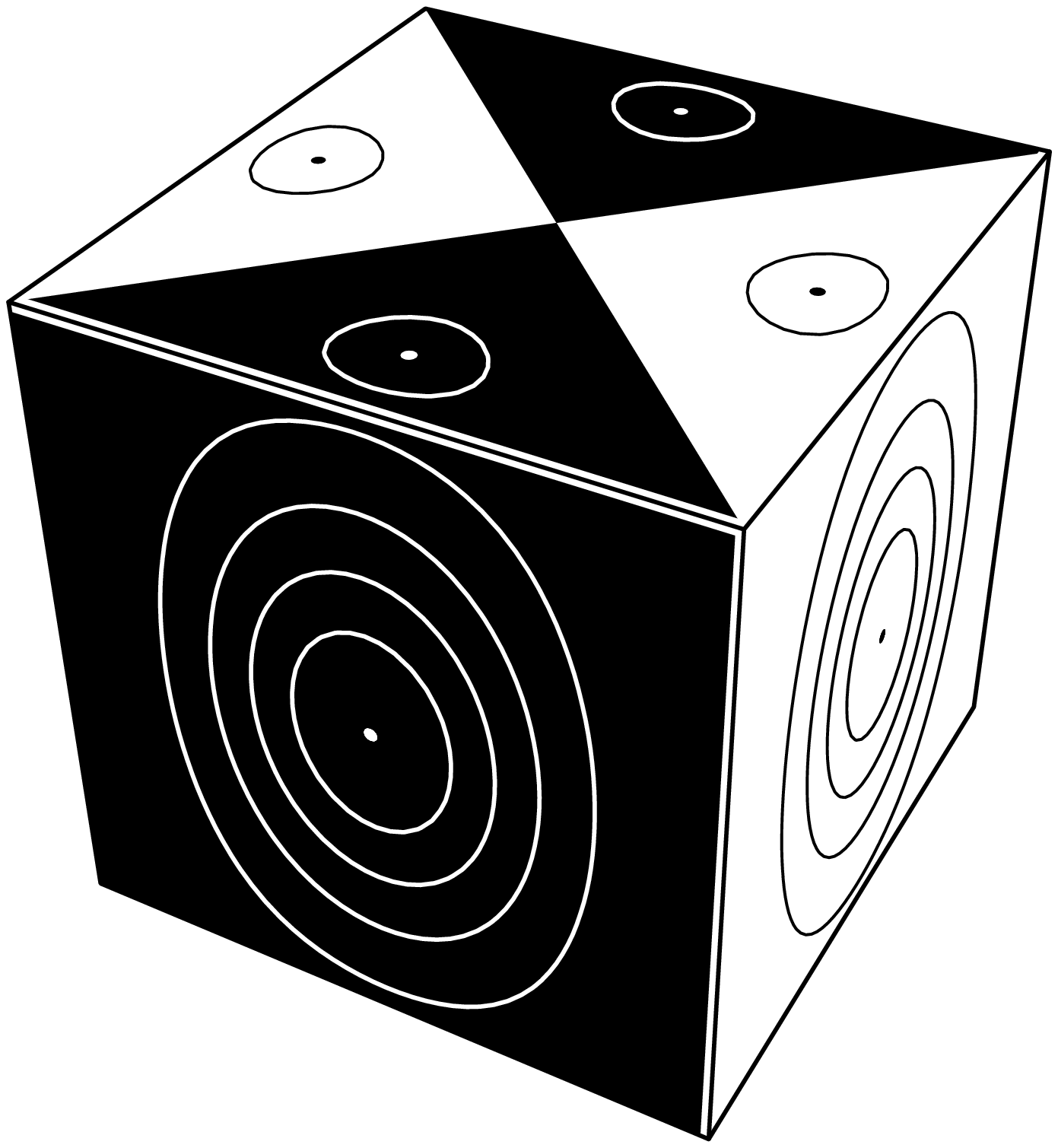}}
\end{tabular}
\caption{
Contour plots of solutions that bifurcate at $s = 11$.  The order of the solutions is the same
as that in Figure~\ref{bifdiag4}, from top to bottom.
The second solution branch intersects the fourth solution branch at an anomaly-breaking bifurcation.
As described in Figure~\ref{bifdiag4}, both have symmetry type $S_{17}$.
The fourth solution's contour plot shows a function which is the negative of the continuation
of the second solution, due to
the ink-saving heuristic that replaced $u$ by $-u$ (see Section~\ref{contour}).
}
\label{bifdiag4c}
\end{figure}

Figure~\ref{bifdiag4} shows the bifurcation diagram of the primary branches that bifurcate
from the fourth
eigenvalue $s = \lam_{3,1,1} = 11$.
This is a bifurcation with $\D_6$ symmetry; the action of $\D_6$ on the critical
eigenspace $\tilde E$ is shown in Figure~\ref{D6bif}.
Contour plots of the primary branches that bifurcate at $s = 11$ are shown in Figure~\ref{bifdiag4c}.

The critical eigenspace is the three-dimensional space
$\tilde E =\spn \{ \psi_{3,1,1} , \psi_{1,3,1} , \psi_{1,1,3} \}$.
The action of $\Gam_0$ on $\tilde E$ satisfies
$$
\Gam_0/\Gam_0'=\langle
(R_{90}, 1) \Gam_0', (R_{120}, 1)\Gam_0', (I_3, -1) \Gam_0' \rangle\cong \D_6.
$$
The action of $\Gam_0/\Gam_0'$ on $\tilde E\cong \R^3$ is isomorphic to the natural action of
$\langle M, R_{120}, -I_3  \rangle$ on $\R^3$, where
$$
M = \left [
\begin{smallmatrix}
0 & 1 & 0\\
1 & 0 & 0 \\
0 & 0 & 1
\end{smallmatrix}
\right ].
$$
Note that $M$ acts as a reflection.
The matrix group
$\langle M, R_{120}, -I_3 \rangle$ is called $\D_{3d}$ in the
Sch\"onflies notation \cite{Tinkham} for crystallographic point groups.

We now describe the symmetries of the solutions shown in Figure~\ref{bifdiag4c}.
For convenience, we will let $\Gamma_i$ denote the symmetry of the solution shown with symmetry type
$S_i$.

There is a one-dimensional fixed-point subspace of $\tilde E$ for the only symmetry in $S_2 = \{\Gamma_2\}$:
$$
\fix(\Gamma_2, \tilde E) = \spn \{ \psi_{3,1,1}+ \psi_{1,3,1}+ \psi_{1,1,3} \}.
$$
This space is the line through the front and back vertices shown as large dots in the first two cubes of
Figure~\ref{D6bif}.
The EBL guarantees that there is a solution with this symmetry;
one such branch is shown in Figure~\ref{bifdiag4}.  The other, negative,
branch is not shown.
Figure~\ref{bifdiag4c} shows a contour map of this solution with $\Gamma_2 \cong \Oh$
symmetry.
The solution has one sign on the shaved cube, as shown, but the sign is opposite at the center of the cube.
The nodal surface has cubic symmetry and is diffeomorphic to a sphere.

Figure~\ref{bifdiag4c} also shows one solution with symmetry
$\Gamma_{18} \in S_{18}$.
The one-dimensional fixed-point subspace of $\tilde E$
for this symmetry is
$$
\fix(\Gamma_{18}, \tilde E) = \spn \{ \psi_{3,1,1}- \psi_{1,3,1} \}.
$$
This fixed-point subspace is the line through the midpoints of two opposite edges, depicted as the thickest
hexagon diagonal in the third cube of Figure~\ref{D6bif}. The two other diagonals of the hexagon are conjugate
fixed-point subspaces.
The $180^\circ$ rotation about each of these diagonals is a symmetry of $\tilde E$.
There is one conjugacy class of branches that bifurcates with symmetry type $S_{18}$ at $s = 11$.
There are 6 branches in this conjugacy class, one of which is shown in Figure~\ref{bifdiag4}.

Figure~\ref{bifdiag4c} shows three solutions with symmetry $\Gamma_{17} \in S_{17}$.
There can be more than one conjugacy class of branches because
the fixed-point subspace of the $\Gamma_{17}$ action on $\tilde E$
is two-dimensional:
$$
\fix(\Gamma_{17}, \tilde E) = \spn \{ \psi_{3,1,1}+ \psi_{1,3,1}, \psi_{1,1,3} \}.
$$
The intersection of this plane with a cube in $\tilde E$ is indicated by dotted lines
in Figure~\ref{D6bif}.
This fixed-point subspace includes the one-dimensional intersection of the AIS 
${\mathcal A}_{1,1,3}$ with $\tilde E$,
$$
\spn\{ \psi_{1,1,3} \} \subseteq \fix(\Gamma_{17}, \tilde E ),
$$
so there is a bifurcating solution branch, which is approximately a multiple of $\psi_{1,1,3}$,
in ${\mathcal A}_{1,1,3}$ (see Equation (\ref{AISpqr})).
It is clear from the contour map that the fourth branch from the top has solutions
that are in ${\mathcal A}_{1,1,3}$.
Note that this branch undergoes an anomaly-breaking bifurcation at $s \approx  3.417$, with a daughter
branch that has the same  $\Gamma_{17}$ symmetry.

\subsection{A bifurcation with $\tet_d$ symmetry}

\begin{figure}
\scalebox{1.0}{\input{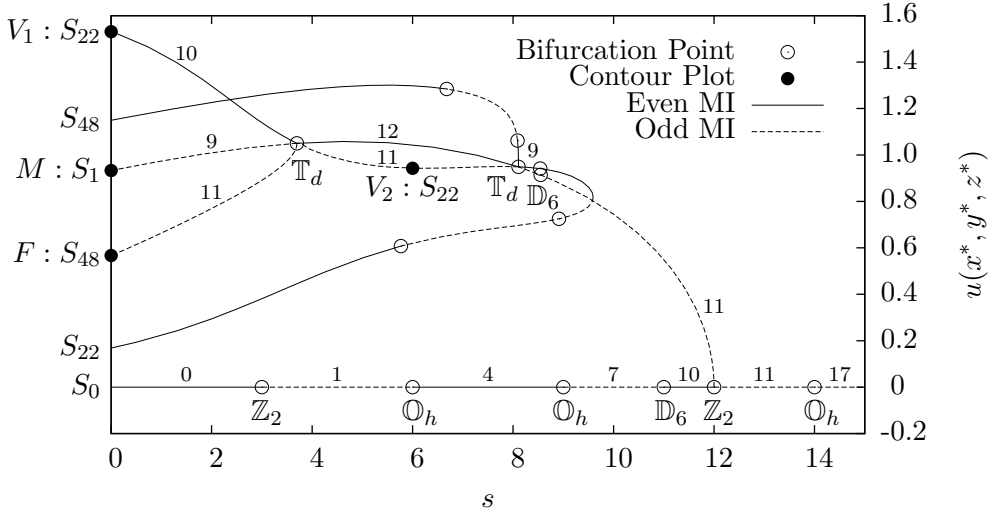}}
\caption{
Partial bifurcation diagram for PDE~(\ref{pde}) on the cube, showing the primary branch
that bifurcates at $s = \lam_{2,2,2} = 12$ and the two bifurcations with $\tet_d$ symmetry
that the primary branch undergoes at $s \approx 3.697$ and $s \approx 8.107$.
There is no bifurcation where the branches appear to cross but no circle is drawn.
The small numerals indicate the MI of the trivial branch, the primary branch,
and the solutions emanating from the bifurcation with $\tet_d$ symmetry on the left.
Contour plots of these solutions are shown in Figure~\ref{T_dBifContours}.
}
\label{bifdiagTd}
\end{figure}

\begin{figure}
\begin{tabular}{cccc}
\raise2cm\hbox{$M$: $S_1$}\hskip1cm
\scalebox{\flagScale}{\includegraphics{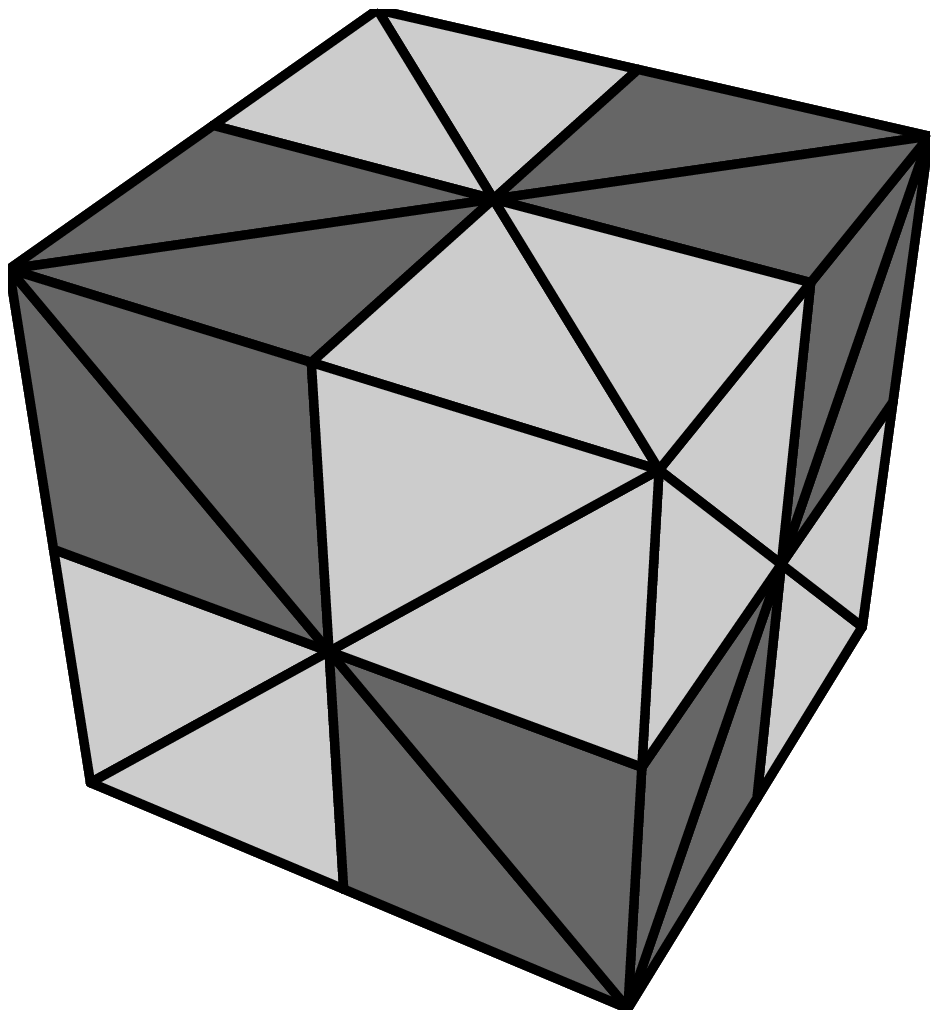}} &
\scalebox{\solScale}{\includegraphics{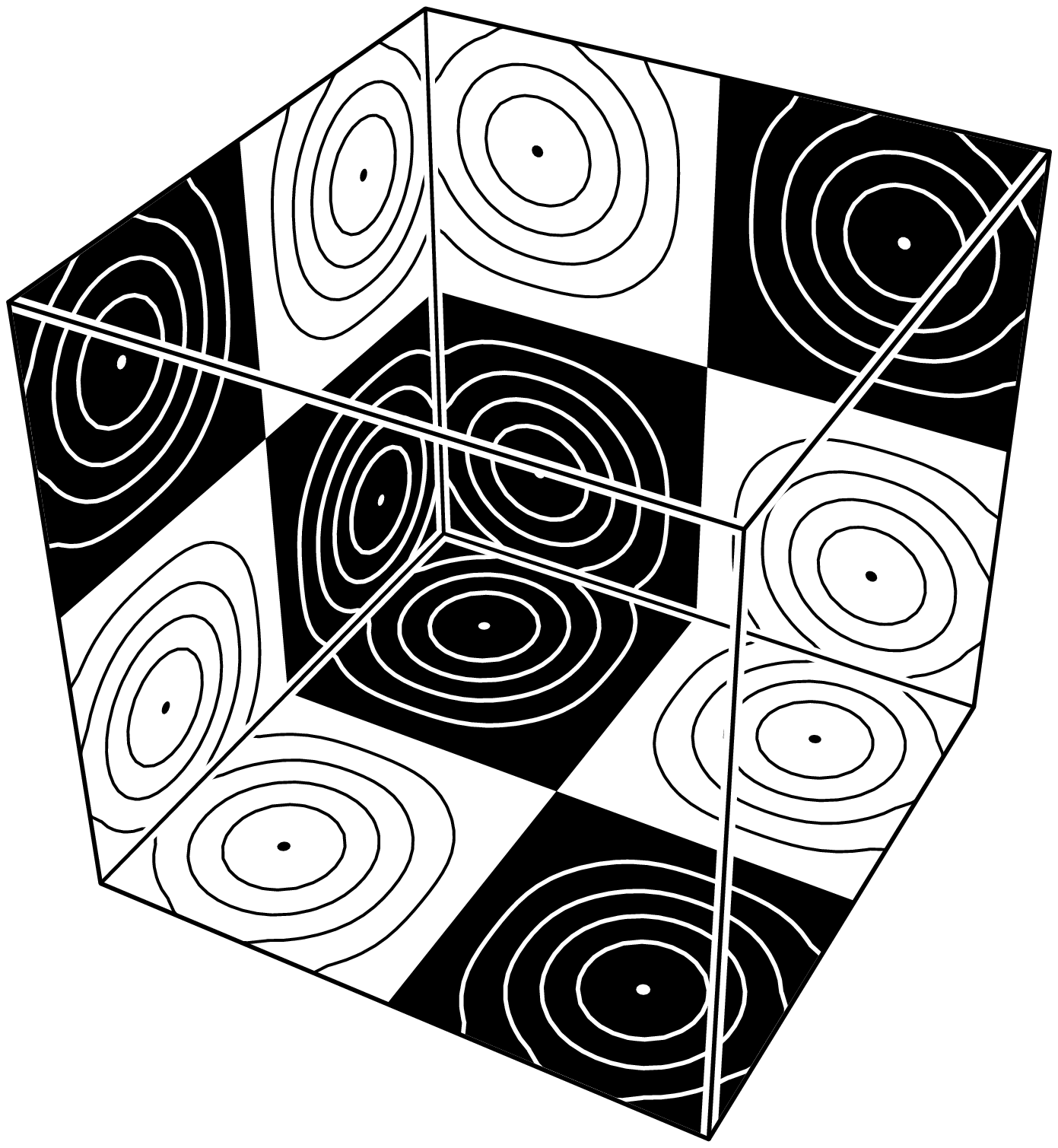}}  &
\scalebox{\solScale}{\includegraphics{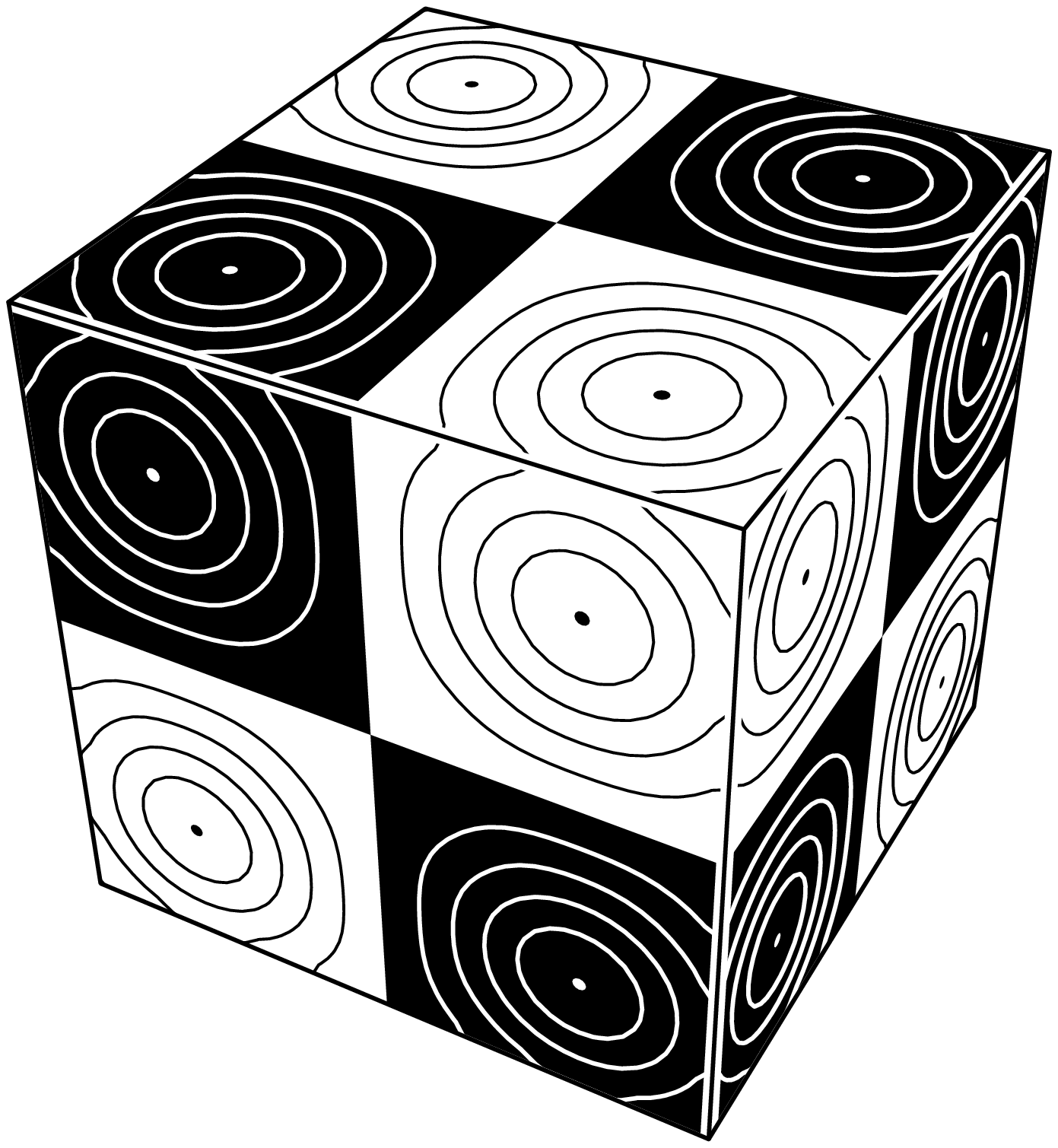}} \\
\raise2cm\hbox{$V_1$: $S_{22}$}\hskip1cm
\scalebox{\flagScale}{\includegraphics{figures/cube/fig.22.eps}} &
\scalebox{\solScale}{\includegraphics{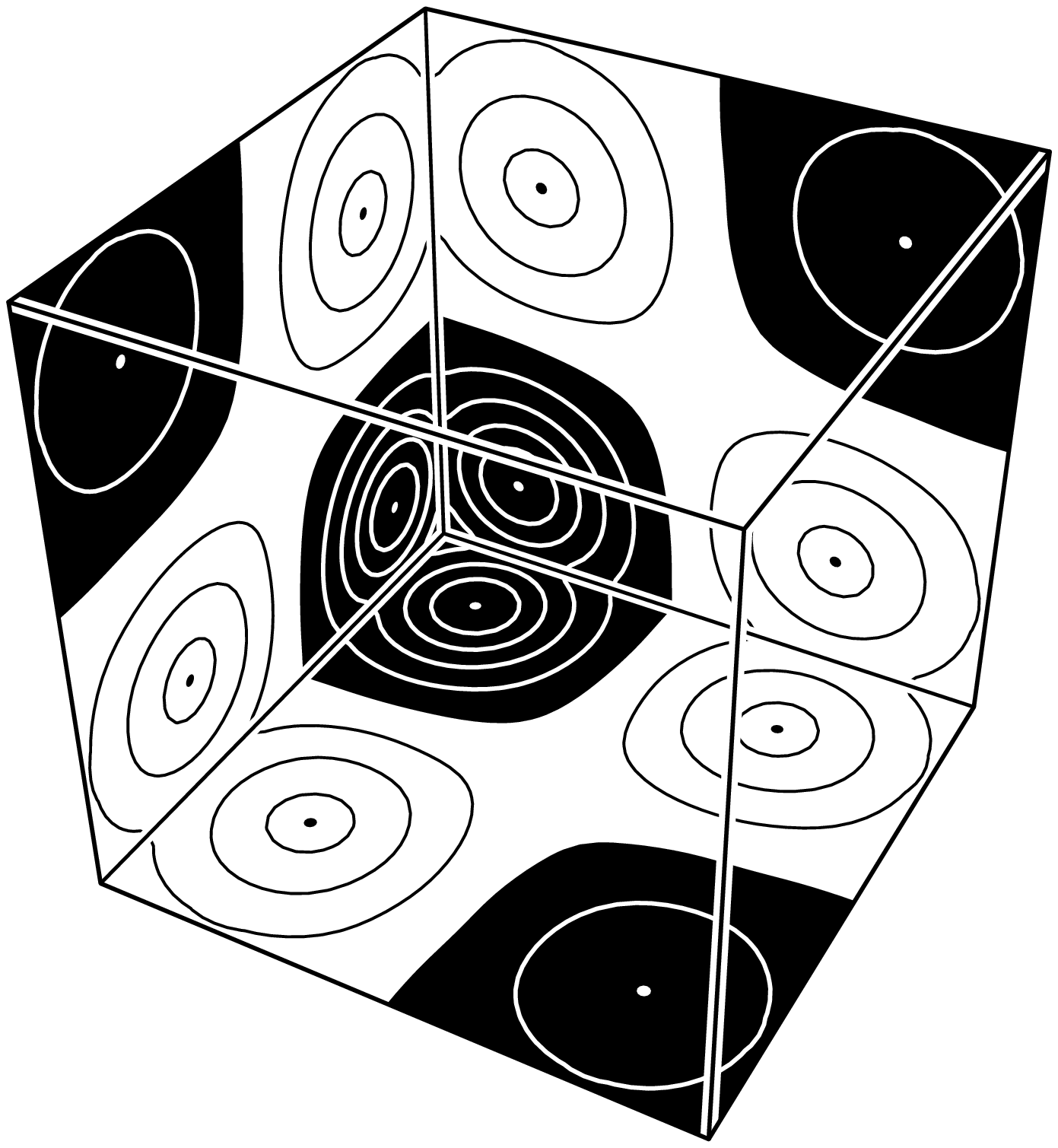}}  &
\scalebox{\solScale}{\includegraphics{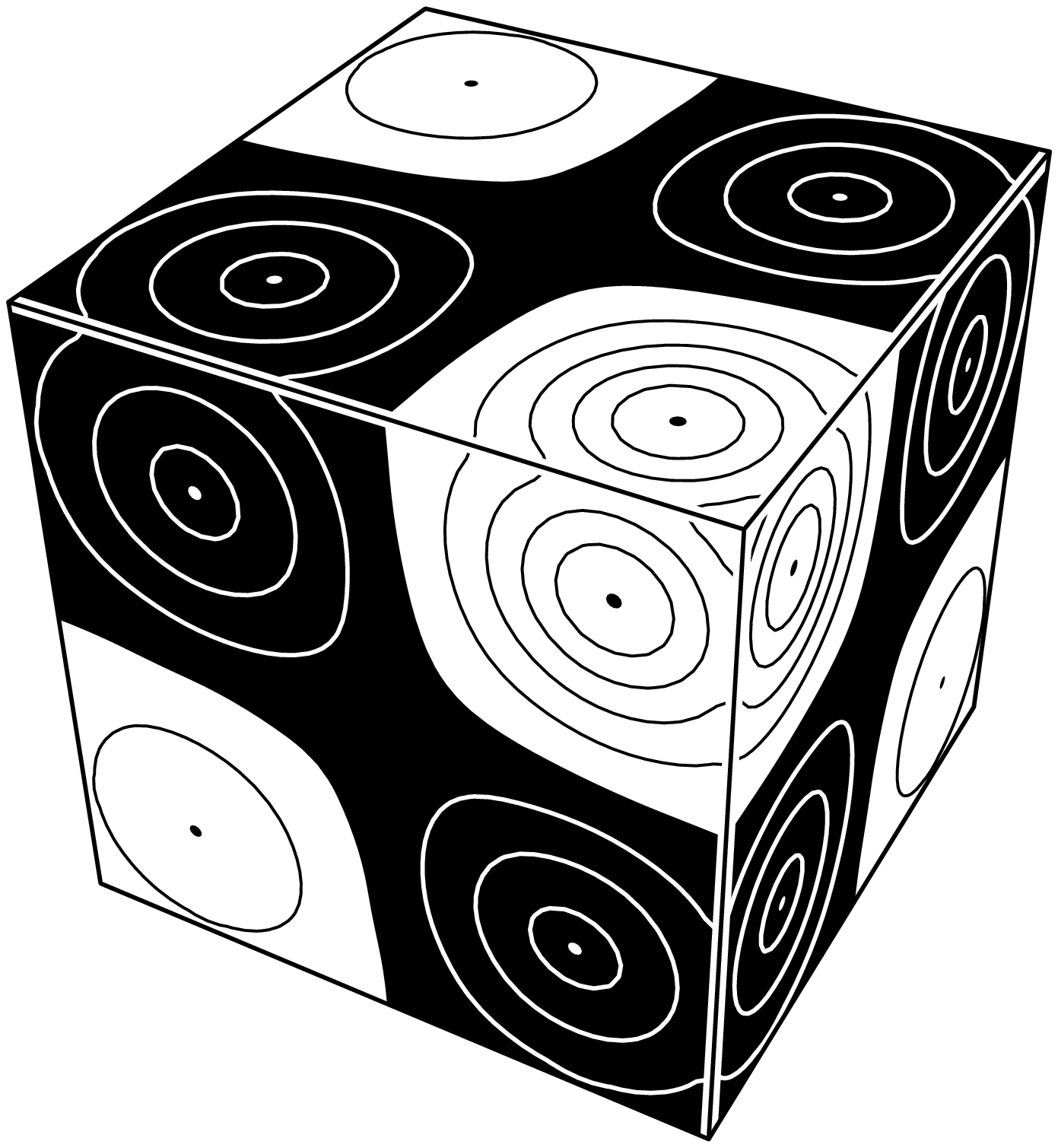}} \\
\raise2cm\hbox{$V_2$: $S_{22}$}\hskip1cm
\scalebox{\flagScale}{\includegraphics{figures/cube/fig.22.eps}} &
\scalebox{\solScale}{\includegraphics{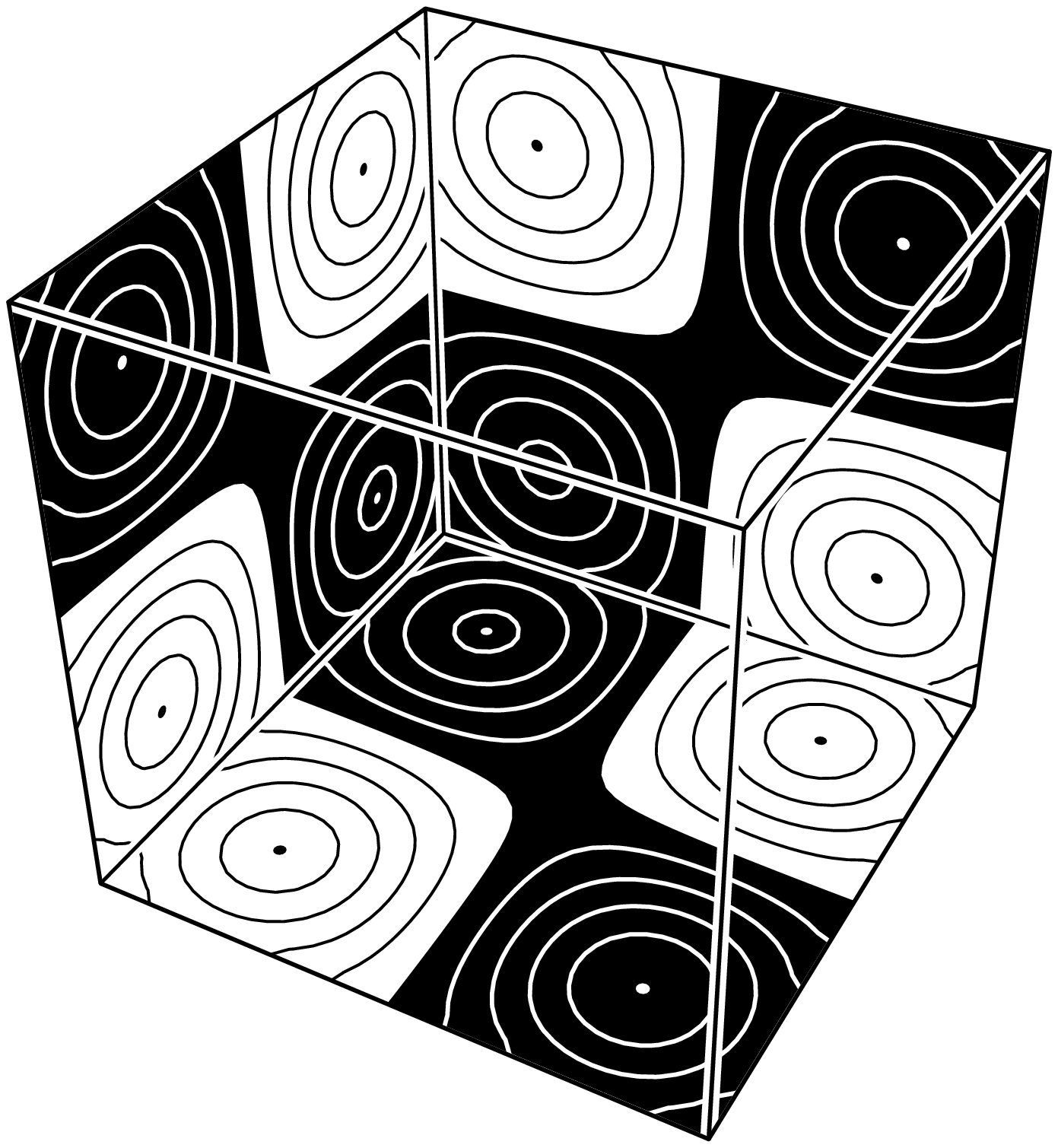}}  &
\scalebox{\solScale}{\includegraphics{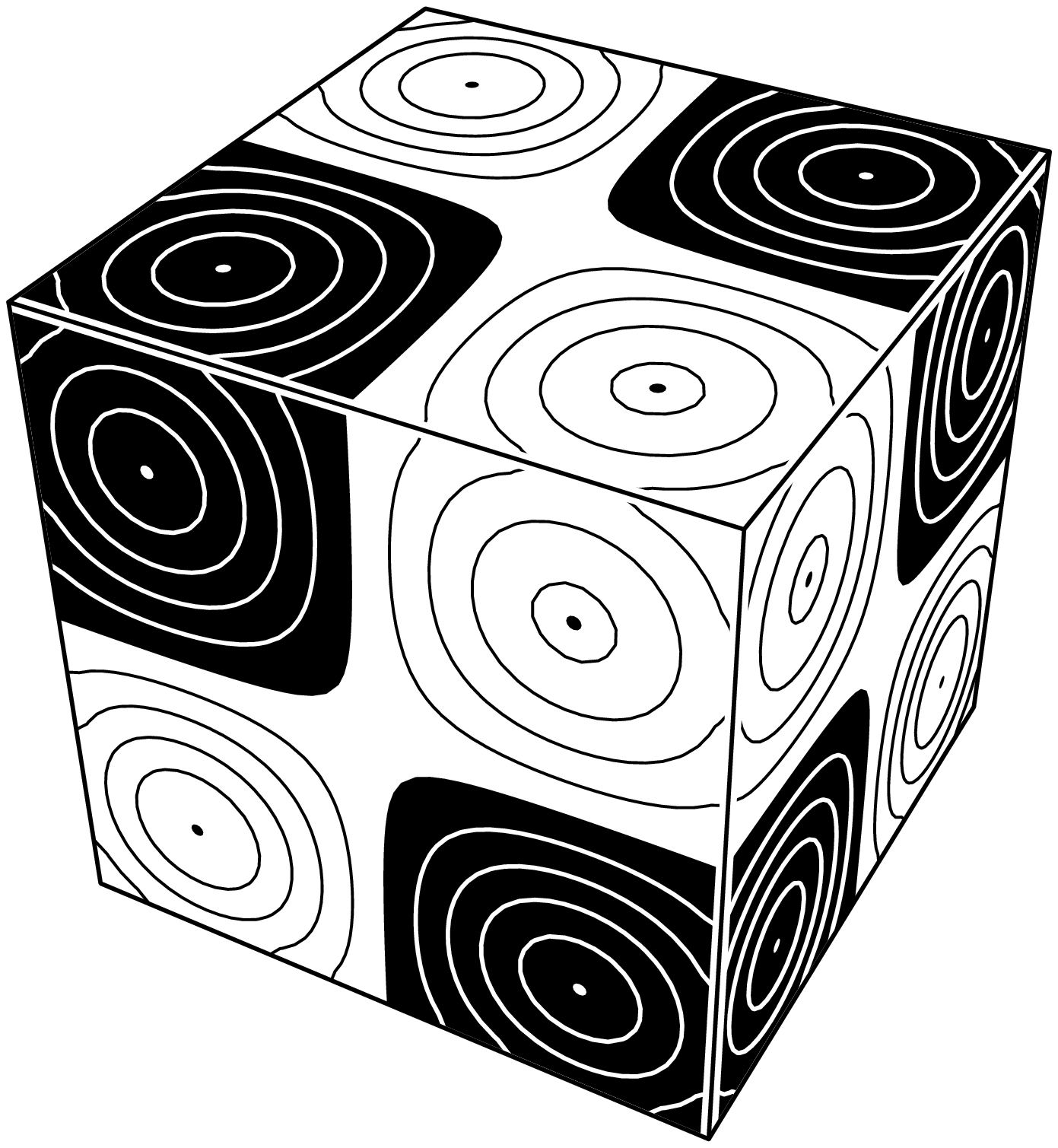}} \\
\raise2cm\hbox{$F$: $S_{48}$}\hskip1cm
\scalebox{\flagScale}{\includegraphics{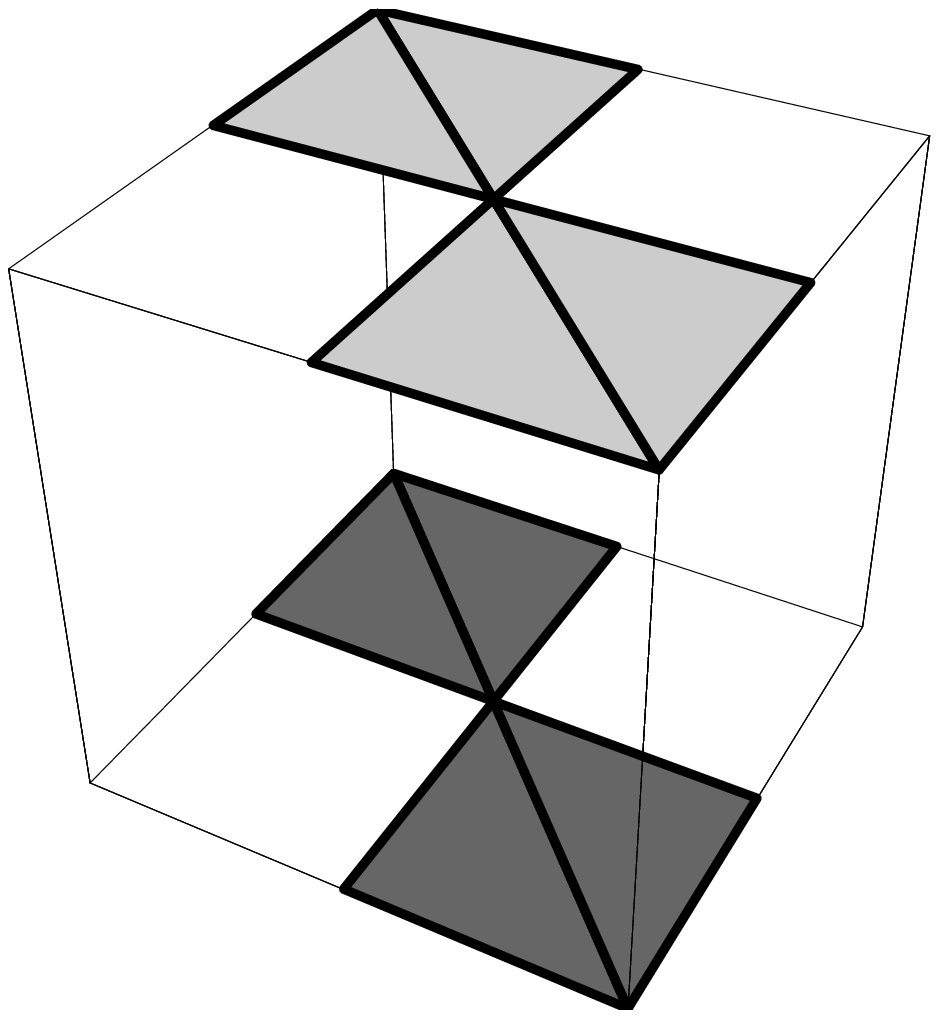}} &
\scalebox{\solScale}{\includegraphics{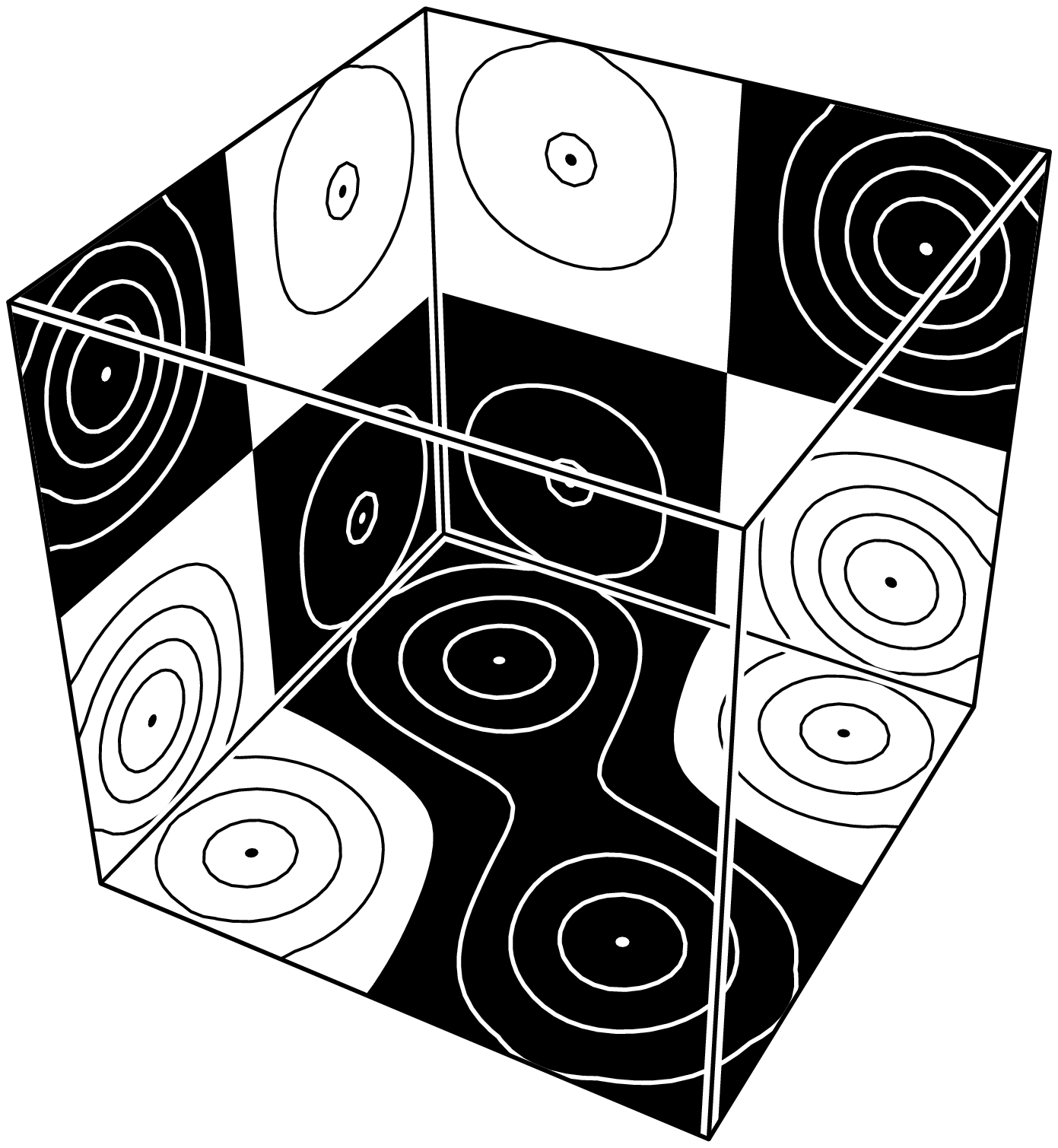}}  &
\scalebox{\solScale}{\includegraphics{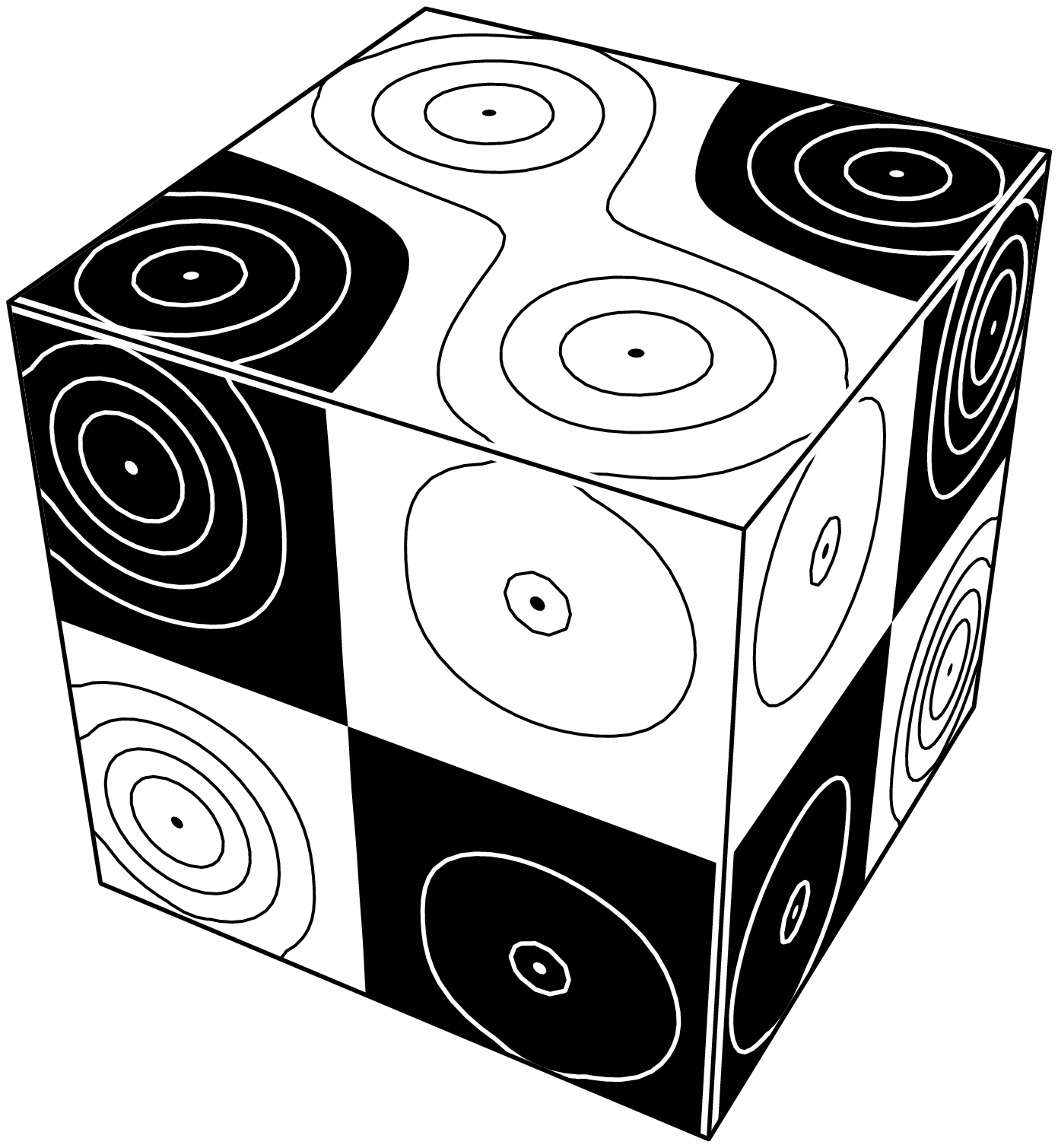}} \\
\end{tabular}
\caption{Contour plot of the mother solution $M$ shown in Figure~\ref{bifdiagTd},
along with two vertex solutions and a face solution born at
a bifurcation with $\tet_d$ symmetry.
The names vertex and face indicate the position in the critical eigenspace,
shown in the middle image of Figure~\ref{irredSpaces3}.
}
\label{T_dBifContours}
\end{figure}

\begin{figure}
\begin{tabular}{cccc}
\raise2cm\hbox{$S_{98}$}\hskip1cm
\scalebox{\flagScale}{\includegraphics{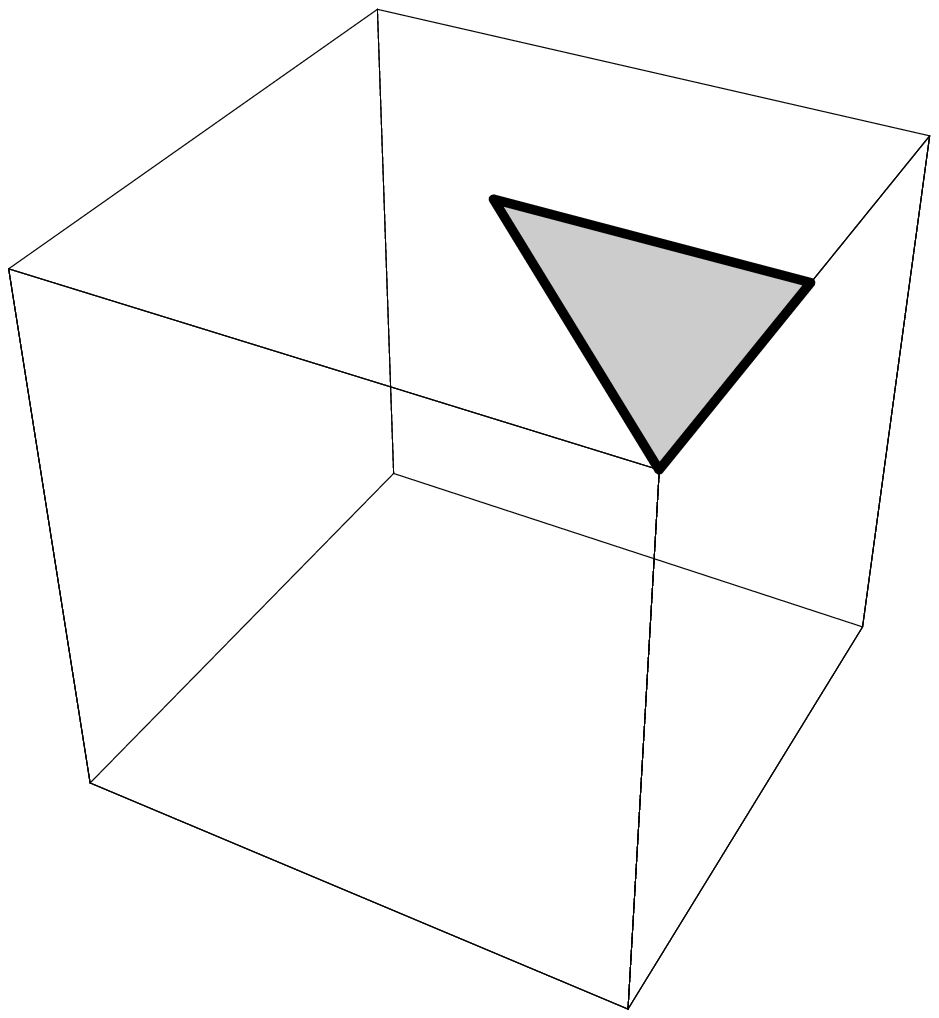}} &
\scalebox{\solScale}{\includegraphics{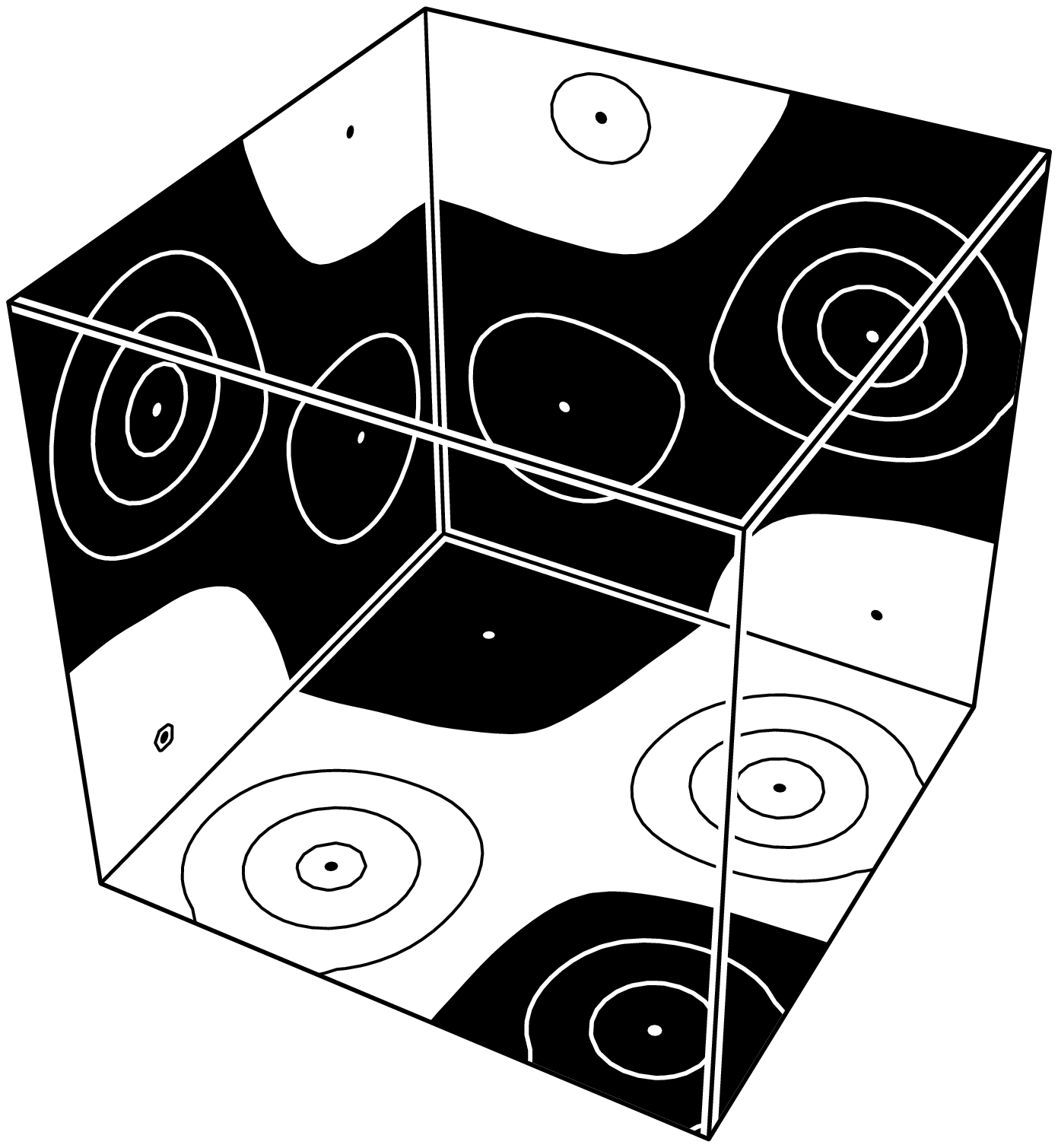}}  &
\scalebox{\solScale}{\includegraphics{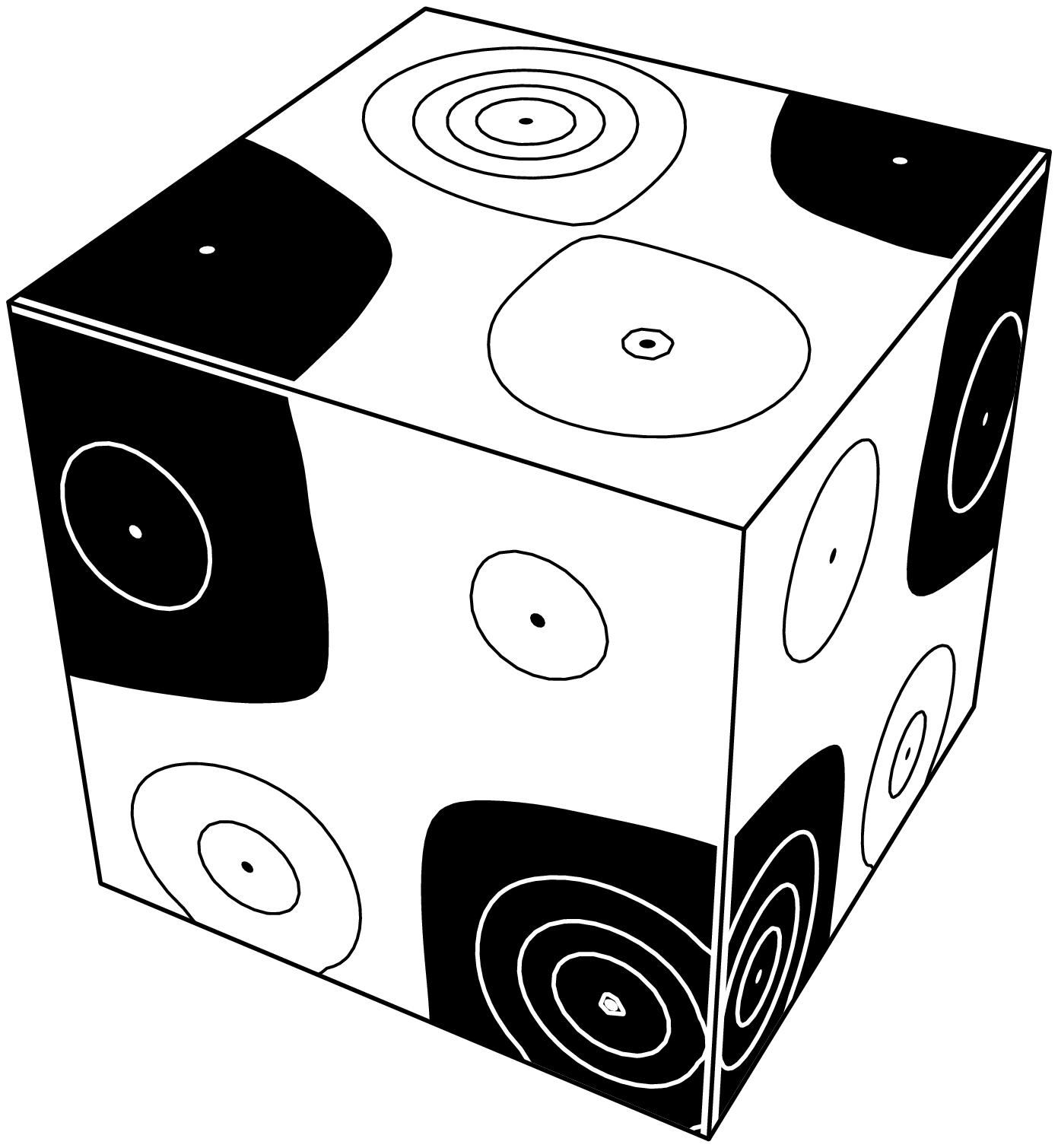}} \\
\end{tabular}
\caption{Contour plot for a solution with trivial symmetry
to PDE~(\ref{pde}) on the cube, at $s = 0$.
This solution is a descendent of the mother branch with symmetry type $S_1$
shown in Figures~\ref{bifdiagTd} and~\ref{T_dBifContours}.
The solution shown is found by following the sequence of bifurcations
$S_0 \xrightarrow{\Z_2}$ $S_1 \xrightarrow{\D_6}$ $S_{37} \xrightarrow{\D_4}$
$S_{91} \xrightarrow{\Z_2}$ $S_{98}$.
}
\label{trivialSymContour}
\end{figure}

Our C++ program can analyze the bifurcations of nontrivial solutions, and follow
all of the daughter branches of most bifurcations.
For example, the primary branch that bifurcates at $s = 12$ undergoes three bifurcations
in the interval $0 < s < 12$, as shown in Figure~\ref{bifdiagTd}.
Two of these bifurcations, at $s \approx 3.687$ and $s \approx 8.107$, are bifurcations
with $\tet_d$ symmetry.
The third bifurcation, at $s \approx 8.547$, is a bifurcation with $\D_6$ symmetry to be discussed later.

We focus on the bifurcation with $\tet_d$ symmetry at $s \approx 3.687$.
The symmetry of the mother solution $u^*$ is
$$
\Gamma_1 = \langle (R_{90}, -1), (R_{120}, 1), (R_{180}, -1), (-I_3, -1) \rangle,
$$
that is, $u^* \in \fix(\Gam_1)$.
For example, a rotation by $90^\circ$ about the $z$-axis, coupled with a sign change, leaves $u^*$
unchanged.
The solution on
the mother branch shown in Figure~\ref{T_dBifContours} looks very much like $\psi_{2,2,2}$.
The critical eigenspace $\tilde E$ has the ordered basis
$$
( \tilde \psi_{2,1,1}, \tilde \psi_{1,2,1}, \tilde \psi_{1,1,2} ),
$$
where $\tilde \psi_{i,j,k}$ is a function with the same symmetry as that of $\psi_{i,j,k}$.
We cannot find $u^*$ or the critical eigenfunctions exactly with pencil and paper, but
we do know the symmetry exactly, and our C++ program is able to use this information.

The action of $\Gamma_1$ on $\tilde E$ satisfies $\Gamma_1' = \langle (-I_3, -1) \rangle$
and the action of $\Gamma_1/\Gamma_1'$ on $\tilde E$ is isomorphic to the natural action
of
$$
\tet_d = \langle -R_{90}, R_{120}, -R_{180} \rangle
$$
on the coordinate space $[\tilde E] = \R^3$.

The symmetry of the reduced vector field $\tilde g$ on $\tilde E$ for this bifurcation
with $\tet_d$ symmetry is shown in Figure~\ref{irredSpaces3}.
The daughter solutions at this bifurcation
can be classified as face solutions or vertex solutions.
Each one-dimensional fixed-point subspace of the $\tet_d$ action on $\tilde E$
is conjugate to one of these two:
$$
\mbox{face }\  [\fix(\Gam_{48}, \tilde E) ] = \{ (0, 0, a) \mid a \in \R \}, \quad \mbox{vertex }\ 
[\fix(\Gam_{22}, \tilde E) ] = \{ (a, a, a) \mid a \in \R \}.
$$
Note that $I_3$, the inversion through the origin, is not in $\tet_d$.
In particular, two antipodal vertex solutions are not conjugate, and there is a transcritical
branch of vertex solutions, as seen in Figure~\ref{bifdiagTd},
leading to the vertex solutions $V_1$ and $V_2$ seen in Figure~\ref{T_dBifContours}.
Note that $V_1$ has two white regions and two black regions on the surface of the cube, whereas
$V_2$ has one white region and one black region.

Figure~\ref{bifdiagTd} shows that the vertex solution $V_2$ is a daughter of
both of the bifurcations with $\tet_d$ symmetry
on the primary branch that bifurcates at $s = 12$.
The third bifurcation on that branch, at
$s \approx 8.547$, is a generic bifurcation with $\D_6$ symmetry.
The MI of the mother branch changes from 11 to 9
as $s$ decreases through that bifurcation.
Unlike the degenerate bifurcation with $\D_6$ symmetry that occurs at $u = 0$, $s = 11$,
generic bifurcations with $\D_6$ symmetry are well-known.
Hence, we do not give the details of the bifurcation at $s \approx 8.547$, except
to mention that one of the two branches created at this bifurcation has a grand-daughter
with trivial symmetry, depicted in Figure~\ref{trivialSymContour}.

\subsection{A six-dimensional critical eigenspace}

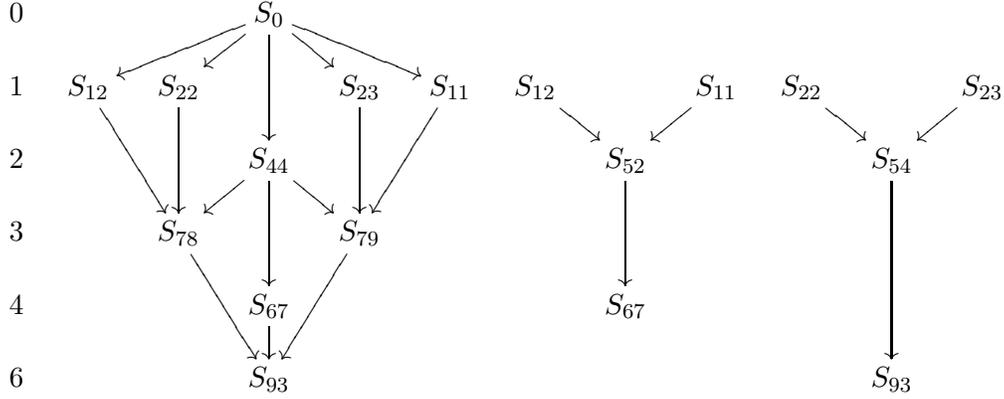
\begin{figure}
\begin{tabular}{cccc}
\xymatrix@=12pt{
0 {\vphantom{S_0}} \\ 
1  {\vphantom{S_0}}\\ 
2 {\vphantom{S_0}} \\ 
3 {\vphantom{S_0}} \\ 
4  {\vphantom{S_0}}\\ 
6  {\vphantom{S_0}}\\ 
}
&
\xymatrix@=12pt{
&   & S_0 \ar[lld] \ar[ld] \ar[dd] \ar[dr] \ar[drr] & & \\ 
 S_{12} \ar[ddr] & S_{22} \ar[dd] & & S_{23} \ar[dd] &  S_{11} \ar[ddl] \\ 
& &   S_{44} \ar[dd] \ar[dl] \ar[dr]&  & \\ 
&  S_{78} \ar[ddr]  &     &   S_{79} \ar[ddl] &\\ 
& &        S_{67} \ar [d] & & \\ 
&       & S_{93} & & 
}
&
\xymatrix@=12pt{
& {\vphantom{S_0}} & \\ 
 S_{12} \ar[rd] & & S_{11} \ar[dl] \\ 
       & S_{52} \ar[dd]& \\ 
& {\vphantom{S_3}} & \\
       & S_{67} & \\ 
& {\vphantom{S_6}} & 
}
&
\xymatrix@=12pt{
& {\vphantom{S_0}} & \\ 
 S_{22} \ar[rd] & & S_{23} \ar[dl] \\
       & S_{54} \ar[ddd] \\
& {\vphantom{S_6}} & \\ 
& {\vphantom{S_6}} & \\ 
       & S_{93} 
}

\end{tabular}

\caption{\label{6Dlattice}
The lattice of isotropy subgroups for the six-dimensional critical
eigenspace $\tilde E$ of the trivial solution at
$s = 14$. For clarity, we display a 3-element partition of the edge set of the Hasse diagram
of the lattice.
The number at the left indicates the dimension $\dim(\fix(\Gam, \tilde E))$ of the fixed-point subspace
for any $\Gam \in S_i$ at that height in the diagram.
}
\end{figure}

\begin{figure}
\begin{tabular}{|c|c|c|c|c|c|c|c|c|c|c|c|c|c|c|}
\hline
$i$ & 0 & 12 & 11 & 22 & 23 & 52 & 54 & 93  & 78 & 79 & 44 & 67 \\
$|\Gamma_i\cdot u|$ & 1 & 6 & 6 & 8 & 8 & 12 & 16 & 48 & 24 & 24 & 12 & 24 \\
MI & 11 & 12 & 14 & 14 & 12 & 15 & 17 & 16 & 14,15 & 13,16 & 12,13,14,15 & 13,14,15,17 \\
\hline
\end{tabular}
\caption{Symmetry, multiplicity and MI at $s=s^-=14-\varepsilon$ of the trivial solution 
and the bifurcating primary solutions at $s=14$. 
The solutions in a given column have symmetry type $S_i$. 
The second row shows the size of the group orbits, while the third row gives the MI
of the solutions in each group orbit. The only local solution at $s=s^+=14+\varepsilon$
is the trivial solution, with $\MI=17$.
Using these MI values, we can make an index theory computation (\ref{degree}) 
to verify that the results are consistent with having obtained all solutions.
}
\label{indexTable}
\end{figure}

\renewcommand{\solScale}{0.18}  
\renewcommand{\flagScale}{0.27}  

\begin{figure}
\begin{tabular}{|ccccc|}
\hline
\multicolumn{1}{|c|}{$S_{12}$} &
\multicolumn{1}{|c|}{$S_{11}$} &
\multicolumn{1}{|c|}{$S_{22}$} &
\multicolumn{1}{|c|}{$S_{23}$} &
$S_{52}$ 
\\
\multicolumn{1}{|c|}{
\scalebox{\flagScale}{\includegraphics{figures/cube/fig.12.eps}}
} &
\multicolumn{1}{|c|}{
\scalebox{\flagScale}{\includegraphics{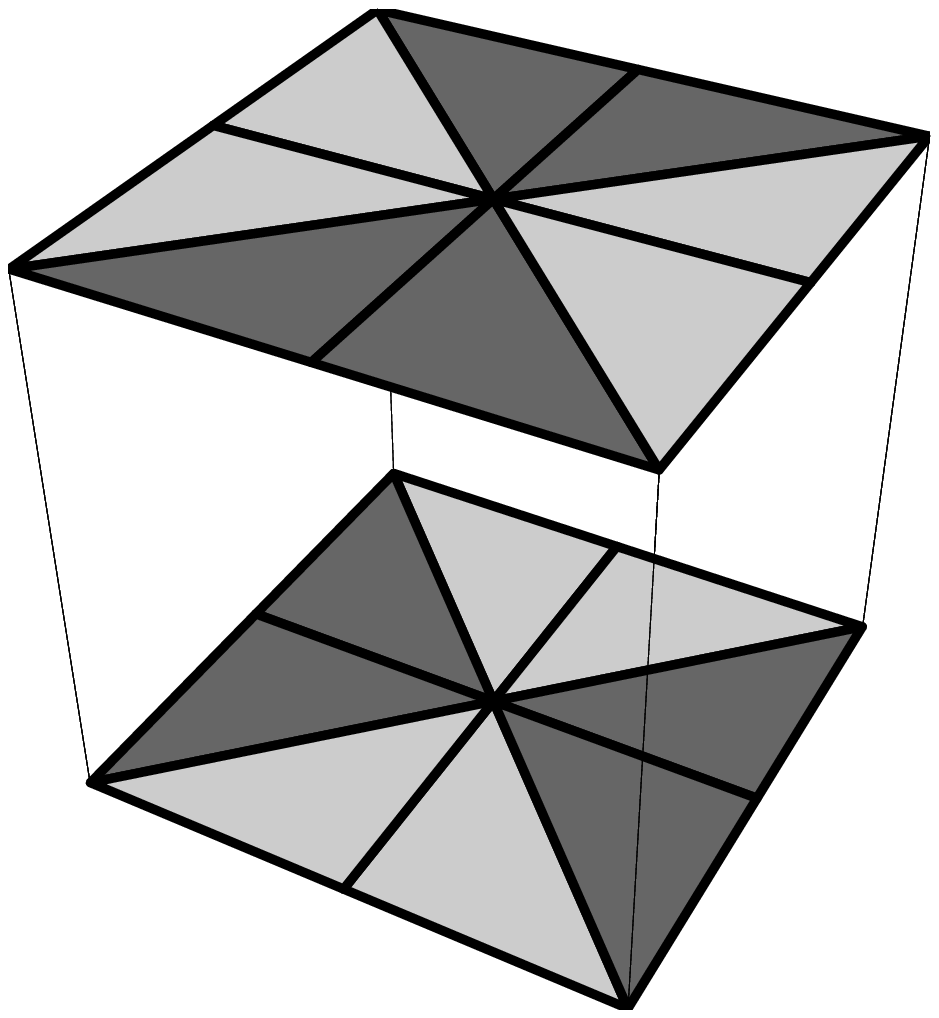}} 
} &
\multicolumn{1}{|c|}{
\scalebox{\flagScale}{\includegraphics{figures/cube/fig.22.eps}} 
} &
\multicolumn{1}{|c|}{
\scalebox{\flagScale}{\includegraphics{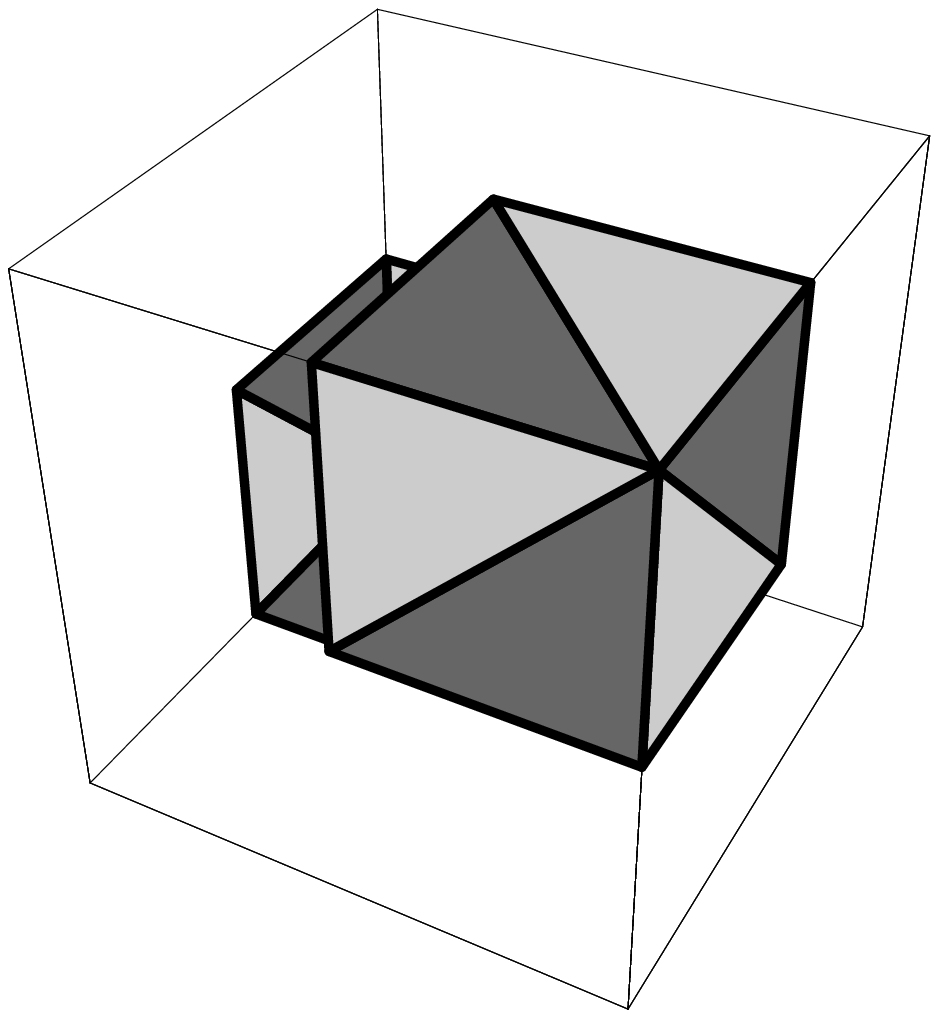}} 
} &
\scalebox{\flagScale}{\includegraphics{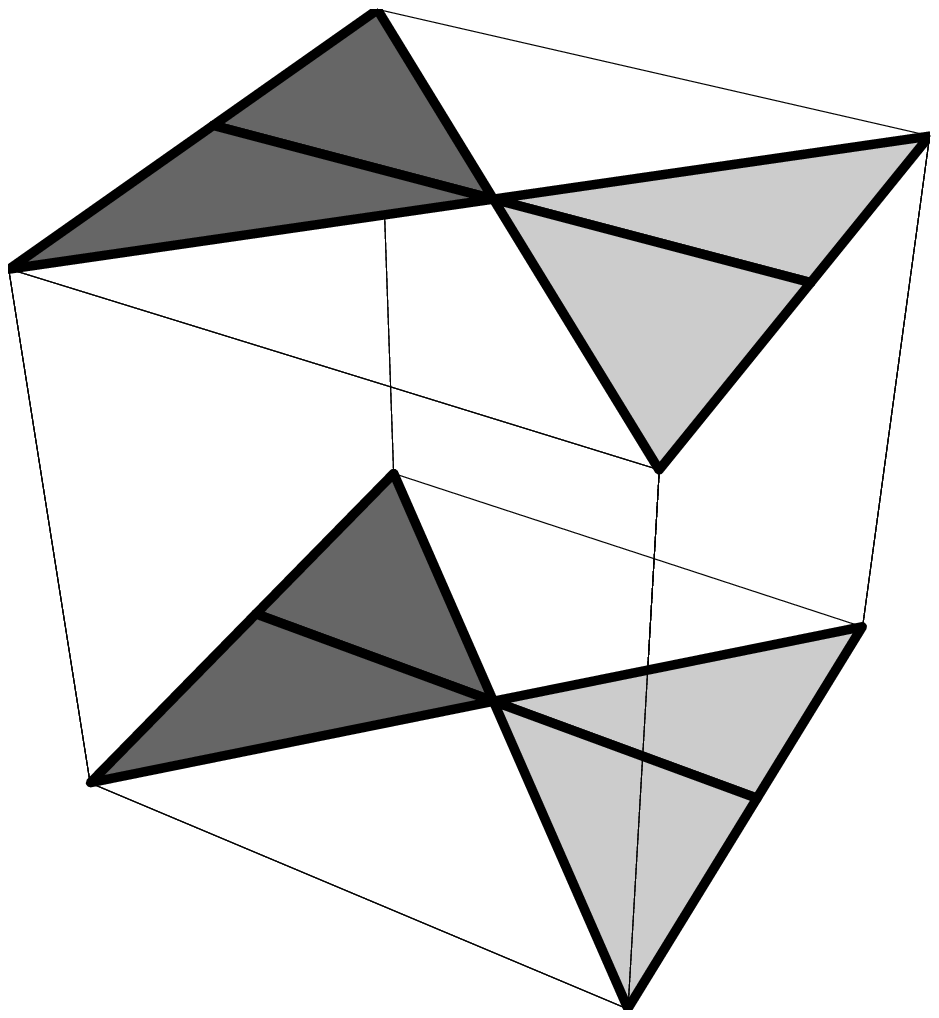}}
\\
\multicolumn{1}{|c|}{
\scalebox{\solScale}{\includegraphics{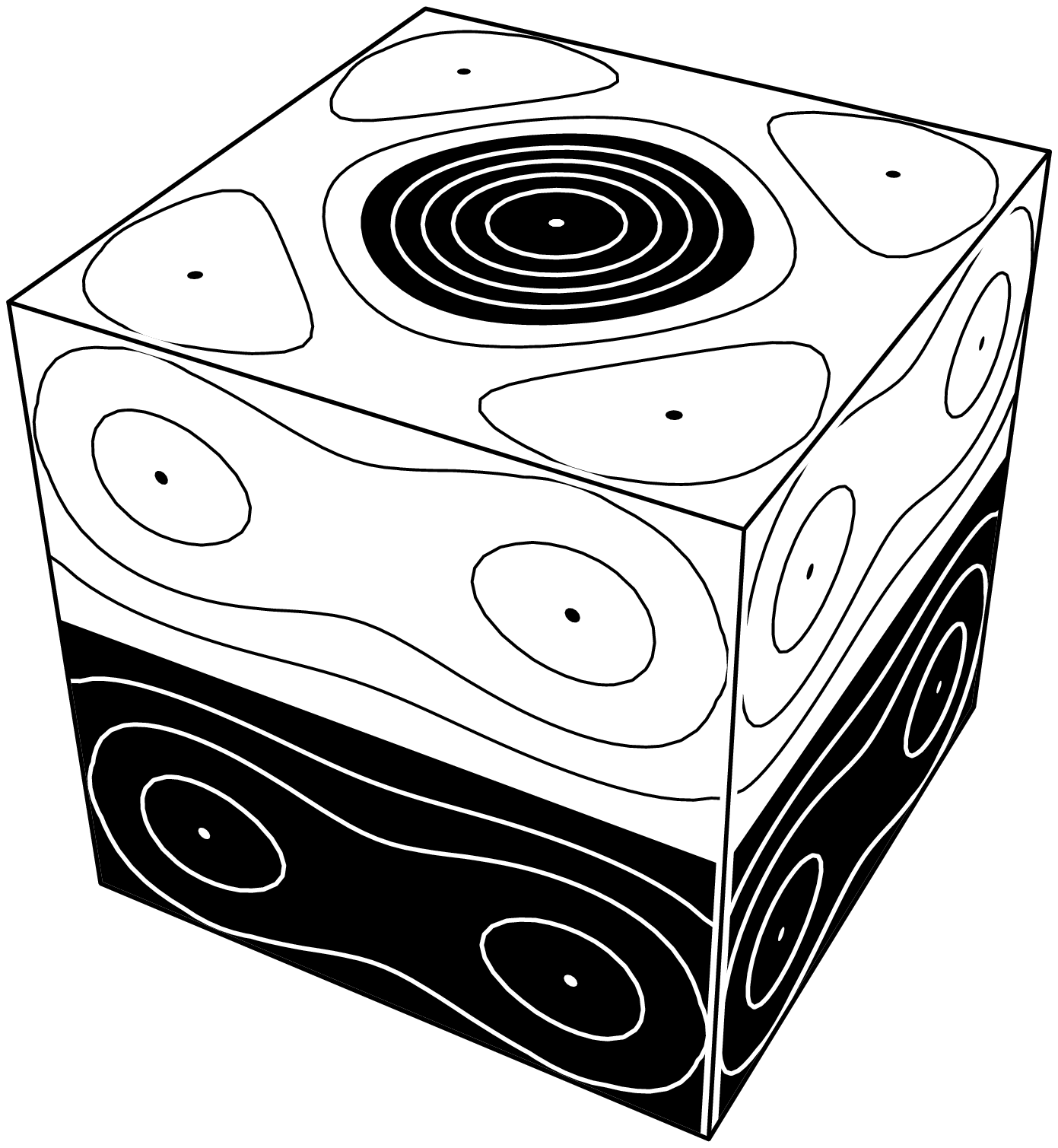}} 
} &
\multicolumn{1}{|c|}{
\scalebox{\solScale}{\includegraphics{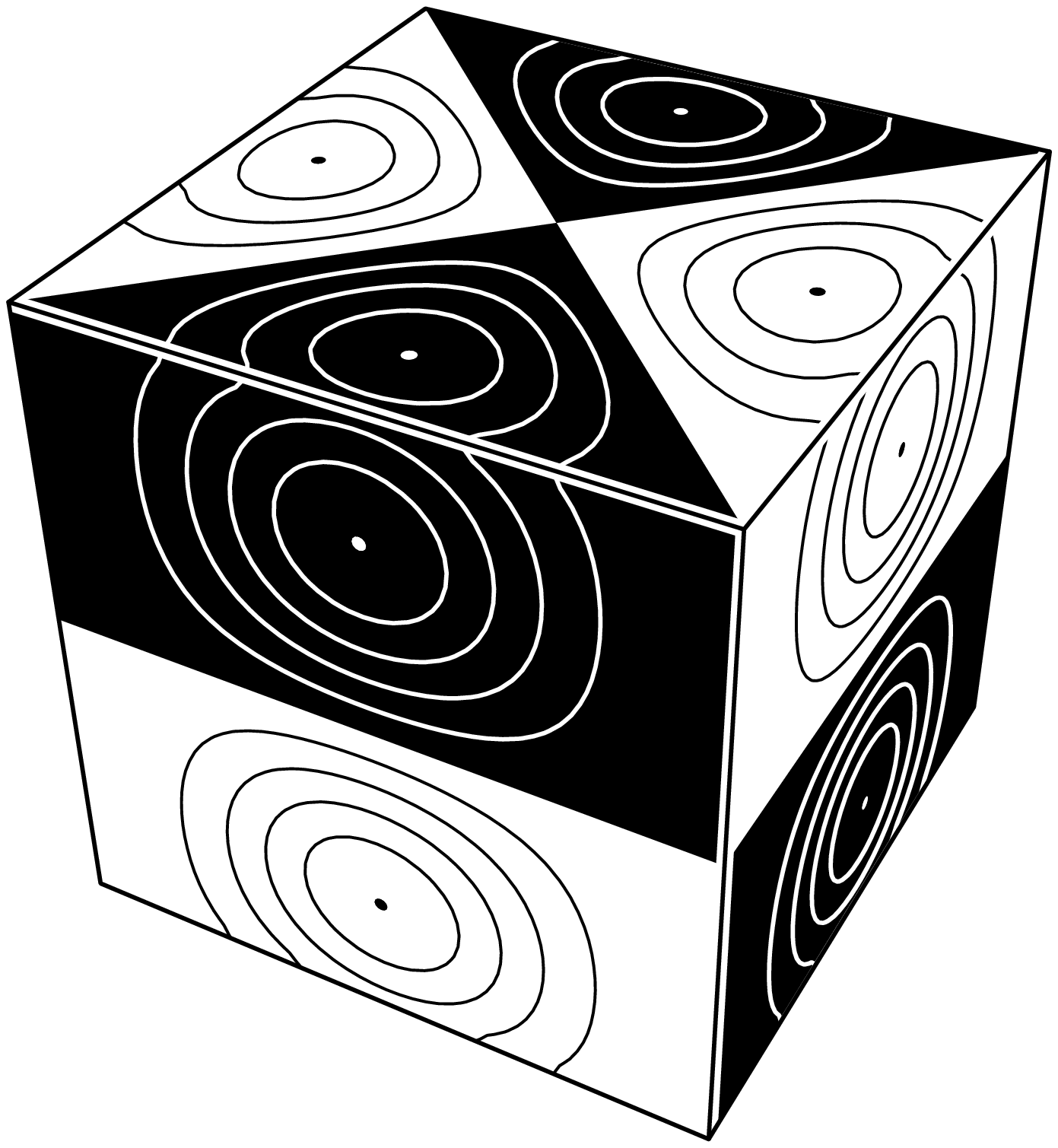}} 
} &
\multicolumn{1}{|c|}{
\scalebox{\solScale}{\includegraphics{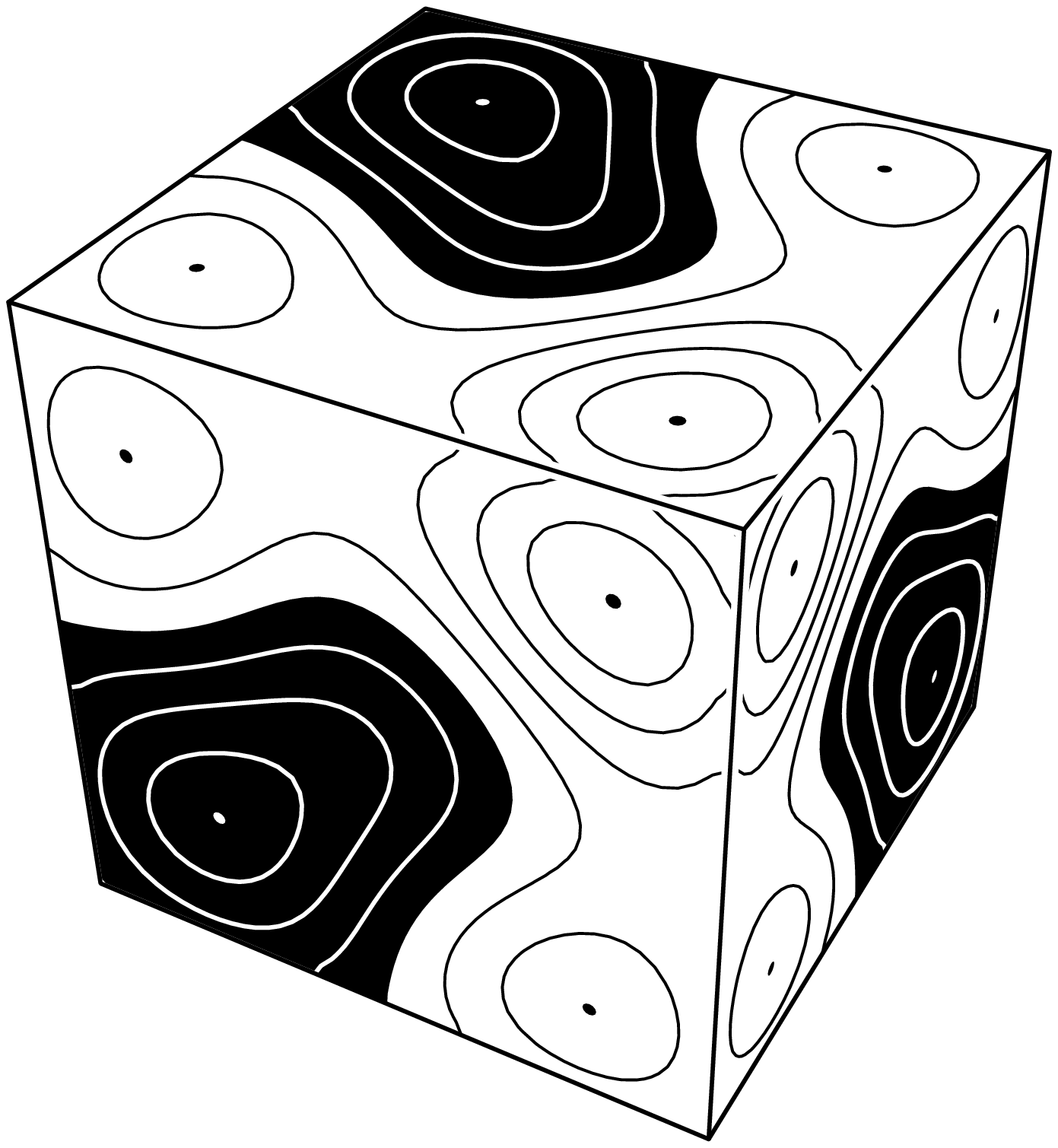}}
} &
\multicolumn{1}{|c|}{
\scalebox{\solScale}{\includegraphics{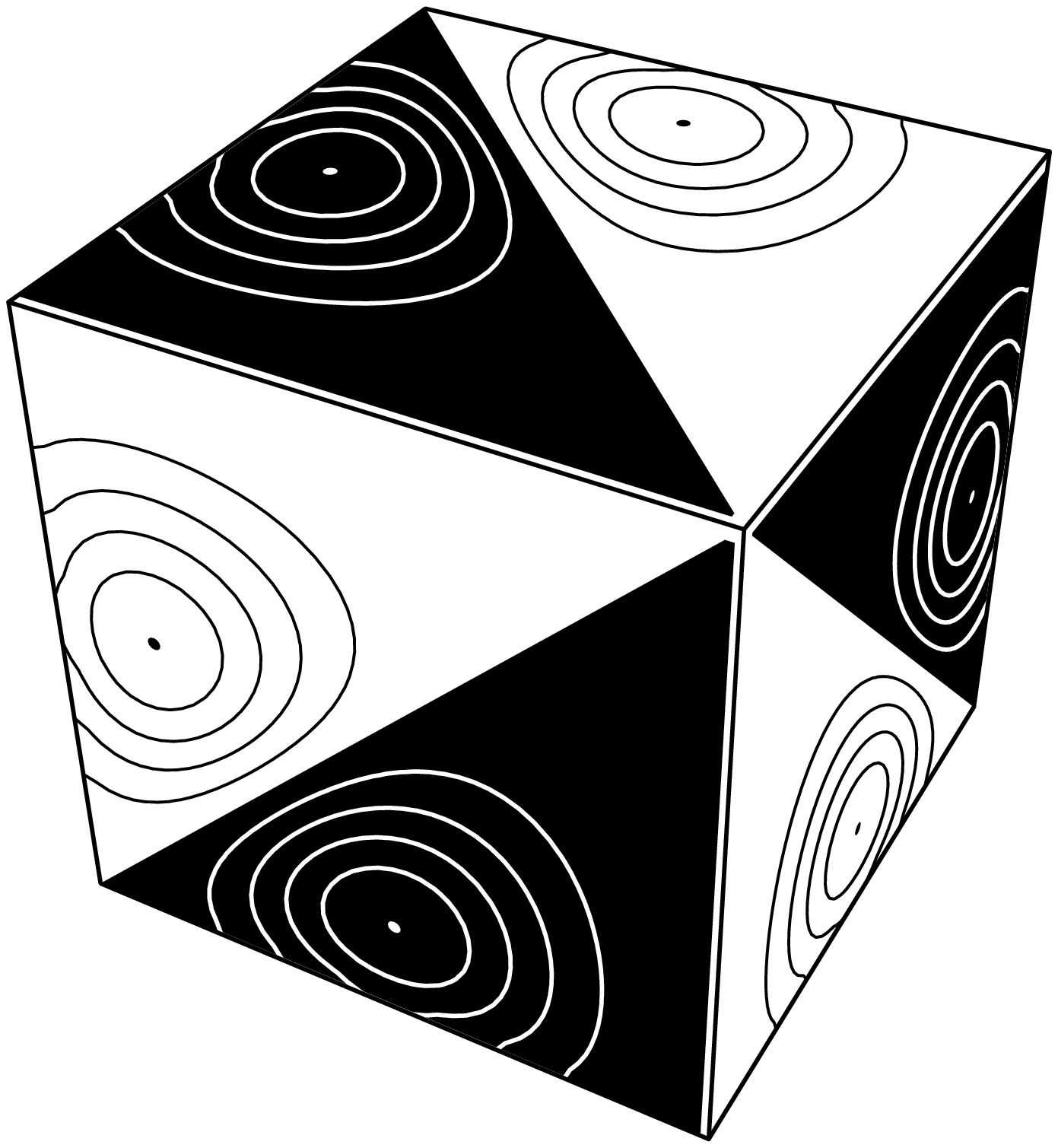}} 
} &
\scalebox{\solScale}{\includegraphics{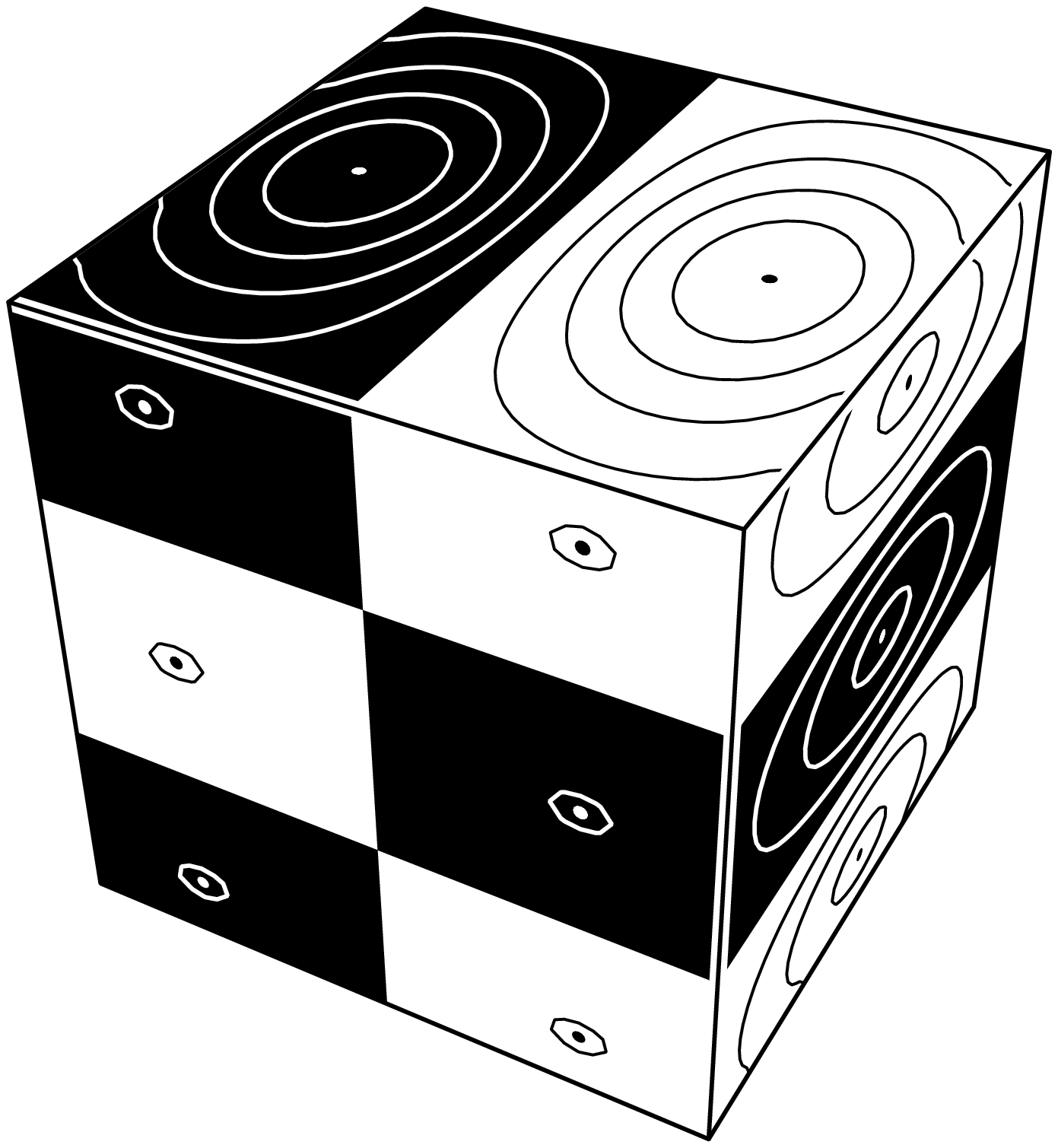}} 
\\
\hline
\multicolumn{1}{|c|}{$S_{54}$} &
\multicolumn{1}{|c|}{$S_{93}$} &
$S_{78} $ & &
\\
\multicolumn{1}{|c|}{
\scalebox{\flagScale}{\includegraphics{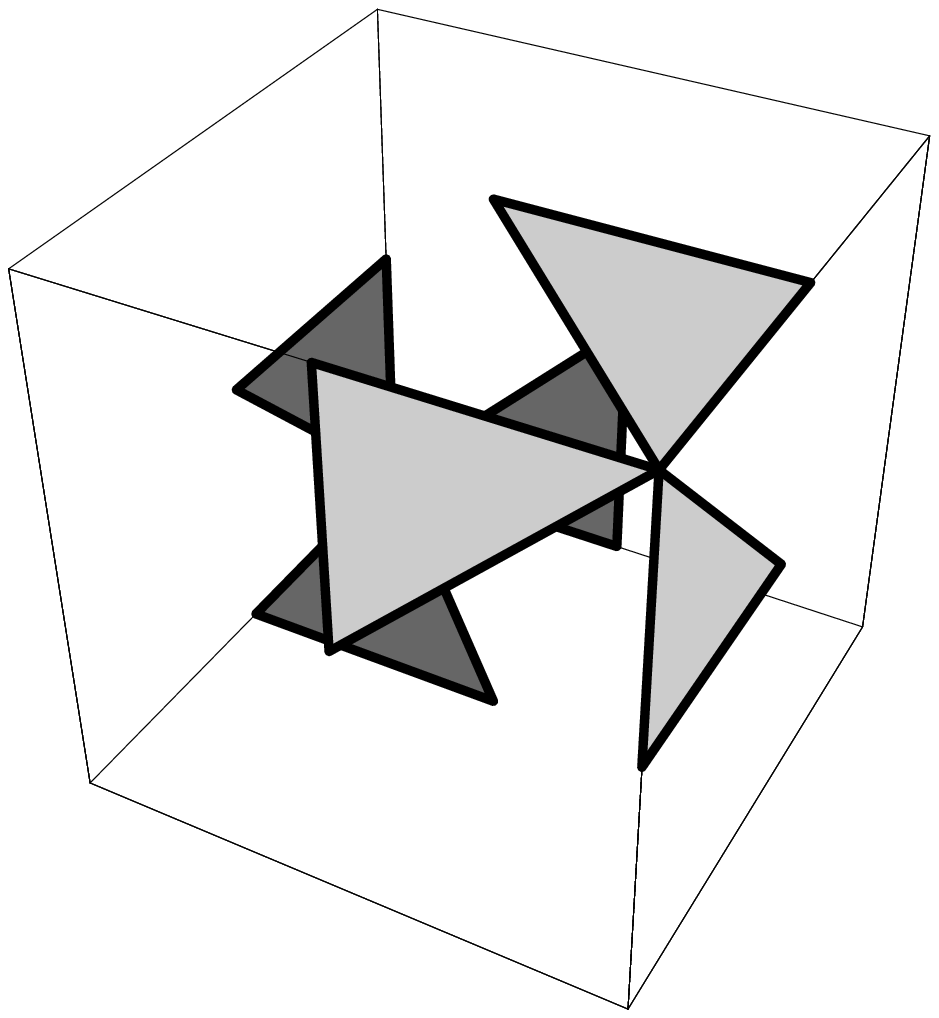}} 
} &
\multicolumn{1}{|c|}{
\scalebox{\flagScale}{\includegraphics{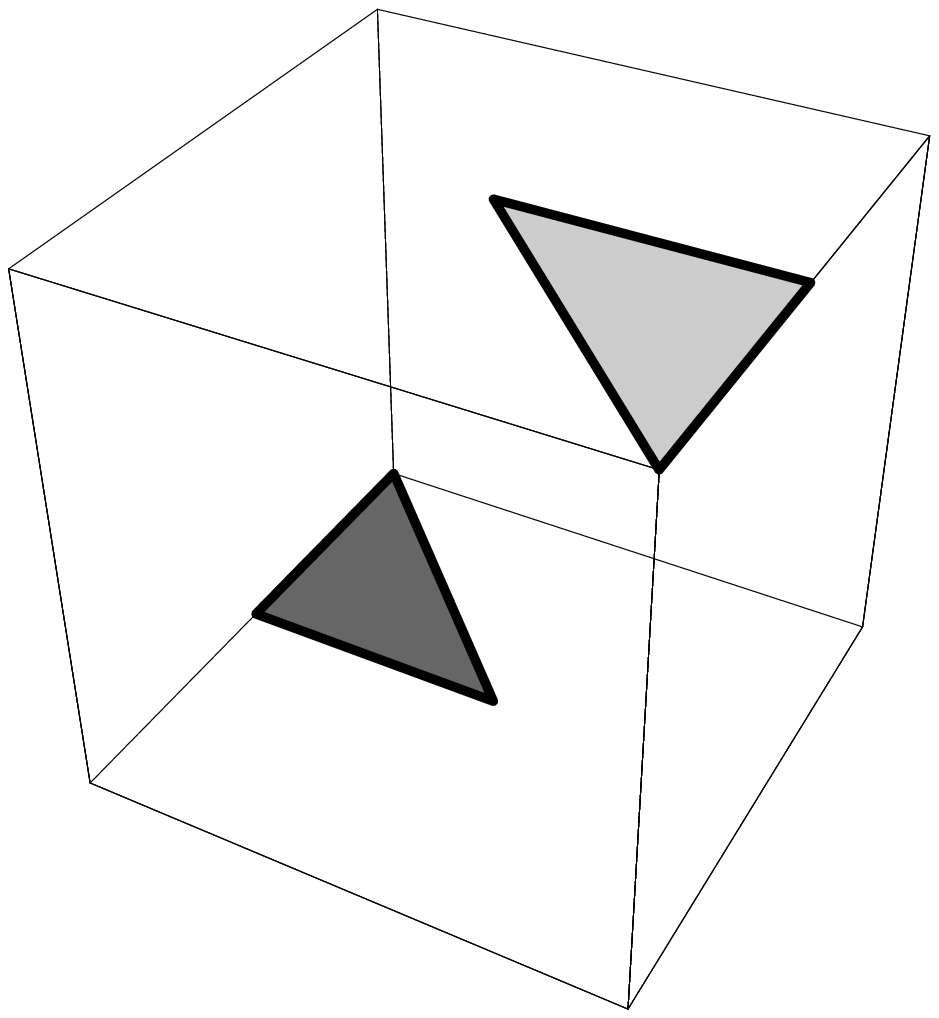}}
} &
\scalebox{\flagScale}{\includegraphics{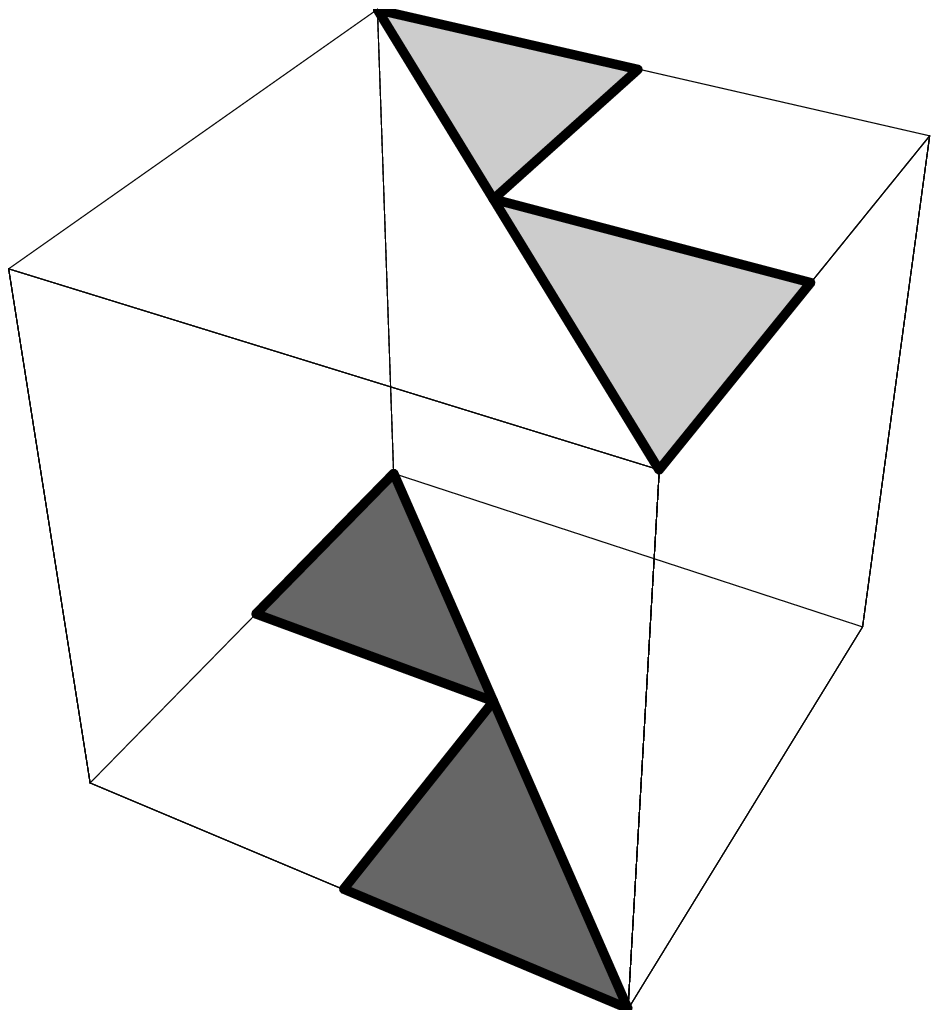}} &
\scalebox{\solScale}{\includegraphics{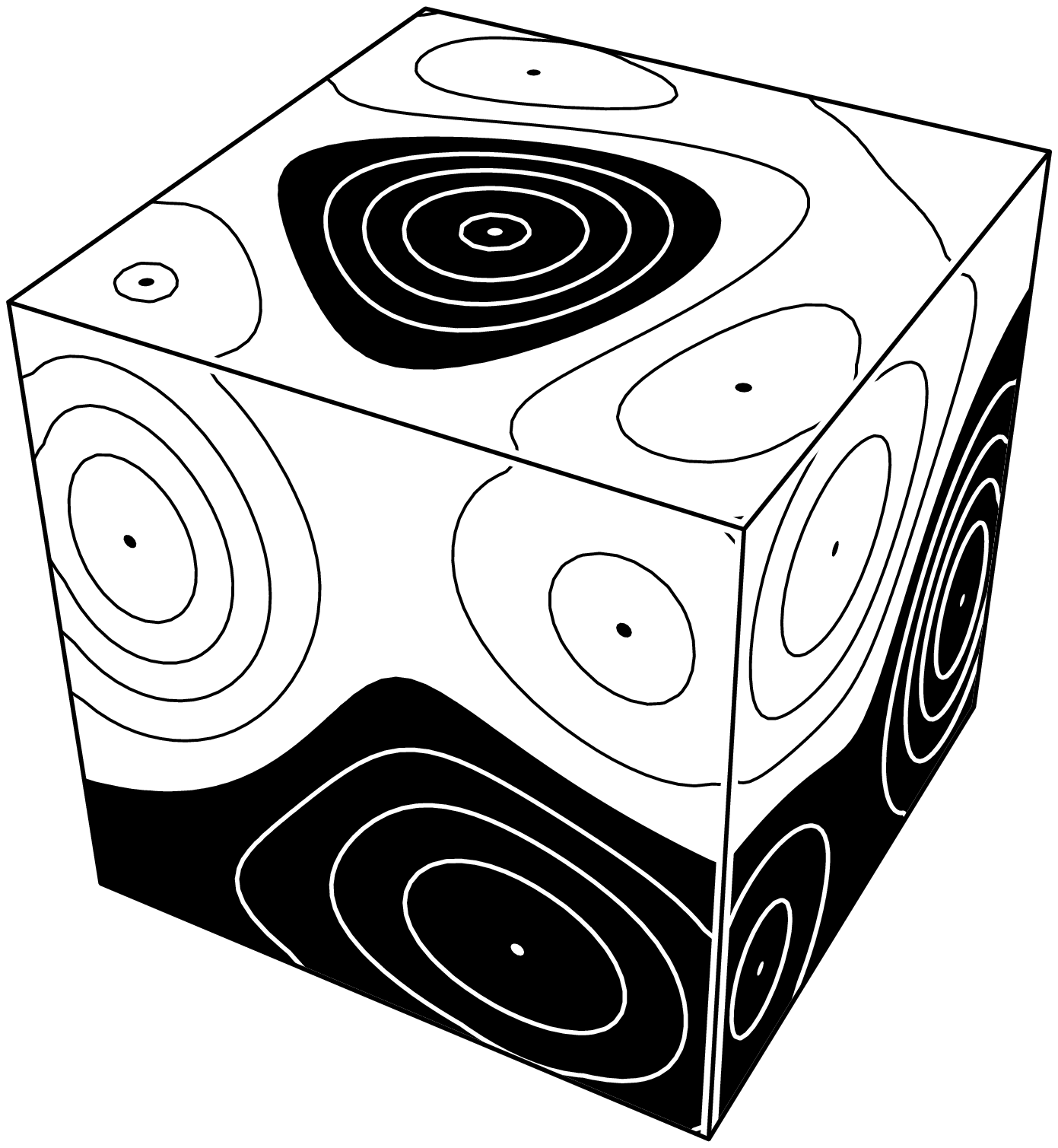}}  &
\scalebox{\solScale}{\includegraphics{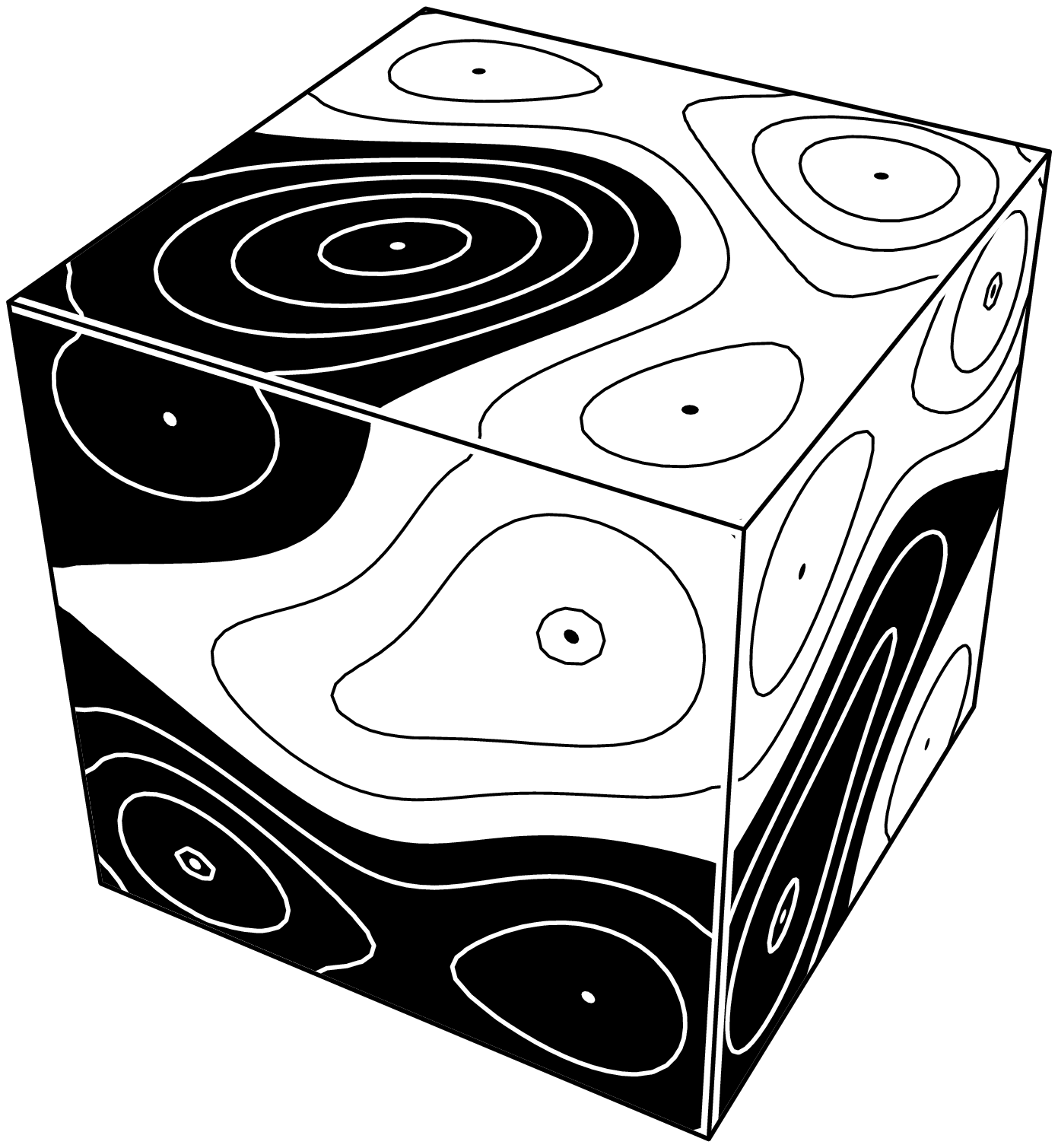}}  \hspace{1pt}
\\
\cline{3-5}
\multicolumn{1}{|c|}{} &
\multicolumn{1}{|c|}{} &
$S_{79}$ &
&
\\
\multicolumn{1}{|c|}{
\scalebox{\solScale}{\includegraphics{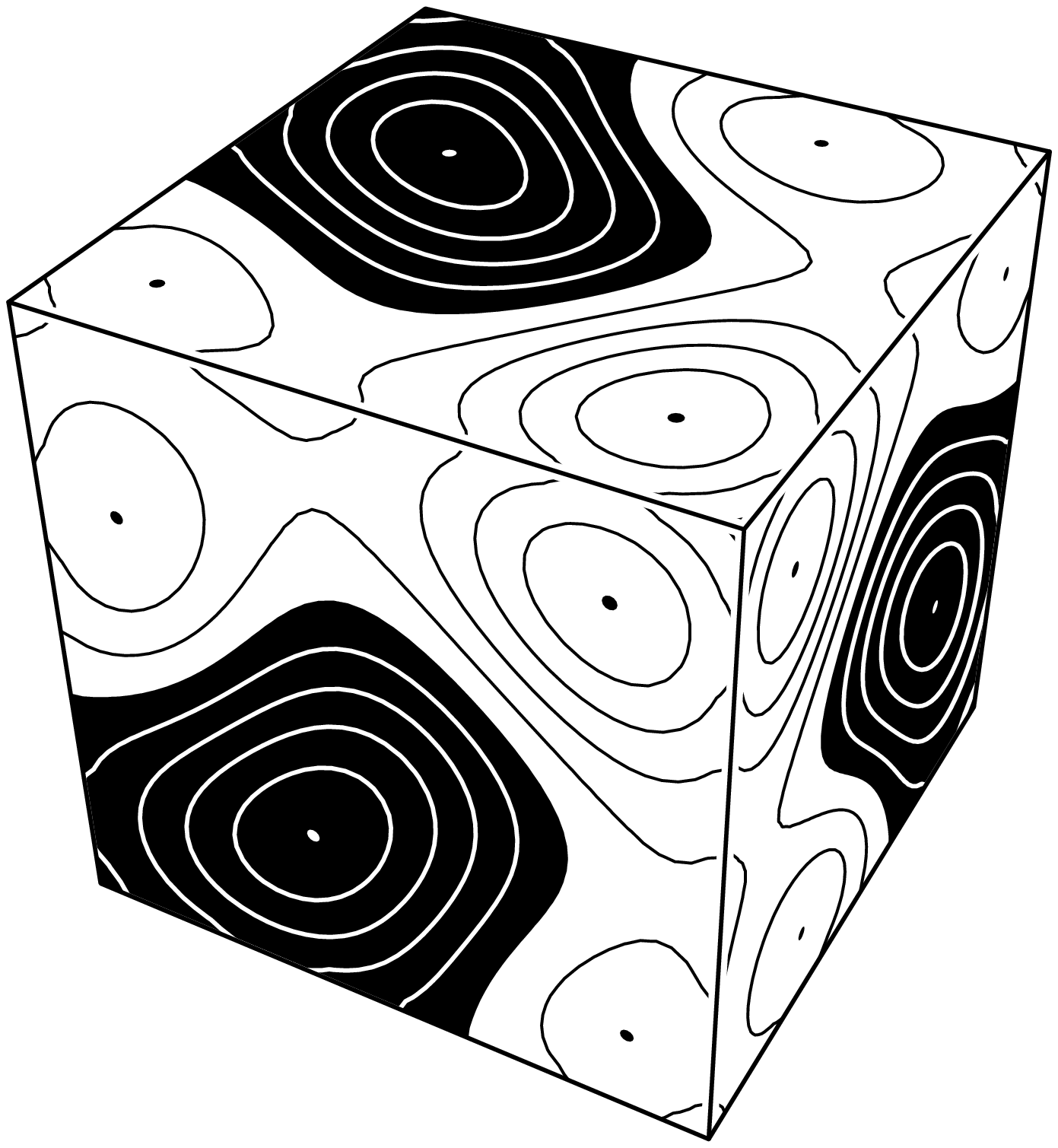}}
} &
\multicolumn{1}{|c|}{
\scalebox{\solScale}{\includegraphics{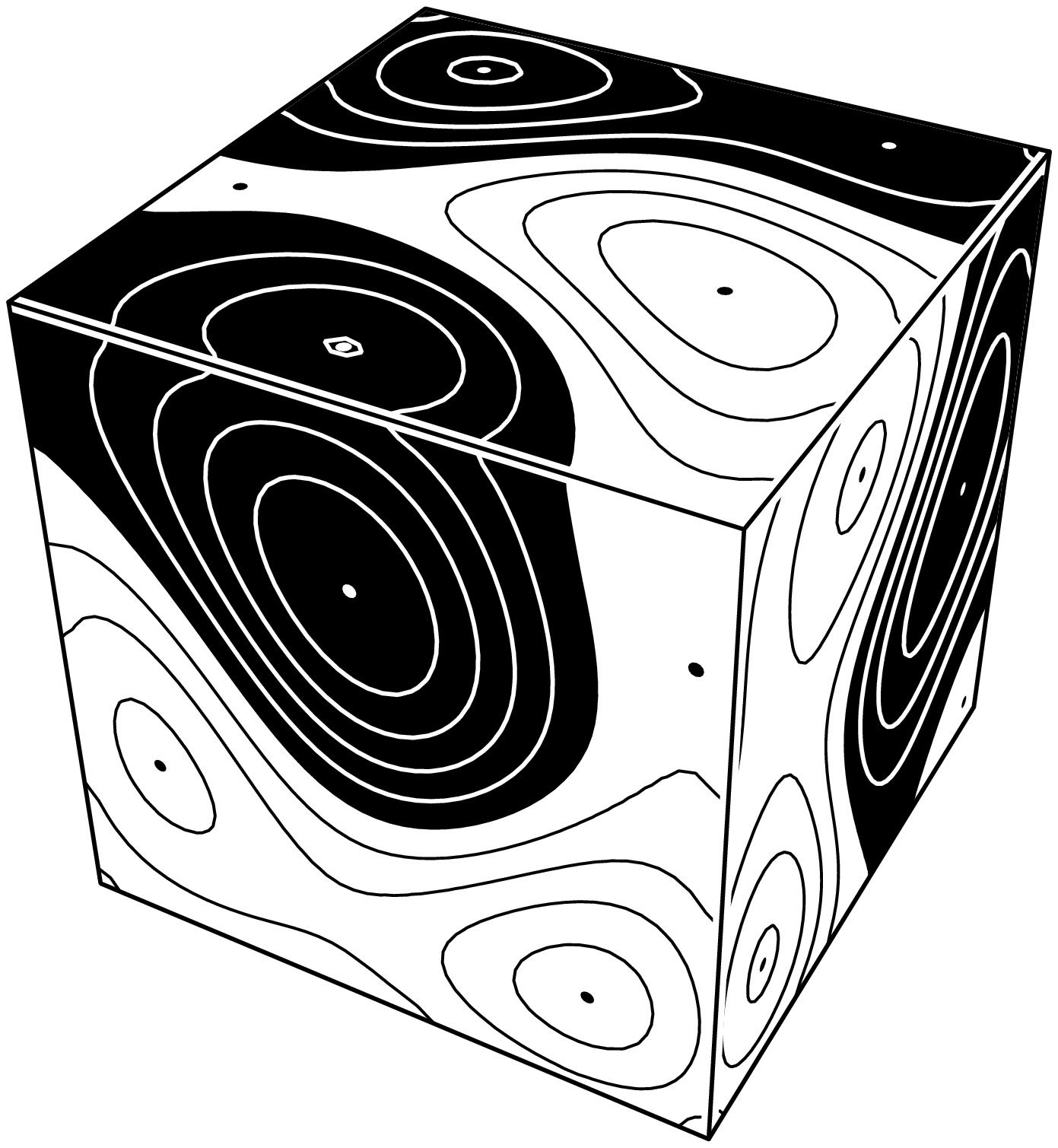}}
} &
\scalebox{\flagScale}{\includegraphics{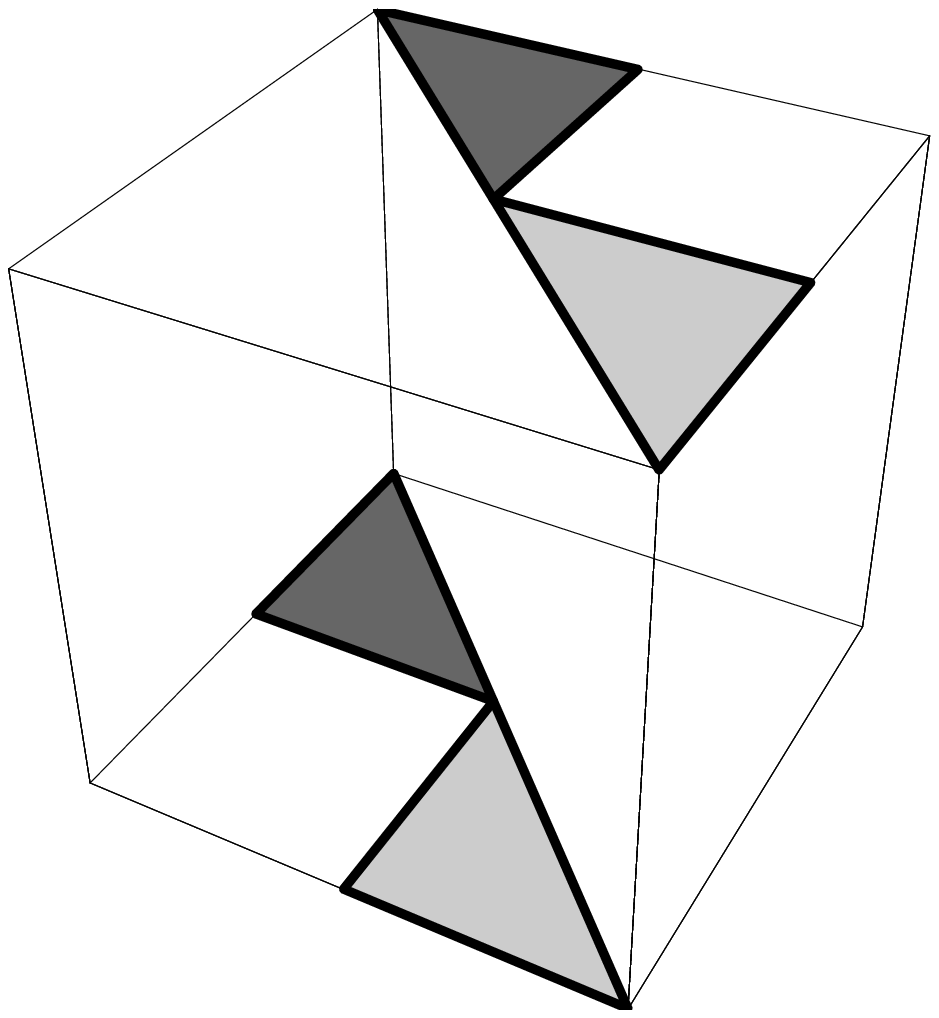}} &
\scalebox{\solScale}{\includegraphics{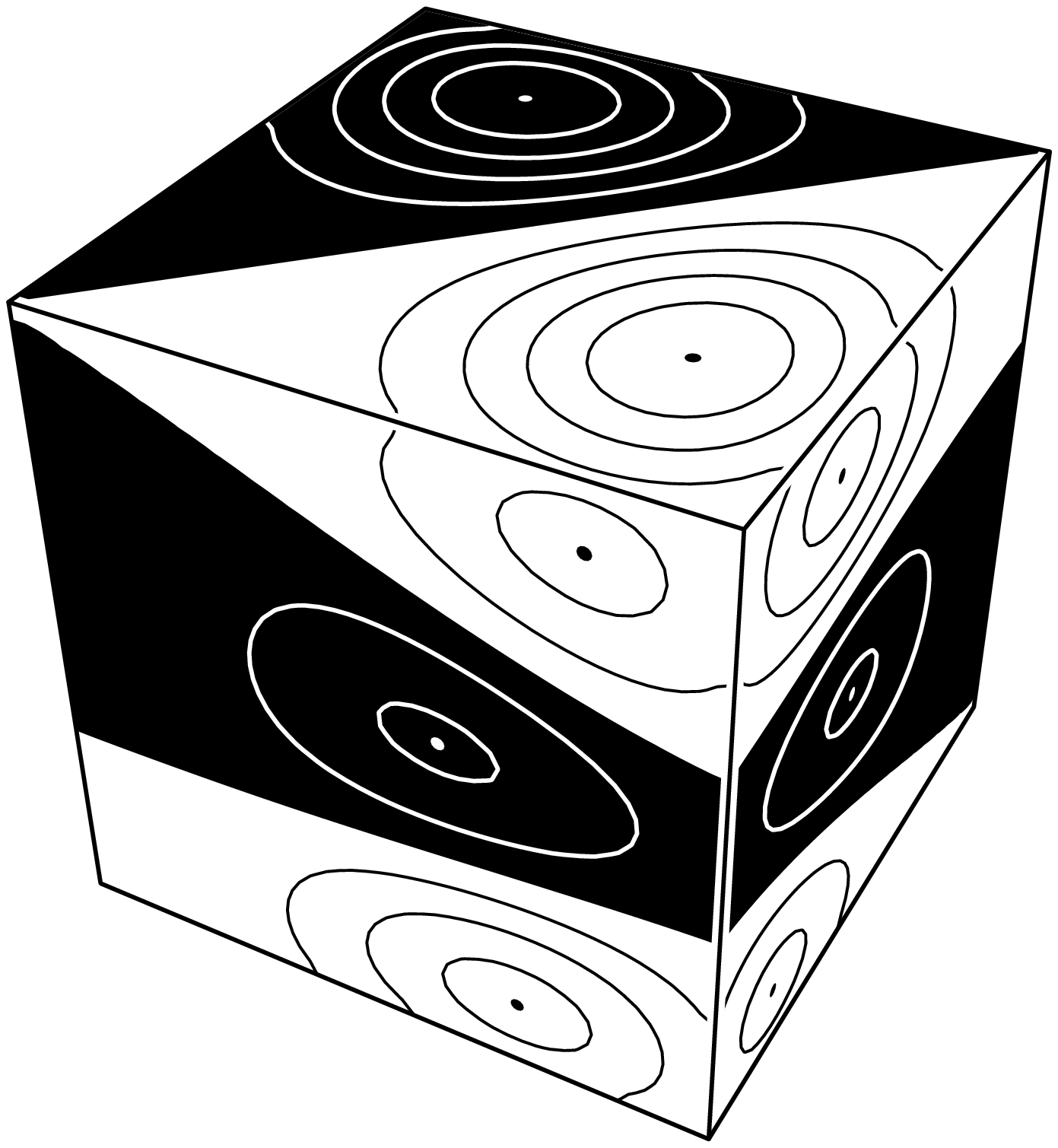}}  &
\scalebox{\solScale}{\includegraphics{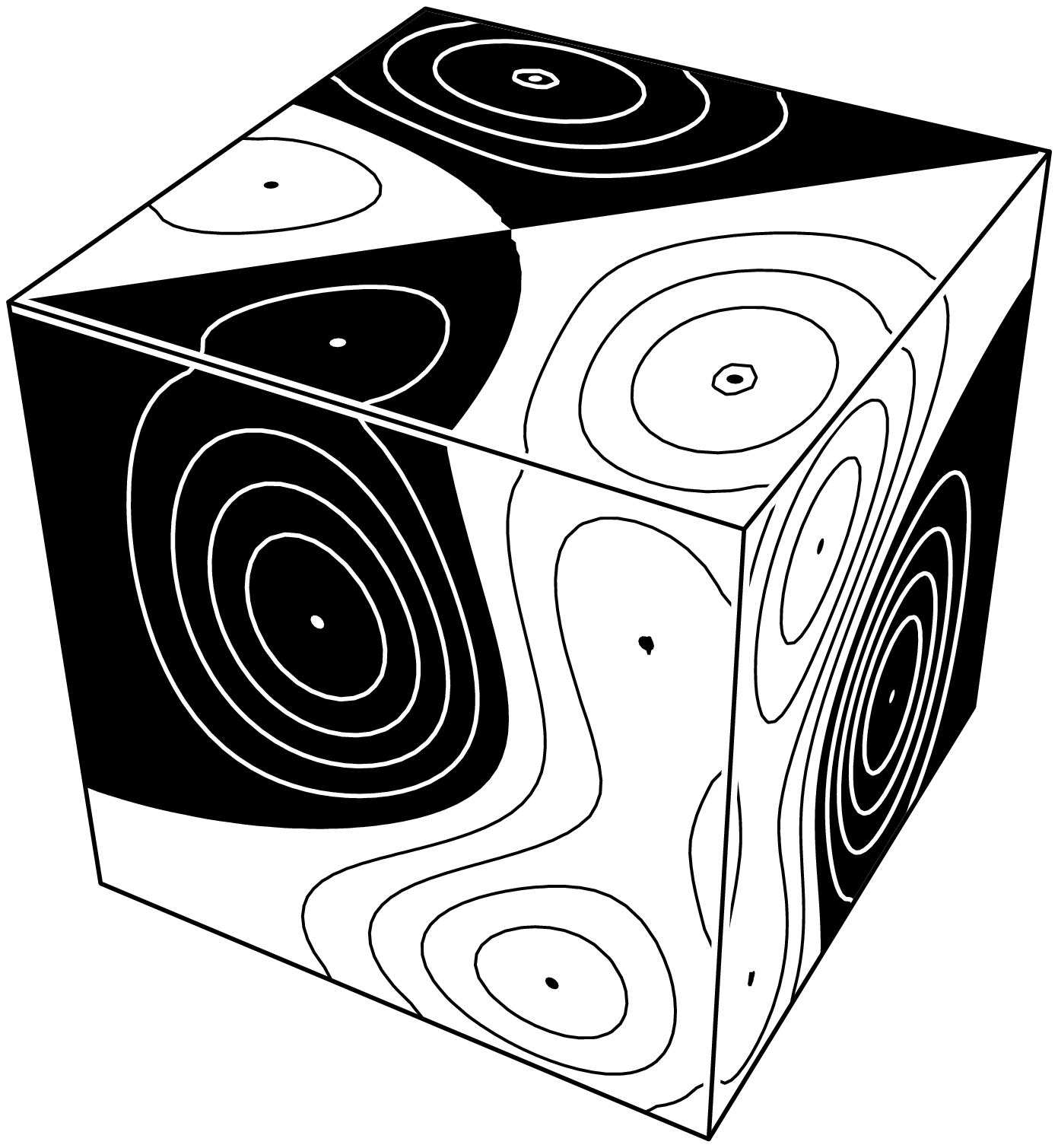}}  
\\
\hline
$S_{44}$ & & & & \\
\scalebox{\flagScale}{\includegraphics{figures/cube/fig.44.eps}} &
\scalebox{\solScale}{\includegraphics{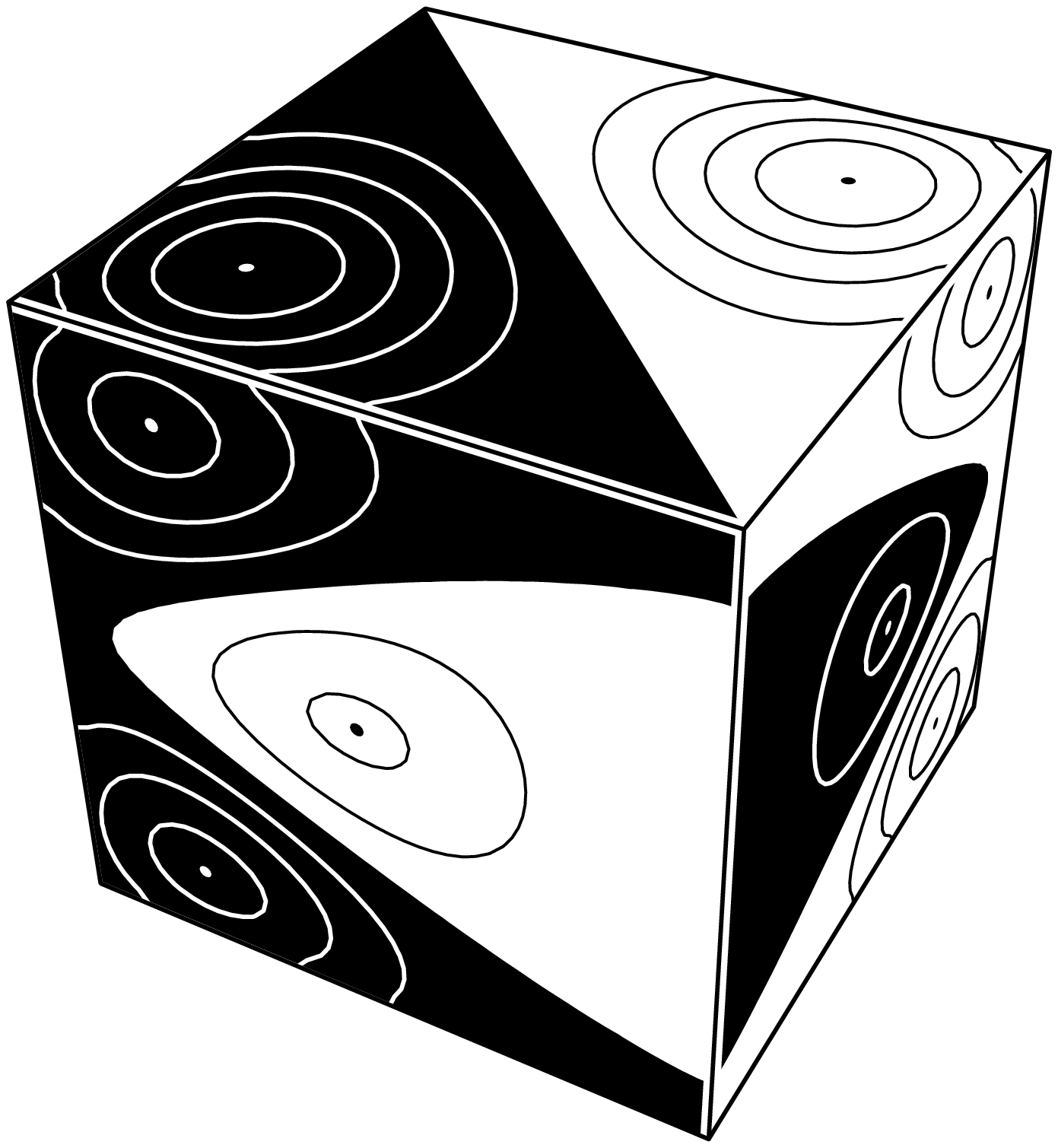}}  &
\scalebox{\solScale}{\includegraphics{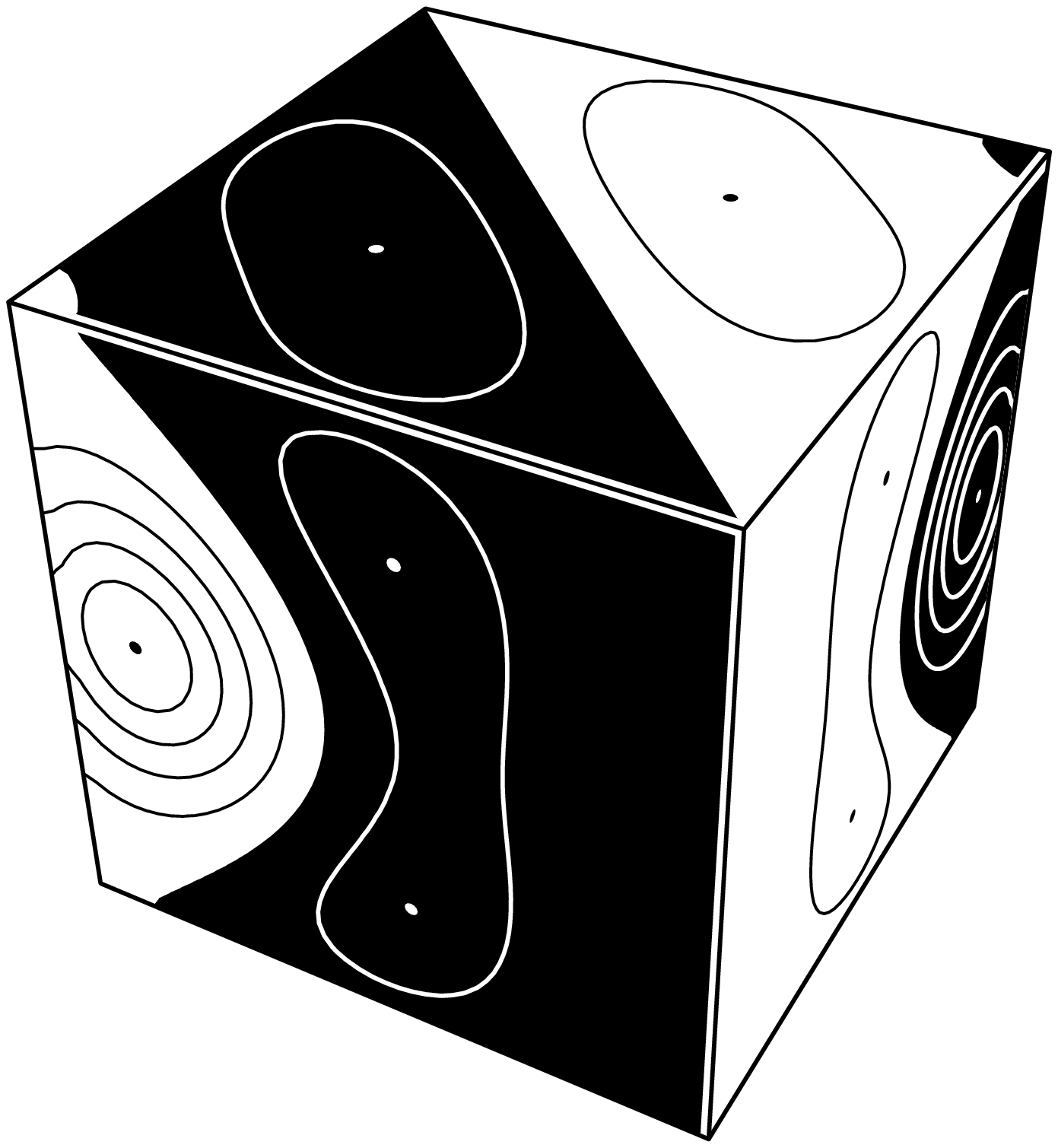}}  &
\scalebox{\solScale}{\includegraphics{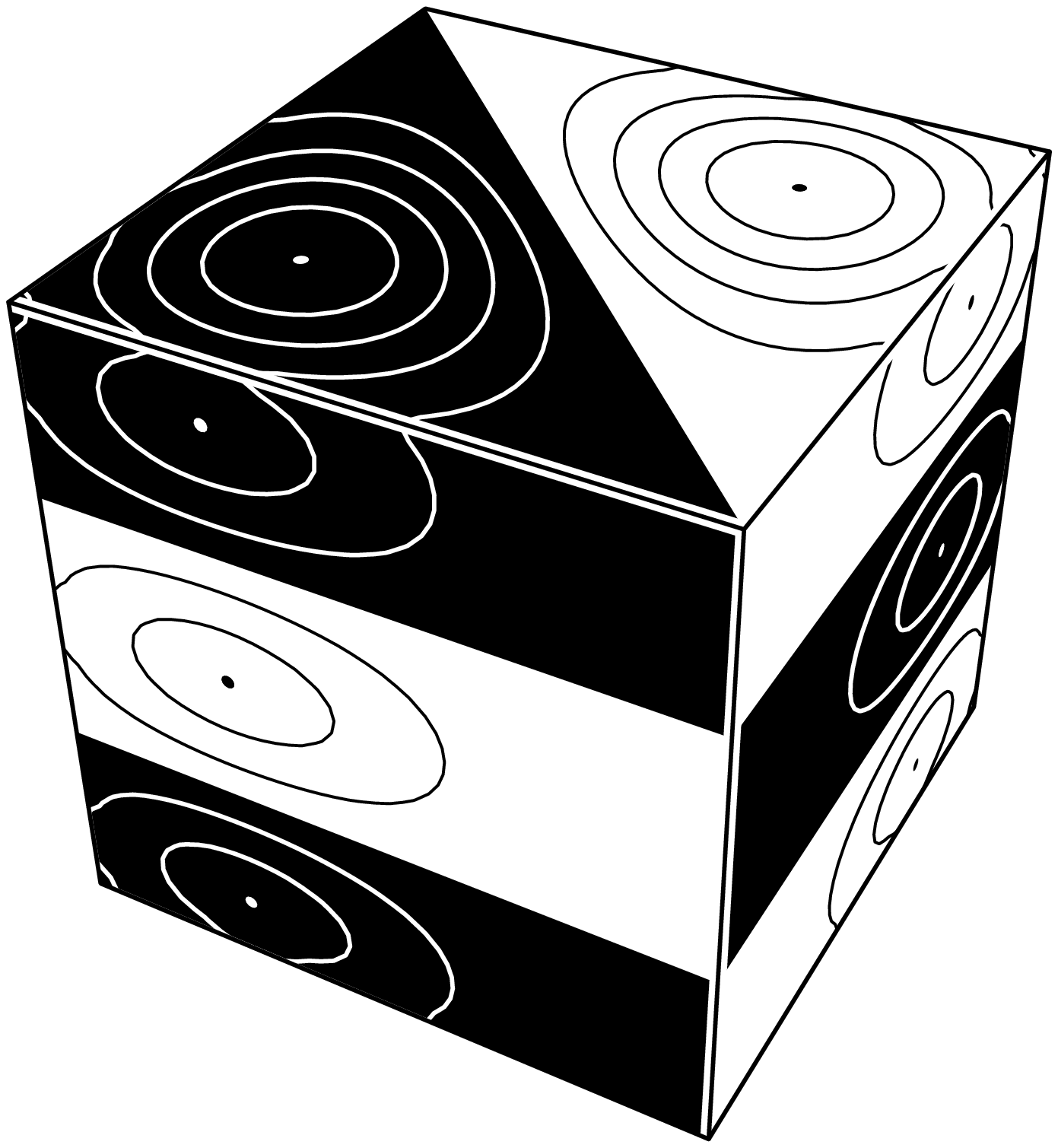}}  &
\scalebox{\solScale}{\includegraphics{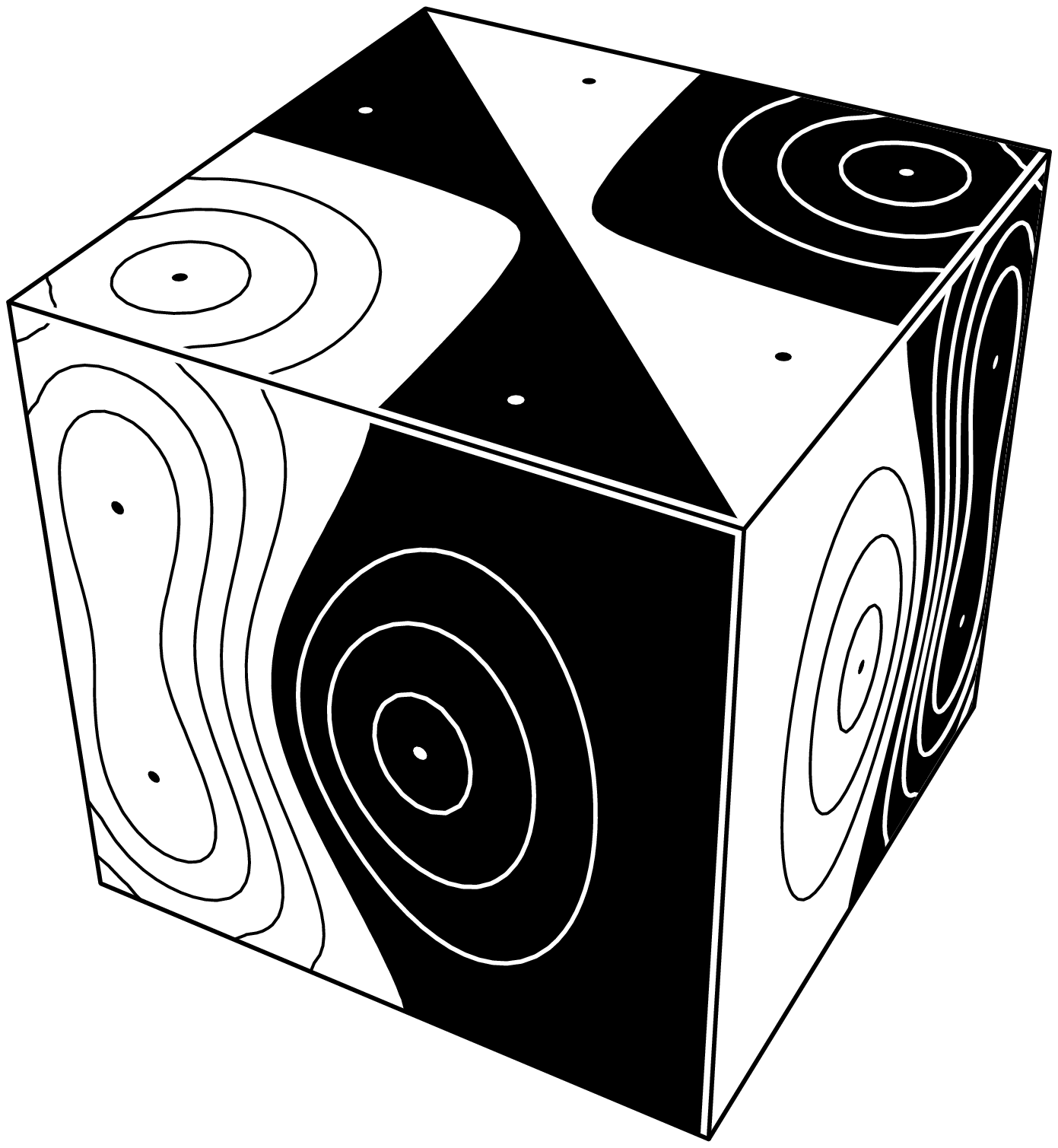}} 
\\
\hline
$S_{67}$ & & & & \\
\scalebox{\flagScale}{\includegraphics{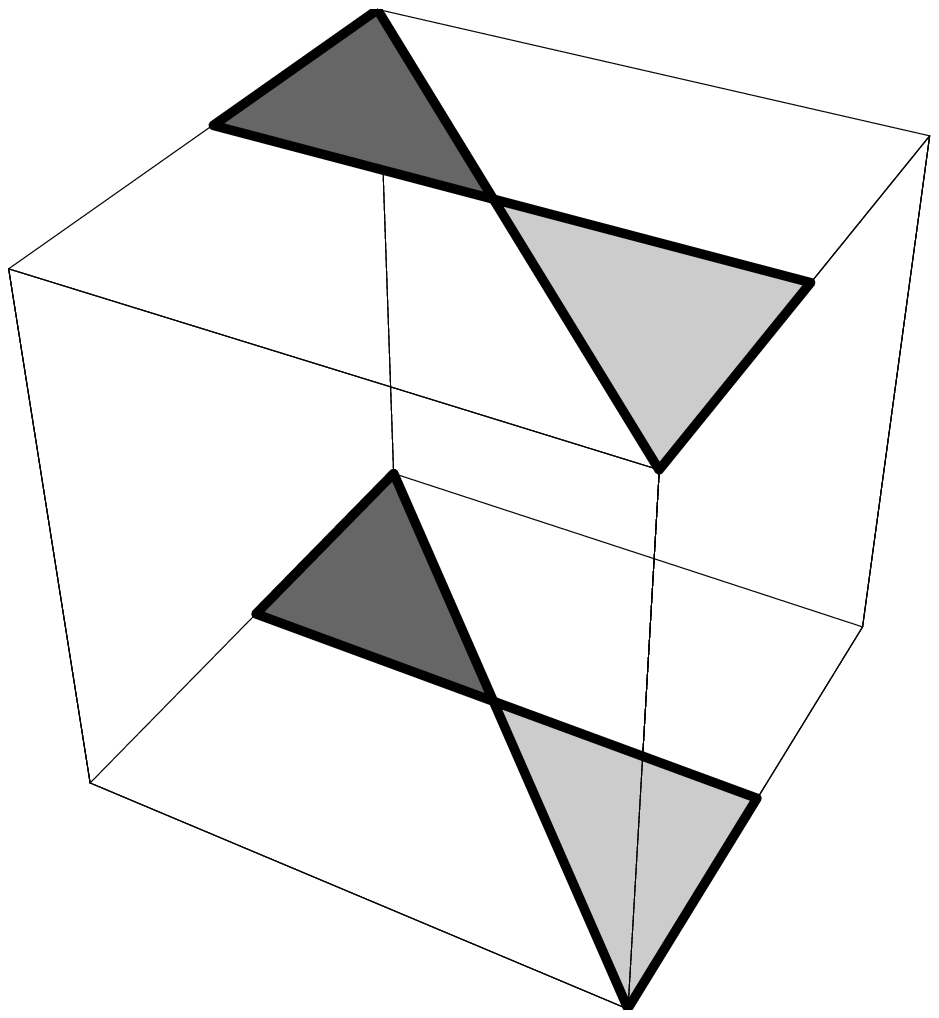}} &
\scalebox{\solScale}{\includegraphics{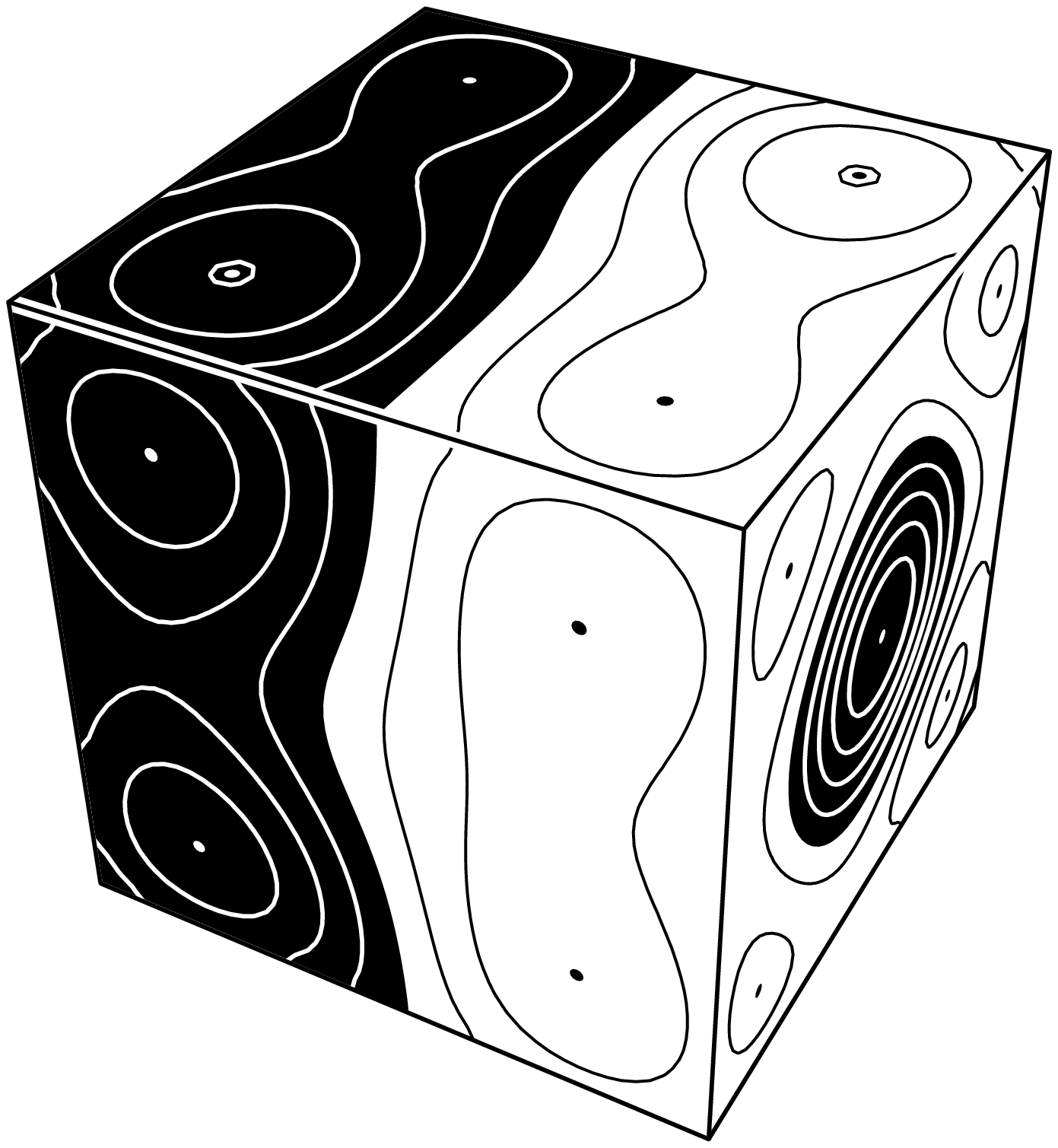}}  &
\scalebox{\solScale}{\includegraphics{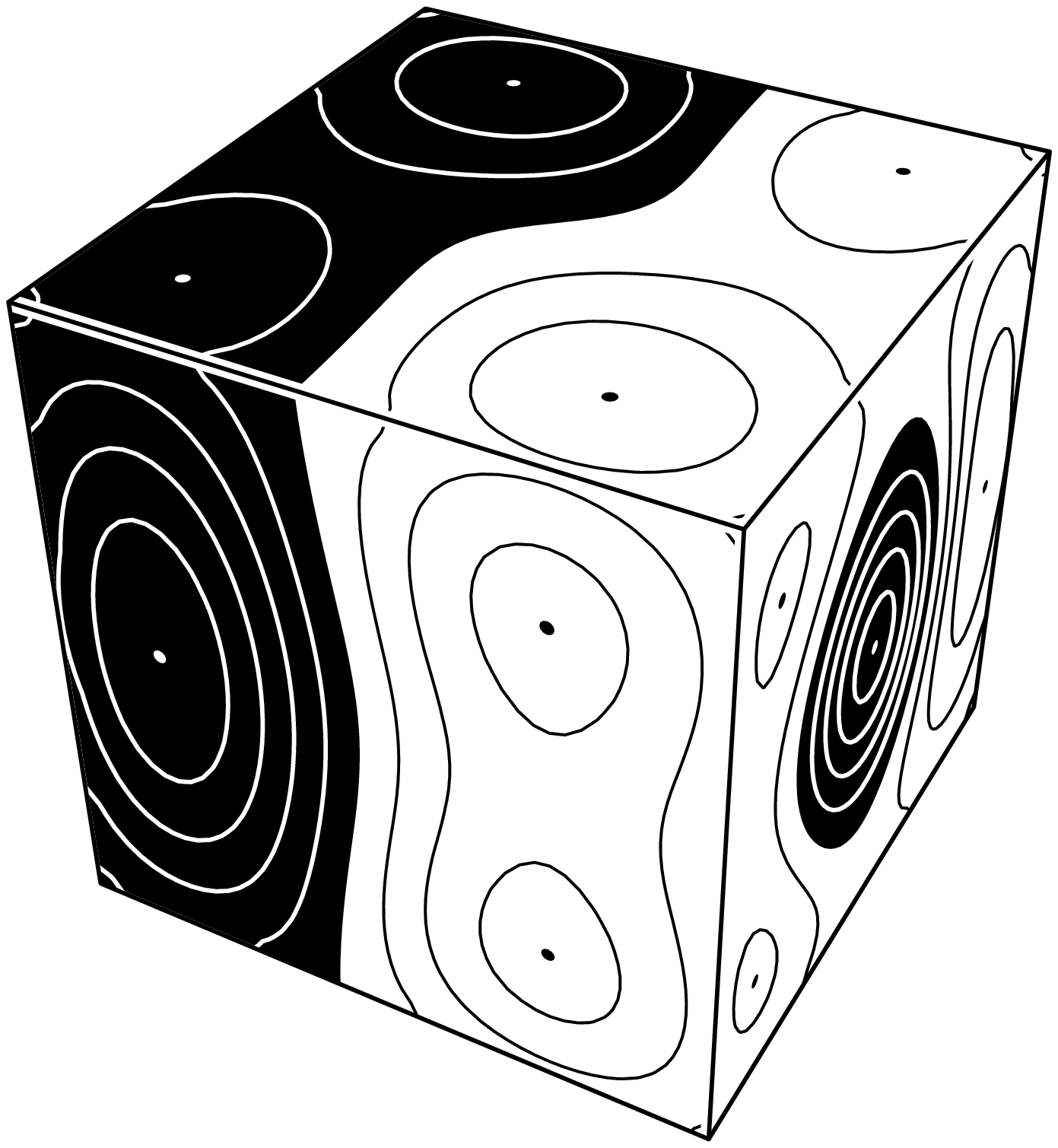}}  &
\scalebox{\solScale}{\includegraphics{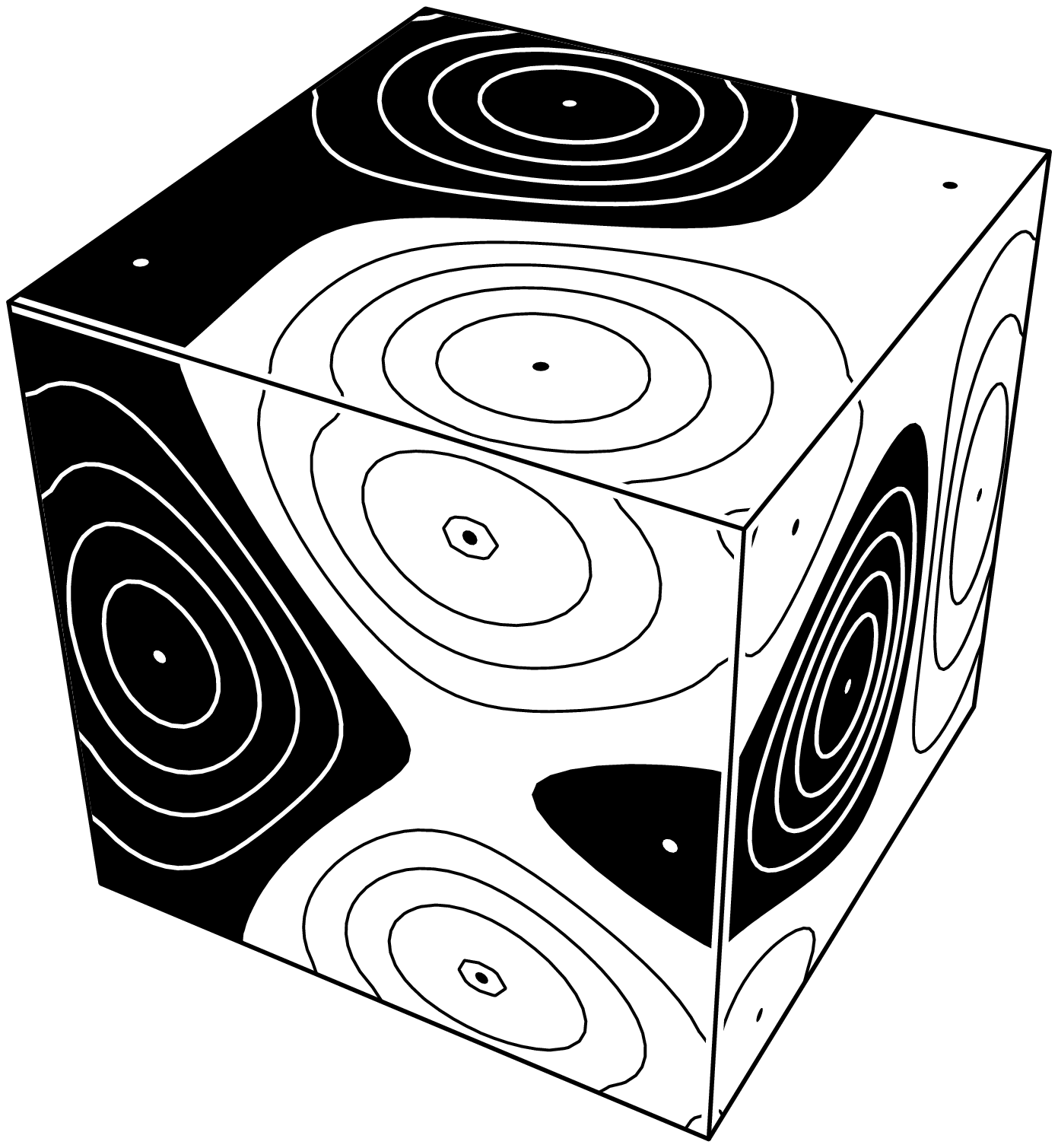}} &
\scalebox{\solScale}{\includegraphics{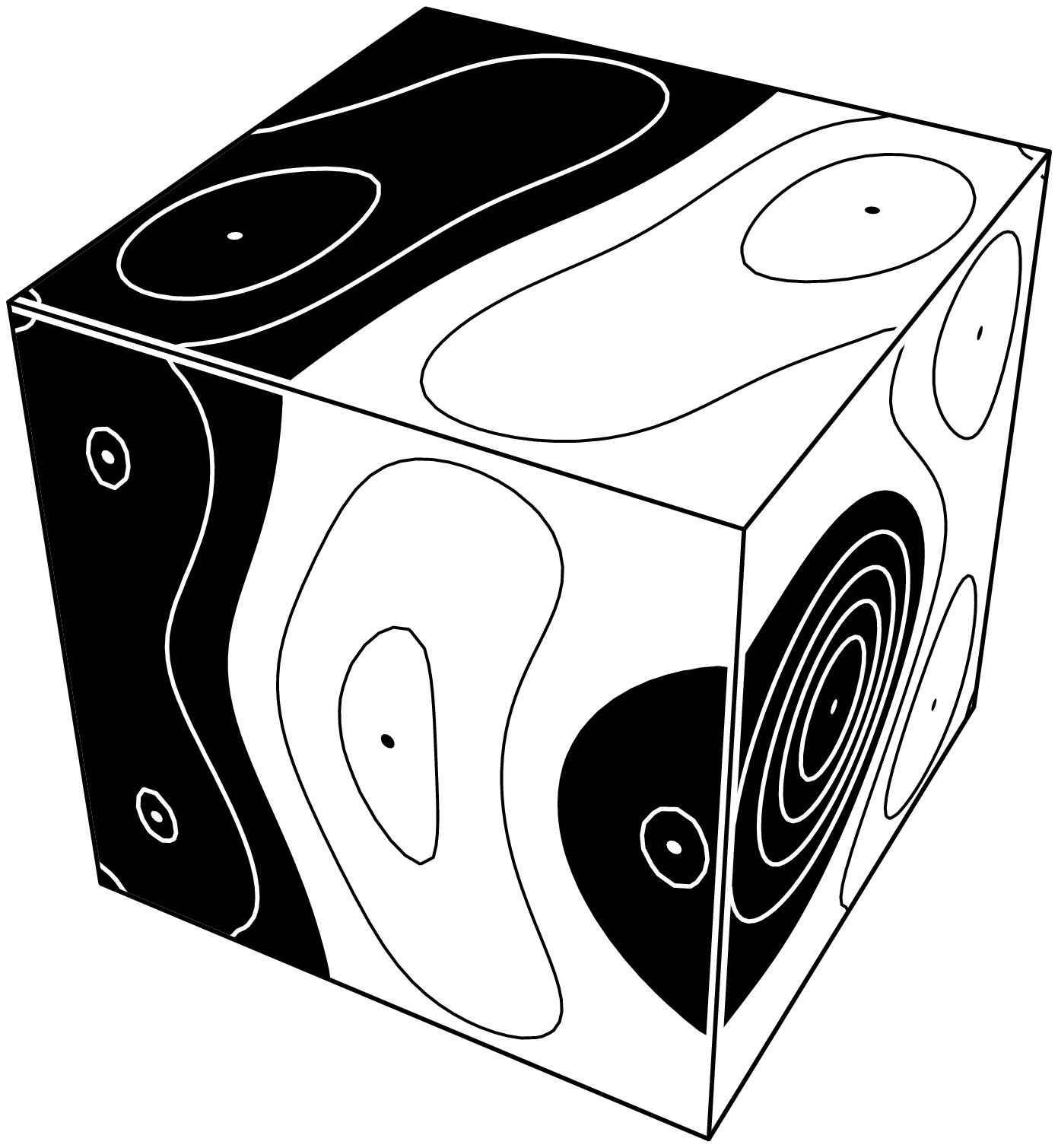}}  
\\
\hline
\end{tabular}
\caption{Contour plots for one solution on each of the  
19 non-conjugate primary branches bifurcating at $s=14$.
The solutions are listed with increasing MI within each symmetry type.
The solutions are shown at $s = 11$,
except for $S_{67}$ with MI 13 and $S_{93}$, which are shown at $s = 13.95$ and $s = 13.69$,
respectively.
We do this because these branches end at $s \approx 13.91$ and $s \approx 13.38$, respectively.
}
\label{s14}
\end{figure}

Figures~\ref{6Dlattice}, \ref{indexTable} and \ref{s14} concern the bifurcation
of the trivial solution at $s =\lambda_{1,2,3}=1^2 + 2^2 + 3^2 = 14$.
The six-dimensional critical eigenspace is
$$
\tilde E=\spn \{\psi_{1,2,3}, \psi_{1,3,2}, \psi_{2,1,3}, \psi_{2,3,1}, \psi_{3,1,2}, \psi_{3,2,1} \}.
$$
The action of $\Gam_0$ on $\tilde E$ satisfies
$\Gam_0'=\langle (-I_3,-1) \rangle$ and $\Gam_0 / \Gam_0' = \Oh$.
The action of $\Gam_0 / \Gam_0'$ on $\tilde E$ is isomorphic to the natural action of
$$
\langle R_{90}\oplus(-R_{90}), R_{120} \oplus R_{120},
R_{180}\oplus(-R_{180}), (-I_3) \oplus (-I_3) \rangle
$$
on  the coordinate space $[\tilde E]=\R^6$ with respect to the ordered basis
$$
( \psi_{213}+\psi_{231},\psi_{321}+\psi_{123},\psi_{132}+\psi_{312},
  \psi_{213}-\psi_{231},\psi_{321}-\psi_{123},\psi_{132}-\psi_{312}
) .
$$
Writing the action in block diagonal form, one sees that
the eigenspace $\tilde E$ is the direct sum of two irreducible spaces.
The trivial subspace of $\tilde E$ has isotropy $S_0$, and $S_{93}$ is the
minimal isotropy subgroup.
Thus, $[\fix(\Gam_{0},\tilde E)] = \{0\}\subseteq \R^6$
and $ [\fix(\Gam_{93},\tilde E)] = \R^6$. The remaining
symmetries in Figures~\ref{6Dlattice} and \ref{s14} in our chosen coordinate space satisfy the following:
\begin{align*}
[\fix(\Gam_{12},\tilde E)] &= \{ (0,0,a,\,0,0,0) \mid a\in \R\},
&
[\fix(\Gam_{11},\tilde E)] &= \{ (0,0,0,\,0,0,a) \mid a\in \R\},
\\
[\fix(\Gam_{22},\tilde E)] &= \{ (a,a,a,\,0,0,0) \mid a\in \R\},
&
[\fix(\Gam_{23},\tilde E)] &= \{ (0,0,0,\,a,a,a) \mid a\in \R\},
\\
[\fix(\Gam_{52},\tilde E)] &= \{ (0,a,0,\,0,b,0) \mid a,b\in \R\},
&
[\fix(\Gam_{54},\tilde E)] &= \{ (a,a,a,\,b,b,b) \mid a,b\in \R\},
\\
[\fix(\Gam_{44},\tilde E)] &= \{ (a,-a,0,\,b,b,0) \mid a,b\in \R\},
&
[\fix(\Gam_{78},\tilde E)] &= \{ (a,-a,b,\,c,-c,0) \mid a,b,c\in \R\},
\\
[\fix(\Gam_{67},\tilde E)] &= \{ (a,b,0,\,c,d,0) \mid a,b,c,d\in \R\},
&
[\fix(\Gam_{79},\tilde E)] &= \{ (a,a,0,\,b,b,c) \mid a,b,c\in \R\}.
\end{align*}
Figure ~\ref{6Dlattice} describes the lattice of isotropy subgroups of $\tilde E$.
Each arrow $S_i \rightarrow S_j$ indicates that some isotropy
subgroup in $S_j$ is a subgroup of some isotropy subgroup in $S_i$.
The arrows generate a partial ordering of the symmetry types.
Note that the lattice of isotropy subgroups is different from
the bifurcation digraph, as explained in~\cite{NSS3}.

To simplify the visual representation, the lattice of
symmetry types for the action of $\Oh$ on two irreducible spaces whose
direct sum is $\tilde E$ are shown on the top row of Figure~\ref{6Dlattice}.
Note that the middle column is the same in each of the top row sub-lattices.
As a result of the presence of $S_{44}$ in both sub-lattices, $\dim(\tilde E \cap \fix(\Gam_{44}) ) = 2$,
whereas  $\dim(\tilde E \cap \fix(\Gam_{i}) ) = 1$ for $i \in \{ 11, 12, 22, 23\}$.
Within these one-dimensional spaces there is a pitchfork bifurcation to an EBL branch,
but the bifurcation to
solutions with symmetry type $S_{44}$ is more complicated.

It is remarkable that there is at least one solution branch bifurcating at $s = 14$ with each
of the symmetry types shown in Figure~\ref{6Dlattice}.  There is even a solution with symmetry 
type $S_{93}$, the lowest symmetry present in $\tilde E$.
The conjugacy class of this branch has a total of 48 branches.
Figure~\ref{s14} shows one solution in each of the
nonconjugate primary branches that bifurcate at the multiplicity six
eigenvalue $s = 14$.
Since each of the solutions in this figure is odd about the center of the cube,
that is $u(x, y, z) = -u(\pi -x, \pi - y, \pi - z)$, we only show the front view of the
contour plot.

The solution in Figure~\ref{s14} with symmetry type $S_{52}$ strongly resembles the eigenfunction
$\psi_{1,2,3}(x,y,z)=\sin(x)\sin(2y)\sin(3z)$.
An analysis of the ``hidden symmetries''
in this problem \cite{Gomes} would explain why the solution with symmetry type $S_{52}$
bifurcates, but it would not explain all of the solutions in Figure~\ref{s14}.
In the space of triply periodic functions on $\R^3$, there is a 12-dimensional
irreducible space spanned by rotations of $\psi_{1,2,3}$ and the similar functions with cosines in place of
sines.

Let $X^-$ and $X^+$ be the set of solutions for $s=s^{-}=14-\varepsilon$ and 
$s=s^+=14+\varepsilon$, 
respectively, that are on branches bifurcating from $(0,14)\in H\times \R$, 
together with solutions on the mother branch,
for a sufficiently small positive $\varepsilon$.
The set $X^-$ contains 345 solutions falling into 20 group orbits
with non-trivial representatives shown in Figure~\ref{s14}. 
Figure~\ref{indexTable} shows multiplicity and MI information for $X^-$.
Since all the bifurcating branches curve to the left, $X^+$ contains only the trivial
solution with $\MI(0,s^+)=17$. 
Thus, one can verify that the Poincar\'e-Hopf Index Theorem of \cite{Arnold} is satisfied since
\begin{equation}
\label{degree}
\sum_{u\in X^{-}} (-1)^{\MI(u,s^-)}=
\sum_i \sum_{u\in X^{-}/\Gamma_i} |\Gamma_i\cdot u|(-1)^{\MI(u,s^-)}  =
-1
= (-1)^{17}
= \sum_{u\in X^{+}} (-1)^{\MI(u,s^+)}.
\end{equation}
This is consistent with our belief that the list of solutions in Figure~\ref{indexTable} is
comprehensive.

\end{section}


\begin{section}{Conclusion}

In this article we have extended the methods from~\cite{NSS2} and~\cite{NSS3}.
In the first paper, the symmetry group was relatively small and a medium-sized grid was used
to produce a reasonable portion of the bifurcation diagram and a selection of contour plots
for a two-dimensional semilinear elliptic PDE.  
In the second article, we completely automated the symmetry analysis 
for investigating the rich symmetries of solutions to 
partial difference equations (PdE) for many interesting low-order graphs.
In the current article, we have shown how to extend these ideas to a three-dimensional problem 
with a large symmetry group, namely the cube.
The large grid and many calculations required the use of a parallel programming environment. 
We developed and employed our own library, MPQueue, 
to implement our branch following and branch switching algorithms 
using self-submitting parallel job queues and MPI.
The new results we have presented here use the symmetry analysis from a low-order graph with the same symmetry group
to generate the corresponding symmetry information for functions discretized over 
the large-sized grid used in the PDE code.
We could not have done this without GAP;
for the cube there are 99 symmetry types 
(and 323 symmetries or isotropy subgroups) with 482 arrows between symmetry types.
This symmetry information is essential to the numerical results in several key ways.
It allows for the efficient construction of block-diagonal Hessians, 
reducing the number of costly integrations required at each Newton step in the presence of symmetry.
It allows us to search for only a single representative of each novel solution type, 
rather than wasting computations on finding many equivalent copies.
By reducing the dimension of a search space, 
symmetry information increases our chance of finding all expected solutions of a given symmetry type 
at each new bifurcation.
The entire suite of programs and new methods for efficiently implementing our algorithms has allowed
us to observe interesting bifurcation symmetries for PDE that we have not previously seen published.
Our procedure demonstrates a robustness for handling degenerate bifurcations, AIS, 
and high-dimensional/reducible critical eigenspaces.
The new contour plots required a number of ideas for efficiently and effectively conveying the
necessary information graphically.
The size of the problem makes it impossible to present a visual representation of the bifurcation digraph
on a single page.
We have constructed a companion website for navigating the digraph, and give examples here
to aid the reader in understanding the digraph and how to use it to interpret our numerical results.

\label{conclusion}
\end{section}

\bibliographystyle{plain}
\bibliography{nss5}

\end{document}